\documentclass[11pt,oneside]{book}

\usepackage{slashed}
\usepackage{empheq}
\usepackage{algorithm}
\usepackage{algpseudocode}
\usepackage{listings}
\usepackage{cancel}

\usepackage{microtype}
\usepackage[T1]{fontenc}
\usepackage[utf8]{inputenc}

\usepackage{autobreak}

\usepackage{epigraph}
\usepackage{datetime}
\newdateformat{monthyeardate}{\monthname[\THEMONTH], \THEYEAR}

\usepackage[usenames,dvipsnames]{xcolor}
\definecolor{light-gray}{gray}{0.95}
\definecolor{red1}{HTML}{921818}
\definecolor{purple1}{HTML}{53047A}
\definecolor{yellow1}{HTML}{edcb52}

\usepackage{graphicx}
\graphicspath{ {Figs/} }

\usepackage{enumitem}

\usepackage[colorinlistoftodos]{todonotes}
\setuptodonotes{fancyline, color=blue!20, figcolor=white}

\usepackage{lettrine}

\usepackage{mathtools}
\usepackage{bm}
\usepackage{amsmath,amsfonts,amssymb}
\allowdisplaybreaks

\usepackage{amsthm}

\theoremstyle{definition}
\newtheorem{definition}{Definition}[chapter]

\usepackage[a4paper]{geometry}

\usepackage{caption}
\usepackage{subcaption}

\usepackage{mfirstuc}

\usepackage{emptypage}
\usepackage{fancyhdr}
\usepackage{titlesec}

\titleformat{\chapter}[display]
{\normalfont\large}{
	\textsc{\chaptertitlename\ \thechapter}\centering
}{0pt}
{\Huge\bfseries\centering}

\titleformat{\section}[display]
{\normalfont}{\textsc{Section \thesection}}{0pt}{\Large\bfseries}

\titleformat{\subsection}
{\normalfont}{\thesubsection}{1em}{\large\bfseries}

\usepackage[square,numbers,sort]{natbib}

\usepackage{setspace}
\usepackage{tcolorbox}
\usepackage{array}
\usepackage{colortbl}
\usepackage{booktabs}

\usepackage{hyperref}
\hypersetup{
	colorlinks=true,
	urlcolor=blue,
	linkcolor=blue,
	citecolor=red,
	pdfdisplaydoctitle=true,
	pdfstartview={XYZ null null 1.00},
}

\usepackage{xparse}

\newcommand{\figref}[1]{Figure (\ref{#1})}
\newcommand{\secref}[1]{Section (\ref{#1})}
\newcommand{\chapref}[1]{Chapter (\ref{#1})}
\newcommand{\appref}[1]{Appendix (\ref{#1})}
\renewcommand{\algref}[1]{Algorithm (\ref{#1})}
\newcommand{\exref}[1]{Example (\ref{#1})}
\newcommand{\tabref}[1]{Table (\ref{#1})}

\newcommand{\supl}{\supbrk{\l}}
\newcommand{\supla}[1]{\supbrk{\l_{#1}}}
\newcommand{\In}{\vec{I} \, \supbrk{n}}
\newcommand{\Iz}{\vec{I} \, \supbrk{0}}
\newcommand{\Jl}{\vec{J} \, \supbrk{\l}}
\newcommand{\Ja}[1]{\vec{J} \, \supbrk{#1}}
\newcommand{\Inm}{\vec{I} \, \supbrk{n,m}}

\newcommand{\Ilnm}{\vec{I} \, \supbrk{\lambda,n,m}}
\newcommand{\Ilzz}{\vec{I} \, \supbrk{\lambda,0,0}}
\newcommand{\Iab}[2]{\vec{I} \, \supbrk{#1,#2}}
\newcommand{\Iabc}[3]{\vec{I} \, \supbrk{#1,#2,#3}}
\newcommand{\Psia}[1]{\Psi \supbrk{#1}}
\newcommand{\Phia}[1]{\Phi \supbrk{#1}}
\newcommand{\PPS}{\PP^{|E|-1}_+}

\newcommand{\qcomma}{\, , \quad}
\newcommand{\pcomma}{, \,}
\newcommand{\plus}{\hspace{0.2cm} + \hspace{0.2cm}}
\newcommand{\minus}{\hspace{0.2cm} - \hspace{0.2cm}}
\newcommand{\quadit}[1]{\quad #1 \quad}
\newcommand{\eq}[1]{\begin{align} #1 \end{align}}

\newcommand{\soft}[1]{\texttt{#1}}
\newcommand{\package}[1]{\textsc{#1}}
\newcommand{\rank}[1]{\mathrm{rank}\!\left(#1\right)}
\newcommand{\spec}[1]{\mathrm{spec}\left(#1\right)}
\renewcommand{\ker}[1]{\mathrm{ker}\left(#1\right)}
\newcommand{\diag}[1]{\soft{Diagonal}\!\left[#1\right]}
\newcommand{\ch}[1]{\mathrm{char}\!\left(#1\right)}
\newcommand{\supp}[1]{\mathrm{supp}\left(#1\right)}
\newcommand{\rowred}[1]{\soft{RowReduce}\!\left[#1\right]}
\newcommand{\jord}[1]{\soft{JordanDecomposition}\!\left[#1\right]}
\renewcommand{\Re}[1]{\mathrm{Re}(#1)}

\newcommand{\Der}{\mathrm{Der}}
\newcommand{\Mons}{\mathrm{Mons}}
\newcommand{\Ext}{\mathrm{Ext}}
\newcommand{\Std}{\mathrm{Std}}
\newcommand{\RStd}{\mathrm{RStd}}
\newcommand{\deRham}{\mathbb{H}^k_A(\b)}

\newcommand{\oq}{\langle \o_q \rangle}
\newcommand{\gfi}{I(\DD_0|\nu)}
\newcommand{\gfic}{c(\DD_0|\nu)}
\newcommand{\newtU}{\mathbf{N}[\mU]}
\newcommand{\newtF}{\mathbf{N}[\mF]}
\newcommand{\baseU}{\mathbf{P}[z_\mU]}
\newcommand{\baseF}{\mathbf{P}[z_\mF]}

\newcommand{\arr}[2]{
        \lrsbrk{
                \begin{array}{#1}
                        #2
                \end{array}
        }
}

\newcounter{ex}[section]
\renewcommand*\theex{\thechapter.\arabic{ex}}
\newenvironment{ex}[1][]
{
        \refstepcounter{ex}\par\medskip\noindent\textbf{Example~\theex.#1} \rmfamily
}
{
        \vspace{0.2cm}
        \hfill
        $\blacksquare$
        \noindent
}

\makeatletter
\newlength{\apb@width}
\newcommand{\autoparbox}[2][c]{\settowidth{\apb@width}{#2}\parbox[#1]{\apb@width}{#2}}
\newcommand{\includegraphicsbox}[2][]{\autoparbox{\includegraphics[#1]{#2}}}
\makeatother

\lstdefinestyle{mystyle}{
backgroundcolor=\color{light-gray},
frame=single,
basicstyle=\scriptsize\ttfamily
}

\DeclareUnicodeCharacter{2212}{-}

\newcommand{\DD}{\mathrm{D}}
\newcommand{\dd}{\mathrm{d}}

\newcommand{\tr}{\text{tr}}
\newcommand{\Int}{\text{int}}
\newcommand{\ext}{\text{ext}}
\newcommand{\Eext}{E_\ext}
\newcommand{\Eint}{E_\Int}
\newcommand{\mzero}{\scalebox{.7}{$\cdot$}}
\newcommand{\Rmat}{\mathsf{R}}
\newcommand{\Bmat}{\mathsf{B}}
\newcommand{\Mmat}{\mathsf{M}}
\newcommand{\zcirc}[1]{\overset{\circ}{z_#1}}
\newcommand{\ft}{\package{Feyntrop}~}
\newcommand{\trop}{^\mathrm{tr}}

\renewcommand{\AA}{\mathbb{A}}
\newcommand{\CC}{\mathbb{C}}
\newcommand{\FF}{\mathbb{F}}
\newcommand{\NN}{\mathbb{N}}
\newcommand{\PP}{\mathbb{P}}
\newcommand{\QQ}{\mathbb{Q}}
\newcommand{\RR}{\mathbb{R}}
\newcommand{\ZZ}{\mathbb{Z}}

\newcommand{\mA}{\mathcal{A}}
\newcommand{\mB}{\mathcal{B}}
\newcommand{\mU}{\mathcal{U}}
\newcommand{\mF}{\mathcal{F}}
\newcommand{\mE}{\mathcal{E}}
\newcommand{\mG}{\mathcal{G}}
\newcommand{\mO}{\mathcal{O}}
\newcommand{\mL}{\mathcal{L}}
\newcommand{\mK}{\mathcal{K}}
\newcommand{\mJ}{\mathcal{J}}
\newcommand{\mD}{\mathcal{D}}
\newcommand{\mM}{\mathcal{M}}
\newcommand{\mN}{\mathcal{N}}
\newcommand{\mI}{\mathcal{I}}
\newcommand{\mR}{\mathcal{R}}
\newcommand{\mV}{\mathcal{V}}
\newcommand{\mP}{\mathcal{P}}

\newcommand{\p}{\partial}

\newcommand{\e}{\epsilon}
\newcommand{\vare}{\varepsilon}

\newcommand{\g}{\gamma}
\renewcommand{\a}{\alpha}
\renewcommand{\l}{\lambda}
\renewcommand{\L}{\Lambda}
\renewcommand{\o}{\omega}
\renewcommand{\O}{\Omega}
\renewcommand{\b}{\beta}
\renewcommand{\d}{\delta}
\newcommand{\s}{\sigma}
\newcommand{\G}{\Gamma}
\renewcommand{\t}{\theta}

\newcommand{\brk}[1]{(#1)}
\newcommand{\lrbrk}[1]{\left(#1\right)}
\newcommand{\bigbrk}[1]{\bigl(#1\bigr)}

\newcommand{\sbrk}[1]{[#1]}
\newcommand{\lrsbrk}[1]{\left[#1\right]}
\newcommand{\bigsbrk}[1]{\bigl[#1\bigr]}

\newcommand{\brc}[1]{\{#1\}}

\newcommand{\supbrk}[1]{^{\brk{#1}}}

\newcommand{\checktoopen}{
	\if@openright\cleardoublepage\else\clearpage\fi
	\ifdef{\phantomsection}{\phantomsection}{}
}

\makeatletter

\def\displaytitle#1{\gdef\@displaytitle{#1}}
\def\subtitle#1{\gdef\@subtitle{#1}}
\DeclareDocumentCommand{\author}{O{} m}{\gdef\@authorURL{#1}\gdef\@author{#2}}
\DeclareDocumentCommand{\supervisor}{O{} m}{\gdef\@supervisorURL{#1}\gdef\@supervisor{#2}}
\DeclareDocumentCommand{\university}{O{} m}{\gdef\@universityURL{#1}\gdef\@university{#2}}
\DeclareDocumentCommand{\school}{O{} m}{\gdef\@schoolURL{#1}\gdef\@school{#2}}
\DeclareDocumentCommand{\group}{O{} m}{\gdef\@groupURL{#1}\gdef\@group{#2}}
\def\degree#1{\gdef\@degree{#1}}
\def\subject#1{\gdef\@subject{#1}}

\newcommand{\abstractname}{Abstract}
\newenvironment{abstract}{
	\checktoopen
	\addcontentsline{toc}{chapter}{\abstractname}	
	\thispagestyle{plain}
	\markboth{\abstractname}{\abstractname}	
	\null
	\vfill
	\begin{center}
		{\huge\itshape\abstractname\par}
	\end{center}
}{
	\vfill
	\null
}

\newcommand{\declarationname}{Author's Declaration}
\newenvironment{declaration}{
	\checktoopen
	\addcontentsline{toc}{chapter}{\declarationname}
	\thispagestyle{plain}
	\markboth{\declarationname}{\declarationname}	
	\begin{center}{\huge\itshape\declarationname\par}\end{center}
	\bigskip
	I hereby declare that}{}

\newcommand{\acknowledgementsname}{Acknowledgements}
\newenvironment{acknowledgements}{
	\checktoopen
	\addcontentsline{toc}{chapter}{\acknowledgementsname}
	\thispagestyle{plain}
	\markboth{\acknowledgementsname}{\acknowledgementsname}	
	\begin{center}{\huge\itshape\acknowledgementsname\par}\end{center}
	\bigskip
}{}

\newcommand{\dedicationname}{Dedication}

\makeatother

\displaytitle{Evaluating Feynman Integrals Using \\ \texorpdfstring{$\mD$-modules}{} and Tropical Geometry}
\title{Evaluating Feynman Integrals Using \texorpdfstring{$\mD$-modules}{} and Tropical Geometry}
\subtitle{}
\author[url]{Henrik Jessen Munch}
\supervisor[url]{Pierpaolo Mastrolia}
\university[url]{University of Padova} 
\school[url]{Something something}
\group[url]{Something something}
\degree{Doctor of Philosophy}
\subject{Physics}

\makeatletter
\AtBeginDocument{
	\hypersetup{pdftitle=\@title} 
	\hypersetup{pdfauthor=\@author} 
}
\makeatother

\usepackage{pdfpages}

\begin{document}
	
	\frontmatter
	
        \includepdf[pages=-]{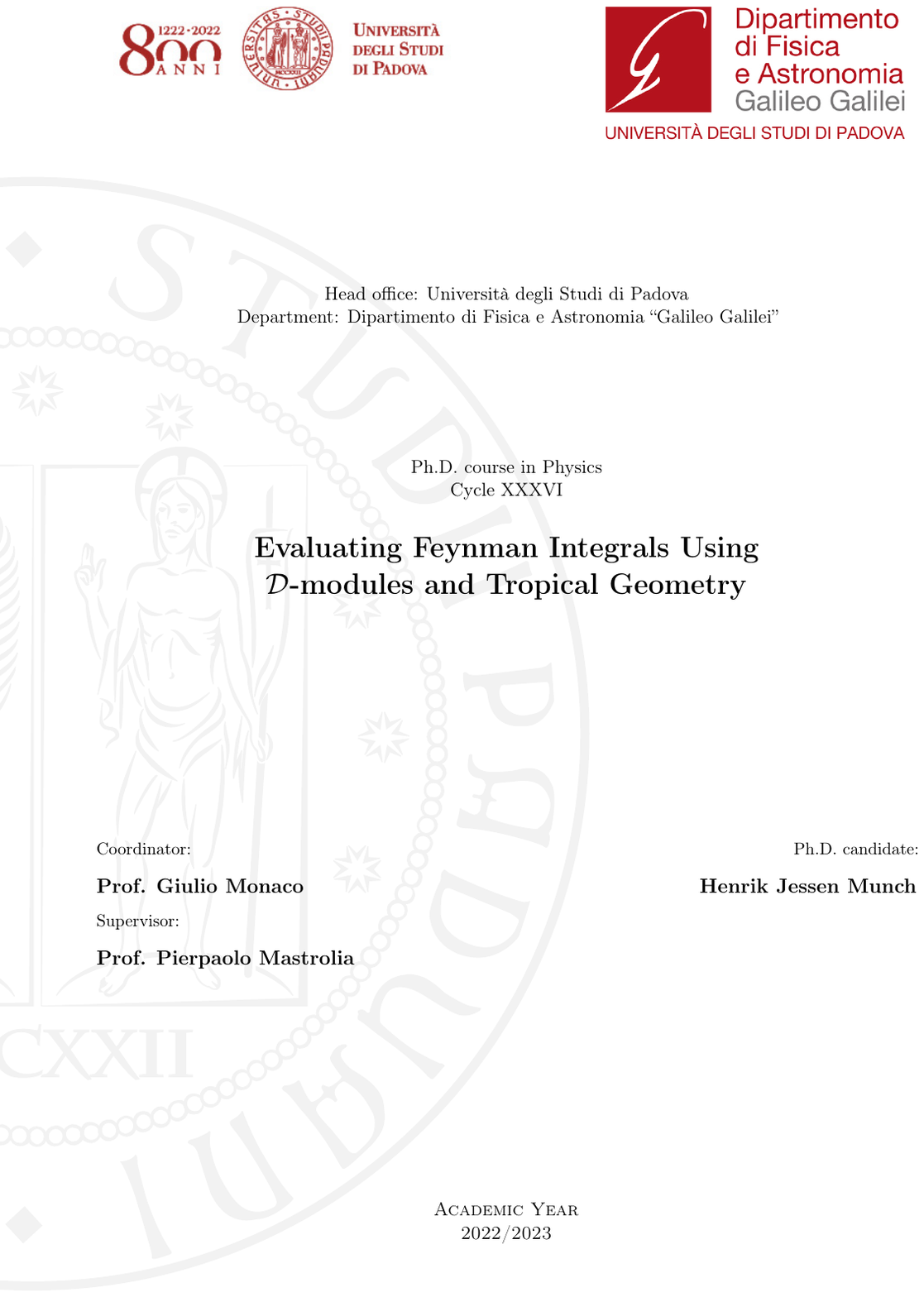}
	
	\begin{abstract}
Feynman integrals play a central role in the modern scattering amplitudes research program.
Advancing our methods for evaluating Feynman integrals will,
therefore,
strengthen our ability to compare theoretical predictions with data 
from particle accelerators such as the Large Hadron Collider.
Motivated by this,
the present manuscript purports to study mathematical concepts related to Feynman integrals.
In particular,
we present both numerical and analytical algorithms for the evaluation of Feynman integrals.

\vspace{0.5cm}

\noindent
The content is divided into three parts.

\textbf{Part I} focuses on the method of DEQs for evaluating Feynman integrals.
An otherwise daunting integral expression is thereby traded for the 
comparatively simpler task of solving a system of DEQs.
We use this technique to evaluate a family of two-loop 
Feynman integrals of relevance for dark matter detection.

\textbf{Part II} situates the study of DEQs for Feynman integrals within the framework of $\mD$-\emph{modules},
a natural language for studying PDEs algebraically.
Special emphasis is put on a particular $\mD$-module called the \emph{GKZ system},
a set of higher-order PDEs that annihilate a generalized version of a Feynman integral.
In the course of matching the generalized integral to a Feynman integral proper,
we discover an algorithm for evaluating the latter in terms of logarithmic series.

\textbf{Part III} develops a numerical integration algorithm.
It combines Monte Carlo sampling with \emph{tropical geometry},
a particular offspring of algebraic geometry that studies "piecewise-linear" polynomials.
Feynman's $i\vare$-prescription is incorporated into the algorithm via contour deformation.
We present a new open-source program named \ft that implements this algorithm,
and use it to numerically evaluate Feynman integrals between 
1-5 loops and 0-5 legs in physical regions of phase space.

\end{abstract}

	
	\begin{declaration}
this thesis is solely authored by me.
The work presented herein was carried out by the author during his Ph.D.~studies at 
Università degli Studi di Padova, Dipartimento di Fisica e Astronomia “Galileo Galilei”.
The following manuscripts were co-authored by me during my Ph.D.

\subsubsection*{Articles}

\begin{enumerate}
        \item
                Raghuveer Garani, 
                Federico Gasparotto, 
                Pierpaolo Mastrolia, 
                Henrik J. Munch, 
                Sergio Palomares-Ruiz and
                Amedeo Primo,
                \emph{Two-photon exchange in leptophilic dark matter scenarios}, 
                JHEP 12 (2021), 
                DOI: 10.1007/JHEP12(2021)212, 
                arXiv: 2105.12116 [hep-ph].
        \item 
                 Vsevolod Chestnov, 
                 Federico Gasparotto, 
                 Manoj K. Mandal, 
                 Pierpaolo Mastrolia, 
                 Saiei J. Matsubara-Heo, 
                 Henrik J. Munch and
                 Nobuki Takayama, 
                 \emph{Macaulay Matrix for Feynman Integrals: Linear Relations and Intersection Numbers}, 
                 JHEP 09 (2022), 
                 DOI: 10.1007/JHEP09(2022)187,
                 arXiv: 2204.12983 [hep-th].
         \item 
                 Michael Borinsky, 
                 Henrik J. Munch and
                 Felix Tellander,
                 \emph{Tropical Feynman integration in the Minkowski regime},
                 COMPHY 292 (2023),
                 DOI: https://doi.org/10.1016/ \\ j.cpc.2023.108874,
                 arXiv: 2302.08955 [hep-ph].
         \item 
                 Vsevolod Chestnov,
                 Saiei J. Matsubara-Heo,
                 Henrik J. Munch and
                 Nobuki Takayama
                 \emph{Restrictions of Pfaffian Systems for Feynman Integrals},
                 arXiv: 2305.01585 [hep-th].
\end{enumerate}

\subsubsection*{Conference proceedings}

\begin{enumerate}
        \item
                Henrik J. Munch,
                \emph{Feynman Integral Relations from GKZ Hypergeometric Systems},
                PoS, LL2022 (2022), 
                DOI: 10.22323/1.416.0042,
                arXiV: 2207.09780 [hep-th].

        \item
                E. Bernardini, 
                M. Carli, 
                M. Y. Elkhashab, 
                A. Ferroglia, 
                M. Fiolhais, 
                L. Gabelli,
                H. J. Munch, 
                D. Krym, 
                P. Mastrolia, 
                G. Ossola, 
                O. Pantano, 
                J. Postiglione,
                J. S. Poveda Correa, 
                C. Sirignano and
                F. Soramel,
                \emph{Using Smartphones to Innovate Laboratories in Introductory Physics Courses},
                GIREP 2022: 
                Effective Learning in Physics from Contemporary Physics to Remote Settings,
                Ljubljana, Slovenia.
\end{enumerate}

\end{declaration}

	\begin{acknowledgements}
\epigraph{\itshape It takes a village to raise a child.}{African proverb}
Let me here express my sincerest gratitude to all those who have so 
wonderfully shaped my Ph.D.~journey over the past three years.

I first and foremost thank my supervisor Pierpaolo Mastrolia for passionately 
introducing me to the subject of scattering amplitudes,
and for allowing me to pursue my research goals.

It is not an exaggeration to say that most of what I learnt throughout my Ph.D.~came 
from interactions with Federico Gasparotto and Seva Chestnov.
I thoroughly enjoyed collaborating with you,
and I cannot thank you enough for sharing your knowledge with me.

Let me give further thanks to all the people associated to the Padova Amplitudes group,
with whom I had many stimulating discussions on topics in physics,
namely
Manoj Kumar Mandal,
Luca Mattiazzi,
Giulio Crisanti,
Giacomo Brunello,
Sid Smith,
and Hjalte Frellesvig;
I thank Hjalte in particular for hosting me at the 
Niels Bohr Institute during the Amplitudes 2021 conference.
I also appreciate all the good times spent with Yousry, Marah, Prajwal and Marco here in Padova.

I feel extremely fortunate to have worked with Felix Tellander and Michael Borinsky,
and I am so proud of the work we achieved together.
I am grateful for the hospitality shown to us three by Nima Arkani-Hamed,
Sebastian Mizera and Aaron Hillman during our visit to the Institute for Advanced Study in Princeton.

Though I was officially enrolled in a physics Ph.D.~program,
my interests lead me into more mathematical areas of research.
In this regard,
I had the pleasure of collaborating with and learning from two brilliant mathematicians,
Saiei-Jaeyeong Matsubara-Heo and Nobuki Takayama.
These interactions had a tremendous impact on my intellectual development,
and instilled in me a great appreciation for the beauty of mathematical structures.

Speaking of mathematics,
I want to thank everyone with whom I engaged with at the 
Max Planck Institute for Mathematics in the Sciences in Leipzig.
I am especially grateful to Bernd Sturmfels for his hospitality,
and for creating such a unique and engaging environment at the institute.
I thank my collaborators Saiei and Claudia Fevola for all the joyous times spent there
diving into the mathematics of hyperplane arrangements and more.

I thank Jonah Stalknecht,
Robert William Moerman and David Damgaard for co-organizing the 
Geometry and Scattering Amplitudes Journal Club with me,
and I appreciate all the speakers and audience members for 
sharing their discoveries within the field of scattering amplitudes.

K\ae mpe tak til mine drenge Skov, Topic og Jeppe for en sindssyg fed tur rundt i Italien.
Det bragte mange gode minder.

Finally,
I wish to thank my mother Ljudmila and my partner Laura for 
their love and support throughout this whole period.

\end{acknowledgements}

	
	\tableofcontents
	
	
	
	
	\mainmatter
	
        \chapter*{List of Acronyms}
\addcontentsline{toc}{chapter}{\protect\numberline{}\hspace{-0.55cm}List of Acronyms}

\begin{tabular}{lll}
& (\textbf{CAS}): & Computer algebra system \\[5pt]
& (\textbf{DM}):  & Dark matter \\[5pt]
& (\textbf{DR}):  & Dimensional regularization \\[5pt]
& (\textbf{FI}):  & Feynman integral \\[5pt]
& (\textbf{GFI}): & Generalized Feynman integral \\[5pt]
& (\textbf{GKZ}): & Gel'fand Kapranov Zelevinsky \\[5pt]
& (\textbf{GP}):  & Generalized permutahedron \\[5pt]
& (\textbf{GPL}): & Generalized polylogarithm \\[5pt]
& (\textbf{HPL}): & Harmonic polylogarithm \\[5pt]
& (\textbf{MC}):  & Monte Carlo \\[5pt]
& (\textbf{ISP}): & Irreducible scalar product \\[5pt]
& (\textbf{LPr}): & Lee-Pomeransky representation \\[5pt]
& (\textbf{MI}):  & Master integral \\[5pt]
& (\textbf{SM}):  & Standard model \\[5pt]
& (\textbf{QCD}): & Quantum chromodynamics \\[5pt]
& (\textbf{QED}): & Quantum electrodynamics \\[5pt]
& (\textbf{QFT}): & Quantum field theory
\end{tabular}

	\chapter{Introduction}

Eugene Wigner famously pondered about the
\emph{unreasonable effectiveness of mathematics in the natural sciences} \cite{Wigner:1960kfi}.
How can it be that esoteric mathematical structures such as 
group theory, calculus and geometry -
often developed for entirely independent reasons -
are able to describe the "laws of Nature" so powerfully?
A prime example is that of quantum field theory (QFT).
This theory combines two pillars of our modern understanding of Nature:
quantum mechanics and special relativity.
Although some intuition can be gained from describing QFT in plain English,
it is only in terms of mathematics that we can properly
(insofar as possible)
\emph{define} the theory and \emph{do} something with it!

The standard model (SM) is the crown jewel of the modern QFT program.
It describes three out of four known forces of Nature:
electromagnetism, the weak and the strong nuclear forces.
Although it does not incorporate gravity (the last, known force) at the quantum level,
the mathematical techniques that were developed to probe the SM are now also being used to study
gravity in the weak field limit \cite{Kosower:2022yvp}.

\begin{figure}[t]
        \includegraphics[scale=0.146]{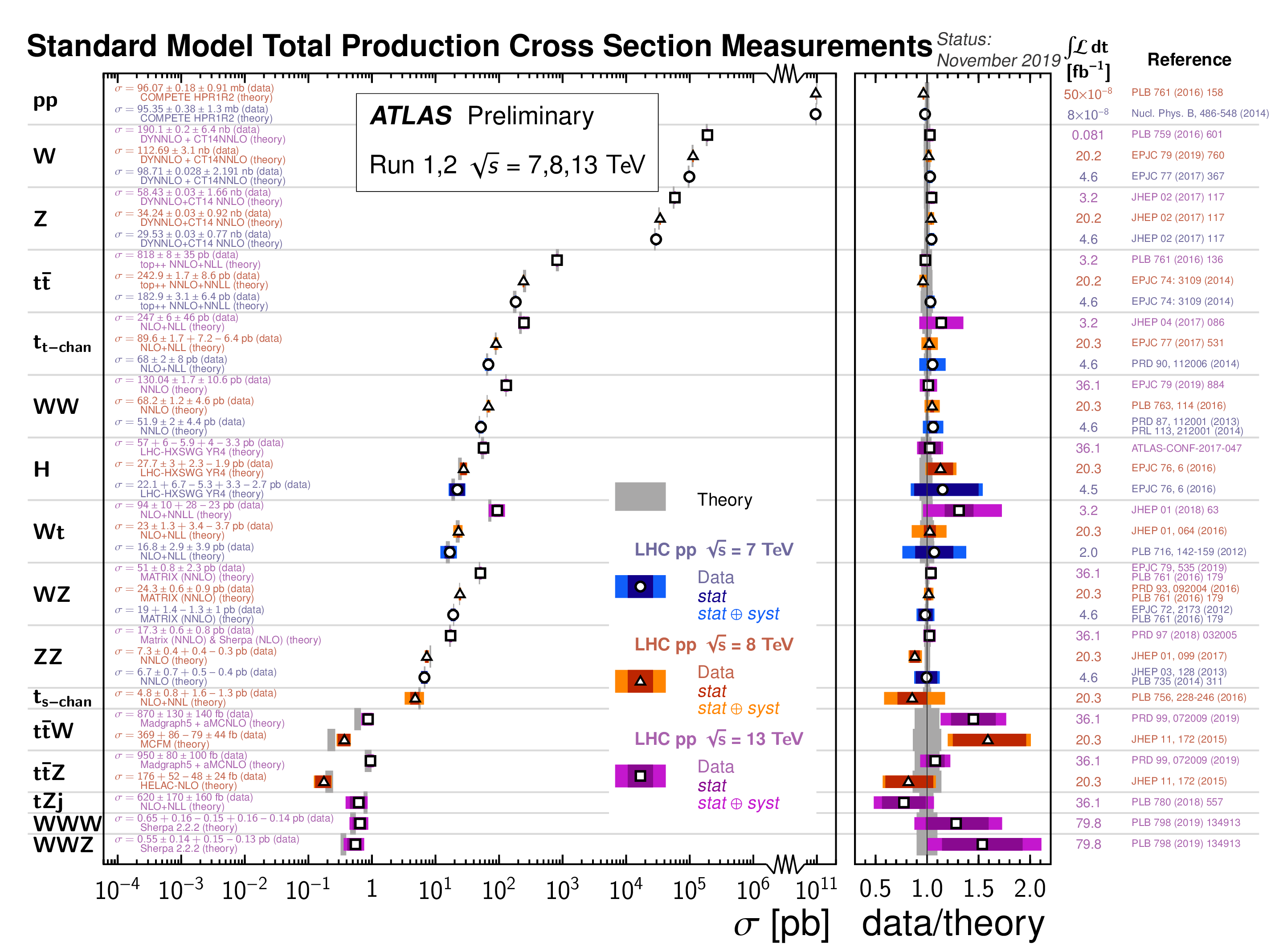}
        \caption{
                Summary plot anno 2019 comparing experimental vs.~theoretical values 
                for Standard Model cross sections \cite{ATL-PHYS-PUB-2019-024}.
        }
        \label{fig:SM_xsections}
\end{figure}

Physicists test the SM in the same way Rutherford first investigated the sub-atomic world a century ago:
by scattering particles together and measuring their cross sections
(roughly speaking the effective range of interaction of the scattered particles).
\figref{fig:SM_xsections} compares a host of experimentally determined total cross sections,
from ATLAS data,
with corresponding theoretical predictions made within the SM.
The level of agreement is striking.

A QFT, such as the SM, is typically defined in the abstract language of Lagrangians,
i.e.~polynomials in quantum fields and derivatives thereof.
How can this be connected to concrete,
observable quantities such as the aforementioned cross sections?
The connection is made through the calculation of \emph{scattering amplitudes}.
These are complex-valued functions
(depending on data such the masses, 
spins and momenta of the particles involved in a given scattering experiment)
which measure the quantum mechanical probability for turning an initial state into a final state.
Given such an amplitude $\mA$,
one obtains a total cross section by integrating $|\mA|^2$ over a relevant region of phase space.

Because the SM Lagrangian contains several small parameters in the form of coupling constants,
it is natural to compute scattering amplitudes as \emph{perturbative} expansions in said constants.
The numerical accuracy of a given amplitude is then determined by how many orders 
of this expansion we are able to compute.
Ever since the work of Feynman \cite{feynman},
physicists have organized these expansions in terms of \emph{Feynman diagrams}.
Schematically,
the perturbative expansion of an amplitude $\mA$ is written as
\eq{
        \includegraphicsbox[scale=0.5]{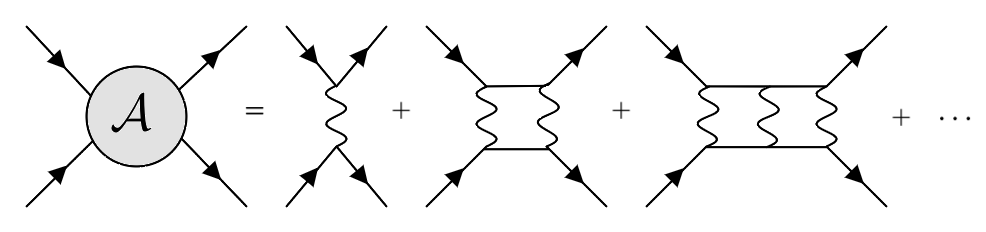}
}
The \emph{Feynman rules},
derived from the Lagrangian of a given theory,
give instructions for how to calculate each diagram.
The leading-order contributions,
written as sums of \emph{tree} diagrams,
are, arguably, now under control,
and so will not be a focus of this thesis.
The subleading-order contributions are written as sums of \emph{loop} diagrams.
In these cases,
the Feynman rules instruct us to compute integrals over the 
momenta of virtual particles running inside the loops.
These \emph{Feynman integrals} (FIs) are the source of nearly all the computational 
complexity that goes into the calculation of perturbative scattering amplitudes beyond leading order.
If we want to probe the SM at increased accuracy in the hopes of 
detecting signs of physics \emph{beyond} the SM
(which we know must exist, 
due to unexplained phenomena such as neutrino masses, 
dark matter and the supposedly "unnatural" value of the Higgs mass),
it thus behooves us to strengthen our mathematical capabilities for computing FIs.
Indeed,
following Wigner,
we should use the unreasonable effectiveness of mathematics to our benefit!

\section*{The context of this thesis}

There is a very long road from writing down a Lagrangian in, say, QCD to 
actually arriving at a cross section at some given order in perturbation theory
(this includes writing down all the relevant Feynman diagrams, 
taking care of tensor structure, 
renormalizing and regularizing UV and IR divergences respectively,
feeding in non-perturbative information via parton distribution functions,
choosing proper values for factorization/renormalization scales,
and much much more).
The modern scattering amplitudes program is,
in a word,
all about finding potential \emph{simplicity} hiding within this highly complex process
in order to lessen the calculational burden \cite{Travaglini:2022uwo}%
\footnote{
        In fact,
        it is now quite common in this field to dispense with Lagrangians entirely,
        as they carry so much physically irrelevant (i.e.~off-shell) information!
}.
More narrowly,
for the purpose of this thesis we may ponder the question:
\emph{do FIs also exhibit any simple mathematical structure?}

\begin{paragraph}{IBPs and DEQs.}
The first sign of "simplicity" was arguably the discovery of the \emph{linear relations} among FIs
called \emph{integration-by-parts} identities (IBPs) \cite{Tkachov:1981wb,Chetyrkin:1981qh,Laporta:2000dsw}.
To mention a concrete application of these relations, 
take the recent 3-loop QCD amplitude computed in \cite{Gehrmann:2023jyv}.
The original expression for the amplitude contains of order $\mO(10^5)$ FIs to be evaluated.
However,
only $\mO(300)$ of these integrals are actually linearly independent!
In other words,
the original $\mO(10^5)$ integrals can all be written as linear combinations of 
$\mO(300)$ \emph{master integrals} (MIs),
where the coefficients in front of the MIs are rational functions in the 
kinematic variables and the dimension of spacetime.
When such linear relations can be generated systematically and efficiently,
it constitutes a significant \emph{simplification} of a given calculation.
There are now many public codes dedicated to this purpose
being used in virtually every current amplitude calculation beyond leading order
\cite{Klappert:2020nbg,Smirnov:2013dia,vonManteuffel:2012np,Wu:2023upw,blade,Lee:2013mka}.
Indeed, 
we shall make frequent use of them in this thesis.

Suppose the many FIs in a given calculation have been reduced to a small set of MIs.
The question is then how to evaluate the MIs themselves.
IBPs once again lend a helping hand,
as they can be used to derive \emph{differential equations} for MIs
\cite{Barucchi:1973zm,Kotikov:1991pm,Remiddi:1997ny,Gehrmann:1999as}.
The second example of "simplicity" that we highlight was discovered by Henn \cite{Henn:2013pwa}:
there is a freedom in choosing which FIs that are promoted to MIs,
and if this choice is made judiciously,
then their DEQs simplify considerably.
Concretely,
the simplification stems from the decoupling, or factorization, of a parameter in the DEQs called $\e$,
which appears because it is related to the dimension $\DD$ of spacetime
(as mentioned above, $\DD$ is brought into the game due to IBPs).
Such a basis is dubbed to be \emph{canonical}.
When the DEQs are brought into this canonical form,
they are \emph{much} easier to solve.
\end{paragraph}

\begin{paragraph}{$\mD$-modules.}
These two notions,
IBPs and DEQs,
thus appear to be central to the (analytical, or symbolic) evaluation of FIs.
To learn more about the mathematical structure of these integrals,
we argue that it is interesting to reformulate these two notions in novel mathematical frameworks%
\footnote{
        A prime example hereof is the recent reformulation of IBPs within the framework of
        \emph{twisted cohomology} \cite{Mastrolia:2018uzb}.
},
as new perspectives often tend to bring new insights.
One such mathematical framework,
which will occupy a large portion of this thesis,
is that of $\mD$-\emph{modules}.
$\mD$-modules study DEQs from an \emph{algebraic} point of view,
as a supplement to the usual tools from complex analysis.
Its mathematical development goes back to the Japanese school 
of the 70s and 80s spearheaded by M. Sato and M. Kashiwara.
(See \cite{schapira2018fifty} for a historical account
and the textbooks \cite{Hotta-Tanisaki-Takeuchi-2008,Borel} for technical details.
Early studies on the relationship between FIs and $\mD$-modules include
\cite{kashiwara_1,kashiwara_2} in the 1970s,
and later \cite{Fujimoto2013GRACEAL,Nasrollahpoursamami:2016,Bitoun:2017nre} in the 2010s.)

One of the goals of this thesis is thus to re-interpret IBPs and DEQs in this language.
To foreshadow the contents of future chapters,
we shall interpret IBPs in terms of so-called \emph{Macaulay matrices},
and solutions to DEQs via \emph{"restrictions" of $\mD$-modules}.
\end{paragraph}

\begin{paragraph}{Numerical integration.}
Alongside the aforementioned developments in the analytical evaluation of FIs,
there is also a tradition for the \emph{numerical} evaluation thereof
(see \cite[Section 3.3]{Heinrich:2020ybq} for a historical account).
However,
because of limited computer resources in the previous century,
this tradition only goes back roughly two decades
(some of the earliest developments include \cite{Passarino:2001jd,Caffo:2002ch}).
But why go down this route?
Although analytic control over FIs is arguably the most favorable situation,
sometimes the symbolic expressions are just too complicated even with current tools!
Nevertheless,
blind numerical evaluation is rarely the most efficient way to proceed.
It is better to perform some kind of symbolic \emph{preprocessing},
so as to reveal an expression that is simple to numerically evaluate.
For instance,
one could still use IBPs to derive DEQs for FIs,
but now with every parameter except one fixed to a number in order to simplify the IBPs.
The DEQs are then written w.r.t.~this single, unfixed parameter,
and are solved through numerical series expansions 
rather than the analytic method of canonical bases.
This approach,
called \emph{auxiliary mass flow},
has proven to be very fruitful in recent years \cite{Liu:2021wks,Liu:2022chg}.
\end{paragraph}

\begin{paragraph}{Tropical geometry.}
With the goal of numerical evaluation in mind,
we shall carry out such a symbolic preprocessing by stepping into the world of \emph{tropical geometry}.
The mathematical discipline of \emph{algebraic geometry} is perhaps already familiar to the reader.
The latter studies the "smooth" geometries associated to the solutions of polynomial equations.
Tropical geometry,
instead,
studies the solutions to equations involving "piecewise-linear" polynomials.
This "piecewise" nature then translates into more "chunky" geometric figures,
more precisely \emph{polytopes},
which are comparatively easier to handle.
(See the textbook \cite{MaclaganSturmfels} for a pedagogical exposition on tropical geometry.)
The connection to FIs comes from representing them as integrals
over certain polynomials raised to rational powers.
By studying the simpler, tropical nature of these integrand polynomials,
one can learn about the UV/IR divergent behaviour of FIs \cite{Arkani-Hamed:2022cqe},
as well as the connection between QFT and string theory amplitudes \cite{Tourkine:2013rda,Eberhardt:2022zay}.

For our purposes,
the central notion will be that of a \emph{tropical approximation} to a polynomial,
envisioned by Panzer \cite{Panzer:2019yxl},
which essentially just replaces $"\!+\!"$ with $"\mathrm{max}"$.
We obtain tropical versions of FIs by applying this operation to the integrands.
Despite their \emph{simplicity} (or perhaps because of it!),
tropical versions of FIs appear to retain an uncanny amount of information about the original integral
\cite[Section 5.1]{Panzer:2019yxl}.
We shall employ this particular symbolic preprocessing, 
i.e.~the tropicalization of the Feynman integrand,
to simplify the numerical evaluation of FIs.
This idea was originally put forward by Borinsky \cite{Borinsky:2020rqs}.
\end{paragraph}

\section*{The content of this thesis}

The above discussion hopefully motivates the main topic of this thesis,
namely the evaluation of FIs.
We have divided the presentation into three parts.

\begin{paragraph}{Part I.}
The first part of the thesis focuses on the evaluation of FIs through the method of DEQs.

In \chapref{ch:feynman_integrals},
starting from a single FI,
we show how to associate to it a whole family of integrals $I_{\nu_1 \ldots \nu_n}$,
where each $\nu_i \in \ZZ$.
There exist linear relations among integrals with different values of $(\nu_1, \ldots, \nu_n)$,
namely the IBPs mentioned above.
We shall describe how to derive these relations,
how to identify a set of MIs $\vec{I}$,
and lastly how set up a system of first-order DEQs
\eq{
        \frac{\p}{\p z_i} \vec{I}(z) = P_i(z) \cdot \vec{I}(z)
        \quadit{\text{for}}
        i = 1, \ldots, N
        \, .
}
$z = (z_1, \ldots, z_N)$ is a collection of kinematic variables,
and $P_i(z)$ are rational matrices.
This is called a \emph{Pfaffian system} of DEQs.
Direct integration of FIs,
often being too hard,
is thereby sidestepped in favor of the comparatively easier task of solving Pfaffian systems.

In \chapref{ch:dark_matter},
we use the method of IBPs and Pfaffian systems to calculate a 
collection of 2-loop 3-point FIs with internally massive propagators.
These integrals are then fed into the form factors for a 
scattering amplitude that is of relevance to dark matter detection.
\end{paragraph}

\begin{paragraph}{Part II.}
The second part of this thesis is much more mathematically formal
(though hopefully still accessible to a physicist audience),
as it dives into the theory of $\mD$-\emph{modules}.

In \chapref{ch:macaulay},
we begin by describing those concepts in the theory of $\mD$-modules 
that are needed for developments in future chapters.
This includes the notion of a \emph{holonomic} $\mD$-module, 
which carries the structure of a finite-dimensional vector space.
This mimics the vector space structure of families of FIs due to IBPs.

In \chapref{sec:macaulay_matrices},
we present a novel algorithm whose output is a Pfaffian system for a holonomic $\mD$-module.
It takes as input a collection of differential operators that annihilate a given function
as well a basis of operators for the associated vector space.
The key concept of the algorithm is that of a \emph{Macaulay matrix},
which encodes relations among differential operators.
This matrix is frequently used in the setting of commutative 
polynomial rings to avoid Gr\"obner basis computations,
as they tend to scale poorly with the complexity of the problem.
Indeed,
one feature of the algorithm presented here is that it too avoids the need for Gr\"obner bases,
now in a non-commutative ring of differential operators.

In \chapref{ch:gkz},
we draw our attention to a specific holonomic $\mD$-module called the \emph{GKZ system},
named after Gel'fand, Kapranov and Zelevinsky \cite{GKZ_1,GKZ_2}.
It is built from a collection of differential operators that can immediately be written down
given the integral representation of an \emph{Euler integral}.
The benefit from employing the GKZ system is two-fold:
1) a complete set of annihilating operators is known, 
and 2) an explicit vector space basis can be computed very fast.

In \chapref{ch:gfi},
the connection between Euler integrals and FIs is established by generalizing 
the parameters present in the \emph{Lee-Pomeransky representation} \cite{Lee:2013hzt} of a FI.
The GKZ system can consequently be used to study this \emph{generalized} FI.
In particular,
we use the GKZ system as input for the Macaulay matrix algorithm
in order to derive Pfaffian systems for a couple of simple FIs.

In \chapref{ch:restrictions},
we address the discrepancy between Euler integrals and FIs:
an Euler integral often has too many parameters compared to the number of kinematic variables in a FI.
To land on the $\mD$-module for the FI proper,
one must take a kind of "limit" of the GKZ system,
called a \emph{restriction} in the language of $\mD$-modules.
We develop two different methods for computing restrictions whose
output is the Pfaffian system for the restricted $\mD$-module.
The first method is based on manipulating Pfaffian systems through 
series expansions and gauge transformations.
The second method simply takes the limits directly at the level of the Macaulay matrix.
The first method turns out to be applicable to the computation 
of logarithmic series solutions to Pfaffian systems.
The second method turns out to allow for restrictions onto 
hyper\emph{surfaces} in addition to just hyper\emph{planes},
which,
to the best of our knowledge,
has so far not been achieved within the theory of $\mD$-modules.

In \chapref{ch:restrictions_examples},
we showcase both restriction algorithms through several FI examples.
The examples concerning logarithmic solutions to Pfaffian systems unfortunately only contain partial results,
and so will have to be completed in future work.
\end{paragraph}

\begin{paragraph}{Part III.}
The third and final part of this thesis shifts focus from 
analytical to numerical methods for evaluating FIs.
The mathematical framework is that of \emph{tropical geometry}.

In \chapref{ch:tropical},
we begin by describing the notion of a \emph{tropical approximation} to a polynomial.
For instance,
given 
\eq{
        p(x) = 2 x_1 x_2 + 3 x_1^2 - 8 x_2^2
        \, ,
}
then its tropical approximation is
\eq{
        p \trop (x) = \mathrm{max}\{x_1 x_2, \, x_1^2, \, x_2^2\}
        \, .
}
We thereafter present a striking observation made by Borinsky \cite{Borinsky:2020rqs}:
for an integral of a polynomial raised to some power $\int p(x)^a \dd^n x$ 
(over a certain integration contour and modulo some technical conditions),
one can build a \emph{probability measure} from $p\trop(x)$.
By sampling points from this measure,
it is thereby possible to perform numerical Monte Carlo integration.
This can be used to evaluate FIs in parametric representation. 
We combine this \emph{tropical integration} algorithm with the 
$\e$-expansion of FIs in the dimensional regularization scheme.
Feynman's $i\vare$-prescription is incorporated via a certain contour deformation $x \to x e^{i f(x)}$,
which maps the Schwinger integration variables $x$ from real line into a path in the complex plane.

In \chapref{ch:feyntrop},
the program \ft is introduced.
It is an implementation of the tropical Monte Carlo integration scheme,
and includes both the $\e$-expansion and the $i\vare$-prescription.
The program can therefore numerically evaluate dimensionally regulated 
FIs for physical kinematics in Minkowski space
(rather than just Euclidean kinematics).
After a tutorial of the program,
we then apply it to several non-trivial examples between 1-5 loops and 0-5 legs.
\end{paragraph}

\vspace{0.5cm}

\noindent
In \chapref{ch:conclusion},
we present the conclusions reached throughout this thesis and speculate on developments for the future.

        \part{Multi-loop Technology}

        \chapter{Feynman Integrals}
\label{ch:feynman_integrals}

This chapter is a summary of basic concepts pertaining to Feynman integrals.
It is far from an exhaustive review 
(for such a text, consult e.g.~\cite{Weinzierl:2022eaz} and \cite{Badger:2023eqz}).

        \section{Definition}

The main object of study in this thesis is the \emph{Feynman integral} (FI) associated to an 
$L$-loop Feynman diagram $G$:
\eq{
        I_G = 
        \lim_{\vare \to 0^+}
        \int_{ \brk{\RR^{1,\DD-1}}^{L} } \,
        \dd^{L \cdot \DD } \ell \,
        \prod_{ e=1 }^{ \Eint } 
        \frac{ 1 }{ \left[- q_e\brk{p,\ell}^2 + m_e^2 - i \vare \right]^{\nu_e} } \, .
        \label{feynman_integral}
}
We often also write $I_{\nu_1 \nu_2 \ldots} = I(\nu_1, \nu_2, \ldots)$.
Let us go through the notation used above.

\subsection{Feynman diagram}
A (scalar) \emph{Feynman diagram} 
$
        G = (E,V)
$ 
is defined by a set of edges
$
        E = \brc{e_1, \ldots, e_{|E|}}
$
and vertices
$
        V = \brc{v_1, \ldots, v_{|V|}}.
$
Given the number of edges and vertices, 
the number of loops is given by
$
        L = |E| - |V| + 1 \, .
$
To each edge $e$, we associate a $\DD$-dimensional momentum vector flowing along the orientation of $e$.
Momentum conservation is imposed at each vertex $v$, 
meaning that all edge momenta flowing into $v$ equal those flowing out of $v$ 
(in analogy with Kirchhoff's law for currents in electrical circuits).

\subsection{The integrand}

Let us partition the set of edges into \emph{external} and \emph{internal} ones, 
i.e.
$
        E = \Eext \, \sqcup \, \Eint
$.
The external edges have one end which is not connected to any other edge,
while the internal edges are connected to at least one other edge on each end.

Consider an edge $e$.
If $e \in \Eext$ is external, then we label its associated momentum vector by
\eq{
        p_e = 
        \bigbrk{
                p_e^0, \ldots, p_e^{\DD-1}
        } 
        \in \RR^{1,\DD-1} \, ,
} 
where $\RR^{1,\DD-1}$ denotes $\DD$-dimensional Minkowski space equipped with the inner product
\eq{
        a \cdot b = 
        a^0 b^0 - 
        a^1 b^1 - \ldots -
        a^{\DD-1} b^{\DD-1} \, .
        \label{minkowski_inner_product}
}
On the other hand, if $e \in \Eint$ is internal,
then we associate to it the vector 
$
        q_e \in \RR^{1,\DD-1}
$
written as a sum of external momenta $p_i$ and \emph{loop momenta} $\ell_j$:
\eq{
        \label{propagator_momentum}
        q_e\brk{p,\ell} =
        \sum_i \pm \, p_i 
        \plus
        \sum_j \pm \, \ell_j
        \, .
}
The signs and summation ranges for $(i,j)$ depend on the chosen \emph{momentum routing} for the graph $G$.
The $L$ loop momenta
$
        \{\ell_1, \ldots, \ell_L\}
$
can be viewed as the leftover degrees of freedom after imposing momentum conservation at each vertex.
In \eqref{feynman_integral}, these loop momenta are integrated against the measure
\eq{
        \dd^{L \cdot \DD} \ell :=
        \dd \ell_1 \supbrk{0}  \wedge
        \cdots \wedge
        \dd \ell_1\supbrk{\DD-1} \wedge
        \cdots \wedge
        \dd \ell_L \supbrk{0}  \wedge
        \cdots \wedge
        \dd \ell_L\supbrk{\DD-1} 
}
over the whole of Minkowski space $\mathbb{R}^{1,\DD-1}$.

To any edge $e$, we additionally assign a scalar mass
$
        m_e \in \RR_{\geq 0}.
$
For an external momentum,
we say that it is \emph{on-shell} if it satisfies Einstein's energy momentum relation:
$p_e^2 = m_e^2$.
An internal momentum is generally not on-shell,
that is $q_e^2 \neq m_e^2$
(apart from special points within the integration range).

The integrand of \eqref{feynman_integral} is constructed by taking a product of 
\emph{Feynman propagators}
\eq{
        \frac{1}{ \left[ - q_e\brk{p,\ell}^2 + m_e^2 - i \vare \right]^{\nu_e} } 
}
for each internal edge.
The integers $\nu_e \in \ZZ$ are called \emph{edge weights} or simply \emph{propagator powers}.
The infinitesimal imaginary part $i \vare$ is introduced to set a particular integration contour.
We shall have more to say about this in \secref{sec:ieps_deformation},
but until then we can safely leave the $i \vare$ as well as the limit 
$
        \vare \to 0^+
$ 
in front of \eqref{feynman_integral} as implicit.
In gauge theories such as QED and QCD,
Feynman propagators become dressed with numerators having some theory-dependent tensor structure.
This additional structure is often factored out in front of scalar FIs
(cf.~\secref{sec:dm_amplitude}).

\begin{ex}
Consider the Feynman diagram
\eq{
        G \qquad = \includegraphicsbox{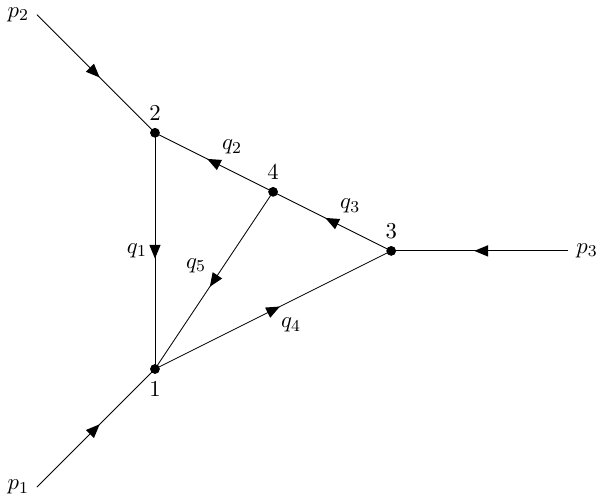} 
        \, .
}
\noindent
It has $L=2$ loops by inspection.
As a sanity check,
we get the same value by writing
$
        L = |E_\Int| - |V_\Int| + 1 = 5 - 4 + 1 = 2,
$
where we just used the internal edges and vertices for counting
(the external lines add three edges but technically
also three vertices, so these cancel in the formula for $L$).

We have chosen an arbitrary momentum routing by drawing arrows on each edge.
The external momenta are denoted by $\{p_1, p_2, p_3\}$,
and the internal ones are $\{q_1, \ldots, q_5\}$.
Imposing momentum conservation at each vertex,
we have
\eq{
        \begin{array}{lll}
                & \text{vertex $1$:} & p_1 + q_1 + q_5 = q_4
                \\
                & \text{vertex $2$:} & p_2 + q_2 = q_1
                \\
                & \text{vertex $3$:} & p_3 + q_4 = q_3
                \\
                & \text{vertex $4$:} & q_3  = q_2 + q_5 
                \, .
        \end{array}
}
There are not enough relations here to completely fix all of the momenta $q_e$.
Given that $L=2$,
there will be two leftover degrees of freedom $\ell_1$ and $\ell_2$.
Arbitrarily choosing to call $q_1 = \ell_1$ and $q_5 = \ell_2$,
then a solution to the linear system above is
\eq{
        \begin{array}{lll}
                q_1 &=& \ell_1
                \\
                q_2 &=& \ell_1 - p_2
                \\
                q_3 &=& \ell_1 + \ell_2 - p_2
                \\
                q_4 &=& \ell_1 + \ell_2 - p_2 - p_3
                \\
                q_5 &=& \ell_2
                \, .
        \end{array}
        \label{FI_example}
}

Suppose the internal masses are given by 
$
        \{m_1, m_2, m_3, m_4, m_5\} = \{m, 0, 0, 0, m\},
$
and that the edge weights are
$
        \{\nu_1, \nu_2, \nu_3, \nu_4, \nu_5\} = \{1, 2, 1, 1, 3\}.
$ 

The FI associated to this data is then
\eq{
        I_G = 
        \int_{ \lrbrk{\RR\supbrk{1,\DD-1}}^2 }
        \frac
        {\dd^{2\cdot\DD} \ell}
        {\lrsbrk{-q_1^2 + m^2} \lrsbrk{-q_2^2}^2 \lrsbrk{-q_3^2} \lrsbrk{-q_4^2} \lrsbrk{-q_5^2 + m^2}^3}
        \, ,
}
where the $q_e$ are given by \eqref{FI_example}.
\end{ex}

\subsection{General features}
\label{sec:general_features}

Having defined our object of study, 
let us now describe some general features of FIs.
The first thing to mention is that they are often \emph{divergent} 
when the loop momenta are small and/or large,
i.e.~when $\ell \to 0$ and/or $|\ell| \to \infty$.
This should not deter us, 
since we can introduce a regularization scheme to make these integrals well-defined.
The most common scheme, and the one we adopt throughout this thesis, is that of 
\emph{dimensional regularization} (DR) \cite{tHooft:1972tcz}.
Although \eqref{feynman_integral} will generically diverge in, say, $\DD=4$ spacetime dimensions,
we can promote $\DD$ to an indeterminate parameter,
thereby considering 
$
        I_G = I_G(\DD)
$ 
as a function of $\DD$.
Divergences then show up as poles in $(\DD-4)$, 
leading us to interpret the FI a Laurent series in $(\DD-4)$.
It is customary to write 
\eq{
        \DD = \DD_0 - 2 \e
}
with $\DD_0 \in \ZZ_{>0}$ and $|\e| \ll 1$,
such that 
\eq{
        I_G(\DD) =
        \sum_{i = i_\text{min}}^\infty
        I_G\supbrk{i} \e^i
        \qcomma
        i_\text{min} \in \ZZ \, .
        \label{epsilon_expansion}
}
The poles in $\e$ happily cancel during the computations of physical observables.
There are two mechanisms for these cancellations: 
UV renormalization and IR subtraction, 
but we shall not dive into these deep topics here.
Note that the mass dimension of the integral measure 
$
        \left[ \dd^{\DD \cdot L} \ell \right] = \DD \cdot L
$ 
changes in DR.
To counterbalance this,
one introduces another arbitrary parameter $\mu$ with mass dimension $[\mu]=1$,
leading to a new measure 
$
        (\mu^2)^{\e L} \, \dd^{\DD \cdot L} \ell.
$
We will often be sloppy with factors of $\mu$ throughout this thesis,
since they are easy to reinstate.

The end goal of a FI computation is to obtain the coefficient functions $I_G\supbrk{i}$.
What do they look like, qualitatively speaking?
Inspecting \eqref{feynman_integral}, 
observe that $I_G\supbrk{i}$ will be a functions of masses $m_e$ 
and $(|\Eext|-1)/2$ scalar products $p_e \cdot p_d$.
The factor of $1/2$ is due to
$
        p_e \cdot p_d = p_d \cdot p_e
$, 
and the factor $(|\Eext|-1)$ comes from momentum conservation 
\eq{
        p_{|\Ext|} = - \sum_{e=1}^{|\Eext|-1} p_e \, ,
}
with the convention that all external momenta are in-coming.
In practical calculations, 
it is useful to trade the dependence on scalar products for a dependence on \emph{Mandelstam variables}
$
        s_{ij} = \brk{p_i + p_j}^2 .
$

So the $I_G\supbrk{i}$ are functions of masses and external momenta, but what kind of functions?
The answer to this question is not yet settled in full generality, 
though much is known at low loop-order \cite{Bourjaily:2022bwx}.
Part of the answer depends on what properties we would \emph{like} a space of functions to have.
Take, for instance, the famous $\Gamma$-function
\eq{
        \Gamma\brk{z} = \int_0^\infty x^{z-1} e^{-x} \dd x 
        \, , \quad
        \Re{z} > 0
        \, .
        \label{gamma_function}
}
Because $\Gamma$ is \emph{defined} by its integral representation, 
we don't think of that integral as being "leftover work" to be done, as opposed to e.g.\
$
        \int_0^z \, x^2 \, \dd x .
$
The integral representation of $\Gamma$ is perfectly fine because 
\begin{enumerate}
        \item 
        It allows for the derivation of many identities,
        such as the famous reflection formula
        $
                \Gamma(1-z)\Gamma(z) = \frac{\pi}{\sin \brk{\pi z}} ,
        $
        which can simplify calculations involving large expressions.
        For instance, there could be two terms like 
        $
                4 \Gamma(1-z)\Gamma(z) - 4 \frac{\pi}{\sin \brk{\pi z}} 
        $
        in some result,
        but it would be wasteful to numerically evaluate this expression since it gives $0$.
        \item 
        It opens the door to other representations which allow for accurate numerical evaluations,
        such as the Stirling series 
        $
                \Gamma(1+z) =
                \sqrt{2 \pi z}
                \lrbrk{z/e}^z
                \exp 
                \lrbrk{
                        \sum_{i=1}^\infty
                        \frac{B_{2i}}{2i(2i-1)z^{2i-1}}
                } 
        $
        written in terms of Bernoulli numbers $B_i$.
        \item
        It is a starting point for determining the analytic continuation of 
        $\Gamma(z)$ 
        to larger regions of the complex plane.
\end{enumerate}
Based on this analogy, 
we claim that "calculating" the $i$th $\e$-coefficient function $I_G\supbrk{i}$ is equivalent to 
rewriting it in terms of functions with "nice" numerical and analytic properties.
A natural choice for such "nice" functions turns out to be the 
\emph{Chen iterated integrals} which we define in \secref{sec:gpls}.

        \section{Integration-by-parts identities}
\label{sec:integration_by_parts_identities}

In state-of-the-art computations at 2- or 3-loop orders,
one typically encounters $\mO(10^4)$ to $\mO(10^6)$ different FIs.
Fortunately,
only a select few of these integrals,
typically of order $\mO(10^2)$,
need to be calculated.
This huge decrease in workload is a consequence of linear relations among FIs called
\emph{integration-by-parts identities} (IBPs) \cite{Tkachov:1981wb,Chetyrkin:1981qh,Laporta:2000dsw}.
These identities express the many integrals appearing in a QFT computation in terms of a small basis of 
\emph{master integrals} (MIs).
The MIs can then be computed via the method of differential equations
\cite{Barucchi:1973zm,Kotikov:1991pm,Remiddi:1997ny,Gehrmann:1999as},
as outlined in \secref{sec:differential_equations}.

IBPs are born from the fact that a total derivative vanishes in DR \cite[Section 6.1]{Weinzierl:2022eaz}:
\eq{
        0 =
        \int \dd^{\DD \cdot L} \ell \,
        \frac{\p}{\p \ell^\mu} \bigsbrk{v^\mu \cdot f\brk{\ell}} \, ,
        \label{vanishing_total_derivative}
}
where $v^\mu$ is an arbitrary vector and $f\brk{\ell}$ is some product of propagators.
Let us go through an example to see why
\eqref{vanishing_total_derivative}
leads to linear relations among FIs.

\begin{ex}
\label{ex:ibp_bubble}
Consider the following 1-loop bubble integral family%
\footnote{
        The solid, thick line has mass $m$, 
        the solid, thin line is massless,
        and the dashed lines have squared momentum $p^2$.
}
\eq{
        I\brk{\nu_1,\nu_2} \ = \
        \includegraphicsbox[scale=0.15]{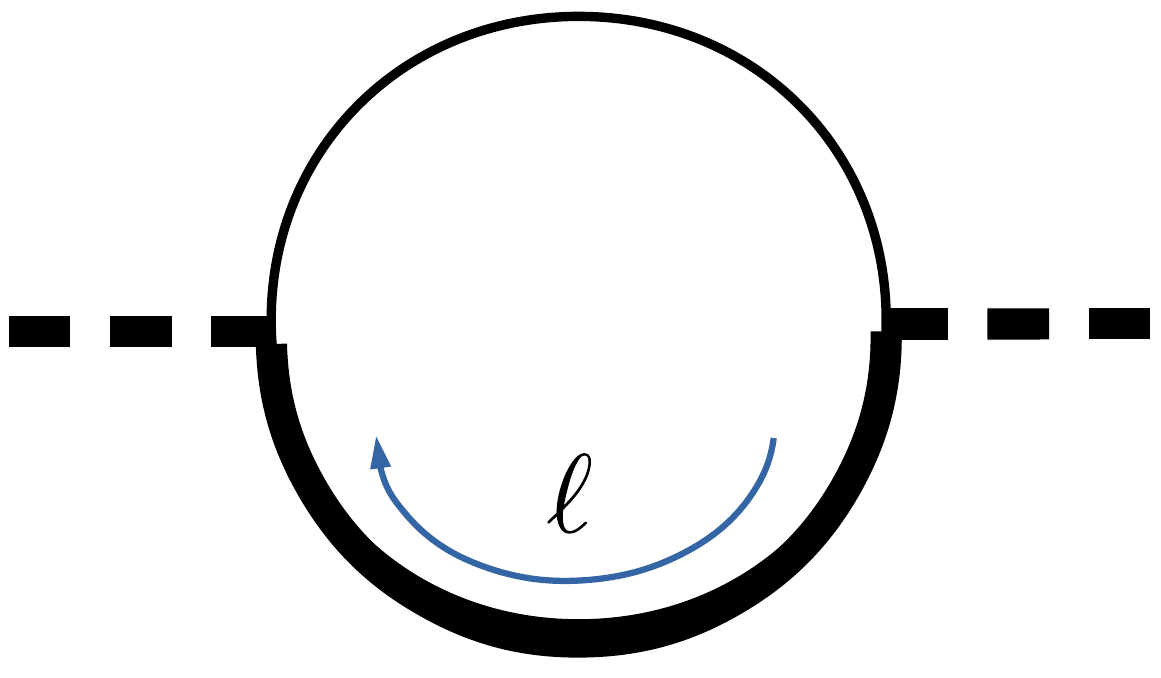} \ = \
        \int
        \frac
        {\dd^\DD \ell}
        { \lrsbrk{-\ell^2 + m^2}^{\nu_1} \lrsbrk{-\brk{\ell+p}^2}^{\nu_2} } \, .
}
Say we choose 
$
        v^\mu = p^\mu
$
in
\eqref{vanishing_total_derivative}.
Setting
$
        D_1 = -\ell^2 + m^2
$
and
$
        D_2 = -\brk{\ell+p}^2,
$
then the chain and product rules yield
\eq{
        \frac{\p}{\p \ell^\mu} \,
        \frac
        {p^\mu}
        { D_1^{\nu_1} D_2^{\nu_2} }
        &=
        \frac{p^\mu}{D_2^{\nu_2}} \,
        \frac{\p}{\p \ell^\mu} \,
        \frac{1}{D_1^{\nu_1}}
        \plus
        \frac{p^\mu}{D_1^{\nu_1}} \,
        \frac{\p}{\p \ell^\mu} \,
        \frac{1}{D_2^{\nu_2}}
        \\[8pt]
        &=
        \frac{p^\mu}{D_2^{\nu_2}} \,
        \frac{ 2 \nu_1 \ell_\mu}{D_1^{\nu_1+1}}
        \hspace{0.33cm} 
        \plus
        \frac{p^\mu}{D_1^{\nu_1}} \,
        \frac{ 2 \nu_2 \brk{\ell_\mu+p_\mu} }{D_2^{\nu_2+1}}
        \label{ibp_bubble_example}
        \, .
}
The combination
$
        2 p^\mu \ell_\mu = 2 p \cdot \ell
$
appears in both terms.
It can be rewritten in terms of the two propagators $D_1$ and $D_2$ as follows:
\eq{
        \begin{array}{ll}
                & D_2 
                = 
                {\color{red} - \ell^2 } - p^2 - {\color{blue} 2 p \cdot \ell}
                =
                {\color{red} D_1 - m^2} - p^2 - {\color{blue} 2 p \cdot \ell}
                \implies
                \\[3pt]
                & {\color{blue} 2 p \cdot \ell}
                =
                D_1 - D_2 - m^2 - p^2
                \, .
        \end{array}
        \label{ibp_bubble_example_2lp}
}
Inserting this into \eqref{ibp_bubble_example} gives
\eq{
        \nu_1
        \frac{D_1 - D_2 - m^2 - p^2}{D_2^{\nu_2} \, D_1^{\nu_1+1}}
        \plus
        \nu_2
        \frac{D_1 - D_2 - m^2 - p^2 + 2 p^2}{D_1^{\nu_1} \, D_2^{\nu_2+1}} \, .
}
Simplifying, 
reinstating the integral sign, 
and equating to zero,
we find the family of IBPs
\eq{
        \nonumber
        0 
        &=
        \brk{\nu_2 - \nu_1}
        I\brk{\nu_1 , \nu_2}
        \\
        \label{ibp_bubble_example_p}
        &\plus
        \nu_1
        \bigsbrk{
                \bigbrk{m^2 + p^2}
                I\brk{\nu_1+1 , \nu_2}
                +
                I\brk{\nu_1+1 , \nu_2-1}
        }
        \\
        &\plus
        \nu_2
        \bigsbrk{
                \bigbrk{m^2 - p^2}
                I\brk{\nu_1 , \nu_2+1}
                -
                I\brk{\nu_1-1 , \nu_2+1}
        } \, .
        \nonumber
}

Choosing instead
$
        v^\mu = \ell^\mu,
$
then a similar calculation yields
\eq{
        \nonumber
        0
        &=
        \brk{\DD - 2\nu_1 - \nu_2} I\brk{\nu_1 , \nu_2}
        \\
        \label{ibp_bubble_example_l}
        &\plus
        2 m^2 \nu_1 I\brk{\nu_1+1 , \nu_2}
        \\
        &\plus
        \nu_2
        \bigsbrk{
                \brk{m^2 - p^2} I\brk{\nu_1 , \nu_2+1}
                -
                I\brk{\nu_1-1 , \nu_2+1}
        }
        \, .
        \nonumber
}
The latter derivation requires the additional identity
\eq{
        \frac{\p}{\p \ell^\mu} \ell^\mu 
        = 
        \brk{\p_0, - \p_i}^\mu \,
        \eta_{\mu \nu} \,
        \brk{\ell_0, \ell_i}^\nu
        =
        \DD.
}

Let's now showcase an \emph{IBP reduction}.
Setting
$
        \nu_1 = \nu_1 = 1
$
in \eqref{ibp_bubble_example_p},
we have
\eq{
        0 
        =
        \brk{m^2 + p^2} I\brk{2,1}
        +
        I\brk{2,0}
        +
        \brk{m^2 - p^2} I\brk{1,2} \, .
        \label{ibp_bubble_example_p_11}
}
Here we set $I\brk{0,2} = 0$.
This is an example of a \emph{scaleless} integral,
i.e.~it does not depend on any variables with mass dimension.
It can be shown that scaleless integrals vanish in DR \cite[Page 40]{Weinzierl:2022eaz}%
\footnote{
        $I\brk{0,2}$ naively does appear to depend on the scale $p$.
        But we can shift 
        $
                \ell \to \ell - p
        $
        without altering the integral since we are integrating over the whole space $\RR^{1,\DD-1}$,
        thereby removing the $p$-dependence.
}.

Setting also 
$
        \nu_1 = \nu_2 = 1
$
in \eqref{ibp_bubble_example_l},
we get
\eq{
        0 =
        \brk{\DD-3} I\brk{1,1}
        +
        2 m^2 I\brk{2,1}
        +
        \brk{m^2-p^2} I\brk{1,2} \, .
        \label{ibp_bubble_example_l_11}
}

The last equation we will need is
\eqref{ibp_bubble_example_l}
for
$
        \nu_1 = 1, \, \nu_2 = 0
$:
\eq{
        0 =
        (\DD-2) I\brk{1,0} 
        +
        2 m^2 I\brk{2,0} \, .
        \label{ibp_bubble_example_l_10}
}
Combining the three equations
\eqref{ibp_bubble_example_p_11},
\eqref{ibp_bubble_example_l_11}
and
\eqref{ibp_bubble_example_l_10},
a few lines of algebra grant
\eq{
        I\brk{2,1} =
        \frac{\DD-2}{2m^2 \brk{p^2-m^2}} 
        I\brk{1,0}
        +
        \frac{\DD-3}{p^2-m^2} 
        I\brk{1,1} \, .
        \label{ibp_bubble_example_reduction}
}
This is an example of an IBP reduction:
we have expressed an integral $I\brk{2,1}$ with "large" denominator powers in terms of two simpler integrals,
$I\brk{1,0}$ and $I\brk{1,1}$.
In fact,
the IBPs
\eqref{ibp_bubble_example_p}
and
\eqref{ibp_bubble_example_l}
allow us to express \emph{any} integral $I\brk{\nu_1,\nu_2}$ in terms of $I\brk{1,0}$ and $I\brk{1,1}$.
These two integrals are thereby dubbed the 
\emph{master integrals} (MIs) for the integral family $I\brk{\nu_1,\nu_2}$.
These MIs are of course not unique.
The MI $I\brk{1,1}$ could for instance be swapped for $I\brk{2,1}$ via \eqref{ibp_bubble_example_reduction}.
\end{ex}

IBP reduction has been automated in many public codes such has
\package{kira} \cite{Klappert:2020nbg},
\package{Fire} \cite{Smirnov:2013dia},
\package{Reduze} \cite{vonManteuffel:2012np},
\package{NeatIBP} \cite{Wu:2023upw},
\package{Blade} \cite{blade},
and
\package{LiteRed} \cite{Lee:2013mka}.

\subsection{Irreducible scalar products}
There is an important detail to deriving IBPs which did not feature in \exref{ex:ibp_bubble},
namely that of \emph{irreducible scalar products} (ISPs).
Recall
\eqref{ibp_bubble_example_2lp},
where we expressed the scalar product $2 p \cdot \ell$ in terms of the propagators $D_1$ and $D_2$.
This is always possible at 1-loop order because there is a 
one-to-one correspondence between scalar products and propagators.
At 2-loop order or higher,
this is no longer the case,
as there will generally be more scalar products than propagators.
These "extra" scalar products are called \emph{irreducible}.

\begin{ex}
Take for instance a 2-loop integral written in terms $3$ propagators:
\eq{
        & I(\nu_1,\nu_2,\nu_2) = 
        \int 
        \frac
        {\dd^{2 \cdot \DD } \ell}
        {D_1^{\nu_1} \, D_2^{\nu_2} \, D_3^{\nu_3} }
        \quadit{\text{where}}
        \\ \nonumber
        & D_1 = -\ell_1^2
        \qcomma
        D_2 = -\ell_2^2
        \qcomma
        D_3 = -(\ell_1+\ell_2+p)^2
        \, ,
}
with $p$ being the only external momentum.
It is possible to form $5>3$ scalar products
\eq{
        \ell_1^2
        \qcomma
        \ell_1^2
        \qcomma
        \ell_1 \cdot \ell_2
        \qcomma
        \ell_1 \cdot p
        \qcomma
        \ell_2 \cdot p 
        \, .
}
Two out of the last three scalar products are irreducible;
say we choose
$
        \ell_1 \cdot \ell_2
        \text{ and }
        \ell_1 \cdot p.
$
The trick is simply to enlarge the original integral family to include two new integrand factors
$D_4 = \ell_1 \cdot \ell_2$ and $D_5 = \ell_1 \cdot p$,
leading to
\eq{
        I\brk{\nu_1, \ldots, \nu_5} =
        \int
        \frac{\dd^{2 \cdot \DD} \ell}{D_1^{\nu_1} \cdots D_5^{\nu_5}} \, .
}
In intermediate steps of an IBP reduction,
$D_4$ and $D_5$ will generally appear in numerators.
This is the reason for letting the edge weights $\nu_e \in \ZZ$ be integral rather than just non-negative.
\end{ex}

        \section{The method of differential equations}
\label{sec:differential_equations}

So IBPs allow us to write a given FI in terms of MIs.
The next question is how to compute the MIs themselves.
Let 
$I(\nu) = I(\nu_1, \nu_2, \ldots)$ 
denote a family of FIs.
This family depends on some kinematic variables,
call them $z = (z_1, z_2, \ldots)$,
such as masses $m_e^2$ and Mandelstam variables $s_{ij}$.
Suppose a vector of MIs $\vec{I}$ has been revealed after IBP reduction.
The \emph{method of differential equations} for computing the MIs consists of the following steps:
\begin{enumerate}
        \item 
                Take derivatives 
                $
                  \p_i \vec{I},
                $ 
                w.r.t.~the kinematic variables $z_i$.
                These derivatives lead to shifts in the edge weights $\nu_e$.
        \item 
                IBP reduce the integrals with shifted $\nu_e$ down to MIs $\vec{I}$.
                The reduction can be factored as 
                $
                        P_i\brk{z} \cdot \vec{I}
                $
                for some matrices $P_i$ whose entries are rational functions of $z$ and $\DD$.
                The resulting system of DEQs is
                \eq{
                        \p_i \vec{I} = P_i(z) \cdot \vec{I}
                        \, .
                }
                This is called a \emph{Pfaffian system},
                and will be a major topic of interest for this thesis.
                The matrices $P_i$ are called \emph{Pfaffian matrices}.
        \item 
                Solve the Pfaffian system in terms of a \emph{path-ordered exponential}.
\end{enumerate}
Let us illustrate the first three steps by an example.
The fourth step will be discussed in the next section.

\newpage

\begin{ex}
\label{ex:deq_bubble}
This is a continuation of \exref{ex:ibp_bubble},
wherein we found two MIs
\eq{
        \vec{I} =
        \arr{c}{
                I\brk{1,0} \\
                I\brk{1,1}
        } 
}
for a 1-loop integral family.

There are two kinematic variables $(z_1,z_2) = (m^2, p^2)$ w.r.t.~which we can take derivatives.
The derivative w.r.t.~$z_1$ works out to
\eq{
        \p_1 \vec{I} 
        &=
        \arr{c}{
                - I\brk{2,0} \\
                - I\brk{2,1}
        } 
        \\[10pt]
        & \hspace{-0.2cm} \overset{ \rm IBP }{=}
        \arr{c}{
                \frac{\DD-2}{2 z_1} \, I\brk{1,0} \\[5pt]
                \frac{\DD-2}{2 z_1 \brk{z_1 - z_2}} \, I\brk{1,0}
                +
                \frac{\DD-3}{z_1 - z_2} \, I\brk{1,1}
        } 
        \\[10pt]
        &=
        \arr{c c}{
                \frac{\DD-2}{2 z_1}                 & 0 \\
                \frac{\DD-2}{2 z_1 \brk{z_1 - z_2}} & \frac{\DD-3}{z_1 - z_2}
        }
        \cdot
        \arr{c}{
                I\brk{1,0} \\
                I\brk{1,1}
        } 
        \\[10pt]
        &=
        P_1 \cdot \vec{I}
        \, .
}
A similar computation can be carried out for $z_2$.
The result is
\eq{
        \p_2 \vec{I} 
        =
        \arr{cc}{
                0                           & 0 \\
                \frac{\DD-2}{2z_2(z_1-z_2)} & \frac{(2-\DD)z_1 + (4-\DD)z_2}{2z_2(z_1-z_2)}
        }
        \cdot \vec{I}
        =
        P_2 \cdot \vec{I}
        \, .
}

Pfaffian matrices such as $P_1$ and $P_2$ satisfy interesting identities.
The \emph{Euler homogeneity condition}
\eq{
        z_1 P_1 + z_2 P_2 = 
        \arr{cc}{\frac{\DD-2}{2} & 0 \\ 0 & \frac{\DD-4}{2}}
}
reflects the fact that the MIs are certain homogeneous functions of $z$.
Furthermore,
the \emph{integrability condition} 
\eq{
        \p_1 P_2 - \p_2 P_1 = [P_1, P_2]
}
follows from the fact that the MIs are $C^2$
(twice continuously differentiable).
\end{ex}

        \subsection{Canonical bases}

Obtaining the Pfaffian system for a basis of MIs is only part of the work,
the next step is to calculate a solution.
In practice,
we are only interested in computing solutions up a specific order in the $\e$-expansion 
\eqref{epsilon_expansion}.
As discovered by Henn \cite{Henn:2013pwa},
it is particularly simple to find this expansion if we first write the Pfaffian system in 
\emph{canonical form}.

The starting point is the Pfaffian system
$
        \p_i \vec{I} = P_i(z|\e) \cdot \vec{I}
$
for $N$ kinematic variables $z = (z_1, \ldots, z_N)$.
Note that we highlight both the $z$- and $\e$-dependence of the matrix $P_i$.
It is more clean to write this system in terms of the total derivative
$
        \dd_z = \dd z_1 \frac{\p}{\p z_1} + \ldots + \dd z_N \frac{\p}{\p z_N}
$
as
\eq{
        \dd_z \vec{I} = P(z|\e) \cdot \vec{I}
        \quadit{\text{where}}
        P = P_1(z|\e) \, \dd z_1 + \ldots + P_N(z|\e) \, \dd z_N
        \, .
        \label{total_derivatie_DEQ}
}
Suppose we would like to change to a new basis of MIs $\vec{J}$.
Due to IBP identities,
there must necessarily exist an invertible matrix $G=G(z|\e)$ such that
$
        \vec{I} = G \cdot \vec{J}.
$
Inserting this expression for $\vec{I}$ into \eqref{total_derivatie_DEQ},
it follows from the chain rule that
\eq{
        \dd_z \vec{J} = Q(z|\e) \cdot \vec{J}
        \quadit{\text{where}}
        Q_i(z|\e) = G^{-1} \cdot \big[ P_i \cdot G - \p_i G\big]
        \, .
        \label{gauge_transformation_G}
}
We say that the new Pfaffian matrices $Q_i$ are obtained from a \emph{gauge transformation} of $P_i$ by $G$.
Now, 
assume that the basis $\vec{J}$ has the following special properties:
\begin{enumerate}
        \item 
                The $\e$-dependence \emph{factorizes} in its associated Pfaffian matrices,
                that is 
                \eq{
                        Q(z|\e) = \e \, Q(z)
                        \, .
                }
        \item 
                The poles coming from the denominators of the Pfaffian matrices all have order 1.
                In particular,
                we can write
                \eq{
                        \e \, Q(z) = \e \sum_{\eta \in A} Q_i \, \dd \log \eta(z)
                        \, ,
                        \label{Q_dlog}
                }
                where
                \eq{
                        \dd \log \eta(z) =
                        \frac{1}{\eta(z)} \frac{\p \eta(z)}{\p z_1} \dd z_1
                        + \ldots +
                        \frac{1}{\eta(z)} \frac{\p \eta(z)}{\p z_N} \dd z_N
                        \, .
                }
                Every matrix entry $(Q_i)_{ab} \in \QQ$ is here a rational number,
                and the
                (possibly algebraic or even transcendental) 
                functions $\eta(z)$ are called \emph{letters}.
                The full set of letters,
                denoted here by $A$,
                is called the \emph{alphabet}.
\end{enumerate}
By virtue of these two properties,
$\vec{J}$ is called a \emph{canonical basis}.
Many calculations in the literature have shown that when condition 1 is satisfied,
then condition 2 follows automatically -
though, to the best of our knowledge, this has yet to be proven rigorously.
In practice,
the effort in searching for canonical bases is hence put into $\e$-factorization.

The $\e$-expansion of the canonical solution vector $\vec{J}$ is now particularly simple:
\eq{
        \vec{J} = 
        \mathcal{P} \exp \left( \e \int_\Gamma Q(z) \right)
        \cdot 
        \vec{J}_0(\e)
        \, ,
}
where the \emph{path-ordered exponential} is given by
\eq{
        \mathcal{P} \exp \left( \e \int_\Gamma Q(z) \right) =
        \mathbf{1} +
        \e 
        \int_0^t Q(\Gamma(t_1))+
        \e^2 
        \int_0^t Q(\Gamma(t_1)) 
        \int_0^{t_1} Q(\Gamma(t_2)) 
        +
        \mO(\e^3)
        \label{path_ordered_exponential}
}
for some chosen integration contour $\G$ parametrized by $t \in \RR$.
The vector of boundary constants also has an $\e$-expansion:
\eq{
        \vec{J}_0(\e) = 
             \vec{J}_0 \hspace{0.005cm} \supbrk{0} +
        \e   \vec{J}_0 \hspace{0.005cm} \supbrk{1} +
        \e^2 \vec{J}_0 \hspace{0.005cm} \supbrk{2} +
        \mO(\e^3)
        \, , 
} 
with each vector $\vec{J}_0 \hspace{0.005cm} \supbrk{n}$ consisting solely of real numbers.
These boundary vectors are often determined by considering special limits of the MIs;
concrete examples are given in \secref{sec:computing_the_FIs}.

The solution above is written somewhat abstractly in terms of a matrix-valued one-form $Q(z)$
(given in \eqref{Q_dlog})
and some unspecified contour $\Gamma$.
In \secref{sec:gpls},
we shall rewrite the iterated integrals 
\eqref{path_ordered_exponential} 
in form that is more suitable for computer implementations.

\subsection{Magnus expansion}
\label{sec:magnus}

With the discussion above in mind,
the key question is thus the following:
given a Pfaffian system
$
        \p_i \vec{I} = P_i \cdot \vec{I},
$
is it possible to algorithmically construct a gauge transformation $G$ such that the vector 
$
        \vec{J} = G^{-1} \cdot \vec{I}
$ 
satisfies a new system in canonical form?
Such an algorithm should preferably terminate in reasonable time,
and be applicable to a large category of FIs.
This question has been,
and is,
subject to intense study
(see e.g.~\cite[Chapter 7]{Weinzierl:2022eaz} and the references therein).
In this section,
we present a method for that calculating canonical forms that is at least effective for the 
class of functions called \emph{generalized polylogarithms} (GPLs),
to be defined in \secref{sec:gpls}.
(There are now also many methods that output canonical forms 
for FIs even when they cannot be evaluated in terms of GPLs \cite{Pogel:2022yat,Frellesvig:2023iwr,Dlapa:2022wdu,Jiang:2023jmk,Gorges:2023zgv,Frellesvig:2021hkr,Giroux:2022wav}.)

The approach described here is based on the 
\emph{Magnus expansion} for solving systems of DEQs \cite{magnus}.
60 years after Magnus' work,
Argeri et al.~\cite{Argeri:2014qva} discovered that his expansion can be used to obtain canonical forms.
The starting point is a Pfaffian system in \emph{pre-canonical form}:
\eq{
        \dd_z \vec{I} = \big[ P\supbrk{0}(z) + \e P\supbrk{1}(z) \big] \cdot \vec{I}
        \, .
}
I.e.~a system linear in $\e$
(the algorithm can be generalized to a polynomial dependence on $\e$ too
\cite[Section 5.2.4]{Schubert-Mielnik:2016cnt}).
By choosing judicial edge weights and simple $\e$-dependent prefactors for the MIs $\vec{I}$,
it is often not too difficult to find a pre-canonical form.

The clever observation is now this:
suppose that there exists a matrix $G = G(z)$,
depending only on $z$,
subject to the matrix DEQ
\eq{
        \dd_z G = P\supbrk{0}(z) \cdot G
        \, .
        \label{matrix_deq}
}
If we change basis to $\vec{I} = G \cdot \vec{J}$,
then the gauge transformation \eqref{gauge_transformation_G} simplifies to
\eq{
        Q(z|\e) 
        &= 
        G^{-1}
        \cdot
        \big[
                \big( P\supbrk{0} + \e P\supbrk{1} \big) \cdot G
                -
                \dd_z G
        \big]
        \\&=
        G^{-1}
        \cdot
        \big[
                \big( P\supbrk{0} + \e P\supbrk{1} \big) \cdot G
                -
                P\supbrk{0} \cdot  G
        \big]
        \\&=
        \e \, G^{-1} \cdot P\supbrk{1} \cdot G
        \, ,
        \label{gauge_transformation_G_magnus}
}
which is indeed $\e$-factorized.

The question is thus how to find the solution $G$ to the matrix DEQ \eqref{matrix_deq}.
It turns out that the solution takes the form of a matrix exponential:
\eq{
        G = e^{ \O \left[ P\supbrk{0} \right] } \cdot G(z=0)
        \, .
        \label{magnus_G_solution}
}
For our purposes,
we can simply take the boundary condition to be the identity matrix, $G(z=0) = \mathbf{1}$.
The $\O$ matrix in the exponent enjoys a \emph{Magnus expansion} 
\eq{
        \O \big[ P\supbrk{0} \big] 
        &=
        \sum_{n=1}^\infty 
        \O_n \big[ P\supbrk{0} \big]
        \, ,
}
with each term computed by integrating iterated commutators:
\eq{
        \nonumber
        \O_1 \big[ P\supbrk{0} \big]
        &=
        \int_\bullet^z \dd z_1 \, P\supbrk{0}(z_1)
        \\[5pt]
        \O_2 \big[ P\supbrk{0} \big]
        &=
        \frac{1}{2} \int_\bullet^z \dd z_1  \int_\bullet^{z_1} \dd z_2 \, 
        \left[ P\supbrk{0}(z_1), \, P\supbrk{0}(z_2) \right]
        \\[5pt]
        \nonumber
        \O_3 \big[ P\supbrk{0} \big]
        &=
        \frac{1}{6} \int_\bullet^z \dd z_1  \int_\bullet^{z_1} \dd z_2 \int_\bullet^{z_2} \dd z_3 \, 
        \Big\{
        \left[ 
                P\supbrk{0}(z_1)
                , \, 
                \lrsbrk{P\supbrk{0}(z_2), \, P\supbrk{0}(z_3)}
        \right]
        + \\& \hspace{5.1cm}
        \left[ 
                P\supbrk{0}(z_3)
                , \, 
                \lrsbrk{P\supbrk{0}(z_2), \, P\supbrk{0}(z_1)}
        \right]
        \Big\}
        \nonumber
        \, .
}
Clean formulas for the $n$th term of this expansion can be found in \cite{arnal2018general}.
Here $P\supbrk{0}$ was assumed to depend on just a single variable $z$.
For multivariate problems,
one can simply apply these formulas one variable at a time \cite[Page 108]{Primo:2017jtw}.

It turns out that the Magnus expansion truncates for FIs that can be evaluated in terms of GPLs,
i.e.~$\O_{n_\mathrm{max}} = 0$ for some $n_\mathrm{max} \in \ZZ_{>0}$.
In such cases,
the matrix $P\supbrk{0}$ is apparently sparse enough to become nilpotent,
meaning that the iterated commutators eventually evaluate to zero.

\begin{ex}
Here we apply the Magnus expansion to the bubble integral from 
Examples \eqref{ex:ibp_bubble} and \eqref{ex:deq_bubble}.
Consider a basis with squared propagators:
\eq{
        \vec{I} = \arr{c}{I(2,0) \\ I(2,1)}
        \, .
}
Using IBPs as in \exref{ex:deq_bubble},
it is possible to obtain two Pfaffian matrices $P_1$ and $P_2$ in the variables $z_1 = m^2$ and $z_2 = p^2$.
Working in units such that $z_1 = 1$,
then $P_1$ can be discarded.
Call $z_2 = z$.
We thence get a pre-canonical system
\eq{
        \frac{\p}{\p z} \vec{I} =
        \Big[ P\supbrk{0} + \e \, P\supbrk{1} \Big]
        \cdot \vec{I}
        \, ,
}
where the constant and linear matrices w.r.t.~$\e$ are 
\eq{
        P\supbrk{0} =
        \arr{cc}{0 & 0 \\ 0 & -\frac{1}{z}}
        \quadit{\text{and}}
        P\supbrk{1} =
        \arr{cc}{
                0 & 0 \\
                \frac{1}{(1-z) z} &
                \frac{1+z}{(1-z)z}
        }
        \, .
}
The Magnus expansion truncates at the first term,
giving
\eq{
        \O\big[P\supbrk{0}\big] = \O_1\big[P\supbrk{0}\big] =
        \int_\bullet^z
        \arr{cc}{0 & 0 \\ 0 & -\frac{1}{z'}} \dd z' =
        \arr{cc}{0 & 0 \\ 0 & -\log(z)} 
        \, ,
}
wherefore
\eq{
        G = e^{ \O\lrsbrk{P\supbrk{0}} } =
        \arr{cc}{1 & 0 \\ 0 & \frac{1}{z} }
        \, .
}
Gauge transforming with $G$ via \eqref{gauge_transformation_G_magnus} yields a new Pfaffian matrix
\eq{
        Q = 
        \e \arr{cc}{0 & 0 \\ \frac{1}{1-z} & \frac{2}{1-z} + \frac{1}{z}}
        \, ,
}
which is indeed canonical.
\end{ex}

\begin{ex}
Every physicist's favorite equation is the DEQ for the harmonic oscillator:
\eq{
        \frac{\dd^2 f(t)}{\dd t^2} + \o^2 f(t) = 0
        \, .
}
The angular frequency $\o$ is taken to be constant.
Let us solve this equation using the Magnus expansion.
The initial step is to rewrite the DEQ as a first-order matrix equation.
Setting $f_1(t) = f(t)$ and $f_2(t) = \frac{\dd f(t)}{\dd t}$,
then the DEQ is equivalent to
\eq{
        \frac{\dd}{\dd t}
        \arr{c}{f_1(t) \\ f_2(t)}
        =
        \arr{cc}{0 & 1 \\ -\o^2 & 0}
        \cdot
        \arr{c}{f_1(t) \\ f_2(t)}
        \, .
}
In addition to solving \emph{matrix} DEQs such as \eqref{matrix_deq},
the Magnus expansion also grants solution \emph{vectors} in the form
\eq{
        \arr{c}{f_1(t) \\ f_2(t)} =
        e^{\O(t)}
        \cdot
        \arr{c}{f_1(0) \\ f_2(0)} 
        \, .
}
The Magnus expansion truncates at the first term because $\o$ is constant.
Thus,
\eq{
        e^{\O(t)} 
        &=
        \exp
        \left(
        \int_0^t
        \arr{cc}{0 & 1 \\ - \o^2 & 0}
        \dd t'
        \right) 
        \\[5pt]&=
        \arr{cc}{
                \cosh \left( \sqrt{-\o^2} t \right) & 
                \frac{ \sinh\left( \sqrt{-\o^2} t \right) }{ \sqrt{-\o^2} } \\ 
                \sqrt{-\o^2} \sinh\left( \sqrt{-\o^2} t \right) & 
                \cosh \left( \sqrt{-\o^2} t \right)
        }
        \\[5pt]&=
        \arr{cc}{
                \cos(\o t) & \frac{\sin(\o t)}{\o} \\
                - \o \sin(\o t) & \cos(\o t)
        }
        \, .
}
We obtain oscillating functions as expected.
Further,
this solution satisfies $f_2(t) = \dot{f}_1(t)$.
\end{ex}

        \section{Generalized polylogarithms}
\label{sec:gpls}

The path-ordered exponential \eqref{path_ordered_exponential} naturally leads to a class of functions
called \emph{Chen iterated integrals} \cite{chen}.
Let us first say what we mean by an integration contour for these integrals.
Suppose that there are $N$ kinematic variables 
$
        z = (z_1, \ldots, z_N)
$ 
which parametrize an $N$-dimensional (smooth) complex manifold $Z$.
We write write a path in this space as
\eq{
        \G : [a,b] \to Z
        \, ,
}
where $\G(a) = z_a \in Z$ and $\G(b) = z_b \in Z$ are the start and end points.
One can think of $t \in [a,b] \subset \RR$ as being a "time" coordinate.

Let $\{\o_1, \ldots, \o_M\}$ denote a collection of differential 1-forms on $Z$.
For example,
they could be the $\dd \! \log$ forms from \eqref{Q_dlog}.
These 1-forms are supposed to be integrated over the contour $\G$.
So instead of parametrizing the $\o_i$ in terms of $z$,
we would like to parametrize them using $t \in [a,b]$.
This is called a \emph{pullback} to the interval $[a,b]$,
written as
\eq{
        \G^* \o_i = f_i(t) \dd t
        \, .
}
This gives a concrete expression for the integral of 1-form along $\G$:
\eq{
        \int_\G \o_i = \int_a^b f_i(t) \dd t
        \, .
}
The question is how to compute the function $f_i(t)$.
Writing
\eq{
        \o_i = \sum_{j=1}^N \o_{ij}(z) \dd z_j
        \quadit{\text{and}}
        \G(t) = \arr{c}{\G_1(t) \\ \vdots \\ \G_N(t)}
        \, ,
}
then the formula for $f_i(t)$ follows from the chain rule:
\eq{
        \G^* \o_i =
        \sum_{j=1}^N 
        \o_{ij} \big( \G(t) \big) \frac{\dd \G_j(t)}{\dd t} \dd t
        \quadit{\implies}
        f_i(t) = 
        \sum_{j=1}^N 
        \o_{ij} \big( \G(t) \big) \frac{\dd \G_j(t)}{\dd t} 
        \, .
        \label{pullback_f}
}

\begin{ex}
To demystify the construction above,
let us compute
\eq{
        \int_\G \o
}
with
\eq{
        \o = \frac{2}{z_2-2} \dd z_1 + \frac{3}{z_1+4} \dd z_2
        \, ,
        \quadit{\text{and}}
        \G : [0,1] \to \CC^2
        \quadit{\text{given by}}
        \G(t) = \arr{c}{t^2 \\ 1-t}
        \, .
        \nonumber
}
Inserting this data into \eqref{pullback_f} gives
\eq{
        f(t) 
        &=
        \o_1 \big(\G(t) \big) \times \frac{\dd \G_1(t)}{\dd t} 
        \plus
        \o_2 \big(\G(t) \big) \times \frac{\dd \G_2(t)}{\dd t} 
        \\&=
        \frac{2}{(1-t)-2} \times 2t
        \plus
        \frac{3}{t^2+4} \times (-1)
        \\&=
        - \frac{4t^3 + 19t + 3}{(1+t)(4+t^2)}
        \, .
}
Using any CAS,
one can then easily integrate
\eq{
        \int_\G \o \ = \ \int_0^1 f(t) \dd t \ \approx \ -1.923 \, .
}
\end{ex}

Given a $t \in [a,b]$,
the \emph{Chen iterated integral} can now be defined as
\eq{
        I_\G(\o_1, \ldots, \o_M | t) 
        &=
        \int_\G \o_1 \wedge \cdots \wedge \o_M 
        \\&=
        \int_a^t f_1(t_1) \dd t 
        \int_a^{t_1} f_2(t_2) \dd t_2 
        \cdots
        \int_a^{t_{M-1}} f_M(t_M) \dd t_M 
        \, .
}
An equivalent,
recursive definition is
\eq{
        I_\G(\,|t) &= 1 
        \\
        I_\G(\o_1, \ldots, \o_M | t) &=
        \int_a^t f_1(t_1) I_\G(\o_2, \ldots, \o_M | t_1) \dd t_1
        \, .
}
Chen proved that $I_\G$ is independent of $\G$,
provided that $\G$ does not hit any singularities nor cross any branch points of the integrand
for $a < t < b$.
If the integrand has singularities at the endpoints $t=a$ or $t=b$,
these are assumed to be integrable singularities 
(caveat: endpoint singularities can also be regularized using so-called shuffle identities \cite{Duhr:2014woa}).

Consider now a special class of 1-forms with a simple pole at $t = z_i$:
\eq{
        \o_\mathrm{GPL}(z_i)  = 
        \frac{\dd t}{t - z_i} = 
        \dd \log(t - z_i) 
        \, .
}
The complex number $z_i$ is treated as a parameter.
Let $\G$ be a path along the non-negative real axis such that $\G(a=0)=0$ and $\G(t) = y \in \RR_{\geq 0}$.
The special class of Chen iterated integrals called \emph{generalized multiple polylogarithms} 
(GPLs) \cite{gpls}
is defined by
\eq{
        G(z_1, \ldots, z_M | y) =
        I_\G\big(\o_\mathrm{GPL}(z_1), \ldots, \o_\mathrm{GPL}(z_M) | y\big)
        \, .
}
Explicitly,
\eq{
        G(\underbrace{0, \ldots, 0}_{M \text{ times }} | y) 
        &= 
        \frac{1}{M!} \log^M(y)
        \\
        G(z_1, \ldots, z_M | y) 
        &=
        \int_0^y \frac{\dd y_1}{y_1-z_1} G(z_2, \ldots, z_M | y_1)
        \, .
}
The $z$-parameters are often called \emph{weights}.
The case $z=0$ above naively constitutes a divergence of the integral representation,
so the first formula above is more of a stipulation than a derivation.

There is a huge literature on these functions,
see \cite{Duhr:2014woa} \cite[Chapters 6,8]{Weinzierl:2022eaz} and references therein.
There are also plenty of powerful codes for symbolically manipulating 
and numerically evaluating these functions:
the \package{Maple} package \cite{Frellesvig:2018lmm},
\package{Hyperint} \cite{Panzer:2014caa},
\package{HPLog} \cite{Gehrmann:2001pz},
\package{HPL} \cite{Maitre:2005uu},
\package{PolyLogTools} \cite{Duhr:2019tlz},
\package{Ginac} \cite{Vollinga:2004sn},
\package{Chaplin} \cite{Buehler:2011ev},
\package{handyG} \cite{Naterop:2019xaf},
and
\package{FastGPL} \cite{Wang:2021imw}.

Now harken back to the path-ordered exponential \eqref{path_ordered_exponential}.
If the functions $\eta(z)$ from \eqref{Q_dlog} are \emph{rational} functions of the kinematic variables $z$,
then the solution for the FI in question can be written in terms of GPLs.
This is indeed a favorable form for the solution because so much is known about the identities among GPLs,
how to analytically continue them,
and how to numerically evaluate them.

One may encounter letters $\eta$ that depend on $z$ in terms of algebraic roots.
In those cases,
one can sometimes find a change of a variables $w = f(z)$ that turns the roots into rational functions.
The package 
\package{RationalizeRoots} \cite{Besier:2019kco} 
automates the procedure for finding such an $f$.

        \chapter{Two-loop Feynman Integrals for Dark Matter Detection}
\label{ch:dark_matter}

We now apply the multi-loop technology from the previous chapter to compute form factors 
for a \emph{leptophillic dark matter} (DM) model.
Form factors are certain building blocks of scattering amplitudes.
For phenomenological implications of these results, 
we refer to the paper \cite{Garani:2021ysl} on which this chapter is based.

        \section{The model}

Consider a DM particle $\chi$ that scatters elastically off a nucleon $A$:
$
        \chi A \to \chi A .
$
This is called a \emph{direct DM search}, 
and is probed by remarkable experiments having large tanks filled with e.g.~oxygen, helium or xenon nucleons.

Assuming that DM does not couple \emph{directly} to SM particles, 
this scattering can only occur via a \emph{mediator} particle $\phi$ that \emph{does} couple to the SM.
Here we consider a \emph{leptophillic} DM model, 
wherein $\chi$ couples to $\phi$, 
which in turn only couples to SM leptons;
the interaction between leptons and the quarks inside the nucleon $A$ is finally mediated via photons.
The mediator $\phi$ is taken to be a singlet scalar (called $\phi_S$) or a pseudo-scalar (called $\phi_P$).
The Lagrangian interaction terms are thus of the form
\eq{
        - \mathcal{L}_S \ \supseteq \
        g_S \, \phi_S \, \bar{\ell} \, \ell
        \ + \
        g_\chi \, \phi_S \, \Gamma_\chi
        \ \quadit{\text{or}}
        - \mathcal{L}_P \, \supseteq \
        i \, g_P \, \phi_P \, \bar{\ell} \, \gamma_5 \, \ell
        \ + \
        g_\chi \, \phi_P \, \Gamma_\chi
        \, ,
        \label{DM_interaction_terms}
}
where
$
        \Gamma_\chi = \chi^\dagger \chi
$ 
for scalar DM and 
$
        \Gamma_\chi = \{\bar{\chi} \, \chi, \, \bar{\chi} \, \gamma_5 \, \chi\}
$
for fermionic DM,
$\{g_S,g_P,g_\chi\}$ are coupling constants,
and $\{\ell, \bar{\ell}\}$ denote a lepton and an anti-lepton.

\begin{figure}
        \centering
        \includegraphicsbox[scale=0.2]{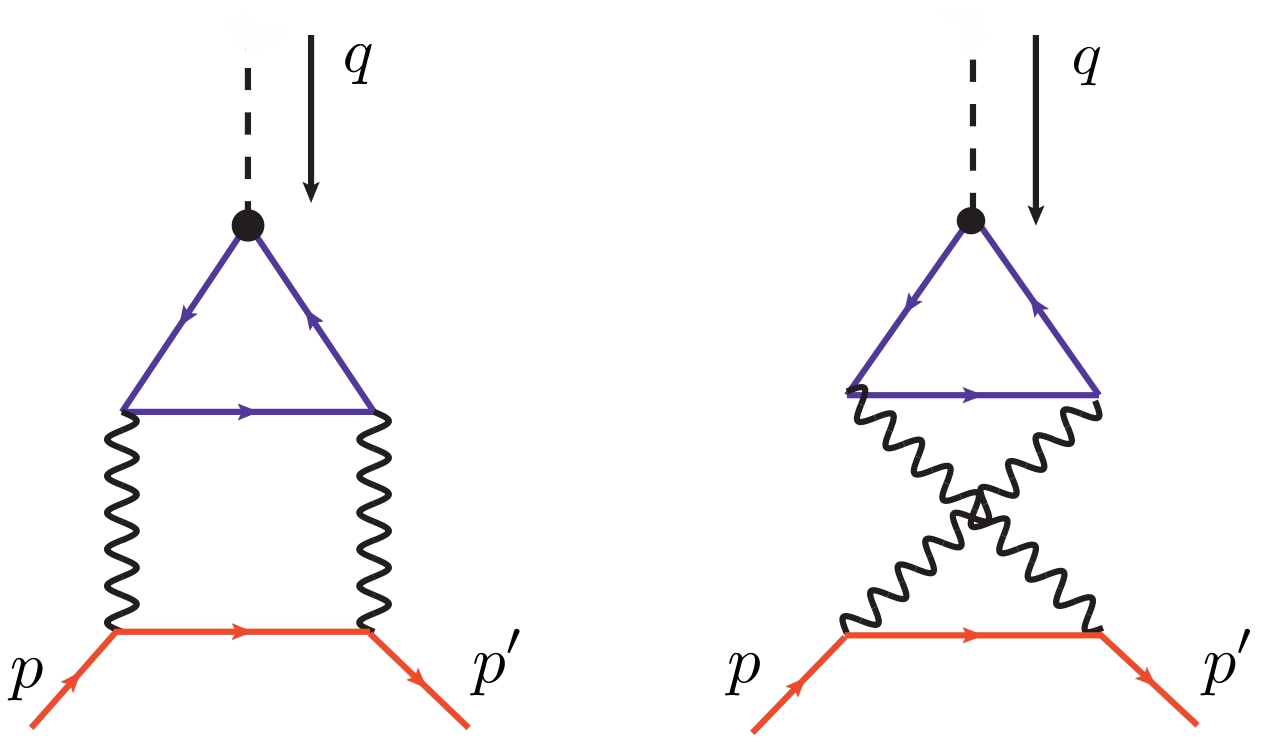}
        \caption{Leading contributions to the scattering amplitude for the leptophillic DM model.}
        \label{fig:DM_diagrams_1}
\end{figure}

Since this model assumes no tree-level couplings to quarks,
the $\chi A$ scattering cross section is purely loop-induced.
It turns out that the leading contributions come from the diagrams%
\footnote{
        These are only the so-called one-body diagrams. 
        The virtual photons can also scatter off two nucleons,
        yielding additional one-loop two-body diagrams.
        See \cite{Garani:2021ysl} for more details.
}
in \figref{fig:DM_diagrams_1}.
This is a $t$-channel process, 
with an incoming mediator $\phi(q)$ (dashed black lines) transferring momentum to a nucleon $A(p)$ 
(leftmost {\color{red} red} lines),
leading to a final state nucleon $A(p')$
(rightmost {\color{red} red} lines).
Momentum conservation dictates that $q = p'-p$.
The {\color{blue} blue} triangle loops represent a lepton which couples to 
(the quarks inside) 
the nucleon via photons (wiggly black lines).

Let $m_A$ denote the nucleon mass, 
meaning that it can either be the proton or neutron mass.
The kinematic invariants for this process are then defined by
\eq{
        t = q^2 = (p'-p)^2
        , \quad
        m_A^2 = p^2 = p'^2
        \, .
}

        \section{The scattering amplitude}
\label{sec:dm_amplitude}

According to the Feynman rules dictated by the Lagrangian terms \eqref{DM_interaction_terms},
the desired scattering amplitude takes the form
\eq{
        \mathcal{A}_{S,P}(t, m_\ell, m_A) =
        i \, g_{S,P} \, Q_A^2
        \sum_{\ell = e, \, \mu, \, \tau}
        Q_{\ell}^2 
        \Big[
                \bar{u}_A(p') 
                \cdot
                \Gamma_{S,P}(t, m_\ell, m_A)
                \cdot
                u_A(p)
        \Big]
        \, .
        \label{DM_amplitude}
}
Let us parse this expression from left to right.
The amplitude $\mathcal{A}$ has an index $S$ or $P$, 
depending on whether the mediator $\phi_{S,P}$ is taken to be scalar or pseudo-scalar.
$\mA$ has kinematic dependence on the momentum transfer variable $t$,
the lepton masses $m_\ell = \{m_e, m_\mu, m_\tau\}$
and the nucleon mass $m_A$.
The prefactor in front of the sum contains the coupling constants $g_{S,P}$ from \eqref{DM_interaction_terms}
and the electric charge of the nucleon, 
where $Q_A = 1$ for the proton and $Q_A = 0$ for the neutron%
\footnote{
        The expression \eqref{DM_amplitude} therefore vanishes when $A$ is the neutron.
        This is an approximation which holds only in the non-relativistic limit,
        which is the regime being probed by DM experiments.
        See \cite{Garani:2021ysl} for more details.
}.
The lepton electric charge inside the sum is $Q_\ell = 1$.
The sum over leptons contains an operator 
$
        \Gamma_{S,P}(t, m_\ell, m_A)
$
sandwiched between spinors $\bar{u}(p')$ and $u(p)$ for the final and initial state nucleons respectively.
This operator stems from the sum of the two diagrams above.
Its analytic expression is
\eq{
        \Gamma_{S,P}(t, m_\ell, m_A) =
        - 32 \pi^2 \alpha_\mathrm{em}
        \int \frac{\dd^\DD \ell_1}{(2\pi)^\DD} \frac{\dd^\DD \ell_2}{(2\pi)^\DD}
        \frac
        {
                g_{\mu \rho} \
                \gamma^\rho \
                (\slashed{\ell}_2 - m_A) \
                g_{\nu \sigma} \
                \gamma^\sigma \
                \mathrm{Tr}_{S,P}^{\mu \nu}
        }
        {D_1 D_2 D_3 D_4 D_5 D_6}
        \, ,
        \label{Gamma_SP}
}
where the electromagnetic coupling is
$
        \alpha_\mathrm{em} =
        e^2/(4 \pi),
$
and the Dirac trace over spinor indices is given by
\eq{
        \mathrm{Tr}_{S,P}^{\mu \nu}
        =
        \mathrm{Tr}
        \left[
                (\slashed{\ell}_1 + \slashed{q} + m_\ell) \cdot
                \Lambda_{S,P} \cdot
                (\slashed{\ell}_1 + m_\ell) \cdot
                \gamma^\mu \cdot
                (\slashed{\ell}_1 + \slashed{\ell}_2 + \slashed{p}' + m_\ell) \cdot
                \gamma^\nu
        \right]
        \, ,
}
with $\Lambda_S = \mathbf{1}$ and $\Lambda_P = \gamma_5$.
The denominators of the momentum space integral \eqref{Gamma_SP} are
\begin{alignat}{6}
        \label{DM_denominators}
        &  D_1 &&= \ell_1^2 - m_\ell^2
        , \quad
        && D_2 &&= (\ell_1+q) - m_\ell^2 
        , \quad
        && D_3 &&= \ell_2^2 - m_A^2 \\
        &  D_4 &&= (\ell_2+p)^2 
        , \quad
        && D_5 &&= (\ell_2+p')^2
        , \quad
        && D_6 &&= (\ell_1+k_2+p')^2 - m_\ell^2
        \, .
        \nonumber
\end{alignat}

It is standard to write such an amplitude containing tensor structure as 
\eq{
        \mathcal{A}_{S,P}(t, m_\ell, m_A) \propto
        \sum_{\ell = e, \mu, \tau}
        \mathcal{F}_{S,P}
}
where the \emph{form factors} $\mathcal{F}_{S,P}$ are related to $\Gamma_{S,P}$ via projection:
\eq{
        \mathcal{F}_{S,P}(t, m_\ell, m_A) =
        \frac{1}{2(p' \pm p)^2}
        \mathrm{Tr}
        \left[
                \Lambda_{S,P} \cdot
                (\slashed{p} \pm m_A) \cdot
                \Gamma_{S,P} \cdot
                (\slashed{p}' \pm m_A)
        \right]
        \, .
}
The signs are $(+1)$ and $(-1)$ for the scalar ($S$) and pseudo-scalar ($P$) cases respectively.
These form factors can in turn be expanded as sums over scalar FIs:
\eq{
        \mathcal{F}_{S,P} =
        \sum_{\vec{\nu}}
        c^{\vec{\nu}}_{S,P}(t, m_\ell, m_A | \e) 
        \times
        I_{\vec{\nu}}(t, m_\ell, m_A | \e)
        , \quad
        \vec{\nu} \in \NN^7
        \, .
        \label{form_factor_definition}
}
The integral family $I_{\vec{\nu}}$ is given by
\eq{
        I_{\nu_1, \ldots, \nu_7}(t, m_\ell, m_A | \e) =
        \int \frac{\dd^\DD \ell_1}{(2\pi)^\DD} \frac{\dd^\DD \ell_2}{(2\pi)^\DD}
        \frac{D_7^{\nu_7}}{D_1^{\nu_1} \cdots D_6^{\nu_6}}
        \, ,
        \label{DM_two_mass_integral_family}
}
with denominators defined in \eqref{DM_denominators} and an ISP chosen to be
\eq{
        D_7 = (\ell_1-p)^2 
        \, .
        \label{DM_ISP}
}
The coefficients
$
        c^{\vec{\nu}}_{S,P}
$ 
are rational functions of the kinematic variables and the DR parameter $\e$,
and stem from carrying out the Dirac algebra%
\footnote{
        The prescription for treating $\gamma_5$ in $\DD \neq 4$ dimensions is taken from 
        \cite{Bernreuther:2004ih}.
}.

It turns out that $\mathcal{F}_{S}$ is a sum of $34$ integrals,
and $\mathcal{F}_{P}$ a sum of $5$ integrals.
The $34$ integrals are not all linearly independent due to IBP relations.
The goal for the rest of this chapter is to compute a linearly independent set of MIs,
which in turn leads to expressions for the form factors $\mathcal{F}_{S,P}$ via 
\eqref{form_factor_definition}.

        \section{Computing the Feynman integrals}
\label{sec:computing_the_FIs}

Although the diagrams in \figref{fig:DM_diagrams_1} are finite in $\DD = 4$ dimensions,
we treat them in DR in order to benefit from IBPs and DEQs.
The limit $\e \to 0$ in \eqref{form_factor_definition} then recovers the desired form factor results.

For the purpose of verification,
we compute MIs for both the two-mass ($m_A \neq m_\ell$) and equal-mass ($m_A = m_\ell$) cases.
The \emph{ab initio} equal-mass calculation can then be compared with the limit
$
        m_A \to m_\ell
$ 
taken on the two-mass form factors.

\subsection{Two-mass case}

\begin{figure}
        \centering
        \includegraphicsbox[scale=0.30]{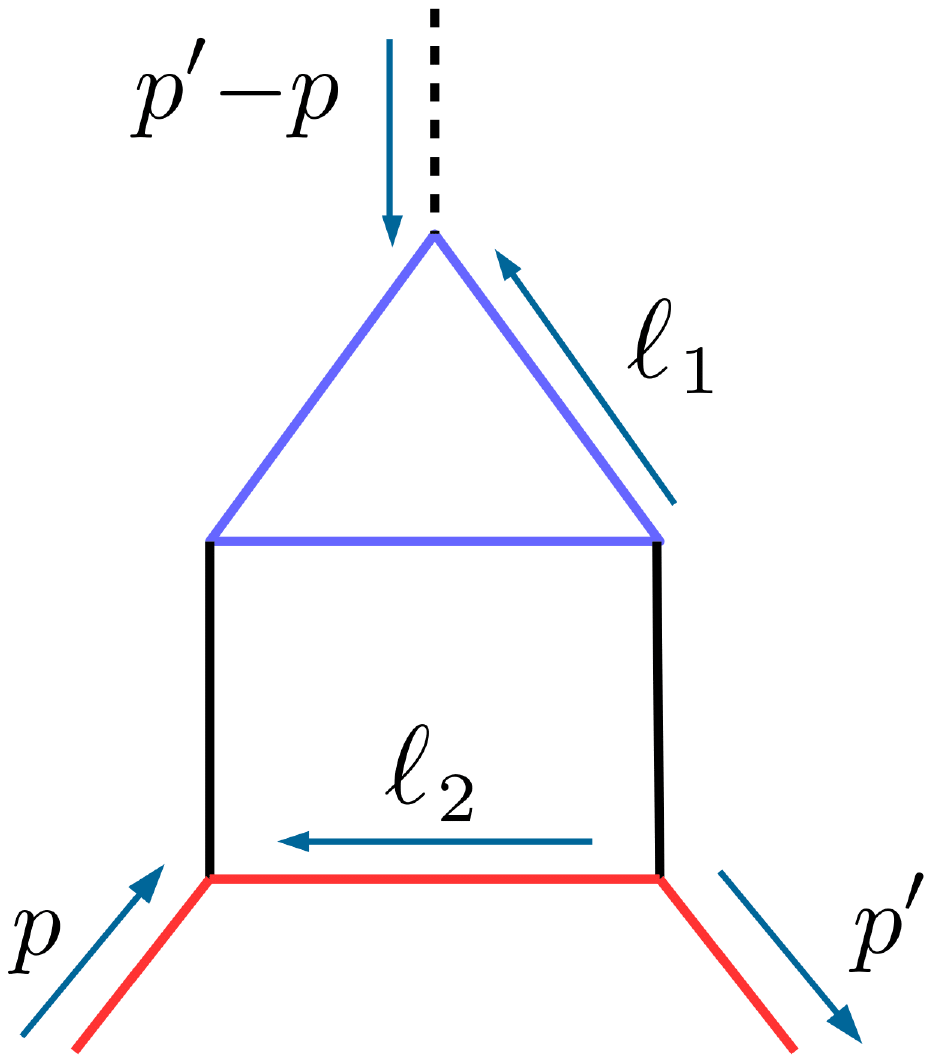}
        \caption{Two-mass Feynman diagram.}
        \label{fig:DM_diagrams_2}
\end{figure}

We begin by computing MIs for the family of FIs associated to \figref{fig:DM_diagrams_2}.
The integral family was defined in the formulas
\eqref{DM_denominators}, \eqref{DM_two_mass_integral_family} and \eqref{DM_ISP}.
As before, 
the kinematics are $p^2 = p'^2 = m_A^2$ and $t = (p-p')^2$.
MIs for this topology have already been computed in \cite{Primo:2018zby} in the context of 
$
        H \to \bar{b} \, b
$
decay;
here we re-compute them independently.

\begin{figure}[H]
        \centering
        \captionsetup[subfigure]{labelformat=empty}
        \subfloat[\hspace{0.5cm}$J_1$]{%
                \includegraphics[width=0.2\textwidth]{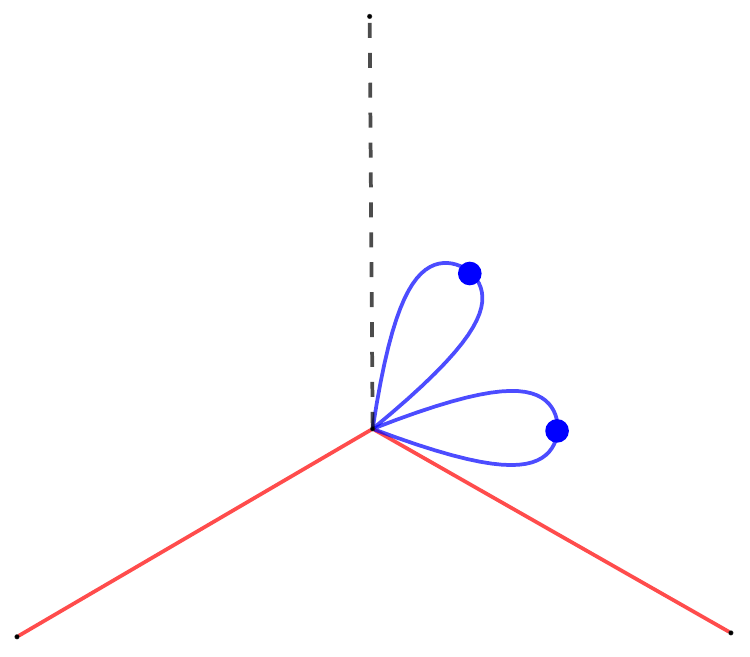}
        }
        \subfloat[\hspace{0.5cm}$J_2$]{%
                \includegraphics[width=0.2\textwidth]{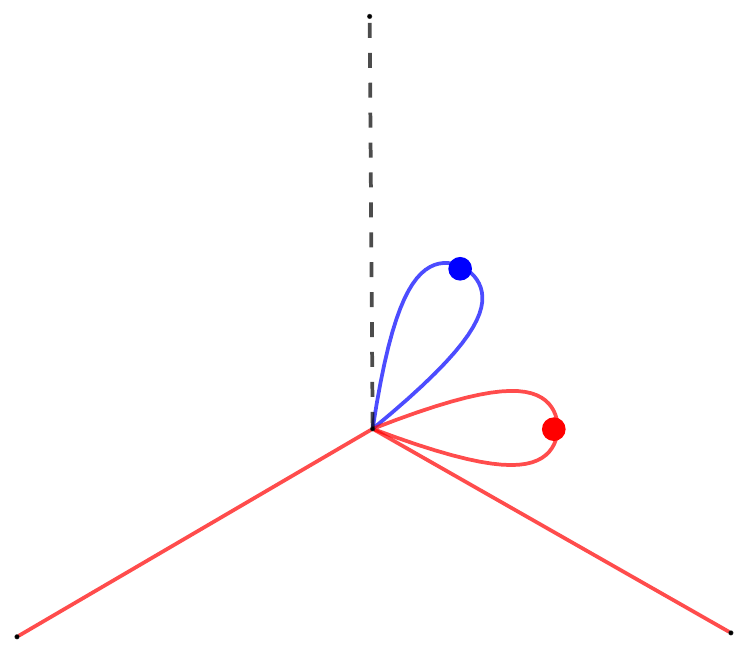}
        }
        \subfloat[\hspace{0.5cm}$J_3$]{%
                \includegraphics[width=0.2\textwidth]{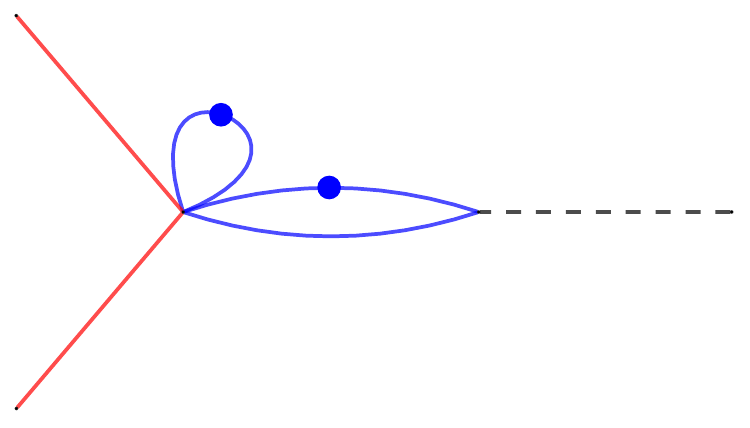}
        }
        \subfloat[\hspace{0.5cm}$J_4$]{%
                \includegraphics[width=0.2\textwidth]{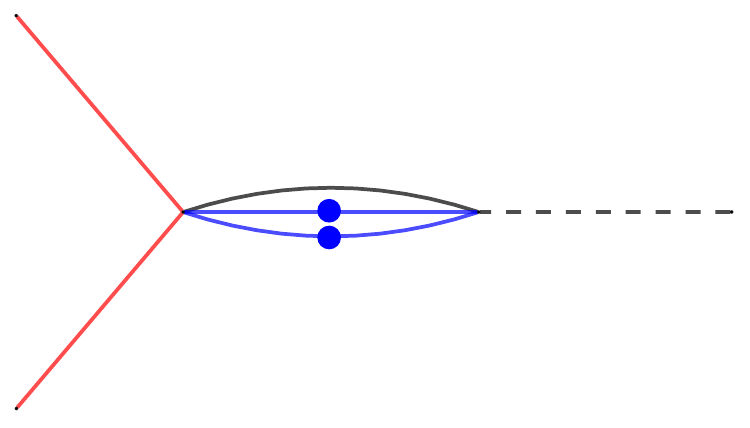}
        }\\
        \subfloat[\hspace{0.5cm}$J_5$]{%
                \includegraphics[width=0.2\textwidth]{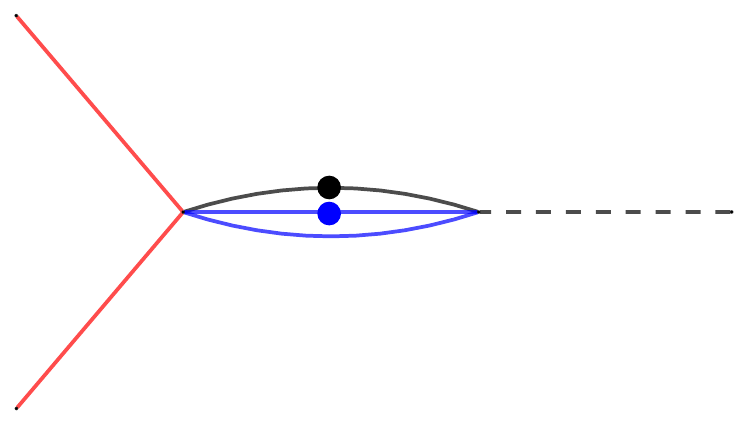}
        }
        \subfloat[\hspace{0.5cm}$J_6$]{%
                \includegraphics[width=0.2\textwidth]{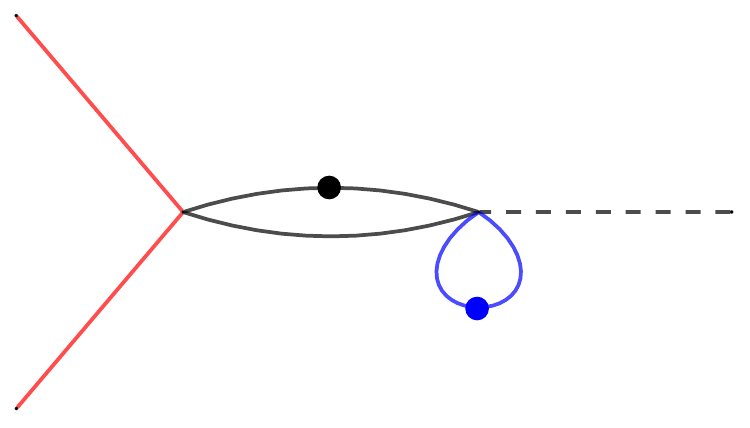}
        }
        \subfloat[\hspace{0.5cm}$J_7$]{%
                \includegraphics[width=0.2\textwidth]{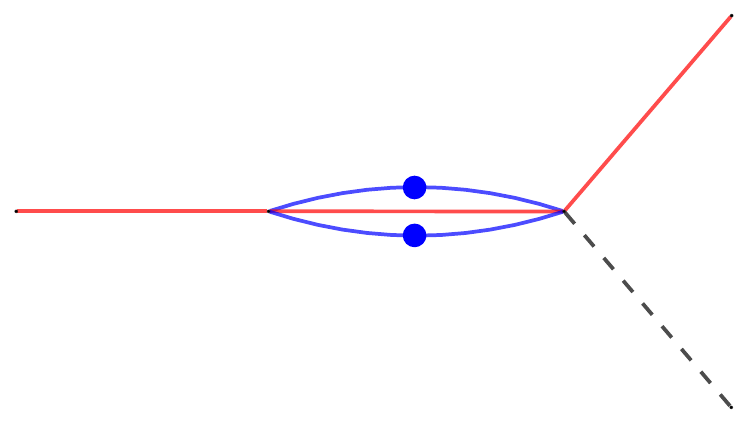}
        }
        \subfloat[\hspace{0.5cm}$J_8$]{%
                \includegraphics[width=0.2\textwidth]{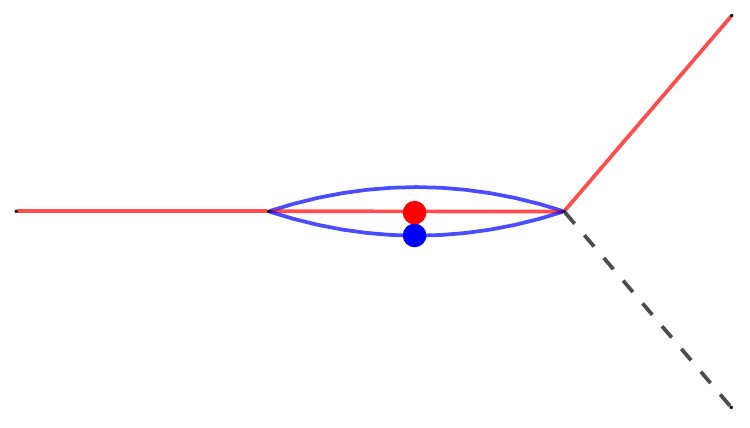}
        }\\
        \subfloat[\hspace{0.5cm}$J_9$]{%
                \includegraphics[width=0.2\textwidth]{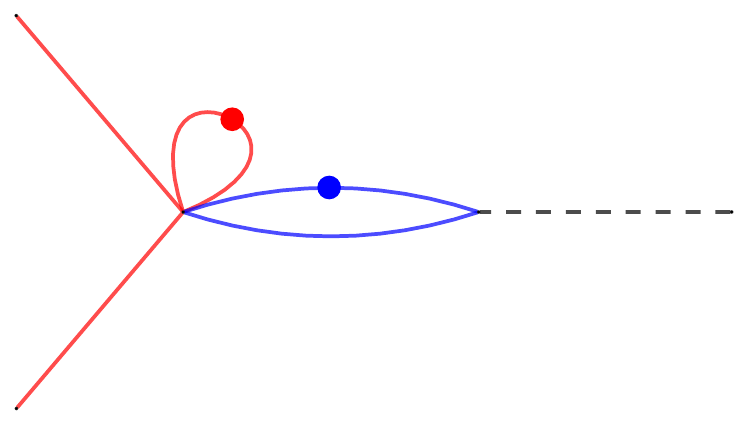}
        }
        \subfloat[\hspace{0.5cm}$J_{10}$]{%
                \includegraphics[width=0.2\textwidth]{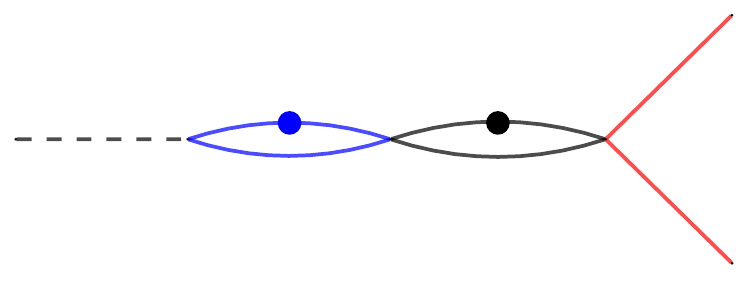}
        }
        \subfloat[\hspace{0.5cm}$J_{11}$]{%
                \includegraphics[width=0.2\textwidth]{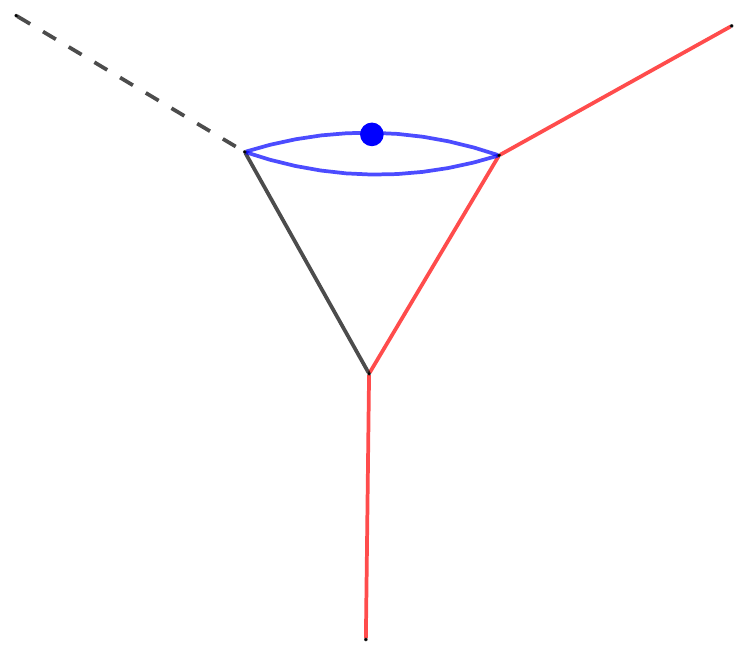}
        }
        \subfloat[\hspace{0.5cm}$J_{12}$]{%
                \includegraphics[width=0.2\textwidth]{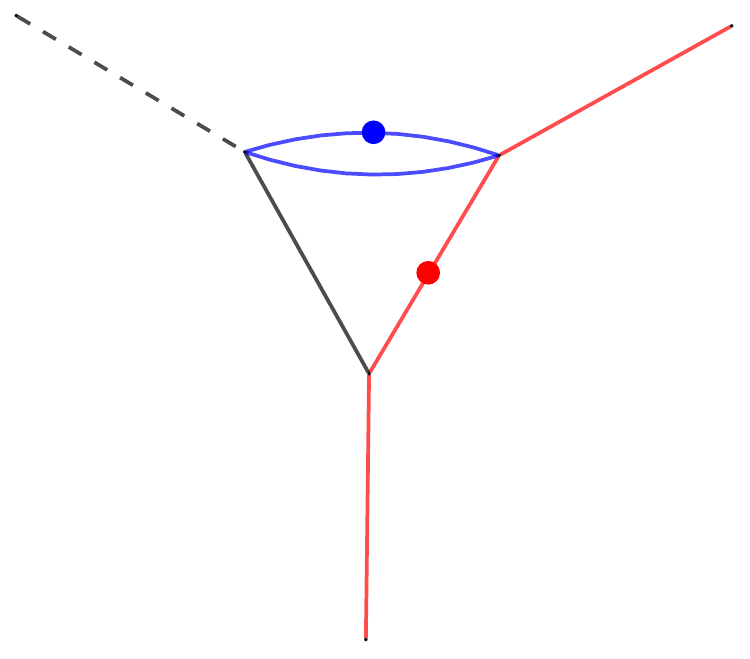}
        }
        \\
        \subfloat[\hspace{0.5cm}$J_{13}$]{%
                \includegraphics[width=0.2\textwidth]{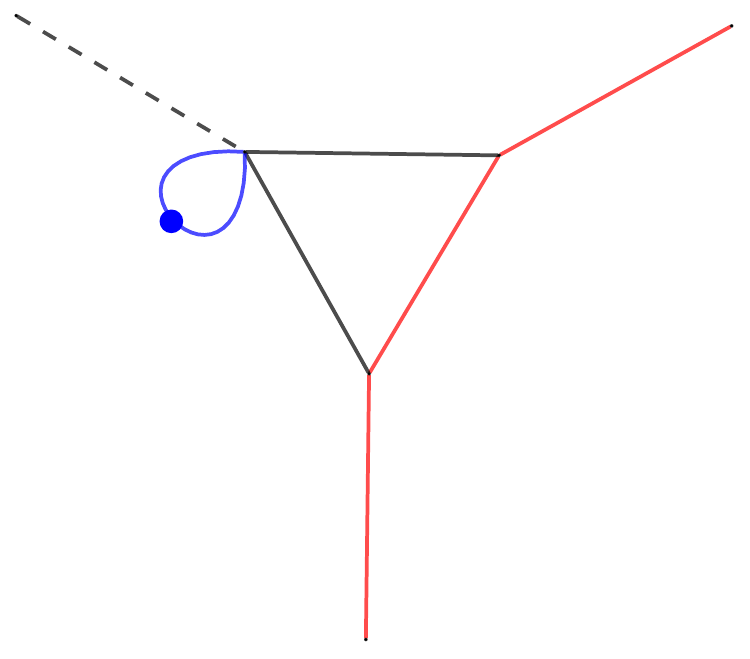}
        }
        \subfloat[\hspace{0.5cm}$J_{14}$]{%
                \includegraphics[width=0.2\textwidth]{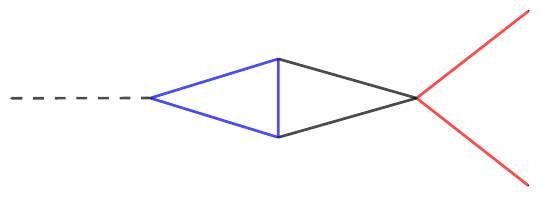}
        }
        \subfloat[\hspace{0.5cm}$J_{15}$]{%
                \includegraphics[width=0.2\textwidth]{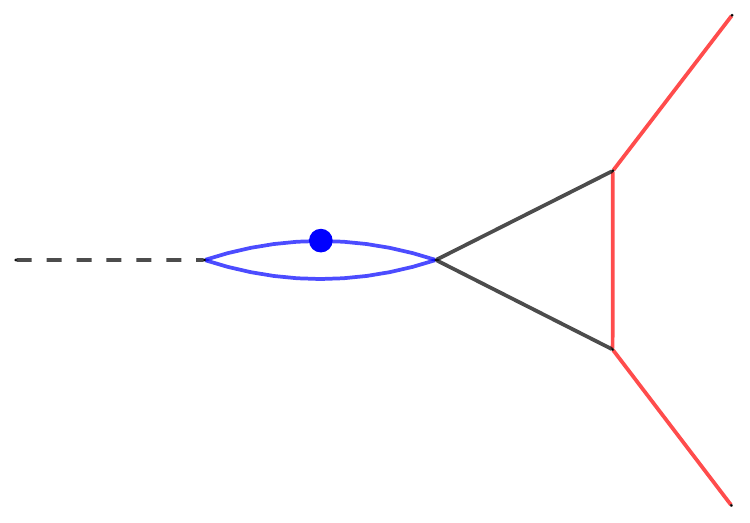}
        }
        \subfloat[\hspace{0.5cm}$J_{16}$]{%
                \includegraphics[width=0.2\textwidth]{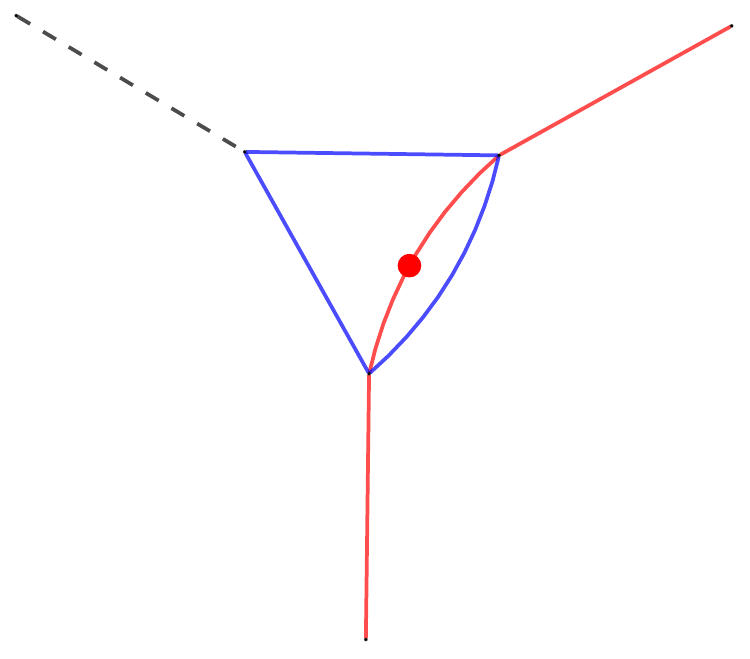}
        }\\
        \subfloat[\hspace{0.5cm}$J_{17}$]{%
                \includegraphics[width=0.2\textwidth]{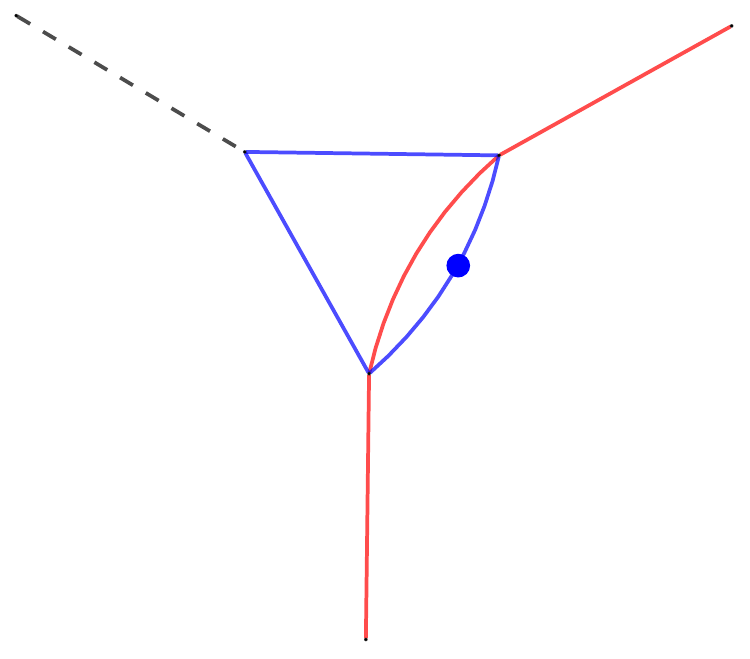}
        }
        \subfloat[\hspace{0.5cm}$J_{18}$]{%
                \includegraphics[width=0.2\textwidth]{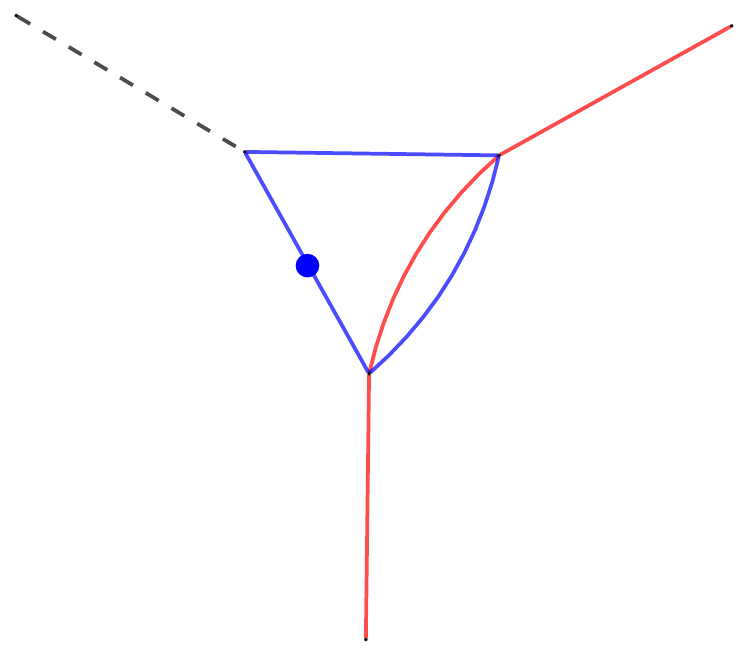}
        }
        \subfloat[\hspace{0.5cm}$J_{19}$]{%
                \includegraphics[width=0.2\textwidth]{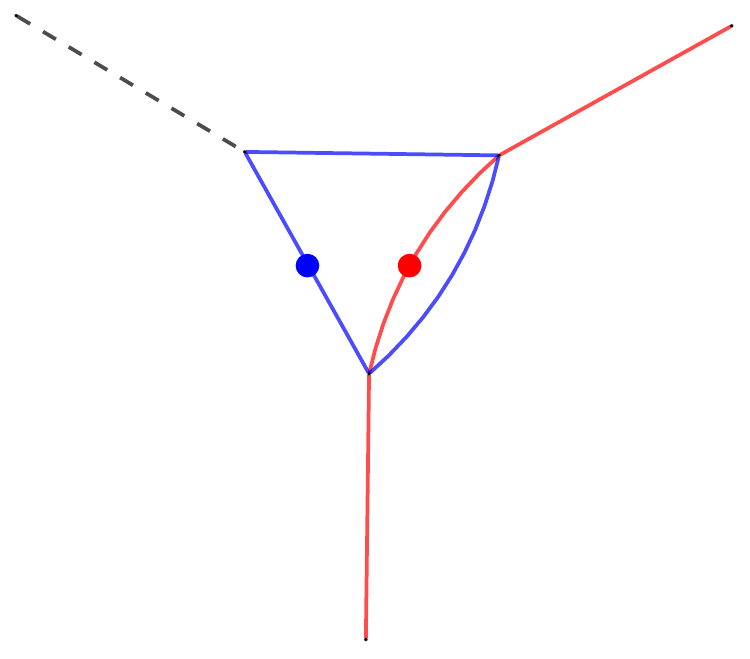}
        }
        \subfloat[\hspace{0.5cm}$J_{20}$]{%
                \includegraphics[width=0.2\textwidth]{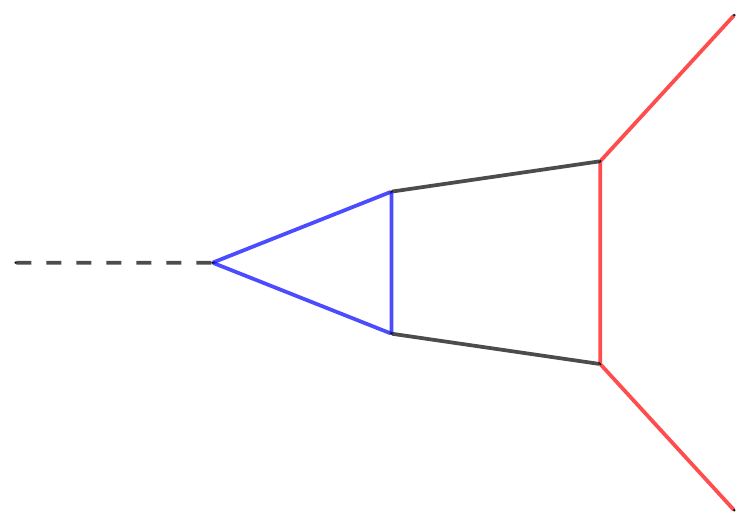}
        }
        \caption{
                The MIs $J_i$ appearing in the canonical basis 
                \eqref{DM_two_mass_master_integrals}.
                Squared propagators are represented by dots.
        }
        \label{fig:DM_two_mass_master_integrals}
\end{figure}

IBP reduction and differentiation of FIs is performed with the codes
\package{Reduze2} \cite{vonManteuffel:2012np} and \package{LiteRed} \cite{Lee:2013mka}.
Using the Magnus expansion from \secref{sec:magnus},
the following set of canonical MIs are identified:

\eq{
        \begin{array}{lllllll}
                & I_1 &=&
                \e^2 J_1
                & I_2 &=&
                \e^2 J_2
                \\ 
                & I_3 &=&
                \e^2 \lambda_\ell J_3
                & I_4 &=&
                - \e^2 t J_4
                \\ 
                & I_5 &=&
                \e^2
                \left[
                        \frac{\lambda_\ell - t}{2} J_4 + \lambda_\ell J_4
                \right]
                & I_6 &=&
                -\e^2 t J_6
                \\ 
                & I_7 &=&
                \frac{\e^2 m_A^2 \rho_A (t + \lambda_\ell)}{\rho_\ell (t + \lambda_A)}
                [J_7 + 2 J_8]
                & I_8 &=&
                \e^2 m_A^2 J_8
                \\ 
                & I_9 &=&
                \e^2 \lambda_\ell J_9
                & I_{10} &=&
                -\e^2 t \lambda_\ell J_{10}
                \\ 
                & I_{11} &=&
                \e^3 \lambda_A J_{11}
                & I_{12} &=&
                \frac{\e^2 \lambda_\ell}{4t}
                \left[
                        (t - \lambda_A) (J_4 + 2 J_5) -
                        4 m_A^2 \lambda_A J_{12}
                \right]
                \\ 
                & I_{13} &=&
                \e^3 \lambda_A J_{13}
                & I_{14} &=&
                \e^3(2\e - 1) t J_{14}
                \\ 
                & I_{15} &=&
                \e^3 \lambda_\ell \lambda_A J_{15}
                & I_{16} &=&
                \e^3 \lambda_A J_{16}
                \\ 
                & I_{17} &=&
                \e^3 \lambda_A J_{17}
                & I_{18} &=& \eqref{two_mass_MI_I18}
                \\ 
                & I_{19} &=& \eqref{two_mass_MI_I19}
                & I_{20} &=& 
                - \e^4 t \lambda_A J_{20}
                \, ,
        \end{array}
        \label{DM_two_mass_master_integrals}
        \raisetag{6.5\baselineskip}
}
where
\eq{
        \label{two_mass_MI_I18}
        I_{18} &=
        \frac{\e^2}{t}
        \left[
                \lambda_\ell (\lambda_A - t) J_9 +
                \e \lambda_A (t - \lambda_\ell) J_{17} +
                (2\e - 1) \lambda_\ell \lambda_A J_{18}
        \right]
}
and
\eq{
        \label{two_mass_MI_I19}
        I_{19} &=
        \frac{\e^2}{2t}
        \Big[
                t (\lambda_\ell - t) J_3 -
                2 t m_\ell^2 (J_7 + 2 J_8) +
                \big( 4 t m_\ell^2 + \lambda_\ell (\lambda_A - t) \big) J_9 +
                \frac{4 t^2 m_A^2}{\lambda_A + t} J_{16} 
                \\ \nonumber & \quad +
                \e \big( \lambda_A (t - \lambda_\ell) 4 t m_\ell^2 \big) J_{17} +
                (2\e - 1) (4 t m_\ell^2 + \lambda_\ell \lambda_A - t^2) J_{18} + 
                2 t^2 (m_\ell^2 - m_A^2) J_{19}
        \Big]
        \, .
}
The integrals $J_i$ are graphically represented in \figref{fig:DM_two_mass_master_integrals}.

\noindent
The canonical integrals contain square root prefactors $\lambda_i$ and $\rho_i$ defined by
\eq{
        \lambda_i = 
        \sqrt{-t} \sqrt{4 m_i^2 - t}
        \, , \quad
        \rho_i =
        \sqrt{ \frac{2 m_i^2 - t - \lambda_i}{m_i^2} }
        \, , \quad 
        i = \ell, A
        \, .
        \label{DM_square_roots}
}
These roots are simultaneously rationalized by the change of variables
\eq{
        t = 
        - m_\ell^2 \frac{(1-x)^2}{x^2}
        \, , \quad
        m_A^2 = m_\ell^2 \frac{(1-x^2)^2 y^2}{(1-y^2)^2 x^2}
        \label{DM_xy_variables}
        \, ,
}
with a corresponding inverse map 
\eq{
        x = \frac{\sqrt{4-\sigma_\ell} - \sqrt{-\sigma_\ell}}{2}
        \, , \quad
        y = \frac{\sqrt{4-\sigma_A} - \sqrt{-\sigma_A}}{2}
        \, , \quad
        \sigma_i = \frac{t}{m_i^2}
        \, .
}
The Euclidean kinematic regime%
\footnote{
        It is standard to solve Pfaffian systems for Euclidean kinematics 
        and then analytically continue the resulting GPLs to physical kinematic regimes.
}
\eq{
        \big\{ t < 0 \big\}
        \quad \bigcap \quad
        \big\{
                0 < m_A^2 < m_\ell^2
                \quadit{\text{or}}
                0 < m_\ell^2 < m_A^2
        \big\}
}
is parametrized by
\eq{
        \big\{ 0 < x < 1 \big\}
        \quad \bigcap \quad
        \big\{ 
                0 < y < x 
                \quadit{\text{or}}
                0 < x < y 
        \big\}
}
in the $(x,y)$ coordinates.
        
The basis \eqref{DM_two_mass_master_integrals} satisfies a canonical Pfaffian system
\eq{
        \dd \vec{I}(x,y | \e) =
        \e \, P(x,y) \cdot \vec{I}(x,y | \e)
        \, , \quad
        P(x,y) = 
        \sum_{i=1}^{12} P_i \ \dd \log \eta_i(x,y)
        \, , \quad
        P_i \in \mathbb{Q}^{20 \times 20}
}
with letters
\begin{alignat}{6}
        \nonumber
        &\eta_1     && = x
        \, , \quad
        &&\eta_2    &&= 1+x
        \, , \quad
        &&\eta_3    &&= 1-x
        \, ,
        \\
        &\eta_4     &&= 1+x^2
        \, , \quad
        &&\eta_5    &&= y
        \, , \quad
        &&\eta_6    &&= 1+y
        \, ,
        \\ 
        \nonumber
        &\eta_7     &&= 1-y
        \, , \quad
        &&\eta_8    &&= 1+y^2
        \, , \quad
        &&\eta_9    &&= x+y
        \, ,
        \\ 
        &\eta_{10}  &&= x-y
        \, , \quad
        &&\eta_{11} &&= 1+xy
        \, , \quad
        &&\eta_{12} &&= 1-xy
        \, .
        \nonumber
\end{alignat}
The general solution to this Pfaffian system is given by the path-ordered exponential 
\eqref{path_ordered_exponential}.
The result takes the form
$
        \vec{I}(x,y|\e) = 
        \left[ \sum_{i=0}^4 M_i(x,y) \, \e^i + O(\e^5) \right] \cdot \vec{I}_0(\e)
        \, ,
$
where the matrices $M_i(x,y)$ contain GPLs built from the letters $\eta_i$,
and $\vec{I}_0(\e)$ is a vector of boundary constants.
The latter is fixed as follows:
\begin{itemize}
        \item 
                By normalizing the integration measure as
                \eq{
                        \left( \frac{m_\ell^2}{\mu^2} \right)^{\e}
                        \int \frac{\dd^\DD \ell_i}{i \pi^{\DD/2} \Gamma(1+\e)}
                        \, , \quad
                        i = 1,2
                        \, ,
                }
                then direct integration (via Feynman parametrization) of the tadpoles $\{I_1, \, I_2\}$ 
                as well as the factorized integral $I_6$ gives
                \eq{
                        \nonumber
                        I_1 &= 1
                        \\
                        I_2 &=
                        \left(
                                \frac{(1-y^2)x^2}{(1-x^2)y^2}
                        \right)^\e
                        \\
                        I_6 &=
                        \left( \frac{x^2}{(1-x^2)^2} \right)^\e
                        \left[
                                1 - \zeta_2 \e^2 - 2 \zeta_3 \e^3 - \frac{9}{4} \zeta_4 \e^4 +
                                O(\e^5)
                        \right]
                        \, ,
                        \nonumber
                }
                where $\zeta_n$ is the $n$th Riemann zeta value.
                These external inputs are fed into the DEQ to give 
                relations among the remaining boundary constants.
        \item 
                Boundary constants for the set of integrals
                $
                        \{
                                I_3, I_4, I_5, I_9, I_{10}, I_{11}, I_{12}, 
                                I_{14}, I_{16}, I_{17}, I_{18}, \allowbreak I_{19}
                        \}
                $       
                are fixed by requiring regularity at the pseudo-threshold limit $t \to 0$.
                In particular for
                $
                        \{
                                I_3, I_4, I_5, I_9, I_{10}, I_{14}, I_{16}, I_{17}
                        \},
                $
                the prefactors in 
                \eqref{DM_two_mass_master_integrals}
                make them vanish in this limit%
                \footnote{
                        The vanishing of $I_{14}$ as $t \to 0$ follows from the results of 
                        \cite{Anastasiou:2006hc}.
                }.
                Further,
                $
                        \{
                                I_{11}, I_{12}, I_{18}, I_{19}
                        \}
                $
                are seen to vanish as $t \to 0$ by analyzing their DEQs.
                Finally,
                the factorized double-bubble integral $I_{10}$ vanishes in the limit because the 
                \emph{massive} bubble factor 
                (drawn in {\color{blue} blue} in \figref{fig:DM_two_mass_master_integrals})
                goes to $0$ at a faster rate than the divergence of the \emph{massless} bubble factor
                (drawn in black).
        \item 
                Boundary constants for
                $
                        \{I_7, I_8\}
                $
                are determined by regularity at $m_A^2 \to 0$.
                Due to the prefactors in \eqref{DM_two_mass_master_integrals},
                both integrals vanish in this limit.

        \item 
                Regularity at the limit 
                $
                        t \to 4 m_A^2
                $
                fixes boundary constants for the final three integrals
                $
                        \{ I_{13}, I_{15}, I_{20} \} .
                $
                Thanks to the prefactors of \eqref{DM_two_mass_master_integrals},
                they all vanish in this limit.
        \item 
                In some of the cases listed above,
                the equations for the boundary constants only led to solutions at very high precision,
                as opposed to analytic expressions in terms of transcendental constants.
                Fortunately, 
                using the implementation of the 
                \soft{PSLQ} algorithm \cite{PSLQ} in \package{PolyLogTools} \cite{Duhr:2019tlz},
                we can reconstruct the analytic constants by fitting to an ansatz.
\end{itemize}

\subsection{Equal-mass case}

\begin{figure}[H]
        \centering
        \includegraphics[scale=0.3]{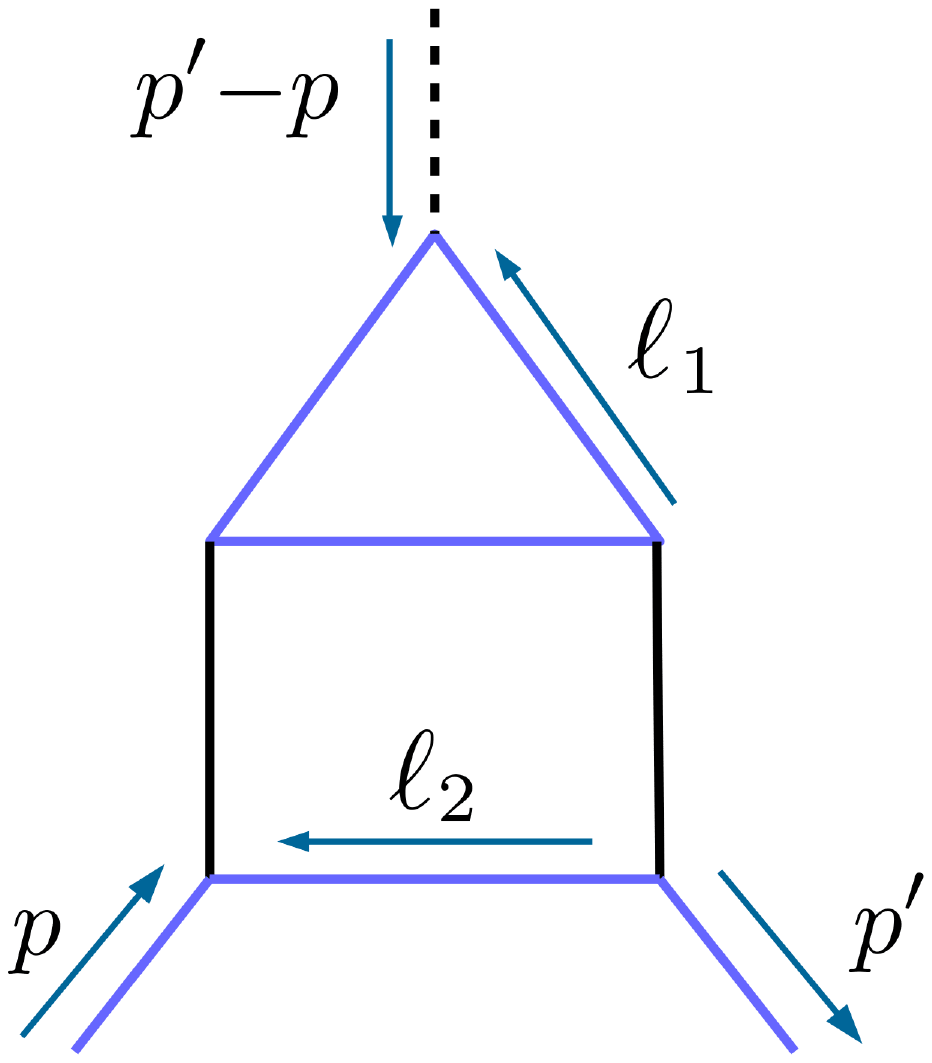}
        \caption{Equal-mass Feynman diagram.}
        \label{fig:DM_diagrams_3}
\end{figure}

The equal-mass topology is shown in \figref{fig:DM_diagrams_3}.
Its associated FI family is given by \eqref{DM_two_mass_integral_family} 
upon setting 
$
        m_A^2 = m_\ell^2 = m^2
$
in
\eqref{DM_denominators} and \eqref{DM_ISP}.

\begin{figure}[H]
        \centering
        \captionsetup[subfigure]{labelformat=empty}
        \subfloat[\hspace{0.5cm}$J_1$]{%
                \includegraphics[width=0.2\textwidth]{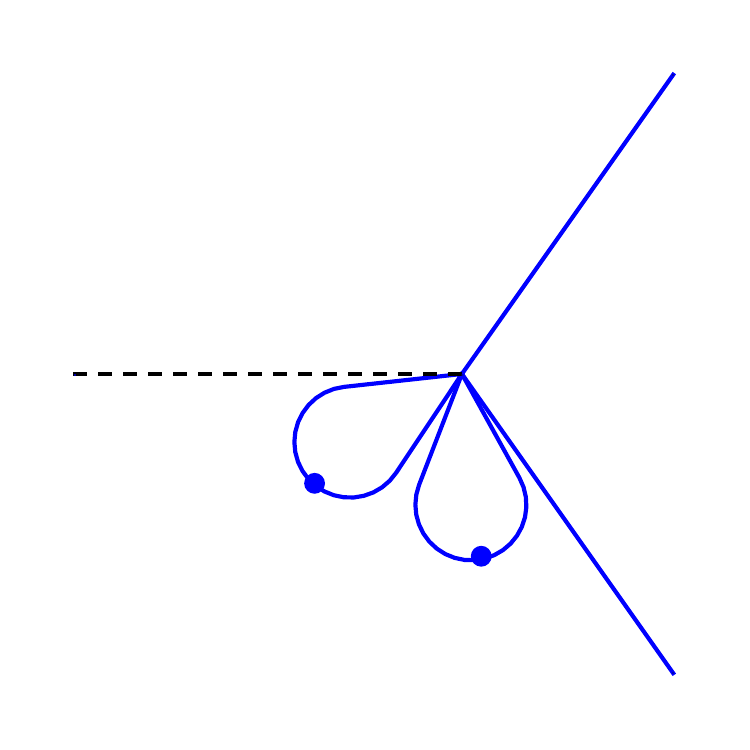}
        }
        \subfloat[\hspace{0.5cm}$J_2$]{%
                \includegraphics[width=0.2\textwidth]{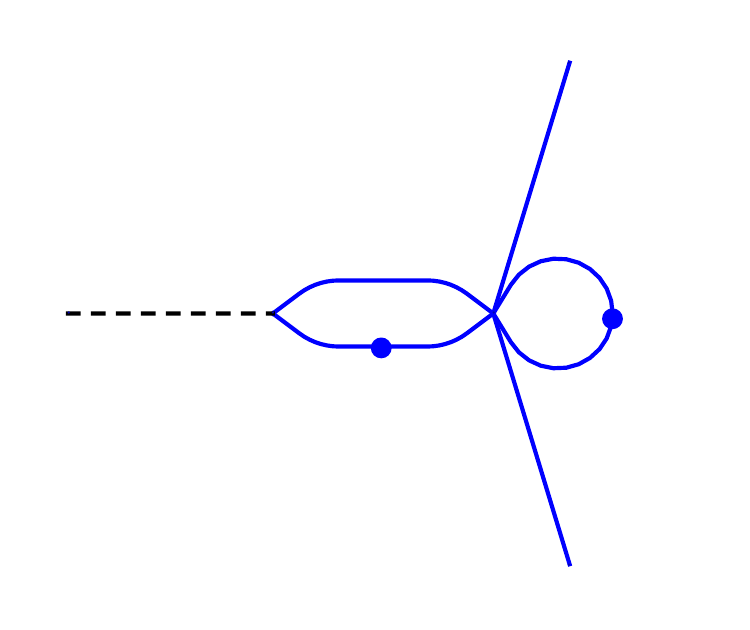}
        }
        \subfloat[\hspace{0.5cm}$J_3$]{%
                \includegraphics[width=0.2\textwidth]{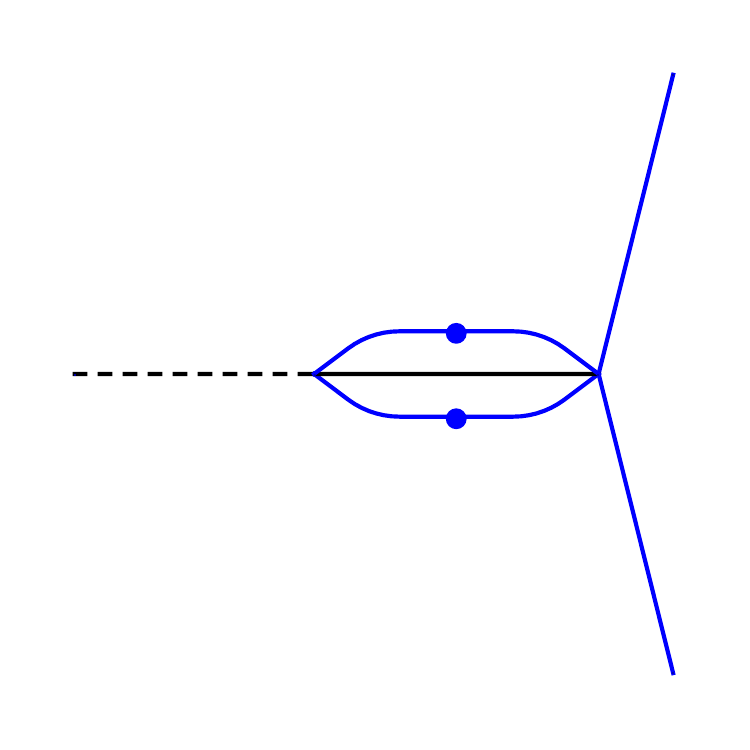}
        }
        \subfloat[\hspace{0.5cm}$J_4$]{%
                \includegraphics[width=0.2\textwidth]{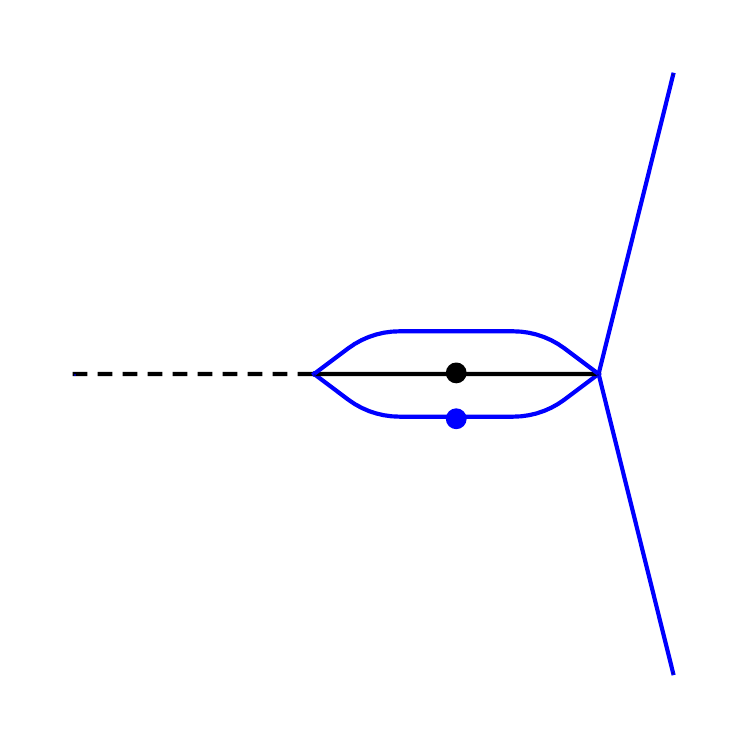}
        }\\
        \subfloat[\hspace{0.5cm}$J_5$]{%
                \includegraphics[width=0.2\textwidth]{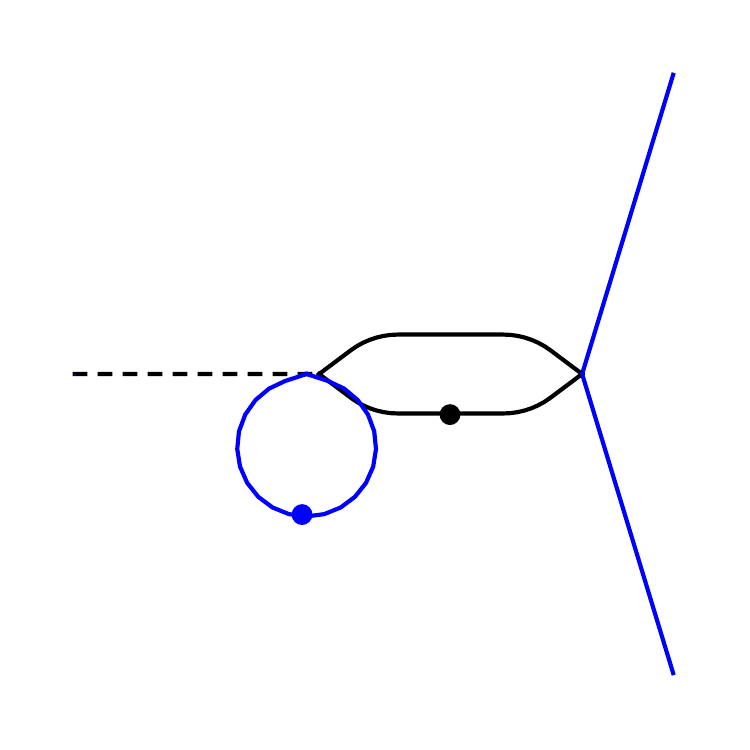}
        }
        \subfloat[\hspace{0.5cm}$J_6$]{%
                \includegraphics[width=0.2\textwidth]{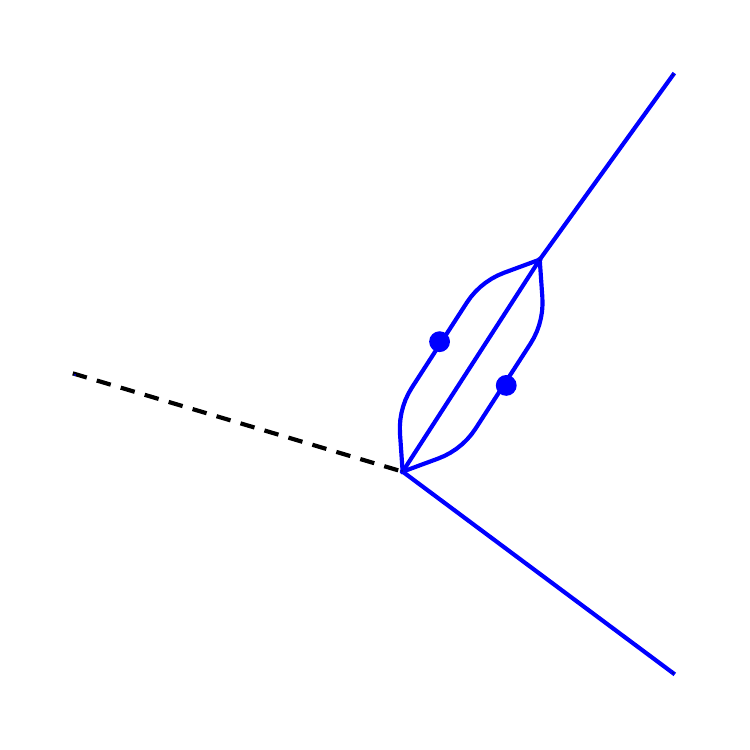}
        }
        \subfloat[\hspace{0.5cm}$J_7$]{%
                \includegraphics[width=0.2\textwidth]{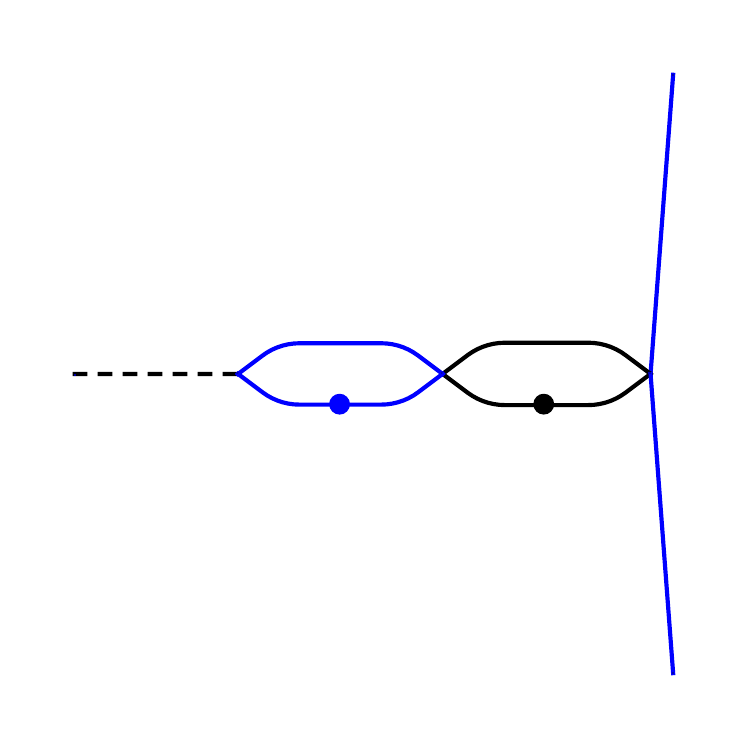}
        }
        \subfloat[\hspace{0.5cm}$J_8$]{%
                \includegraphics[width=0.2\textwidth]{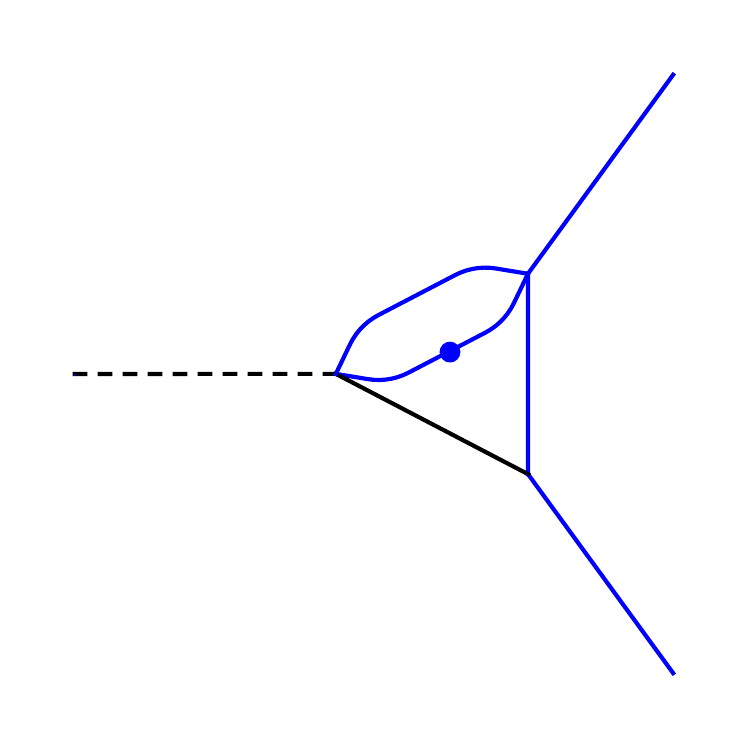}
        }\\
        \subfloat[\hspace{0.5cm}$J_9$]{%
                \includegraphics[width=0.2\textwidth]{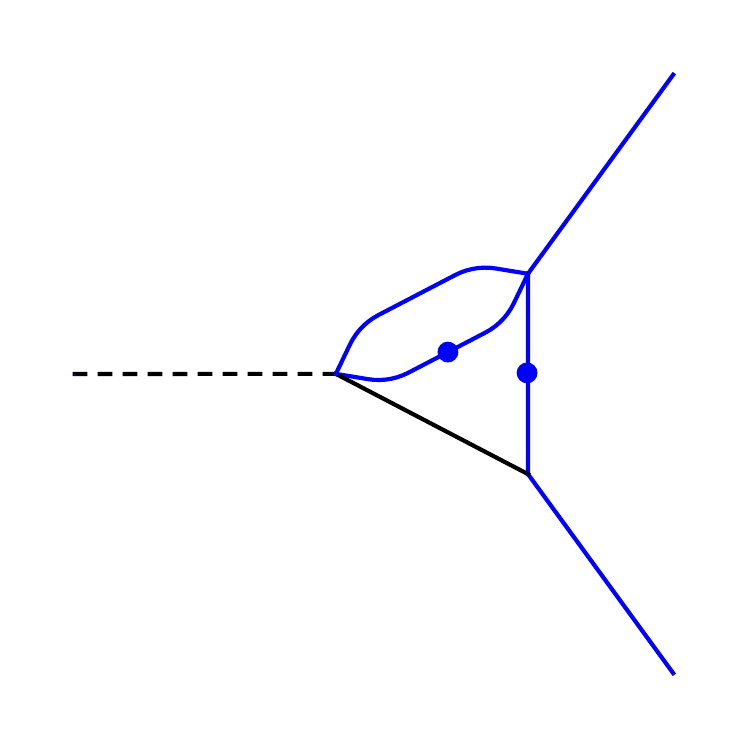}
        }
        \subfloat[\hspace{0.5cm}$J_{10}$]{%
                \includegraphics[width=0.2\textwidth]{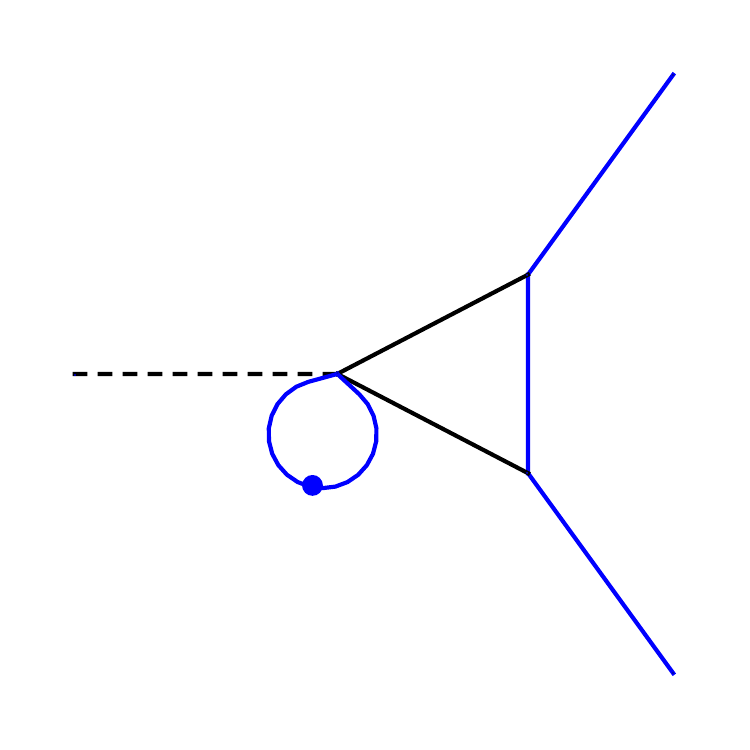}
        }
        \subfloat[\hspace{0.5cm}$J_{11}$]{%
                \includegraphics[width=0.2\textwidth]{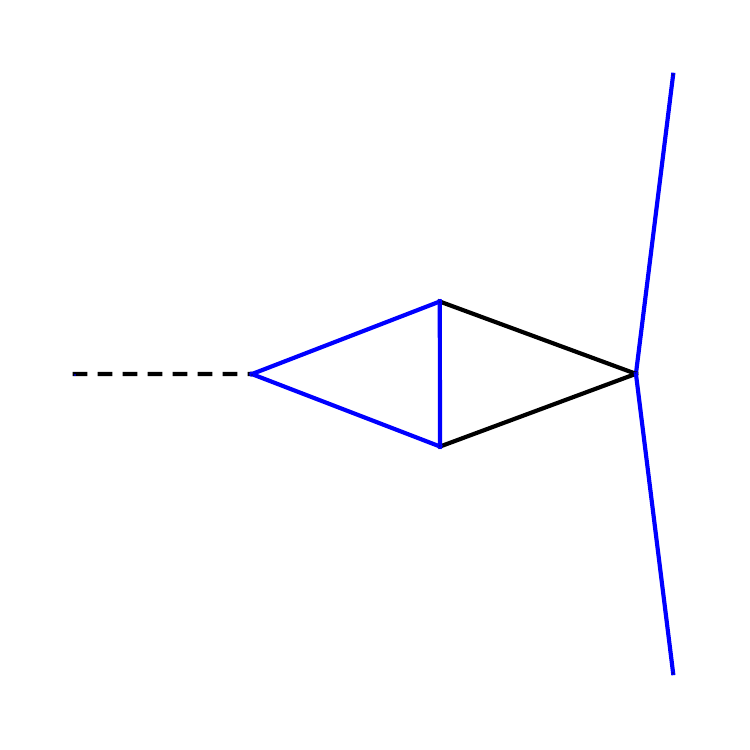}
        }
        \subfloat[\hspace{0.5cm}$J_{12}$]{%
                \includegraphics[width=0.2\textwidth]{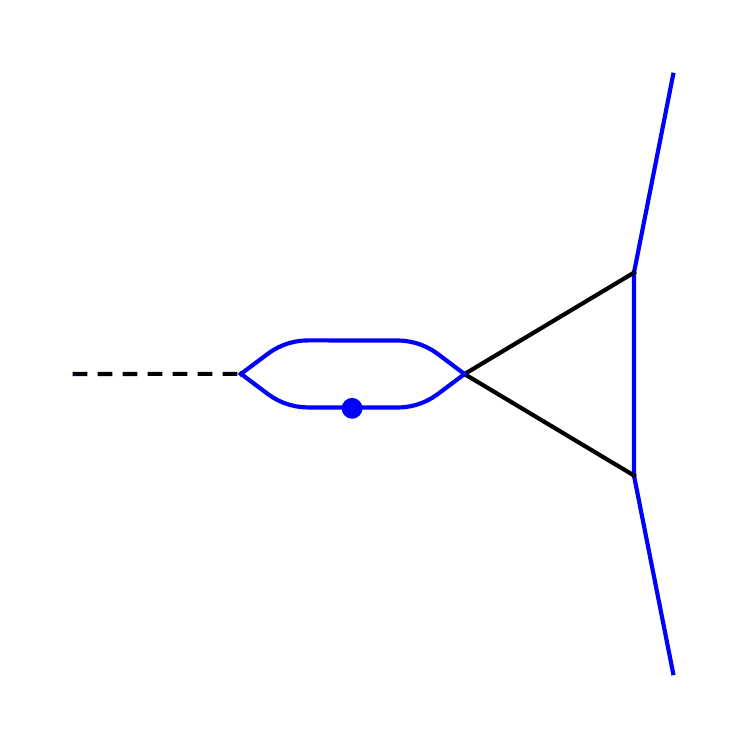}
        }
        \\
        \subfloat[\hspace{0.5cm}$J_{13}$]{%
                \includegraphics[width=0.2\textwidth]{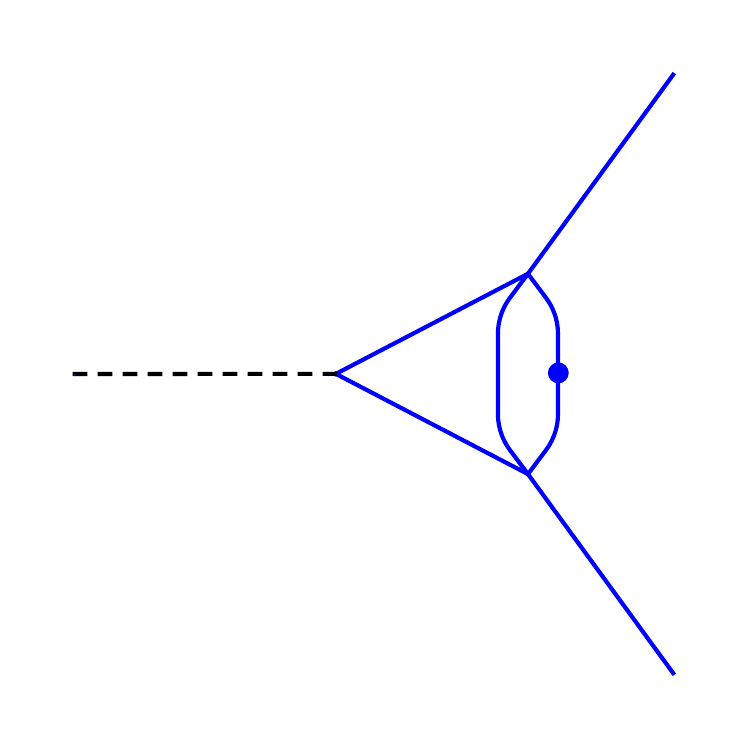}
        }
        \subfloat[\hspace{0.5cm}$J_{14}$]{%
                \includegraphics[width=0.2\textwidth]{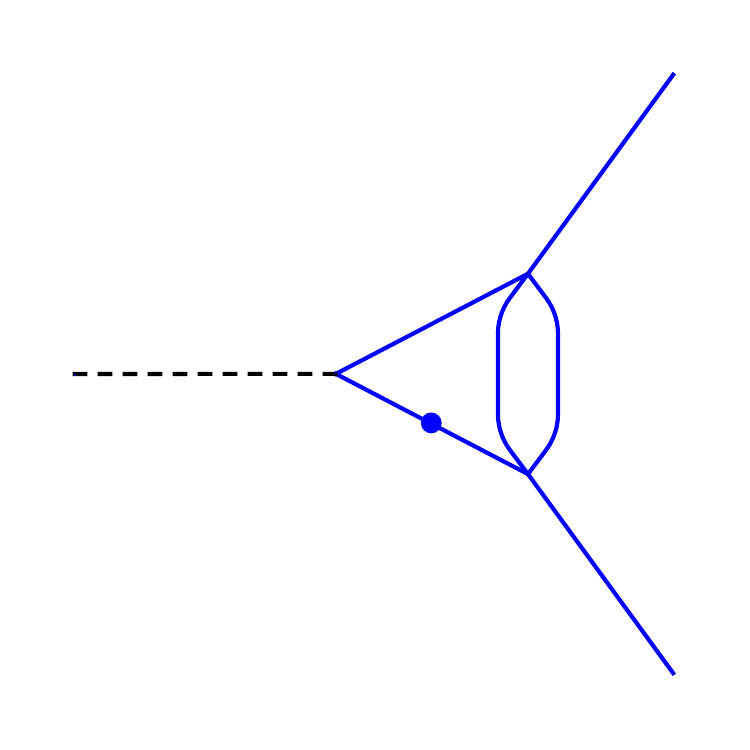}
        }
        \subfloat[\hspace{0.5cm}$J_{15}$]{%
                \includegraphics[width=0.2\textwidth]{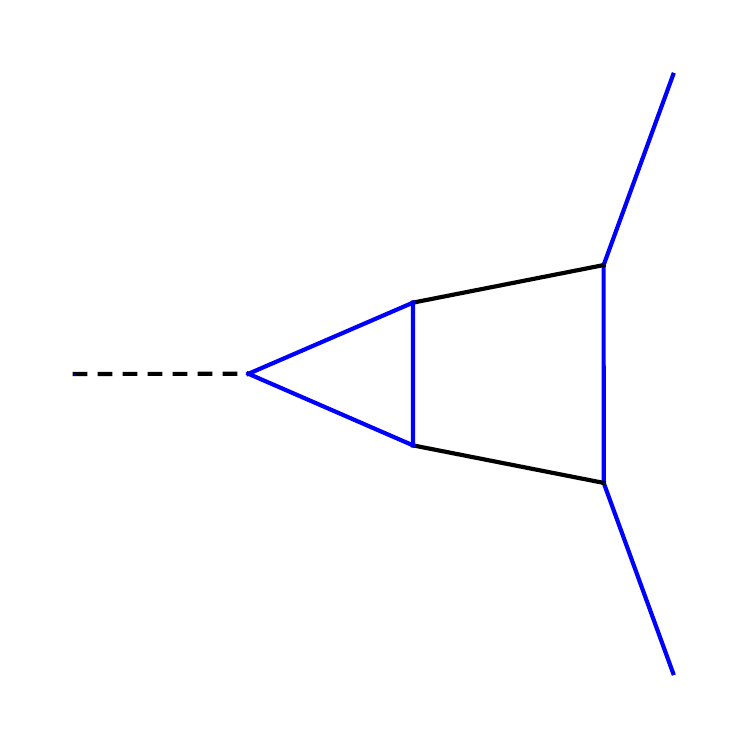}
        }
        \caption{
                The MIs $J_i$ appearing in the canonical basis 
                \eqref{DM_equal_mass_master_integrals}.
                Squared propagators are represented by dots.
        }
        \label{fig:DM_equal_mass_master_integrals}
\end{figure}

\noindent
IBP reduction reveals 15 MIs.
The Magnus expansion then yields the canonical basis
\eq{
        \begin{array}{lllllll}
                & I_1 &=&
                \e^2 J_1
                & I_2 &=&
                \e^2 \lambda_m J_2
                \\ 
                & I_3 &=&
                -\e^2 t J_3
                & I_4 &=&
                \e^2
                \left[
                        \frac{\lambda_m + t}{2} J_3 + \lambda_m J_4 
                \right]
                \\ 
                & I_5 &=&
                -\e^2 t J_5
                & I_6 &=&
                \e^2 m^2 J_6
                \\ 
                & I_7 &=&
                -\e^2 t \lambda_m J_7
                & I_8 &=&
                \e^3 \lambda_m J_8
                \\ 
                & I_9 &=&
                \e^2
                \left[
                        \frac{4m^2 - \lambda_m - t}{4}(J_3 + 2 J_4) +\
                        m^2 (4m^2 - t) J_9
                \right]
                & I_{10} &=&
                \e^3 \lambda_m J_{10}
                \\ 
                & I_{11} &=&
                \e^3(1 - 2\e) t J_{11}
                & I_{12} &=&
                \e^3 t (t - 4m^2) J_{12}
                \\
                & I_{13} &=&
                \e^3 \lambda_m J_{13}
                & I_{14} &=& \eqref{dm_equal_mass_MI_I14}
                \\
                & I_{15} &=&
                -\e^4 \lambda_m t J_{15}
                \, ,
        \end{array}
        \label{DM_equal_mass_master_integrals}
        \raisetag{4.3\baselineskip}
}
with
\eq{
        \label{dm_equal_mass_MI_I14}
        I_{14} =
        \e^2
        \left[
                (4m^2 - \lambda_m - t)(J_2 - \e J_{13}) + 
                (2\e - 1)(4m^2 - t) J_{14}
        \right]
        \, .
}
Diagrams for the integrals $J_i$ can be seen in \figref{fig:DM_equal_mass_master_integrals}.

\noindent
The canonical integrals contain an algebraic function $\l_m$ given by
\eq{
        \lambda_m =
        \sqrt{-t} \sqrt{4m^2 - t}
        \, .
        \label{DM_equal_mass_square_root}
}
Both of these roots are simultaneously rationalized by the coordinate change
\eq{
        t = - m^2 \frac{(1-w)^2}{w}
        \, ,
}
whose inverse map is given by
\eq{
        w = 
        \frac
        {\sqrt{4m^2 - t} - \sqrt{-t}}
        {\sqrt{4m^2 - t} + \sqrt{-t}}
        \, .
}
The Euclidean kinematic regime
\eq{
        \big\{ t < 0 \big\}
        \quad \bigcap \quad
        \big\{ m^2 > 0 \big\}
}
is then given by
\eq{
        0 < w < 1
        \, .
}

The canonical Pfaffian system for the basis \eqref{DM_equal_mass_master_integrals} takes the form
\eq{
        \dd \vec{I}(w, \e) =
        \e \, P(w) \cdot \vec{I}(w,\e)
        \, , \quad
        P(w) = \sum_{i=1}^3 P_i \, \dd \log \eta_i(w)
        \, , \quad
        P_i \in \mathbb{Q}^{15 \times 15}
        \, ,
}
where the three letters are
\eq{
        \eta_1 = w
        \, , \quad
        \eta_2 = 1+w
        \, , \quad
        \eta_3 = 1-w
        \, .
}
This means that the Pfaffian system has a solution in terms 
\emph{harmonic polylogarithms} (HPLs) denoted by $H(\vec{z}|w)$,
i.e.~GPLs with weights $z_i \in \{-1,0,1\}$ \cite{Remiddi:1999ew}.

Boundary constants are determined as follows:
\begin{itemize}
        \item 
                The integrals
                $
                        \{I_1, I_5, I_6\}
                $
                can be directly integrated,
                and are provided as external input.
                The first integral is normalized to unity by the measure,
                and the second integral is swiftly computed with Feynman parameters,
                giving
                \eq{
                        I_1 &= 1
                        \\
                        I_5 &= 
                        \left( \frac{w}{(1-w)^2} \right)^\e
                        \left[
                                1 - \zeta_2 \e^2 - 2 \zeta_3 \e^3 - \frac{9}{4} \zeta_4 \e^4 + 
                                O(\e^5)
                        \right]
                        \, .
                }
                The equal-mass sunrise integral $I_6$ is computed by following the strategy of 
                \cite{Argeri:2002wz}.
                Since the square of its external momentum equals its internal masses,
                $p^2 = m^2$,
                this integral evaluates to a number rather than a function,
                wherefore it cannot immediately be computed from a Pfaffian system.
                The trick is to change two of the propagator masses to $M^2$ rather than $m^2$,
                and then solve a Pfaffian system in the auxiliary variable $x = m^2/M^2$.
                Boundary constants are fixed by regularity at $x = 0$. 
                The desired integral is obtained by setting $x = 1$,
                giving the result
                \eq{
                        \label{DM_equal_mass_I_6}
                        I_6 =& -
                        \frac{\zeta_2}{2} \e^2 +
                        \left[ 3 \zeta_2 \log(2) - \frac{7}{4} \zeta_3 \right] \e^3 
                        \\ & +
                        \left[
                                \frac{31}{4} \zeta_4 - 6 \zeta_2 \log^2(2) - 
                                \frac{\log^4(2)}{2} - 12 \mathrm{Li}_4(\tfrac{1}{2})
                        \right] \e^4 +
                        O(\e^5)
                        \, .
                        \nonumber
                }

        \item 
                Boundary values for the integrals
                $
                        \{ I_2, I_3, I_4, I_7, I_8, I_9, I_{11}, I_{13}, I_{14} \}
                $
                are fixed by regularity at the pseudo-threshold $t \to 0$.
                In particular, due to the prefactors in \eqref{DM_equal_mass_master_integrals},
                the integrals
                $
                        \{ I_2, I_3, I_4, I_8, I_{11}, I_{13} \}
                $
                go to $0$ in this limit.
                By analyzing the Pfaffian system as $t \to 0$,
                we obtain the relations
                $
                        I_9 |_{t \to 0} = 3 I_6
                $
                and
                $
                        I_{14} |_{t \to 0} = 6 I_6,
                $
                which fixes boundary values for $I_9$ and $I_{14}$ because $I_6$ is known by the formula
                \eqref{DM_equal_mass_I_6}.
                The integral $I_7$ vanishes because the \emph{massive} canonical bubble factor
                goes to zero faster than the divergence of \emph{massless} bubble factor.
        \item 
                For $I_{10}$,
                we use regularity at the pseudo-threshold $t \to 4 m^2$.
                Thanks to the prefactor in \eqref{DM_equal_mass_master_integrals},
                the integral vanishes in this limit.
        \item 
                Boundary constants for the integral $I_{12}$ follow from comparing the 
                left- and right-hand sides of 
                $
                        I_{12} = I_2 \times I_{10},
                $
                since $I_2$ and $I_{10}$ are known.
                This relation can be gleaned from \figref{DM_equal_mass_master_integrals} by
                recalling that the double-tadpole $I_1$ is normalized 
                to unity due to the integration measure.
        \item 
                We are able to determine boundary constants for the top sector integral $I_{15}$ 
                by comparing our result with \cite{Bonciani:2003hc}.
\end{itemize}
The final results for the canonical MIs are quite compact.
For instance,
the top sector integral reads
\eq{
        I_{15} =
        \e^4 &
        \Big[
                2 \zeta_3 H(0|w) +
                4 \zeta_2 H(0,0|w) +
                8 H(0,0,-1,0|w) + 
                H(0,0,0,0|w) 
                \\ & +
                6 H(0,0,0,1|w) + 
                4 H(0,1,0,0|w) +
                3 \zeta_4
        \Big]
        +
        O(\e^5)
        \, .
        \nonumber
}

\subsection{Soft limit}
\label{sec:soft_limit_integrals}

\begin{figure}[H]
        \centering
        \includegraphics[scale=0.2]{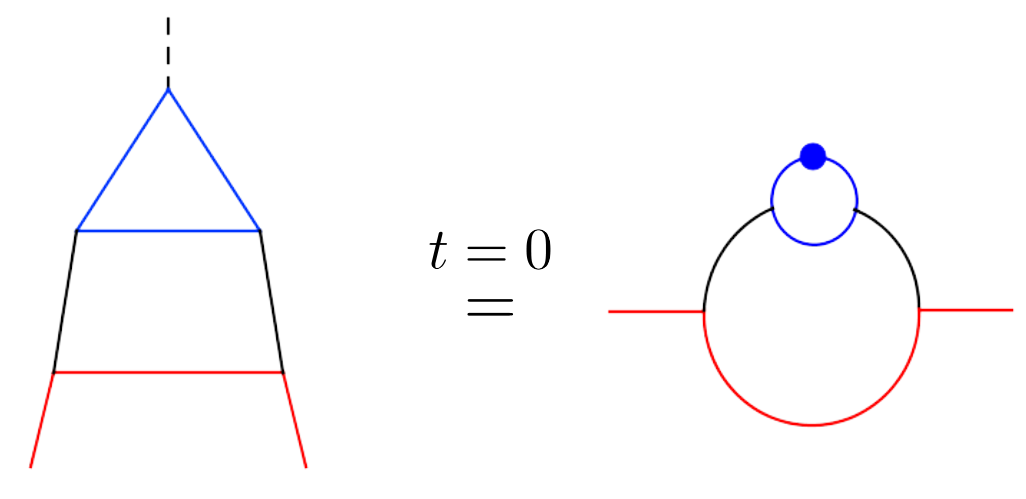}
        \caption{Diagrammatic representation of the soft limit.}
        \label{fig:DM_soft_limit}
\end{figure}

The soft limit $t \to 0$ of the form factors in 
\secref{sec:form_factor_soft_limit} is of phenomenological interest.
For this reason,
we now compute the two-mass MIs in this limit.
The original two-mass topology simplifies in this limit,
as depicted in \figref{fig:DM_soft_limit}.

\noindent
The reason for this simplification is as follows.
Recall that $t = q^2$,
where $q^\mu$ is the momentum of the mediator particle.
Setting $q^\mu = 0$ in the integral family \eqref{DM_two_mass_integral_family} implies that
$
        D_2 \to D_1 \text{ and } D_5 \to D_4,
$
wherefore we only need to consider IBPs and DEQs for the simplified integral family
\eq{
        I_{\nu_1 \nu_3 \nu_4 \nu_6 \nu_7} =
        \int \frac{\dd^\DD \ell_1}{(2\pi)^\DD} \frac{\dd^\DD \ell_2}{(2\pi)^\DD}
        \frac{D_7^{\nu_7}}{D_1^{\nu_1} D_3^{\nu_3} D_4^{\nu_4} D_6^{\nu_6} }
        \label{DM_soft_limit_integral_family}
}
having four denominators
\eq{
        \begin{array}{lllllll}
                & D_1 &=& \ell_1^2 - m_\ell^2 
                \, ,
                & D_3 &=& \ell_2^2 - m_A^2 
                \, ,
                \\
                & D_4 &=& (\ell_2+p)^2 
                \, ,
                & D_6 &=& (\ell_1+\ell_2+p)^2 - m_\ell^2
                \, ,
        \end{array}
}
and an ISP $D_7 = (\ell_1-p)^2$.
Note that $p'$ has been set equal to $p$ due to momentum conservation.

\begin{figure}[H]
        \centering
        \captionsetup[subfigure]{labelformat=empty}
        \subfloat[\hspace{0.5cm}$J_1$]
        {%
          \includegraphics[width=0.2\textwidth]{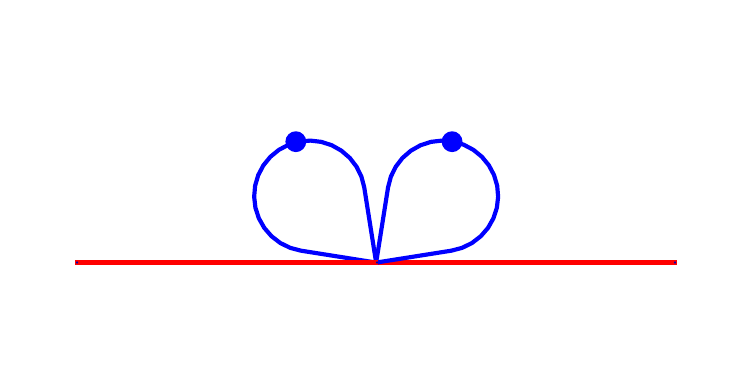}
        }
        \subfloat[\hspace{0.5cm}$J_2$]{%
          \includegraphics[width=0.2\textwidth]{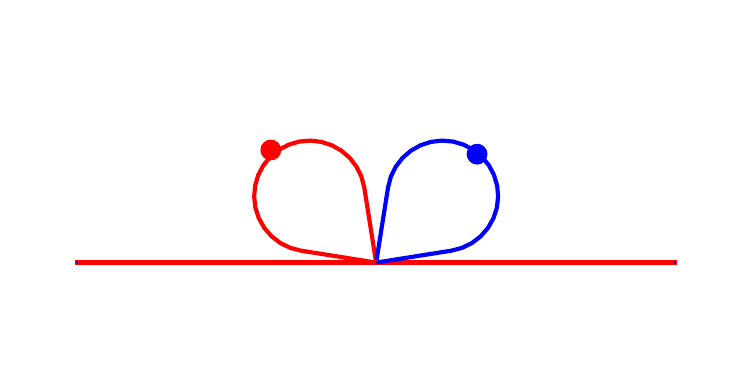}
        }
        \subfloat[\hspace{0.5cm}$J_3$]{%
          \includegraphics[trim={0 2cm 0 0},clip,  width=0.2\textwidth]{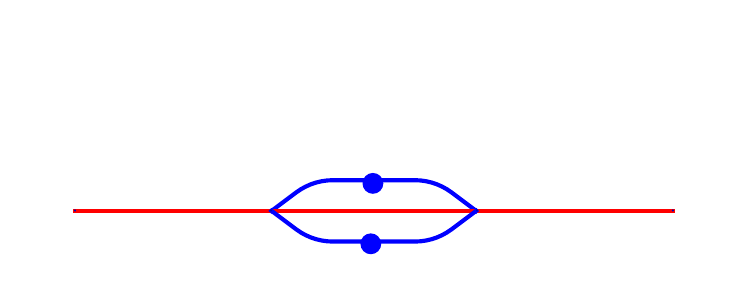}
        }
        \subfloat[\hspace{0.5cm}$J_4$]{%
          \includegraphics[trim={0 1.8cm 0 0},clip,width=0.2\textwidth]{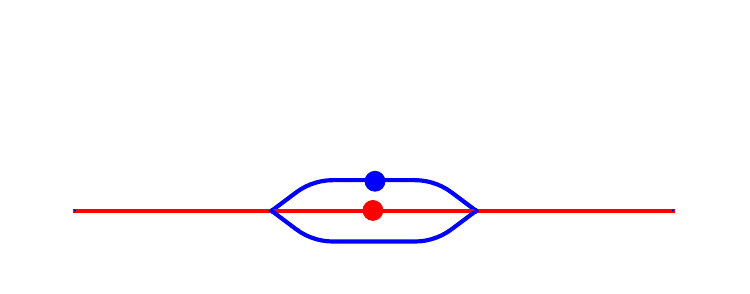}
        }
        \caption{
                The integrals $J_1$ appearing in the canonical basis \eqref{DM_soft_limit_canonical_basis}.
                Squared propagators are depicted by dots.
        }
        \label{fig:DM_soft_limit_diagrams}
\end{figure}

IBP reduction for the family \eqref{DM_soft_limit_integral_family} reveals 4 MIs.
By the Magnus exponential, 
we obtain the canonical basis
\eq{
        \begin{array}{lllllll}
                & I_1 &=& \e^2 J_1
                \, , 
                & I_2 &=& \e^2 J_2
                \, , \\ 
                & I_3 &=& \e^2 m_\ell m_A \big[J_3 + 2 J_4\big]
                \, , 
                & I_4 &=& \e^2 m_A^2 J_4 
                \, ,
        \end{array}
        \label{DM_soft_limit_canonical_basis}
}
in terms of the integrals $\{J_1, J_2, J_3, J_4\}$ shown in \figref{fig:DM_soft_limit_diagrams}.

\noindent
Defining the mass-ratio variable
\eq{
        z = \frac{m_A}{m_\ell}
        \, ,
        \label{DM_z_variable}
}
the canonical Pfaffian system takes the form
\eq{
        \dd \vec{I}(z|\e) = \e \, P(z) \cdot \vec{I}(z|\e)
        \, , \quad
        P(z) = \sum_{i=1}^3 P_i \, \dd \log \eta_i(z)
        \, , \quad
        P_i \in \QQ^{4 \times 4}
        \, ,
}
where the $\eta_i$ are HPL letters
\eq{
        \eta_1 = z
        \, , \quad
        \eta_2 = 1+z
        \, , \quad
        \eta_3 = 1-z
        \, .
}

Boundary constants for the solution vector are fixed as follows:
\begin{itemize}
        \item 
                The two double-tadpoles $\{I_1,I_2\}$ can be directly integrated to
                \eq{
                        I_1 = 1 
                        \quadit{\text{and}}
                        I_2 = z^{-2\e}
                        \, .
                }
        \item 
                Boundary values for the two sunrises $\{I_3,I_4\}$ are fixed by regularity at $m_A \to 0$
                as in \cite{Argeri:2002wz}.
                In particular, 
                due to the prefactors of \eqref{DM_soft_limit_canonical_basis},
                both integrals vanish in this limit.
\end{itemize}

        \section{Form factor results}

We now present form factor results for the following cases:
\begin{itemize}
        \item 
                Two masses ($m_A \neq m_\ell$): scalar and pseudo-scalar.
        \item 
                Equal-mass ($m_A = m_\ell$): scalar and pseudo-scalar.
        \item 
                Soft limit ($t \to 0$): scalar and pseudo-scalar.
\end{itemize}
Let us emphasize again that we only include results for one-body interactions.
The analytic form factors in this section are all rescaled as
$
        \mathcal{F}_{S,P} \to \frac{\pi^2}{\alpha_{\mathrm{em}}} \mathcal{F}_{S,P}
$
for the sake of readability. 

\subsection{Two-mass case}

\begin{figure}
        \centering
        \begin{subfigure}[b]{0.55\textwidth}
                \includegraphicsbox[scale=0.33]{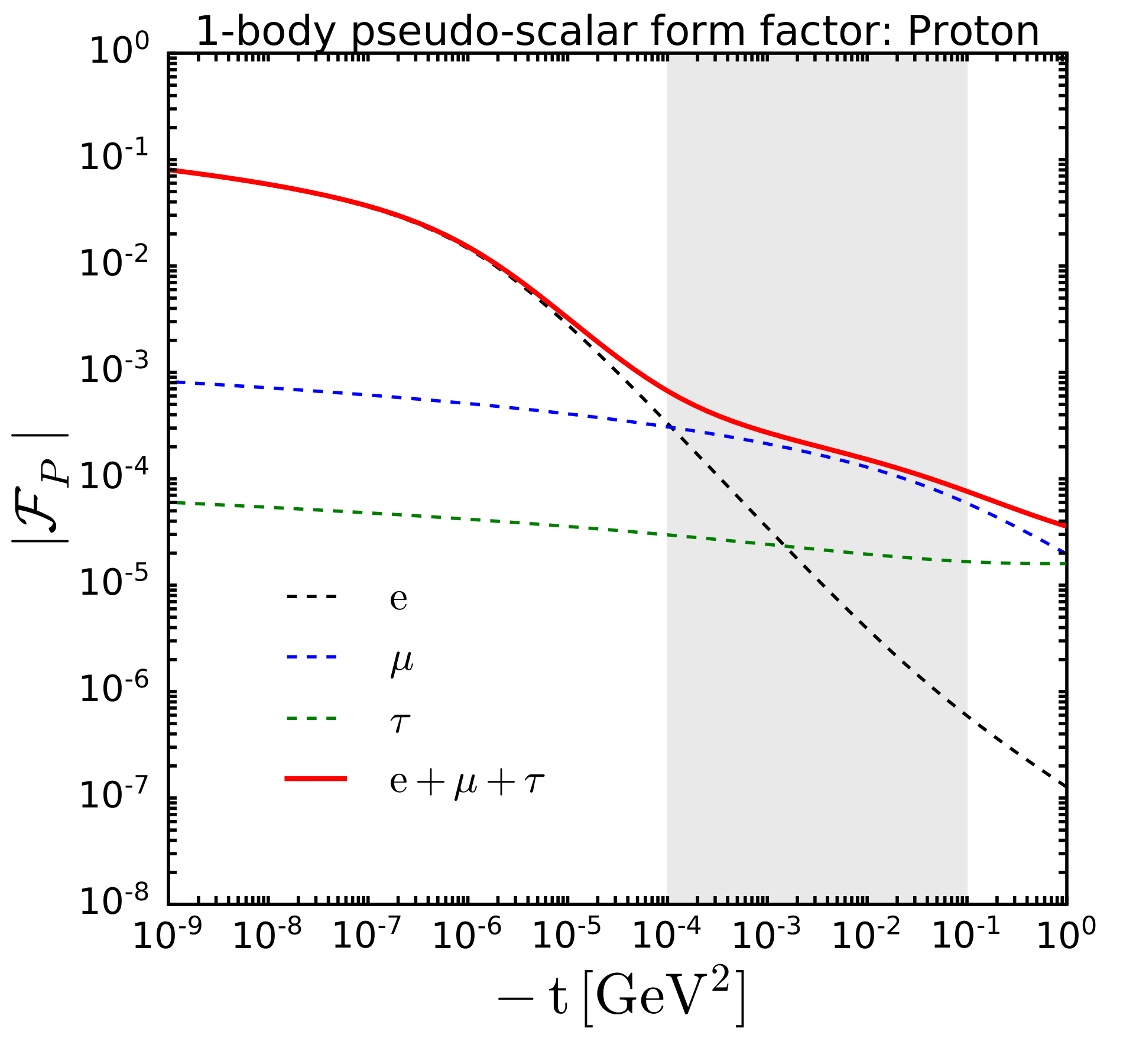}
                \caption{
                        Two-mass \textbf{pseudoscalar} form factor as a function of momentum transfer $t$.
                        The nucleon mass $m_A$ is set equal to the proton mass.
                        The dashed lines represent individual contributions from the electron ($e$),
                        the muon ($\mu$),
                        and the tau lepton ($\tau$).
                        The solid {\color{red} red} line shows the sum of all contributions.
                        The grey shaded band indicates the range of $t$ that is expected to be relevant for
                        DM particles arriving from our local galaxy.
                }
                \label{fig:form_factor_pseudoscalar}
        \end{subfigure}

        \vspace{0.15cm}

        \begin{subfigure}[b]{0.55\textwidth}
                \includegraphicsbox[scale=0.33]{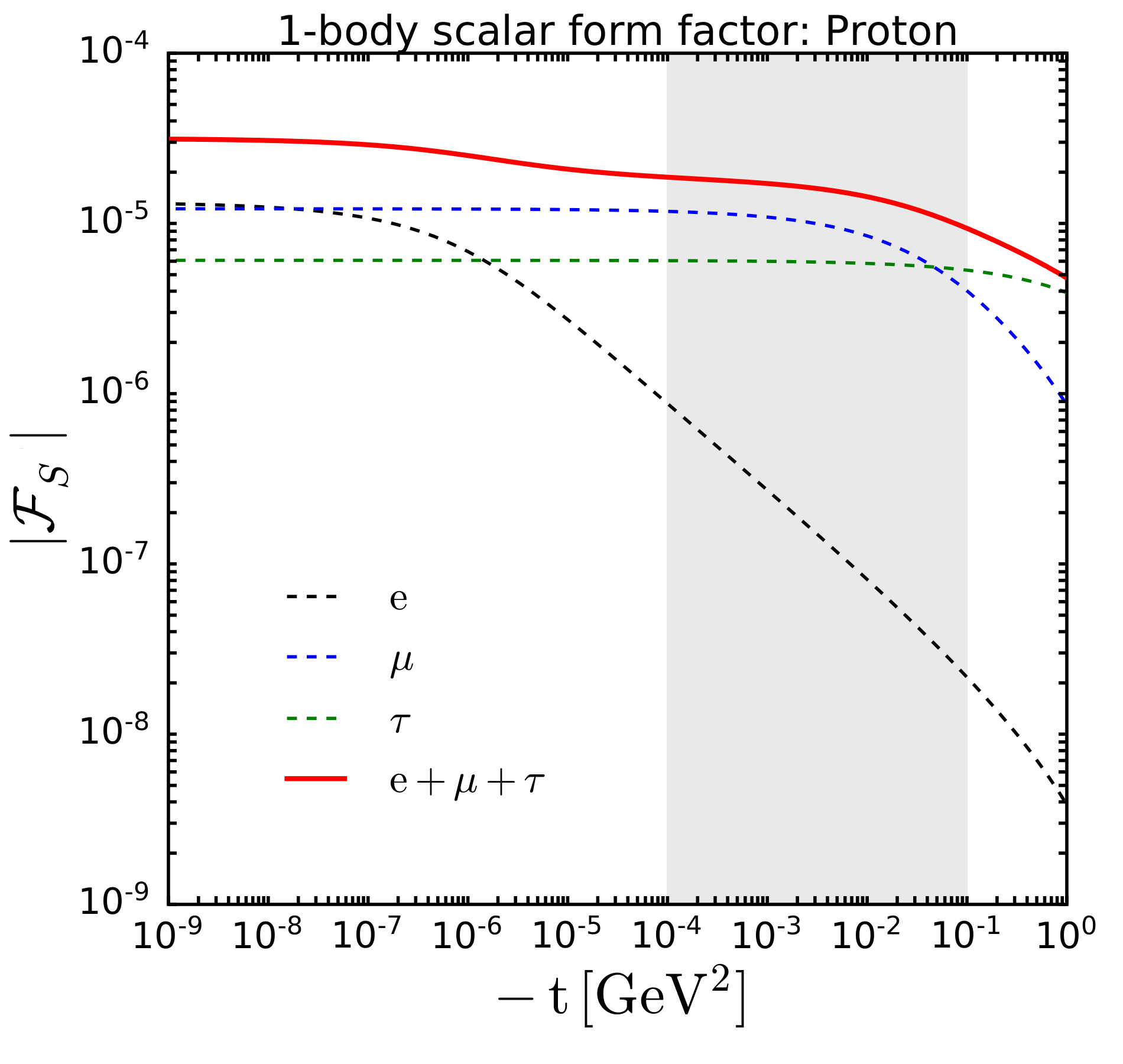}
                \caption{
                        Two-mass \textbf{scalar} form factor as a function of momentum transfer $t$.
                        See the caption of \figref{fig:form_factor_pseudoscalar} for additional details.
                }
                \label{fig:form_factor_scalar}
        \end{subfigure}
\end{figure}

The analytic expressions for the two-mass form factors
$
        \mathcal{F}_{S,P}(t, m_\ell, m_A)
$
would span several pages,
so we omit them here.
Instead, 
we showcase the results as numerical plots in Figures
\eqref{fig:form_factor_pseudoscalar} and \eqref{fig:form_factor_scalar}.

\subsection{Equal-mass case}
Define the shorthand notation 
$
        G(\vec{z}|w) := G_{\vec{z}}(w).
$

\subsubsection{Scalar form factor}

The analytic expression for the equal-mass form factor with a scalar mediator particle is
{\small\eq{
& \mathcal{F}_S(w) =  {-} \frac{ (20 \, w^2 {+} 4 \, w)}{ \, (w{-}1) \, (w{+}1)^2} \,  G_0(w)
\; {+} \; \frac{ w}{ \, (w{-}1) \, (w{+}1)} \, \big[24 \, G_{{-}1,0}(w) {+} 4 \, G_{1,0}(w) \big] 
\nonumber\\
& {+} \frac{ w}{ \, (w{-}1)^2 \, (w{+}1)^3} [{-}12 \, w^3 {-} 38 \, w^2 {+} 8 \, w {+} 2] \, G_{0,0}(w)
\; {-} \; \frac{w}{\pi^2 \, (w{-}1)^2} \, \big[2 \, G_{0,0,0}(w) {+} 4 \, \zeta_3\big] 
\nonumber\\
& {+} \frac{ w \, \zeta_3}{ \, (w{-}1)^3 \, (w{+}1)} \, [5 \, w^2 {-} 6 \, w {+} 5] \, G_0(w)
\; {+} \; \frac{w}{\pi^2 \, (w{+}1)^2} \, \left[12 \, G_{0,0,1}(w) {-} \frac{4 \, \pi^2}{3} \, G_0(w) {+} 16 \, G_1(w) \right] 
\nonumber\\
& {+} \frac{ w}{ \, (w{-}1)^2 \, (w{+}1)^2} \, \Big[({-}16 \, w^2 {+} 64 \, w {-} 16) \, G_{0,{-}1,0}(w) {+} (8 \, w^2 {-} 48 \, w {+} 8) \, G_{1,0,0}(w) {+} 16 \, w \, G_{0,1,0}(w) \Big] 
\nonumber\\ &
{+} \frac{ w}{ \, (w{-}1) \, (w{+}1)^3} \Bigg[({-}14 \, w^2 {+} 4 \, w {-} 14) \, G_{0,1}(w) {+} \left(\frac{w^2}{3} {+} 2 \, w {+} \frac{1}{3}\right) \, G_{1,0}(w) \, \pi^2 
\\
& {+} (2 \, w^2 {+} 12 \, w {+} 2) \, \big(G_{1,0,0,0}(w) \, {-} \, G_{0,0,1,0}(w)\big) {+} \left(2 \, w^2 {-} \frac{20 \, w}{3} {+} 2 \right) \pi^2 \Bigg]
\nonumber\\
& {+} \frac{ w}{ \, (w{-}1)^3 \, (w{+}1)^3} \Bigg[\left(\frac{5 \, w^4}{6} {-} 2 \, w^3 {+} \frac{23 \, w^2}{3} {-} 2 \, w {+} \frac{5}{6}\right) \, G_{0,0}(w) \, \pi^2
\nonumber\\
& {+} \left(8 \, w^4 {-} 32 \, w^3 {+} 112 \, w^2 {-} 32 \,  w {+} 8\right) \, G_{0,0,{-}1,0}(w) {+} \left(\frac{3 \, w^4}{2} {-} 2 \, w^3 {+} 9 \, w^2 {-} 2 \, w {+} \frac{3}{2}\right) \, G_{0,0,0,0}(w)
\nonumber\\
& {+} \left({-}9 \, w^4 {+} 12 \, w^3 {-} 54 \, w^2 {+} 12 \, w {-} 9\right) \, G_{0,0,0,1}(w) {+} \left({-}2 \, w^4 {+} 24 \, w^3 {-} 76 \, w^2 {+} 24 \, w {-} 2\right) \, G_{0,1,0,0}(w)
\nonumber\\
& {+} \left(\frac{w^4}{18} {-} \frac{2 \, w^3}{45} {+} \frac{11 \, w^2}{45} {-} \frac{2 \, w}{45} {+} \frac{1}{18} \right) \, \pi^4 \Bigg].
\nonumber
}}

\noindent
This expression can be turned into an $s$-channel form factor by crossing symmetry.
Then it is related to a form factor for Higgs decay into two heavy quarks \cite{Bernreuther:2005gw}.

\subsubsection{Pseudo-scalar form factor}

The equal-mass form factor for the pseudoscalar reads 
{\small\eq{
& \mathcal{F}_P(w) =  {-} \frac{ w^2}{ \, (w{-}1)^3 \, (w{+}1)} \, \left[ \frac{2\pi^2}{3} \, G_0(w) {+} 4 \, G_{0,0,0}(w) \right] 
\nonumber\\
& {+} \frac{ w}{ \, (w{-}1)^2} \, \left[ G_{0,0}(w) {-} 4 \, G_{0,0,1}(w) {-} 4 \, G_{0,1,0}(w) {+} 8 \, G_{1,0,0}(w) {+} 12 \, \zeta_3 {+} \frac{\pi^2}{3} \right] 
\\
& {+} \frac{ w}{ \, (w{-}1) \, (w{+}1)} \, \left[ {-} 3 \, \zeta_3 \, G_0(w) {-} \frac{\pi^2}{6} \, G_{0,0}(w) {-} \frac{\pi^2}{3} \, G_{1,0}(w) {-} \frac{1}{2} \, G_{0,0,0,0}(w) \right. 
\nonumber \\
& \hspace{3.15cm} \left. {+} 3 \, G_{0,0,0,1}(w) {+} 2 \, G_{0,0,1,0}(w) {-} 2 \, G_{0,1,0,0}(w) {-} 2 \, G_{1,0,0,0}(w) {-} \frac{\pi^4}{45} \right].
\nonumber
}}

\noindent
Using crossing symmetry, 
this form factor is related to the $s$-channel QCD form factor dubbed $A_R(s,m,m)$ in 
\cite{Bernreuther:2004ih}.

\subsection{Soft limit}
\label{sec:form_factor_soft_limit}

The soft limit $t \to 0$ is important for phenomenological applications,
as the DM particles are expected to scatter elastically.
It is therefore useful to have simple expressions for the form factors in this limit
that can be easily numerically evaluated.

\subsubsection{Scalar form factor}

The scalar form factor is finite at $t=0$,
so the result 
$
        \mathcal{F}_S(t=0, m_\ell, m_A)
$ 
will only depend on the mass ratio $z = m_A/m_\ell$.
Writing the form factor as an expansion in terms of FIs as in \eqref{form_factor_definition},
we then IBP reduce those integrals to the soft limit MIs from \secref{sec:soft_limit_integrals}.
The smooth $4$-dimensional limit $\e \to 0$ finally yields the result
\eq{
        \mathcal{F}_S(t=0, m_\ell, m_A) &=
        -\frac{2}{z}
        \left[
                1 - \frac{\log(z)}{2} + f_S(z) + f_S(-z)
        \right]
        \\[5pt] \nonumber
        \text{with}
        \quad
        f_S(z) &= \frac{4 + 3z + z^3}{4 z^2}
        \big[
                \log |z| \log(1+z) + \mathrm{Li}_2(-z)
        \big]
        \, .
}

\subsubsection{Pseudo-scalar form factor}

The pseudo-scalar form factor is logarithmically divergent as $t \to 0$,
so we compute an asymptotic expansion in terms of the variables $(x,y)$ from \eqref{DM_xy_variables}.
Keeping $m_{A,\ell}$ fixed, 
the soft limit then corresponds to $(x,y) \to (1,1)$.

\newpage

Our strategy for computing the asymptotic expansion is as follows:
\begin{enumerate}
        \item
                Staring with the full expression for the form factor 
                $
                        \mathcal{F}_P(t, m_\ell, m_A)
                $
                in terms of GPLs,
                we use \package{PolyLogTools} to rewrite every $G(\vec{z} \, | \, v)$ such that 
                $
                        \vec{z} \in \ZZ^M
                $
                and $v = x$ or $v = y$.
                The weight vectors $\vec{z}$ will thereby not contain any $x$'s or $y$'s,
                which makes it easier to expand the GPLs.
        \item 
                \package{PolyLogTools} has a built-in function for asymptotically expanding GPLs.
                However,
                it only supports an expansion around $v \sim 0$,
                not $v \sim 1$.
                Introducing a dummy variable $v'$ by $v = 1-v'$,
                we hence seek to rewrite every $G(\vec{z}\, |\, v)$ in terms of $G(\vec{z} \, | \, v')$.
                This is done by employing the well-known recursion 
                \eq{
                        G(z_1, \ldots, z_M | v) &=
                        G(z_1, \ldots, z_M | 1-v') \\&=
                        G(z_1, \ldots, z_M | 1) +
                        \int_0^{v'} \frac{\dd t}{t - (1-z_1)} G(z_2, \ldots, z_M |1 -t)
                        \, ,
                        \nonumber
                }
                which produces GPLs having argument $v'$ due to the upper integration bound 
                in the second term.
                Experimentally,
                we find that this recursion admits a closed-form solution
                \eq{
                        \label{GPL_1_minus_z_recursion}
                        G(z_1, \ldots, z_M|1-v') =
                        \sum_{i=0}^M 
                        & G(1-z_1, \ldots, 1-z_i | v') 
                        \\ \times
                        & G(z_{i+1}, \ldots, z_M | 1) \, \big|_{\text{shuffle reg.}}
                        \, ,
                        \nonumber
                }
                where we emphasized the need for \emph{shuffle regularization} 
                to deal with divergent GPLs \cite[Section 5.3]{Duhr:2014woa}.
                We use the convention
                $
                        G(1-z_1, \ldots, 1-z_i | v') = 1
                $
                when $i = 0$.

                As an aside,
                although it is not a deep result,
                we have not been able to locate this closed-form solution elsewhere in the literature.
        \item 
                \package{PolyLogTools} can now expand the terms
                $
                        G(1-z_1, \ldots, 1-z_i | v')
                $
                from \eqref{GPL_1_minus_z_recursion} around $v' \sim 0$ 
                up the desired order
                $
                        \mO\big((1-x)^n,(1-y)^m \big).
                $
                The latter notation means that we first expand in $(1-x)$ up to $n$th order,
                and then in $(1-y)$ up to $m$th order.
                The reversed expansion order is written as
                $
                        \mO\big((1-y)^n,(1-x)^m \big).
                $
                The maximal powers
                $(n,m)$ 
                are chosen so as to give sufficiently accurate numerical results in comparison with
                the full form factor
                $
                        \mathcal{F}_P(t, m_\ell, m_A)
                $
                when the latter is evaluated for small $t$.
\end{enumerate}

\newpage

A double asymptotic expansion in $(x,y)$ will depend on the orderings
$
        x<y \text{ or } y<x,
$
so we treat these two cases separately.
The orderings correspond to the mass hierarchies
$
        m_\ell < m_A \text{ and } m_A < m_\ell
$
respectively.
By the steps outlined above,
we obtain the following results for the pseudoscalar form factor in the soft limit:
\begin{itemize}
        \item 
                Mass hierarchy $m_\ell < m_A \, ( x < y)$:
        \eq{
                \nonumber
                \mathcal{F}_P(t \sim 0, \, m_\ell < m_A) =
                -& 2 \zeta_2 
                + \frac{(1-x)}{(1-y)}
                \big[
                        - 2 \log(1-x) - 2 \log(2) + 3
                \big]
                \\ \nonumber
                +& \frac{(1-x)^2}{(1-y)}
                \left[
                        - \log(1-x) - \log(2) + \frac{1}{2}
                \right]
                \\ 
                +& (1-x)^2 
                \left[
                        \frac{\log(1-x)}{2} + \frac{\log(2)}{2} + \frac{37 \zeta_2}{60} - \frac{1}{4}
                \right]
                \\ \nonumber
                +& \frac{6(1-x)}{\zeta_2}
                \left[
                        \log(1-x) + \log(2) + \frac{3 \zeta_2}{2} - \frac{3}{2}
                \right]
                \\ \nonumber
                +& \mO\big((1-y), \, (1-x)^3 \big)
                \, .
        }
        \item 
                Mass hierarchy $m_A < m_\ell \, ( y < x)$:
        \eq{
                \mathcal{F}_P(t \sim 0, \, m_A < m_\ell) =
                & \frac{(1-x)}{(1-y)}
                \left[
                        -\frac{3 \log(1-x)}{2} - \frac{(1-y)}{2} - 2 \log(2) + \frac{17}{8}
                \right]
                \\ \nonumber
                +& (1-x)
                \left[
                        \frac{3 \log(1-x)}{4} + 
                        \frac{\log(1-y)}{4} + 
                        \log(2) + 
                        \frac{3\zeta_2}{2} - 
                        \frac{21}{16}
                \right]
                \\ \nonumber
                +& \mO\big((1-x)^2, \, (1-y) \big)
                \, .
        }
\end{itemize}
These formulas differ by roughly $1\%$ in comparison with the full analytic results evaluated at
$t = - 10^{-7} \text{GeV}^2$.

        \part{\texorpdfstring{$\mD$-modules}{}}
        
        \chapter{Introduction to \texorpdfstring{$\mD$}{}-modules}
\label{ch:macaulay}

The preceding chapters of this thesis dealt with tried and tested methods for evaluating FIs.
We now venture into somewhat uncharted territory from the perspective of the physics literature.

Based on the papers \cite{Chestnov:2022alh,Chestnov:2023kww},
the next six chapters of this thesis study FIs within the framework of $\mD$-\emph{modules}.
A $\mD$-module can,
informally,
be thought of as a vector space of linear differential operators that annihilate a given function.
As a supplement to the tools coming from complex \emph{analysis},
$\mD$-modules provide \emph{algebraic} machinery to uncover properties of PDEs.
For instance,
one can learn the dimension of the solution space,
construct bases of differential operators,
obtain relations between basis elements,
and more.
We will only scratch the surface of this vast topic,
so we refer to the classic textbooks 
\cite{Hotta-Tanisaki-Takeuchi-2008} and \cite{Borel}
for exhaustive discussions on $\mD$-modules.

Our overarching goal in \textbf{Part II} of this thesis is  
to build a bridge between these two independently well-studied topics,
namely FIs and $\mD$-modules.
But,
as it so often happens in blue skies research,
practical applications will spring out as well.

        \section{What is a \texorpdfstring{$\mD$}{}-module?}
\label{sec:what_is_a_D_module}

How might one represent a function?
Take $\sin(x)$ as an example.
It can be represented in several ways,
each one having different advantages.
For instance,
it can be analytically represented as a power series,
or geometrically via the unit circle.
In the philosophy of $\mD$-modules,
$\sin(x)$ is represented by the set of linear PDEs which annihilate the function;
in this case simply
$
        \{\p^2 + 1\}.
$

We note that this philosophy would not work for non-linear PDEs, 
since a differential operator $L$ cannot be "pulled out" in the form 
$
        L \bullet f = 0
$
if there is a term like $f(x)^2$.

\newpage

\begin{ex}
A striking example of this philosophy comes from calculating roots of an $N$th order polynomial
\eq{
        z_0 + z_1 x + z_2 x^2 + \cdots + z_{N-1} x^{N-1} + z_N x^N =
        (x-r_1) \cdots (x-r_N)
        \, .
}
Is there an equation for the roots 
$
        r_k = r_k(z_1, \ldots, z_N)
$
in terms of the indeterminate polynomial coefficients $z_k$?
There is,
for instance,
a well-known formula for 
$
        r_{1,2}(z_0,z_1,z_2)
$
in the quadratic case $N = 2$.

Alas, 
Galois theory teaches us that no such formula exists in terms of radicals when $N \geq 5$.
But if we settle for an infinite series representation of $r_k$, 
then such a formula \emph{does} exist!

Fixing $k$, 
it turns out that $r_k$ is annihilated,
indeed \emph{represented},
by the following set of partial differential operators:
\eq{
        \nonumber
        \frac{\p^2}{\p z_i \p z_j} -
        \frac{\p^2}{\p z_k \p z_l} 
        \quad &\text{whenever} \quad
        i+j = k+l
        \, ,
        \\
        \sum_{i=0}^N i \, z_i \frac{\p}{\p z_i} + 1
        \quad & \hspace{0.4cm} \text{and} \hspace{0.4cm} \quad
        \sum_{i=0}^N z_i \frac{\p}{\p z_i}
        \, .
        \label{GKZ_system_for_roots}
}
Many features of polynomial roots can then be learned from studying these differential operators.
For one,
it is possible to construct power series solutions of the form
$
        r_k = 
        \sum_{i_0 \ldots i_N}^\infty f_k(i_0, \ldots, i_N) z_0^{i_0} \cdots z_N^{i_N}
        ,
$
where the $f_k$'s are various combinations of factorials \cite{STURMFELS2000171}.
\end{ex}

So,
what is a $\mD$-module?
It is a module over a ring of $\mD$ifferential operators.
Let us spell this out more formally.

Let $R$ be a ring with a unit $"1"$,
and $M$ an abelian group w.r.t.~addition $"+"$.
Given $r \in R$ and $m \in M$,
we assume the existence of a map
\eq{
        \nonumber
        &\psi: R \times M \to M
        \\
        &\psi(r,m) = r \bullet m
        \, ,
}
where $r \bullet m$ denotes the \emph{action} of $r$ on $m$.
For instance, 
if $R = \CC$ and $M = \CC^n$,
then $\psi$ could represent the scalar multiplication of a complex number $r$ on a complex vector $m$.

\begin{definition}
\label{def:R_module}

An abelian group $M$ becomes a \emph{left}%
\footnote{
        \emph{Right} $R$-modules are defined by the same axioms with $r$'s acting from the right,
        i.e.~$m \bullet r$.
}
$R$-\emph{module}
when the following three axioms are fulfilled%
\footnote{
        We omit $"+"$ subscripts of kind
        $(r_1 +_R r_2) \bullet m = r_1 \bullet m +_M r_2 \bullet m$
        for notational cleanliness.
}:
\begin{enumerate}
        \item
                For any $r \in R$ and $m_i \in M$,
                we have
                $
                        r \bullet (m_1 + m_2) = r \bullet m_1 + r \bullet m_2.
                $
        \item 
                For any $r_i \in R$ and $m \in M$,
                we have
                $
                        (r_1 r_2) \bullet m = r_1 \bullet (r_2 \bullet m)
                $
                and
                $
                        (r_1 + r_2) \bullet m = r_1 \bullet m + r_2 \bullet m.
                $
        \item 
                For any $m \in M$, 
                we have 
                $
                        1 \bullet m = m.
                $
\end{enumerate}
\end{definition}

\noindent
If $R$ is a commutative field such as $\CC$,
then a left $R$-module is nothing but an $R$-vector space.

The $N$th \emph{Weyl algebra} 
\eq{
        \mD_N = \CC[z_1, \ldots, z_N] \langle \p_1, \ldots, \p_N \rangle
        \label{Weyl_algebra}
}
is generated by symbols $z_i$ and $\p_i$ subject to the commutation relations
\eq{
        [z_i, z_j] = 0
        \, , \quad
        [\p_i, \p_j] = 0
        \, , \quad
        [\p_i, z_j] = \d_{ij}
        \, .
        \label{Weyl_algebra_commutators}
}
We shall often suppress the subscript in $\mD_N$ when the number of variables is understood.
The square brackets $[\ldots]$ in 
\eqref{Weyl_algebra} 
mean that the $z_i$ are thought of as coefficients of the generators $\p_i$.
Note that the last commutator of 
\eqref{Weyl_algebra_commutators} 
encodes Leibniz' rule of differentiation when we insert a dummy function $f(z)$:
$
        \frac{\p}{\p z}[z f(z)] - z \frac{\p}{\p z} f(z) = f(z).
$
We therefore interpret $\mD$ as the collection of all differential operators having polynomial coefficients.
For instance,
$
        (z_1^2 + z_3 + 1) \p_1^2 \in \mD_3.
$
We emphasize that $\mD$ has the structure of a ring.

We also have the $N$th \emph{rational} Weyl algebra
\eq{
        \mR_N = \CC(z_1, \ldots, z_N) \langle \p_1, \ldots, \p_N \rangle
        \, ,
}
wherein the coefficients are promoted to rational functions.
We shall often shorten $\mR_N$ to $\mR$.
In this case,
the commutators are extended to
\eq{
        [z_i, z_j] = 0
        \, , \quad
        [\p_i, \p_j] = 0
        \, , \quad
        [\p_i, r(z)] = \frac{\p r(z)}{\p z_i}
        \, ,
        \label{rational_Weyl_algebra_commutator}
}
for a rational function 
$
        r(z) \in \CC(z_1, \ldots, z_N).
$
The rational Weyl algebra is also endowed with a ring structure.

\begin{definition}
        A $\mD$-module is a left $R$-module 
        (see Definition \eqref{def:R_module}) 
        with the specific choice of ring $R = \mD$.
        An $\mR$-module is defined similarly for $R = \mR$.
        \label{def:D_module}
\end{definition}

We shall be careful in distinguishing between ring \emph{multiplication} and ring \emph{action} $"\bullet"$,
because there is a difference between e.g.
\eq{
        \p_i z_i = z_i \p_i + 1 
        \quadit{\text{and}} 
        \p_i \bullet z_i = 1
        \, .
}

Definition \eqref{def:D_module} can
be instantiated in many different ways because it only specifies the ring $R$.
\emph{Particular} $\mD$- and $\mR$-modules follow from further specifying the 
abelian group $M$ and the action map $\psi$.
Let us list some examples.
 
\begin{ex}
The Weyl algebras $\mD$ and $\mR$ are themselves $\mD$- and $\mR$-modules respectively.
The abelian group $M$ is formed from the addition of differential operators.
The algebras acts on themselves according to 
\eqref{Weyl_algebra_commutators} in the case of $\mD$
and 
\eqref{rational_Weyl_algebra_commutator} in the case of $\mR$,
which specifies $\psi$.
\end{ex}
\begin{ex}
\label{ex:function_space_as_D_module}
Many function spaces are equipped with the structure of a $\mD$-module.
Take $M$ to be some suitable space of holomorphic functions on a domain of $\CC^n$.
Assume that $M$ is closed under differentiation,
meaning that if $f(z) \in M$ then also $\frac{\p f}{\p z_i} \in M$.
The addition of such functions indeed forms an abelian group.
The action map
$
        \psi: R \times M \to M
$
is defined by
\eq{
        \psi(\p_i, f) = \p_i \bullet f := \frac{\p f}{\p z_i} 
        \, , \quad
        \psi(z_i, f) = z_i \bullet f := z_i \cdot f
        \, ,
}
where $"\cdot"$ in the last equation stands for ordinary scalar multiplication.
\end{ex}
\begin{ex}
The most important example of a $\mD$-module (or $\mR$-module) for the study of PDEs 
is arguably that of the \emph{quotient} ring $\mD/\mI$ (or $\mR/\mI$),
where $\mI$ is a so-called \emph{left ideal} to be defined shortly.
The abelian group $M$ is given by the addition of operators,
and the action is that of the Weyl algebra.

To motivate the notion of a left ideal $\mI$, 
consider a collection of differential operators 
$
        \{L_1, L_2, \ldots\}
$
that annihilate some function $f$:
$
        L_1 \bullet f = L_2 \bullet f = \ldots = 0 .
$
Note that we still get zero if we act with any other operator $L \in \mD$ from the left:
$
        L L_1 \bullet f = L L_2 \bullet f = \ldots = 0 .
$
The definition%
\footnote{
        This is more of a "physicist's definition".
        Formally,
        a left ideal $\mI$ of a ring $R$ satisfies 
        1) it is a subgroup under addition,
        and 2) for $r \in R$ and $L \in \mI$ we have $r \, L \in \mI$.
}
of a left ideal $\mI$ in the Weyl algebra encapsulates this fact:
\eq{
        \mI =
        \langle L_1, L_2, \ldots \rangle =
        \{L L_1, \, L L_2, \ldots  \ | \ \forall \  L \in \mD\} 
        \, .
}
The notation $\langle L_1, L_2, \ldots \rangle$ means that the $L$'s \emph{generate} $\mI$.

The quotient ring
$
        \mD / \mI
$
is then given by all operators $L \in \mD$ modulo the equivalence relation $\sim$ determined by
\eq{
        L \sim K \iff L = K + K_\mI
        \quad \text{where} \quad
        K \in \mD
        \, , \
        K_\mI \in \mI
        \, .
        \label{modulo_I}
}
Take,
for instance, 
the left ideal generated by
$
        \mI = \langle \p_1 \p_2 \rangle
        \, .
$
Define two operators by
\eq{
        \nonumber
        L \ &= \ \p_2 \quadit{\text{and}} \\
        K \ &= \
        \p_1 z_1 \p_2 \ \overset{\eqref{Weyl_algebra_commutators}}{=} \
        (1 + z_1 \p_1)\p_2 \ = \
        \p_2 + z_1 \p_1 \p_2                
        \, .
}
$L$ and $K$ represent \emph{the same} element inside 
$
        \mD_2 / \mI
$
because they only differ by a term $z_1 \p_1 \p_2 \in \mI$.
We say that $L$ and $K$ lie in the same \emph{equivalence class},
written as 
$
        [L] = [K].
$

We shall see that the algebraic structure of the $\mD$-module $\mD/\mI$ 
carries a wealth of information about the functions that are annihilated by the ideal $\mI$.
This $\mD$-module hence embodies the intuition,
argued for in the beginning of this section,
that a function can be represented by the differential operators that annihilate it.
\end{ex}

\section{Holonomic \texorpdfstring{$\mD$}{}-modules}

A $\mD$-module $\mD / \mI$ has particularly nice features when it is \emph{holonomic}.
Holonomic $\mD$-modules describe FIs,
trigonometric functions,
various probability distributions,
hypergeometric functions,
Bessel functions,
and GPLs.
It is necessary to introduce a few concepts before we can define the notion of holonomicity
(we refer to the seminal text \cite{SST} for many more details).
\begin{itemize}
        \item
                Using the Weyl algebra commutation relations,
                any operator $L \in \mD$ can be uniquely expressed in the \emph{normally ordered form}
                \eq{
                        L \quadit{=}
                        \sum_{(p,q) \, \in \, \supp{L}}
                        c_{p,q} \, z^p \, \p^q
                        \, , \quad
                        c_{p,q} \in \CC
                        \, .
                }
                The summation range $\supp{L}$ denotes a finite set of non-zero multi-indices 
                $
                        (p,q) = (p_1, \ldots, p_N, q_1, \ldots, q_N),
                $
                and we introduced the multi-index exponent notation
                \eq{
                        z^p = z_1^{p_1} \cdots z_N^{p_N}
                        \quadit{\text{and}}
                        \p^q = \p_1^{q_1} \cdots \p_N^{q_N}
                        \, ,
                }
                which will be used frequently throughout this text.
        \item 
                For a given a vector $(u,v) \in \RR^{2N}$,
                the \emph{$(u,v)$-weight} of $z^p \, \p^q$ is defined by
                $
                        u \cdot p + v \cdot q.
                $
        \item 
                The differential operator $\mathrm{in}_{(u,v)}(L)$ is constructed by only
                keeping the largest $(u,v)$-weight terms in $L$.
        \item 
                The \emph{initial form} is defined by $\mathrm{in}_{(u,v)}(L)$ for the specific choice
                $
                        u = (0,\ldots,0) = \mathbf{0}
                $
                and
                $
                        v = (1, \ldots, 1) = \mathbf{1}.
                $
                I.e.~one assigns weight $0$ to all $z$'s,
                and weight $1$ to all $\p$'s.
                The standard notation for the initial form is
                \eq{
                        \mathrm{in}_{(\mathbf{0}, \mathbf{1})}(L) =
                        \sum_{\substack{
                                (p,q) \, \in \, \supp{L}, \\
                                |q| \text{ is max}
                        }}
                        c_{p,q} \, z^p \, \xi^q
                        \, ,
                        \label{principal_symbol}
                }
                where 
                $
                        |q| = q_1 + \ldots + q_N.
                $
                The $\p$'s have here been replaced by commuting $\xi$ variables,
                but this replacement may only be done \emph{after} writing $L$ in normally ordered form.
                As an example,
                take the normally ordered operator
                \eq{
                        L = z_1^2 + 2 \p_1^2 \p_2 + z_2^3 \p_1 \p_2 + z_1 z_2 \p_1^3 .
                }
                From left to right,
                the terms have weights
                $
                        (0,3,2,3).
                $
                So we only keep the 2nd and 4th terms,
                giving
                \eq{
                        \mathrm{in}_{(\mathbf{0}, \mathbf{1})}(L) =
                        2 \xi_1^2 \xi_2 + z_1 z_2 \xi_1^3
                        \, .
                }
        \item 
                Given a left ideal $\mI$ in $\mD_N$,
                we set
                \eq{
                        \mathrm{in}_{(\mathbf{0}, \mathbf{1})}(\mI) =
                        \CC \cdot
                        \{
                                \mathrm{in}_{(\mathbf{0}, \mathbf{1})}(L) 
                                \ | \
                                L \in \mI
                        \}
                        \, ,
                }
                i.e.~the ideal generated by the highest 
                $
                        (\mathbf{0},\mathbf{1})\text{-weight}
                $
                terms in $\mI$.
                It is a \emph{polynomial} ideal since it is written 
                in terms of commuting variables $(z,\xi)$.

        \item 
                The \emph{characteristic variety} of an ideal is defined as
                \eq{
                        \ch{\mI} =
                        \left\{
                                (z,\xi) \ | \
                                p(z,\xi) = 0
                                \ \text{ for all } \
                                p \, \in \, 
                                \mathrm{in}_{(\mathbf{0}, \mathbf{1})}(\mI)
                        \right\}
                        \ \subset \
                        \CC^{2N}
                        \, .
                }
                Namely,
                it the zero-set of all the polynomials in the ideal
                $
                        \mathrm{in}_{(\mathbf{0}, \mathbf{1})}(\mI).
                $
                The computation of $\ch{\mI}$ typically requires knowing a
                \emph{Gr\"obner basis} for $\mI$ 
                (to be defined in the next section).
\end{itemize}
With all of these notions in place,
we can finally state 
\begin{definition}
        An ideal $\mI$ is \emph{holonomic} when its characteristic variety $\ch{\mI}$ has dimension%
        \footnote{
                According to a result called Bernstein's inequality,
                all connected components of $\ch{\mI}$ have dimension at least $N$.
                An ideal is hence holonomic when its characteristic variety 
                has the smallest possible dimension
                (the maximal dimension would be $2N$).
        }
        $N$.
        A $\mD$-module $\mD / \mI$ is holonomic if and only if $\mI$ is holonomic%
        \footnote{
                There is actually an independent definition of holonomicity for $\mD$-modules,
                depending on so-called Bernstein filtrations.
                The fact that holonomic $\mD$-modules coincide with 
                holonomic ideals is then a non-trivial theorem.
                See \cite[Chapter 6]{dojo} for details.
        }.
        The \emph{holonomic rank} is defined as the following vector space dimension over $\CC(z)$:
        \eq{
                \nonumber
                \rank{\mI} 
                &= 
                \dim_{\CC(z)}
                \big(
                        \CC(z)[\xi] \, / \, \CC(z)[\xi] \cdot \ch{\mI}
                \big)
                \, .
        }
\end{definition}

The above definition of $\rank{\mI}$ relies on the following important fact:
a holonomic $\mD$-module carries the structure of a \emph{finite-dimensional} vector space \cite{dojo,SST}.
It follows that there exists a \emph{basis} of differential operators for the vector space $\mD / \mI$
having exactly $\rank{\mI}$ many elements.
Quite remarkably,
$\rank{\mI}$ also counts the number of independent holomorphic solutions to the system of PDEs
defined by $\mI$.
When $\mI$ is the annihilating ideal for a FI,
then $\rank{\mI}$ is indeed the number of MIs!

\subsection{Regular holonomic}
\label{sec:regular_holonomic}

Throughout the rest of this thesis,
when we say an ideal $\mI$ is holonomic we shall always mean \emph{regular holonomic}.
The extra adjective "regular" signifies,
loosely speaking,
that the solutions to the PDEs dictated by $\mI$ only have mild singularities.

Suppose that $z=0$ is a singularity of a solution $f(z)$ in the univariate case.
It is a \emph{regular singularity} if there exists a $k \in \ZZ_{>0}$ such that $|z|^k f(z) \to 0$ 
as $z \to 0$ at a fixed angle $\mathrm{arg}(z)$ in the complex plane.

The definition of regular holonomic is much more complicated in the multivariate case.
We refer to \cite[Section 2.4]{SST} \cite[Chapter 6]{Hotta-Tanisaki-Takeuchi-2008} for details.
A working definition for our purposes is to say that the solutions to $\mI$ admit
logarithmic series expansions near the singular hypersurfaces.
Equivalently,
the Pfaffian system born from $\mI$,
to be discussed in \secref{sec:Pfaffian_systems},
will have at most simple poles
(higher-order poles can appear,
but they are "fake" in the sense they can be turned into simple poles via a gauge transformation).

\section{Gr\"obner bases}

How does one compute a basis for a holonomic $\mD$- or $\mR$-module?
The answer,
as with almost any computational question in commutative and non-commutative algebra,
is via \emph{Gr\"obner bases}.
A couple of extra concepts are needed before the definition can be stated.
\begin{itemize}
        \item 
                Given two monomials $\p^p$ and $\p^{q}$,
                a \emph{term order} $\prec$ discerns which of them is the largest one.
                We refer to \cite{dojo} for the technical definition,
                and content ourselves with giving an example,
                namely the \emph{graded reverse lexicographic order}%
                \footnote{
                        The speed with which Gr\"obner bases are calculated 
                        can heavily depend on the chosen term order
                        (of which there are many).
                        The graded reverse lexicographic order tends to be the fastest.
                }.
                Setting $|p| = p_1 + \cdots + p_N$,
                then
                $
                        \p^p \prec \p^q
                $
                iff
                \begin{align*}
                        |p| < |q|
                        \quadit{\text{or}}
                        |p| = |q|
                        \ \text{and the \emph{last} non-zero entry of} \ 
                        q-p \in \ZZ^N
                        \ \text{is \emph{negative}.}
                \end{align*}
                So the monomial with the largest degree wins,
                unless there is a tie,
                in which case the one with the smallest variable index wins.
                In the case of $N=3$ variables and degree less than 3,
                we have
                \eq{
                        1 
                        \prec 
                        \p_3
                        \prec 
                        \p_2
                        \prec
                        \p_1
                        \prec
                        \p_3^2 
                        \prec 
                        \p_2 \p_3 
                        \prec 
                        \p_1 \p_3 
                        \prec
                        \p_2^2
                        \prec
                        \p_1 \p_2
                        \prec
                        \p_1^2
                        \, .
                }
        \item  
                Consider an operator
                $
                        L = 
                        \sum_{q \in \supp{L}}
                        c_q(z) \, \p^q
                        \in \mR
                $
                in normally ordered form.
                Fixing a term order,
                the \emph{initial monomial}%
                \footnote{
                        The similarity between this notation and $\mathrm{in}_{(u,v)}$ from 
                        \eqref{principal_symbol} is not coincidental.
                        There is indeed a strong connection between comparing monomials according to weights 
                        $
                                (u,v) \in \RR^{2N}
                        $
                        and according to term orders $\prec$ \cite[Chapter 1]{SST}.
                }
                $\mathrm{in}_\prec(L)$ is defined as the largest monomial $\p^q$
                (stripped of its rational function $c_q(z) \in \CC(z)$)
                w.r.t.~$\prec$.
        \item 
                The \emph{initial ideal} of $\mI$ is the ideal generated by its initial monomials:
                \eq{
                        \label{initial_ideal}
                        \mathrm{in}_\prec(\mI) =
                        \CC \cdot \big\{
                                \mathrm{in}_\prec(L) 
                                \ | \
                                L \in \mI
                        \big\}
                        \ \subset \
                        \CC[\p_1, \ldots, \p_N]
                        \, .
                }
\end{itemize}

\begin{definition}
        A finite subset 
        $
                G = \{g_1, \ldots, g_M\} \in \mR
        $
        is called a \emph{Gr\"obner basis} for the ideal $\mI$ w.r.t.~the term order $\prec$ iff
        \begin{enumerate}
                \item 
                        $G$ generates $\mI$,
                        i.e.~$\langle g_1, \ldots, g_M \rangle = \mI$.
                \item 
                        $
                                \mathrm{in}_\prec(\mI) =
                                \langle
                                        \mathrm{in}_\prec(g_1)
                                        ,\ldots,
                                        \mathrm{in}_\prec(g_M)
                                \rangle.
                        $
        \end{enumerate}
        A Gr\"obner basis containing the minimal number of generators $g_i$ is called \emph{reduced},
        and is unique for a given term order.
        A monomial $\p^q$ is called a \emph{standard monomial} w.r.t.~$G$ 
        when it does \emph{not} belong to $\mathrm{in}_\prec(\mI)$.
\end{definition}

Absent any other methods,
the standard monomials only become known at the end of a Gr\"obner basis computation.

We can now answer the question posed in the very beginning of this section:
there are exactly $\rank{\mI}$ many standard monomials,
and they form an operator basis for the $\mD$-module $\mD / \mI$!

Gr\"obner basis computations are typically performed using Buchberger's algorithm%
\footnote{
        Bruno Buchberger was in fact himself the inventor of Gr\"obner bases,
        but he named them after his doctoral supervisor Wolfgang Gr\"obner as a sign of respect.
} \cite{dojo}.
The intensity of such calculations grows significantly with the number of variables.
For this reason,
the main theme of the next chapter will be to develop a more efficient algorithm that avoids Gr\"obner bases.

\section{Pfaffian systems}
\label{sec:Pfaffian_systems}

Let $\mI$ be a holonomic ideal born from a system of PDEs,
and suppose $f=f(z)$ is a solution to this system.
Assume a basis of standard monomials is known for the $\mR$-module $\mR/\mI$.
It is here denoted by
\eq{
        \Std = \{\p^{q_1},  \p^{q_2}, \ldots, \p^{q_R}\}
        \, ,
}
where $R := \rank{\mI}$ and each $q_i \in \ZZ^N_{\geq 0}$.
Without loss of generality, 
we may set $\Std_1 = 1$.
A basis of \emph{functions} for the PDE solution space is then given by the action of each
standard monomial on $f$:
\eq{
        e = 
        \arr{c}{
                \Std_1 \bullet f = f \\
                \Std_2 \bullet f     \\
                \vdots            \\
                \Std_R \bullet f
        }
        \, .
        \label{e_basis}
}
We are interested in finding a set of first-order PDEs satisfied by $e$.
To this end,
consider the Weyl algebra element $\p_i \Std_j$.
Since the standard monomials form a basis for $\mR / \mI$,
it must be possible to write
\eq{
        \p_i \Std_j =
        \sum_{k=1}^R p_{ij}^k(z) \, \Std_k
        \quad
        \text{mod} \ \mI
        \, ,
        \quad p_{ij}^k(z) \in \CC(z)
        \, .
        \label{pi_sj}
}
The notation $"\text{mod} \ \mI"$ means that this equation holds modulo the ideal $\mI$,
in the sense of \eqref{modulo_I}.
Most of the time we leave the $"\text{mod} \ \mI"$ as implicit.

Combining \eqref{e_basis} with \eqref{pi_sj},
we then have
\eq{
        \p_i \bullet e 
        =
        \arr{c}{
                (\p_i \Std_1) \bullet f \\
                (\p_i \Std_2) \bullet f \\
                \vdots               \\
                (\p_i \Std_R) \bullet f
        } 
        =
        P_i(z) \cdot e
        \quad
        \text{mod} \ \mI
        \, , \quad
        i = 1, \ldots, N
        \, ,
        \label{pfaffian_system}
}
where $P_i(z)$ is an $R \times R$ matrix whose $(j,k)$th entry equals the rational function $p_{ij}^k(z)$.
The set of first-order PDEs \eqref{pfaffian_system} is dubbed a \emph{Pfaffian system}.
By manipulating the standard monomials,
it can be shown \cite{dojo} that the \emph{Pfaffian matrices} $P_i$ satisfy the 
\emph{integrability condition}
\eq{
        \p_i \bullet P_j + P_i \cdot P_j =
        \p_j \bullet P_i + P_j \cdot P_i
        \, .
        \label{integrability}
}

The next chapter will be dedicated to deriving Pfaffian systems from the point of view of $\mD$-modules.
Such a calculation requires two steps:
\begin{enumerate}
        \item 
                Find a set of standard monomials $\Std$.
        \item 
                Derive the decomposition \eqref{pi_sj}.
\end{enumerate}
Both steps are possible when a Gr\"obner basis is known
(the second step is then performed via an algorithm analogous to that of polynomial division).
However,
as mentioned earlier,
we shall seek to avoid Gr\"obner bases as their computation often requires huge resources.
As an aside,
if a Pfaffian system is known by other means,
then one can immediately construct a Gr\"obner basis too \cite[Appendix B]{Chestnov:2022alh}.

We end this section with the following observation:
if $f$ were a single FI,
then \eqref{pfaffian_system} is exactly the kind of Pfaffian systems 
that we studied in \textbf{Part I} of this thesis! 
The decomposition of $\p_i \Std_j$ in \eqref{pi_sj} is then analogous to an IBP reduction.

        \chapter{Macaulay Matrices}
\label{sec:macaulay_matrices}

Based on \cite{Chestnov:2022alh},
this chapter presents a novel algorithm for the computation of Pfaffian systems.
It avoids the use of Gr\"obner bases in favor of linear algebra methods,
making it quite efficient.

Throughout this chapter we assume that
\eq{\text{
        a set of standard monomials $\Std = \{\p^{q_i}\}_{i=1,\ldots,R}$ is given,
}}
where the cardinality $|\Std| = R$ equals the holonomic rank of a given left ideal $\mI$.
While this is a strong assumption,
it will actually be fulfilled for the so-called GKZ system that we study in \chapref{ch:gkz}.

\section{From Pfaffian to Macaulay matrix}

Let 
$
        \mI = \langle d_1, \ldots, d_D \rangle
$
be a holonomic ideal generated by a collection of $D$ operators $d_i \in \mR$.
In this section we suppose to know the Pfaffian matrices $P_i$ associated to $\mI$.
It holds that
\eq{
        \p_i \Std_j =
        \sum_{k=1}^R (P_i)_{jk} \, \Std_k
        \quadit{\text{in}}
        \mR/ \mI
        \, .
        \label{pi_sj_in_R/I}
}
This expression can be lifted from the quotient ring $\mR / \mI$ to the whole ring $\mR$ by writing
\eq{
        \p_i \Std_j =
        \sum_{k=1}^R (P_i)_{jk} \, \Std_k 
        \plus
        \sum_{k=1}^D \Delta_{jk} \, d_k
        \quadit{\text{in}}
        \mR
        \, ,
        \label{pi_sj_in_R}
}
where each $\Delta_{jk} \in \mR$.
Our short-term goal is to aptly rewrite the second sum above, 
which will culminate in \eqref{C_M_Mons}.
We warn that this will require some index gymnastics,
but the final result does simplify.
To ease the notation in intermediate expressions,
we shall keep the index $i$ fixed
(otherwise we would have to write $\Delta_{ijk}$ etc.).
The index $i$ will be reinstated at the very end.

To start, 
we bring the operator $\Delta_{jk}$ into normally ordered form:
\eq{
        \label{Delta_jk}
        \Delta_{jk} \hspace{0.2cm} =
        \sum_{q \, \in \, \supp{\Der}}
        \d_{jkq}(z) \, \p^q
        \, , \quad
        \p^q \in \Der
        \, ,
}
where 
$
        \d_{jkq} \in \CC(z)^{R \times D \times |\Der|}
$
is a newly introduced rank-3 tensor of rational functions.
The set $\Der$ contains all the derivative monomials appearing on the RHS.

If we substitute the operators $\Delta_{jk}$ from 
\eqref{Delta_jk} into the second sum of \eqref{pi_sj_in_R},
then the monomials $\p^q \in \Der$ will act on the generators $d_i$ 
(which potentially contain rational functions in $z$ as prefactors).
In normally ordered form,
this action is written as
\eq{
        \label{pq_Lk}
        \p^q \, d_k =
        \sum_{p \, \in \, \supp{\Mons}}
        M_{qkp}(z) \, \p^p
        \, , \quad
        \p^p \in \Mons
        \, .
}
Here we introduced a rank-3 tensor of rational functions
$
        M_{qkp} \in \CC(z)^{|\Der| \times D \times |\Mons|}
$
and defined a set $\Mons$ whose elements consists of all the monomials $\p^p$ appearing on the RHS.
Upon concatenation of the first two indices $q$ and $k$, 
the rank-3 tensor $M_{qkp}$ turns into the \emph{Macaulay matrix} \cite{ohara2015pfaffian}
\eq{
        \label{Macaulay_matrix_definition}
        M_{op}(z) := M_{(qk) p}(z) \in \CC^{(|\Der| \cdot D) \times |\Mons|}
        \, .
}
The combined index $o = (qk) = (kq)$ runs over the Cartesian product of sets
$
        \supp{\Der} \times \{1,\ldots,D\}.
$

Plugging equations
\eqref{Delta_jk}, \eqref{pq_Lk} and \eqref{Macaulay_matrix_definition} 
into the second sum of \eqref{pi_sj_in_R},
this term now takes the form of a triple matrix product:
\eq{
        \sum_{k=1}^D \Delta_{jk} \, d_k 
        \quadit{=}
        \sum_{o, \, p}
        \d_{jo}(z) \, M_{op}(z) \, \p^p 
        \quadit{=}
        \big(
                C \cdot M \cdot \Mons
        \big)_j
        \, .
        \label{C_M_Mons}
}
$
        C \in \CC(z)^{R \times (|\Der| \cdot D)}
$
stands for the coefficient matrix 
$
        C_{jo} = \d_{j (kq)}(z),
$
when $\Der$ is regarded as a column vector in \eqref{Delta_jk}.
The set $\Mons$ is here also regarded as a column vector.

Some of the monomials in $\Mons$ may be standard,
but all need not be.
The non-standard monomials in $\Mons$ shall henceforth be called \emph{exterior monomials}:
\eq{
        \Ext = \Mons \setminus \Std
        \, .
}
The disjoint union
\eq{
        \Mons = \Ext \ \sqcup \ \Std
}
then induces a partition on the columns of the Macaulay matrix:
\eq{
        \label{M_Std_Ext}
        M = 
        \Big[ \ M_\Ext & \ \Big | \ M_\Std \ \Big]
        \, ,
        \quadit{\text{with}}
        \\[8pt]
        M_\Ext \in \CC(z)^{(|\Der| \cdot D) \times |\Ext|}
        \qquad & \text{and} \qquad
        M_\Std \in \CC(z)^{(|\Der| \cdot D) \times |\Std|}
        \nonumber
}
Thinking of the set $\Std$ as a column vector,
we further decompose the LHS of \eqref{pi_sj_in_R} as
\eq{
        \label{C_Std_Ext}
        &
        \p_i \Std 
        \quad = \quad  
        \Big[ \ C_\Ext \ \Big | \ C_\Std \ \Big]
        \cdot \Mons 
        \quad = \quad 
        C_\Ext \cdot \Ext \plus C_\Std \cdot \Std
        \, ,
        \quadit{\text{with}}
        \\[8pt] & 
        \hspace{2.2cm}
        C_\Ext \in \{0,1\}^{|\Std| \times |\Ext|}
        \qquad \text{and} \qquad
        C_\Std \in \{0,1\}^{|\Std| \times |\Std|}
        \, .
        \nonumber 
}
In other words,
$C_\Ext$ and $C_\Std$ are two binary matrices that encode whether the monomials in
$
        \p_i \Std
$
are exterior or standard.

Finally,
inserting equations
\eqref{C_M_Mons}, \eqref{M_Std_Ext} and \eqref{C_Std_Ext}
into the Pfaffian equation \eqref{pi_sj_in_R},
we arrive at
\eq{
        (C_\Ext - C \cdot M_\Ext) \plus
        (C_\Std - C \cdot M_\Std) \quad = \quad
        P_i \cdot \Std
        \, .
}
But, 
because $\Ext$ and $\Std$ are linearly independent in $\mR$,
this is equivalent to the system
\begin{empheq}[box=\fbox]{align}
        \label{MM_Ext_eq}
        C_\Ext\supbrk{i} - C\supbrk{i} \cdot M_\Ext &= 0
        \\[5pt]
        C_\Std\supbrk{i} - C\supbrk{i} \cdot M_\Std &= P_i \, .
        \label{MM_Std_eq}
\end{empheq}
Here we also reinstated the dependence on the index $i$,
as promised earlier in this section.
These two matrix equations are central to the algorithm for computing Pfaffian matrices
that we propose in the next section.

\section{From Macaulay to Pfaffian matrix}

Now we reverse the logic in the derivation leading to equations 
\eqref{MM_Ext_eq} and \eqref{MM_Std_eq}.
Although we started by assuming to know the Pfaffian system \eqref{pi_sj_in_R/I},
we now regard it as the goal.
Supposing that
$
        C_\bullet\supbrk{i} 
        \ \text{and} \ 
        M_\bullet
$
in \eqref{MM_Ext_eq} and \eqref{MM_Std_eq} are easily calculable ($\bullet = \Ext, \, \Std$),
then $C\supbrk{i}$ and $P_i$ are the only unknowns.
The purported strategy is this:
\begin{align*}
        \text{solve \eqref{MM_Ext_eq} for the rational matrix $C\supbrk{i}$}
        \implies
        \text{insert $C\supbrk{i}$ into \eqref{MM_Std_eq} to obtain $P_i$.}
\end{align*}
Let us formalize this strategy into a proper algorithm.
To begin,
define the set of all derivative monomials bounded by a degree $Q \in \ZZ_{\geq 0}$:
\eq{
        \Der_Q =
        \{ \p^q \}_{|q| \leq Q}
        \, .
}
In analogy with $\Mons$ from \eqref{pq_Lk},
we let $\Mons_Q$ denote the derivative monomials appearing in the normally order form of the set
$
        \big\{ \p^q \, d_k \big\}_{\p^q \, \in \, \Der_Q}^{k = 1, \ldots, D} .
$
The Macaulay matrix $M_Q$ of degree $Q$ is defined as in \eqref{Macaulay_matrix_definition},
but with $\Der$ and $\Mons$ replaced with $\Der_Q$ and $\Mons_Q$.
It thus holds that
\eq{
        \label{Macaulay_matrix_degree_Q}
        \Big\{ \p^q \, d_k \Big\}_{\p^q \, \in \, \Der_Q}^{k = 1, \ldots, D} 
        \quad &= \quad
        \sum_{p \, \in \, \supp{\Mons_Q}} 
        (M_Q)_{(qk)p} \, \p^p 
        \\[4pt]
        \quad &= \quad
        M_Q \cdot \Mons_Q
        \, .
        \label{Macaulay_matrix_degree_Q_2}
}
The set on the LHS of \eqref{Macaulay_matrix_degree_Q} 
is here viewed as a column vector of operators with rational function coefficients,
and $M_Q=M_Q(z)$ is the coefficient matrix of the monomial derivatives $\Mons_Q$.
The set $\Der_Q$ can be thought of as "seeding" the Macaulay matrix.

The idea is now to adjust the degree $Q$ until \eqref{MM_Ext_eq} admits a solution for $C\supbrk{i}$:
\begin{algorithm}
    \begin{tabbing}
        \underline{Input}: 
        Standard monomials $\Std$, 
        generators $\{d_1, \ldots d_D\}$ of a holonomic ideal $\mI$, 
        direction $i$.
        \\[6pt]
        \underline{Output}: 
        Pfaffian matrix $P_i$.
    \end{tabbing}
    \begin{algorithmic}[1]
        \State $ Q = 0 $
        \State $ M = M_0 = \big[ \ M_\Ext \ \big | \ M_\Std \ \big] $
        \State $\Mons = \Mons_0$
        \While{
            $C_\Ext\supbrk{i} = 
            C\supbrk{i} \cdot M_\Ext$ 
            is not solvable w.r.t.~$C\supbrk{i}$
        }
            \State $ Q++ $
            \State $ M \leftarrow M_Q $
            \State $ \Mons \leftarrow \Mons_Q $
        \EndWhile
        \State Solve $ C_\Ext\supbrk{i} = C\supbrk{i} \cdot M_\Ext $ for $C\supbrk{i}$
        \\
        \Return $ P_i = C_\Std\supbrk{i} - C\supbrk{i} \cdot M_\Std $
    \end{algorithmic}
    \caption{: Pfaffian matrix by the Macaulay matrix method}
    \label{alg:Pfaffian_by_MM}
\end{algorithm}

\noindent
There are a couple of immediate optimizations to this algorithm.
First,
notice that \eqref{MM_Ext_eq} has a solution when the rows of $C_\Ext\supbrk{i}$ 
lie in the row space of $M_\Ext$.
The condition in the \textbf{while} statement in line 4 of \algref{alg:Pfaffian_by_MM} 
then translates into the requirement
\eq{
        \mathrm{rank}
        \arr{c}{M_\Ext \\ \hline C_\Ext\supbrk{i}}
        =
        \mathrm{rank}
        \big[ M_\Ext \big]
        \, .
}
It is quick to check this equality by setting every parameter 
in the problem to a number in $\QQ$ or a finite field 
$
        \FF_p = \ZZ / p \ZZ
$
where $p$ is prime.

Second,
to solve the system \eqref{MM_Ext_eq},
it often suffices to pick only a subset of the rows in $M_\Ext$.
The reason is that some rows may be linearly dependent.
Such a subset can be identified by replacing parameters with numbers and row reducing.

Lastly,
since every matrix in the problem is defined over the rational function field $\CC(z)$,
it is vastly more efficient to employ \emph{rational reconstruction} 
algorithms rather than Gaussian elimination.
These algorithms reconstruct analytic expressions 
for rational functions by repeated evaluations over $\FF_p$.
One need not even reconstruct the intermediate "dummy matrix" $C\supbrk{i}$,
as we are ultimately only interested in the Pfaffian matrix $P_i$.
I.e.,
rational reconstruction can be postponed until all the matrix operations have been carried out in
\eqref{MM_Std_eq}.
This thesis employs the rational reconstruction software \package{FiniteFlow} \cite{Peraro:2019svx},
because it allows one to easily chain together several operations 
(such as matrix multiplication and addition of matrices),
with reconstruction postponed until the final step.
See \cite{Klappert:2019emp,Klappert:2020aqs,Magerya:2022hvj,Liu:2023cgs} for alternative implementations.

\begin{ex}
Consider the holonomic ideal
\eq{
        \mI = \langle z \p - \b \rangle
        \label{MM_example_1}
        \, ,
}
where 
$
        z_1 = z, \, \p_1 = \p
$
and $\b \in \CC$ is a parameter.
Although this example is too simple for an interesting application of the Macaulay matrix method,
we include it to showcase the many definitions in the above text.

To build the Macaulay matrix,
we must first choose the seeds $\Der_Q$ defined in \eqref{Macaulay_matrix_degree_Q}.
It turns out that $Q=0$ is a sufficient degree,
but for the sake of illustration let us pick $Q=1$.
Then
$
        \Der_Q = \{1, \p\}.
$
Following \eqref{Macaulay_matrix_degree_Q}, 
we apply the monomials in $\Der_Q$ to the generator of $\mI$:
\eq{
        & 1 (z \p - \b) = z \p - \b
        \\[5pt]
        & \p (z \p - \b) =
        \p z \p - \b \p =
        (1 + z \p) \p - \b \p =
        z \p^2 + (1 - \b) \p 
        \, .
}
The set $\Mons_Q$ is formed from all the derivative monomials appearing above:
\eq{
        \Mons_Q = 
        \{\p^2, \p, 1 \} = 
        \Ext \ \sqcup \ \Std = 
        \{\p^2, \p\} \ \sqcup \ \{1\}
        \, .
}
The Macaulay matrix can now be constructed as
\eq{
        \arr{c}{1 (z \p - \b) \\ \p (z \p - \b)} &=
        \arr{c|c}{M_\Ext & M_\Std}
        \cdot
        \Mons_Q
        \\ &=
        \arr{cc|c}{
                0 & z & -\b \\
                z & 1-\b & 0
        }
        \cdot
        \arr{c}{\p^2 \\ \p \\ 1}
        \, .
}
Next we build the two matrices $C_\Ext$ and $C_\Std$ as per \eqref{C_Std_Ext}:
\eq{
        \p \Std &=
        \p \big[ 1 \big] =
        \big[ \p \big] \\&=
        \arr{c|c}{C_\Ext & C_\Std} 
        \cdot
        \Mons_Q \\&=
        C_\Ext \cdot \Ext
        \plus
        C_\Std \cdot \Std \\&=
        \arr{cc}{0 & 1} 
        \cdot 
        \arr{c}{\p^2 \\ \p}
        \plus
        \big[ 0 \big]
        \cdot
        \big[ 1 \big]
        \, .
}
At this point we have all ingredients present in \eqref{MM_Ext_eq} and \eqref{MM_Std_eq}.
The first equation is
\eq{
        C_\Ext - C \cdot M_\Ext =
        \arr{cc}{0 & 1} -
        \arr{cc}{C_{11} & C_{12}} \cdot
        \arr{cc}{0 & z \\ z & 1-\b} =
        \arr{cc}{0 & 0} 
        \, .
}
Its solution is
$
        C = \arr{cc}{\frac{1}{z} & 0} 
        \, .
$
Inserting this into \eqref{MM_Std_eq},
we get
\eq{
        C_\Std - C \cdot M_\Std =
        \big[ 0 \big] -
        \arr{cc}{\frac{1}{z} & 0} \cdot
        \arr{c}{-\b \\ 0 } = 
        \left[ \frac{\b}{z} \right] =
        P
        \, ,
}
where $P$ is the Pfaffian matrix 
(albeit just a scalar in this case).
In other words,
we have derived that
\eq{
        \p \Std = 
        P \cdot \Std 
        \implies
        \big[ \p \big] =
        \left[ \frac{\b}{z} \right]
        \, ,
}
which is exactly the relation dictated by the ideal \eqref{MM_example_1}!
So we have obtained a trivial result,
but we can at least be satisfied that everything is self-consistent.
\end{ex}

\begin{ex}
\label{ex:2_F_1_macaulay}
Let
$
        \mI = \langle d_1, d_2, d_3, d_4 \rangle
$
be the ideal generated by
\eq{
        \nonumber
        \label{2_F_1_generators}
        d_1 &= z_1 \p_1 + z_2 \p_2 + z_3 \p_3 + z_4 \p_4 - \b_1
        \\ 
        d_2 &= z_2 \p_2 + z_4 \p_4 - \b_2
        \\
        \nonumber
        d_3 &= z_3 \p_3 + z_4 \p_4 - \b_3
        \\
        d_4 &= \p_1 \p_4 - \p_2 \p_3
        \, .
        \nonumber
}
In \secref{sec:integrand_rescaling},
we show that this is the annihilating ideal for Gauss' ${}_2F_1$ hypergeometric function.
The $\mD$-module $\mD/\mI$ turns out to have a basis of standard monomials given by
$
        \Std = \{\p_4, 1\}
$
(later on, in \secref{sec:basis_of_stds}, we explain how to easily find such a basis).
The goal of this example is to find the Pfaffian system in direction $z_4$:
\eq{
        \p_4 \Std =
        \p_4 \arr{c}{\p_4 \\ 1} =
        \arr{c}{\p_4^2 \\ \p_4} =
        P_4 \cdot \arr{c}{\p_4 \\ 1}
        \, .
}
To compute $P_4$,
we need to resolve the higher-order monomial $\p_4^2$ in terms of $\p_4$ and $1$.
But $\p_4^2$ does not appear in the generators $d_i$ from \eqref{2_F_1_generators},
so the seeding set $\Der_{Q=0} = \{1\}$ is insufficient.
It turns out that the degree $Q=1$ monomials
$
        \Der_1 = \{1, \p_1, \p_2, \p_3, \p_4\}
$ 
lead to a solvable system.

It is rather tedious to compute the Macaulay matrix by hand given this seed.
Instead,
we use the newly developed package \package{mt\_mm} written by 
Nobuki Takayama and Saiei J. Matsubara-Heo in the CAS called \package{asir}.
\package{asir} can be built from source by cloning the \package{git} repository
\href
{https://github.com/openxm-org/OpenXM}
{\texttt {\color{blue} https://github.com/openxm-org/OpenXM}}.
Source code and documentation for \package{mt\_mm} can be found in the directories
\begin{verbatim}
        OpenXM/src/asir-contrib/packages/src/mt_mm
        OpenXM/doc/asir-contrib/en.
\end{verbatim}
For the case at hand,
the Macaulay matrix is obtained by entering the following commands into \package{asir}%
\footnote{
        We call the generators \soft{Di} rather than \soft{di}
        since most lower-case objects are name-protected in \package{asir}.
}:
\begin{lstlisting}[style=mystyle]
D1 = z1*dz1 + z2*dz2 + z3*dz3 + z4*dz4 - b1;
D2 = z2*dz2 + z4*dz4 - b2;
D3 = z3*dz3 + z4*dz4 - b3;
D4 = dz1*dz4 - dz2*dz3;
Ideal = [D1, D2, D3, D4];
Std = [dz4, 1];
mt_mm.find_macaulay(Ideal, Std, [z1, z2, z3, z4]);
\end{lstlisting}
The output is a list
$
        \{M_\Ext, M_\Std, \Ext, \Std\}.
$
Explicitly,
we obtain the following data:
\newpage
{\footnotesize
\begin{equation*}
\rotatebox{-90}{$
M_\Ext =
\left[
\begin{array}{c||ccccccccccccccccccccc}
\Ext \to & \p_1 \p_2 \p_3 & \p_2^2 \p_3 & \p_2 \p_3^2 & \p_1^2 \p_4 & \p_1 \p_2 \p_4 & \p_1 \p_3 \p_4 & \p_2 \p_3 \p_4 & \p_1 \p_4^2 & \p_1^2 & \p_1 \p_2 & \p_2^2 & \p_1 \p_3 & \p_2 \p_3 & \p_3^2 & \p_1 \p_4 & \p_2 \p_4 & \p_3 \p_4 & \p_4^2 & \p_1 & \p_2 & \p_3 \\ \hline \hline
 1 d_1 & \mzero & \mzero & \mzero & \mzero & \mzero & \mzero & \mzero & \mzero & \mzero & \mzero & \mzero & \mzero & \mzero & \mzero & \mzero & \mzero & \mzero & \mzero & z_1 & z_2 & z_3 \\ \hline
 1 d_2 & \mzero & \mzero & \mzero & \mzero & \mzero & \mzero & \mzero & \mzero & \mzero & \mzero & \mzero & \mzero & \mzero & \mzero & \mzero & \mzero & \mzero & \mzero & \mzero & z_2 & \mzero \\ \hline
 1 d_3 & \mzero & \mzero & \mzero & \mzero & \mzero & \mzero & \mzero & \mzero & \mzero & \mzero & \mzero & \mzero & \mzero & \mzero & \mzero & \mzero & \mzero & \mzero & \mzero & \mzero & z_3 \\ \hline
 1 d_4 & \mzero & \mzero & \mzero & \mzero & \mzero & \mzero & \mzero & \mzero & \mzero & \mzero & \mzero & \mzero & 1 & \mzero & -1 & \mzero & \mzero & \mzero & \mzero & \mzero & \mzero \\ \hline \hline
 \p_1 d_1 & \mzero & \mzero & \mzero & \mzero & \mzero & \mzero & \mzero & \mzero & \mzero & \mzero & \mzero & \mzero & \mzero & \mzero & z_1 & z_2 & z_3 & z_4 & \mzero & \mzero & \mzero \\ \hline
 \p_1 d_2 & \mzero & \mzero & \mzero & \mzero & \mzero & \mzero & \mzero & \mzero & \mzero & \mzero & \mzero & \mzero & \mzero & \mzero & \mzero & z_2 & \mzero & z_4 & \mzero & \mzero & \mzero \\ \hline
 \p_1 d_3 & \mzero & \mzero & \mzero & \mzero & \mzero & \mzero & \mzero & \mzero & \mzero & \mzero & \mzero & \mzero & \mzero & \mzero & \mzero & \mzero & z_3 & z_4 & \mzero & \mzero & \mzero \\ \hline
 \p_1 d_4 & \mzero & \mzero & \mzero & \mzero & \mzero & \mzero & 1 & -1 & \mzero & \mzero & \mzero & \mzero & \mzero & \mzero & \mzero & \mzero & \mzero & \mzero & \mzero & \mzero & \mzero \\ \hline \hline
 \p_2 d_1 & \mzero & \mzero & \mzero & \mzero & \mzero & \mzero & \mzero & \mzero & \mzero & \mzero & \mzero & z_1 & z_2 & z_3 & \mzero & \mzero & z_4 & \mzero & \mzero & \mzero & 1-\b_1 \\ \hline
 \p_2 d_2 & \mzero & \mzero & \mzero & \mzero & \mzero & \mzero & \mzero & \mzero & \mzero & \mzero & \mzero & \mzero & z_2 & \mzero & \mzero & \mzero & z_4 & \mzero & \mzero & \mzero & -\b_3 \\ \hline
 \p_2 d_3 & \mzero & \mzero & \mzero & \mzero & \mzero & \mzero & \mzero & \mzero & \mzero & \mzero & \mzero & \mzero & \mzero & z_3 & \mzero & \mzero & z_4 & \mzero & \mzero & \mzero & 1-\b_3 \\ \hline
 \p_2 d_4 & \mzero & \mzero & 1 & \mzero & \mzero & -1 & \mzero & \mzero & \mzero & \mzero & \mzero & \mzero & \mzero & \mzero & \mzero & \mzero & \mzero & \mzero & \mzero & \mzero & \mzero \\ \hline \hline
 \p_3 d_1 & \mzero & \mzero & \mzero & \mzero & \mzero & \mzero & \mzero & \mzero & \mzero & z_1 & z_2 & \mzero & z_3 & \mzero & \mzero & z_4 & \mzero & \mzero & \mzero & 1-\b_1 & \mzero \\ \hline
 \p_3 d_2 & \mzero & \mzero & \mzero & \mzero & \mzero & \mzero & \mzero & \mzero & \mzero & \mzero & z_2 & \mzero & \mzero & \mzero & \mzero & z_4 & \mzero & \mzero & \mzero & 1-\b_3 & \mzero \\ \hline
 \p_3 d_3 & \mzero & \mzero & \mzero & \mzero & \mzero & \mzero & \mzero & \mzero & \mzero & \mzero & \mzero & \mzero & z_3 & \mzero & \mzero & z_4 & \mzero & \mzero & \mzero & -\b_3 & \mzero \\ \hline
 \p_3 d_4 & \mzero & 1 & \mzero & \mzero & -1 & \mzero & \mzero & \mzero & \mzero & \mzero & \mzero & \mzero & \mzero & \mzero & \mzero & \mzero & \mzero & \mzero & \mzero & \mzero & \mzero \\ \hline \hline
 \p_4 d_1 & \mzero & \mzero & \mzero & \mzero & \mzero & \mzero & \mzero & \mzero & z_1 & z_2 & \mzero & z_3 & \mzero & \mzero & z_4 & \mzero & \mzero & \mzero & 1-\b_1 & \mzero & \mzero \\ \hline
 \p_4 d_2 & \mzero & \mzero & \mzero & \mzero & \mzero & \mzero & \mzero & \mzero & \mzero & z_2 & \mzero & \mzero & \mzero & \mzero & z_4 & \mzero & \mzero & \mzero & -\b_3 & \mzero & \mzero \\ \hline
 \p_4 d_3 & \mzero & \mzero & \mzero & \mzero & \mzero & \mzero & \mzero & \mzero & \mzero & \mzero & \mzero & z_3 & \mzero & \mzero & z_4 & \mzero & \mzero & \mzero & -\b_3 & \mzero & \mzero \\ \hline
 \p_4 d_4 & 1 & \mzero & \mzero & -1 & \mzero & \mzero & \mzero & \mzero & \mzero & \mzero & \mzero & \mzero & \mzero & \mzero & \mzero & \mzero & \mzero & \mzero & \mzero & \mzero & \mzero 
\end{array}
\right]
$}
\end{equation*}}
\newpage

\begin{equation*}
M_\Std =  
\left[
\begin{array}{c||cc}
 \Std \to & \p_4 & 1 \\ \hline \hline
 1 d_1 & z_4 & -\b_1 \\ \hline
 1 d_2 & z_4 & -\b_3 \\ \hline
 1 d_3 & z_4 & -\b_3 \\ \hline
 1 d_4 & \mzero & \mzero \\ \hline \hline
 \p_1 d_1 & 1-\b_1 & \mzero \\ \hline
 \p_1 d_2 & 1-\b_3 & \mzero \\ \hline
 \p_1 d_3 & 1-\b_3 & \mzero \\ \hline
 \p_1 d_4 & \mzero & \mzero \\ \hline \hline
 \p_2 d_1 & \mzero & \mzero \\ \hline
 \p_2 d_2 & \mzero & \mzero \\ \hline
 \p_2 d_3 & \mzero & \mzero \\ \hline
 \p_2 d_4 & \mzero & \mzero \\ \hline \hline
 \p_3 d_1 & \mzero & \mzero \\ \hline
 \p_3 d_2 & \mzero & \mzero \\ \hline
 \p_3 d_3 & \mzero & \mzero \\ \hline
 \p_3 d_4 & \mzero & \mzero \\ \hline \hline
 \p_4 d_1 & \mzero & \mzero \\ \hline
 \p_4 d_2 & \mzero & \mzero \\ \hline
 \p_4 d_3 & \mzero & \mzero \\ \hline
 \p_4 d_4 & \mzero & \mzero 
\end{array}
\right]
\, .
\end{equation*}

\noindent
The 21 columns in $M_\Ext$ are labeled by the exterior monomials $\Ext$.
The two columns in $M_\Std$ are labeled by $\Std$.
The $20$ rows in $M_\Ext$ and $M_\Std$ are labeled by the seeds $\Der_1$ acting on the generators $d_i$.

Rows of the matrix $M_\Ext$ dictate relations among the external monomials.
One notices,
however,
that some of the rows are linearly dependent on each other
(this stems from an overzealous choice of the set 
$
        \big\{ \p^q d_k \big\}       
$ 
from \eqref{Macaulay_matrix_degree_Q},
as only a subset of the seeds would suffice).
Row reduction of $M_\Ext$ reveals that the $17$ rows numbered by
$
        \{1, \ldots, 14, 16, 17, 20\}
$
are independent.
Let $M_\Ext'$ be the $17 \times 21$ matrix built from these rows.
The $17 \times 2$ matrix $M_\Std'$ is defined in a similar manner.
This row reduction can naturally be performed by fixing parameters to numbers for more complicated examples.

Next we construct the two matrices $C_\Ext\supbrk{4}$ and $C_\Std\supbrk{4}$ according to \eqref{C_Std_Ext},
finding that
\eq{
        C_\Ext\supbrk{4} &=
        \arr{ccccccccccccccccccccc}{
                \mzero & \mzero & \mzero & \mzero & \mzero & \mzero & \mzero & \mzero & 
                \mzero & \mzero & \mzero & \mzero & \mzero & \mzero & \mzero & \mzero & 
                \mzero & 1 & \mzero & \mzero & \mzero \\
                \mzero & \mzero & \mzero & \mzero & \mzero & \mzero & \mzero & \mzero & 
                \mzero & \mzero & \mzero & \mzero & \mzero & \mzero & \mzero & \mzero & 
                \mzero & \mzero & \mzero & \mzero & \mzero \\
        }
        \\
        C_\Std\supbrk{4} &=
        \arr{cc}{
                \mzero & \mzero \\
                1 & \mzero
        }
        \, .
}
Now we can employ the central equations \eqref{MM_Ext_eq} and \eqref{MM_Std_eq}:
The two-step process is
\begin{enumerate}
        \item 
                Define an unknown matrix $C\supbrk{4}$ of size $R \times S$
                where
                $
                        R = |\Std| = 2
                $
                and 
                $
                        S = \#\{\text{rows in } M_\Ext'\} = 17.
                $
                This matrix is determined by solving
                \eq{
                        C_\Ext\supbrk{4} - C\supbrk{4} \cdot M_\Ext' = 0
                        \, ,
                }
                which can be done instantaneously with e.g.~\package{Mathematica}.
        \item 
                The solution for $C\supbrk{4}$ is inserted into 
                \eq{
                        C_\Std\supbrk{4} - C\supbrk{4} \cdot M_\Std' = P_4
                }
                to obtain the Pfaffian matrix $P_4$.
\end{enumerate}

\noindent
The resulting Pfaffian system in direction $z_4$ works out to
\eq{
        \p_4 \arr{c}{\p_4 \\ 1} =
        \left[
        \begin{array}{cc}
                \frac{-\b_1+\b_3+\b_3-1}{z_4} + \frac{\b_1 z_1}{z_1 z_4-z_2 z_3} & 
                \frac{\b_3 \b_3 z_1}{z_4(z_2 z_3-z_1 z_4)} \\ 
                1 & 0 \\ 
        \end{array}
        \right]
        \cdot \arr{c}{\p_4 \\ 1}
        \, .
}
In particular, 
the first row of this Pfaffian equation says that
\eq{
        \p_4^2 = 
        \frac{z_2 z_3 (\b_1-\b_3-\b_3+1)+z_1 z_4 (\b_3+\b_3-1)}{z_4 (z_1 z_4-z_2 z_3)} \p_4
        -
        \frac{\b_3 \b_3 z_1}{z_4 (z_1 z_4-z_2 z_3)} 1
        \quad \text{mod} \
        \mI \, ,
        \nonumber
}
which,
in physicist terms,
could be thought of as an "IBP reduction" of $\p_4^2$ in terms of the "MIs" $\p_4$ and $1$.

Pfaffian matrices in the directions $\{z_1, z_2, z_3\}$ are found in a similar manner.
\end{ex}

        \chapter{GKZ Systems}
\label{ch:gkz}

\epigraph
{\itshape 
        Among the Euler type integrals ... there are the integrals 
        $\int \Pi P_i(t_1,\ldots,t_n)^{\a_i} \ \times$ 
        $t_1^{\b_1} \ldots t_n^{\b_n} \, \dd t_1 \ldots \dd t_n$,
        where $P_i$ are polynomials, i.e.~practically all integrals which arise in quantum field theory.
}
{
        Gel'fand, Kapranov, Zelevinsky ('89) \cite{GKZ_1}
}

\noindent
A GKZ system is a particular $\mD$-module $\mD / H_A(\b)$,
where $H_A(\b)$ is an annihilating ideal built from an integer matrix $A$ and a collection of
complex parameters $\b_i$.
We have secretly already shown examples of generators for GKZ systems in equations
\eqref{GKZ_system_for_roots}, \eqref{MM_example_1} and \eqref{2_F_1_generators}.

Why study this specific $\mD$-module?
As indicated by the quote above,
GKZ systems can be related to FIs when they are written in 
parametric representations rather than in momentum space
(the connection between GKZ systems and FIs will be explained in the next chapter).
What's more,
it is easy to determine the annihilating ideal $H_A(\b)$ as well as a basis of standard monomials,
which is computationally advantageous.

        \section{Euler integrals}

To define the GKZ system,
let us begin by considering the class of functions annihilated by the generators of $H_A(\b)$.
These functions are called \emph{Euler integrals} \cite{Matsubara-Heo:2023ylc},
written as%
\footnote{
        The most general definition contains several polynomials $\{g_1, \ldots, g_l\}$ in the integrand,
        but for our inquiry a single polynomial $g(z|x)$ will suffice.
}
\eq{
        \label{Euler_integral}
        f_\b(z) = 
        \int_\Gamma 
        g(z|x)^{\b_0} x_1^{-\b_1} \cdots x_n^{-\b_n}
        \frac{\dd x}{x}
        \quadit{\text{with}}
        \frac{\dd x}{x} :=
        \frac{\dd x_1 \wedge \cdots \wedge \dd x_n}{x_1 \cdots x_n}
        \, .
}
Let us dwell on the building blocks of this integral.
\begin{enumerate}
        \item
                $\Gamma$ is some chosen integration cycle in $\CC^n$.
                For the purpose of this thesis we may pick
                $
                        \Gamma = (0,\infty)^n
                $
                (though to be rigorous one should also specify a Riemann sheet data, 
                as the integrand is multivalued;
                see \cite[Section 3.1]{Matsubara-Heo:2023ylc}).
        \item 
                $g(z|x)$ is a Laurent polynomial in the integration variables
                $
                        x = (x_1, \ldots, x_n)
                $
                with monomial coefficients
                $
                        z = (z_1, \ldots, z_N).
                $
                In multivariate exponent notation
                \eq{
                        g(z|x) =
                        \sum_{i=1}^N z_i \, x^{\a_i}
                        \, , \quad
                        \a_i \in \ZZ^n
                        \, .
                        \label{g_Euler_integrand}
                }
                Crucially,
                each monomial coefficient $z_i$ is regarded is an 
                \emph{independent} variable of the function $f_\b(z)$.
        \item 
                The exponents
                $
                        \b = (\b_0, \ldots, \b_n) \in \CC^{n+1}
                $
                are complex parameters.
                We generally assume that they are \emph{generic} real numbers,
                meaning that they should not be integral.
                This will be formulated more precisely as a \emph{non-resonance} condition later on.
\end{enumerate}
\noindent

By appending%
\footnote{
        This is called the \emph{Cayley trick} or the \emph{Cayley configuration}.
}
a "1" in front of each multi-index $\a_i$ from \eqref{g_Euler_integrand},
we define a collection of $N$ column vectors
$
        a_i = \big[1, \a_i \big]^T \in \ZZ^{n+1}.
$
These vectors are assumed to span $\ZZ^{n+1}$.
Further,
they give rise to an $(n+1) \times N$ dimensional matrix
\eq{
        A = 
        \arr{c|c|c}{
                a_1 & \ldots & a_N
        }
        \, .
        \label{GKZ_A_matrix}
}
The polynomial $g$ in the integrand of \eqref{Euler_integral} thus specifies the integer matrix $A$,
and vice versa.
The (left) kernel of $A$ is defined as
\eq{
        \ker{A} =
        \left\{
                (u_1, \, \ldots, \, u_N) \in \ZZ^N
                \ \Big | \
                \sum_{i=1}^N u_i \, a_i = \textbf{0}
        \right\}
        \, .
}
This can alternatively be viewed as a linear space spanned by finitely many integer vectors $u$ satisfying
$
        A \cdot u = \mathbf{0}
$.
Given a vector $u \in \ker{A}$,
we shall decompose it into positive and negative components by writing
$
        u = u^+ - u^-
$
for
$
        u^+, \, u^- \in \ZZ^N_{\geq 0}.
$

        \section{GKZ \texorpdfstring{$\mD$-module}{}}

The main characters of this chapter are the following operators parametrized by
the integer matrix $A$ and the complex parameters $\b$:
\eq{
        \label{GKZ_operator_Euler}
        E \quadit{&=} A \cdot \theta - \b
        \\[5pt]
        \Box_u \quadit{&=} \p^{u^+} - \p^{u^-}
        \, , \quad
        \forall \, u \in \ker{A}
        \, .
        \label{GKZ_operator_toric}
}
The vector
$
        \theta = \big[ \theta_1, \, \ldots, \, \theta_N \big]^T
$
is a collection of \emph{Euler operators} 
$
        \theta_i = z_i \, \p_i
        \, .
$
The $n+1$ operators $E_i$ are often called the \emph{homogeneity operators}
(cf.~\secref{sec:integrand_rescaling}).
The ideal generated by the operators $\Box_u$ is called the \emph{toric ideal}%
\footnote{
        The name stems from its relation to \emph{toric geometry},
        a field of algebraic geometry that studies varieties containing a torus as an open dense subset
        \cite{SST}.
}.

Quite remarkably,
$E$ and $\Box_u$ can be shown \cite{GKZ_2} to be the complete set of annihilating operators 
for the Euler integral \eqref{Euler_integral}:
\eq{
        \label{E_i_annihilator}
        E_i \bullet f_\b(z) \quadit{&=} 0
        \, , \quad
        i = 0, \ldots, n
        \\[5pt]
        \Box_u \bullet f_\b(z) \quadit{&=} 0
        \, , \quad
        \forall \, u \in \ker{A}
        \, .
        \label{Box_u_annihilator}
}
The \emph{GKZ system} \cite{GKZ_1,GKZ_book} 
associated to an Euler integral is defined as the $\mD$-module $\mD / H_A(\b)$,
where $H_A(\b)$ is the annihilating ideal generated by $E$ and $\Box_u$:
\eq{
        H_A(\b) 
        \quad = \quad
        \sum_{i=0}^{n} \mD \, E_i
        \ \ \plus
        \sum_{u \, \in \, \ker{A}} \mD \, \Box_u
        \, .
}

Given some complicated integral,
it might well be very hard to find the complete set of annihilating operators.
But for the case of Euler integrals having indeterminate $z$ variables,
we can apparently immediately write down all of those operators!
In the $\mD$-module philosophy,
we may thus study the GKZ system as a proxy for studying the Euler integral itself.

The GKZ system is holonomic \cite{Adolphson},
so it is endowed with the structure of a finite-dimensional vector space.
This means that we can use the Macaulay matrix method to construct first-order 
Pfaffian systems for Euler integrals given the higher-order operators that generate $H_A(\b)$.
This will be topic of \chapref{ch:gfi}.

\begin{ex}
Consider the Euler integral
\eq{
        f_\b(z) =
        \int_\Gamma
        \big(
                z_1 x_1 + z_2 x_2 + z_3 x_3 + z_4 x_1 x_2 + z_5 x_3^2
        \big)^{\b_0}
        x_1^{-\b_1}
        x_2^{-\b_3}
        x_3^{-\b_3}
        \frac{\dd x}{x}
        \, .
        \label{gkz_example_integral}
}
We seek to construct its GKZ system.
The first step is to input the exponent multi-indices
(with $1$'s appended)
into the columns of the $A$-matrix \eqref{GKZ_A_matrix}:
\eq{
        A =
        \arr{ccccc}{
                1 & 1 & 1 & 1 & 1 \\ \hline 
                1 & 0 & 0 & 1 & 0 \\
                0 & 1 & 0 & 1 & 0 \\
                0 & 0 & 1 & 0 & 2 
        }
        \, .
}
The 4th column,
for instance,
represents the monomial 
$
        z_4 x^{(1,1,0)} = z_4 x_1 x_2.
$
The operators $E_i$ from \eqref{GKZ_operator_Euler} can now be written as
\eq{
        \label{gkz_example_homogenous}
        E =
        A \cdot \theta - \b =
        \arr{c}{
                \theta_1 + \theta_2 + \theta_3 + \theta_4 + \theta_5 - \b_0 \\
                \theta_1 + \theta_4 - \b_1 \\
                \theta_2 + \theta_4 - \b_2 \\
                \theta_3 + 2 \theta_5 - \b_3 \\
        }
        \, .
}
The toric part \eqref{GKZ_operator_toric} of the GKZ system requires knowing $\ker{A}$.
This can be found by solving a linear system,
or by using one's favorite CAS
(e.g.~the command \soft{NullSpace} in \package{Mathematica}).
One finds that
$
        \ker{A} = \mathrm{span}(u),
$
where $u$ is a vector that we decompose as
\eq{
        u 
        \quadit{=}
        \arr{c}{
                1 \\ 1 \\ -2 \\ -1 \\ 1
        } 
        \quadit{=}
        u^+ - u^-
        \quadit{=}
        \arr{c}{
                1 \\ 1 \\ 0 \\ 0 \\ 1
        } 
        -
        \arr{c}{
                0 \\ 0 \\ 2 \\ 1 \\ 0
        } 
        \, .
}
The toric operator is therefore
\eq{
        \label{gkz_example_toric}
        \Box_u = \p^{u^+} - \p^{u^-} = \p_1 \p_2 \p_5 - \p_3^2 \p_4 
        \, .
}

In summary,
the annihilating ideal for the Euler integral \eqref{gkz_example_integral} is generated as 
$
        H_A(\b) = \langle E, \Box_u \rangle,
$
where $E$ and $\Box_u$ are given by \eqref{gkz_example_homogenous} and \eqref{gkz_example_toric}.
\end{ex}

        \section{Twisted de Rham cohomology}

The GKZ system can be related to another useful vector space for studying Euler integrals,
namely the \emph{twisted de Rham cohomology group}.
The elements of this group are,
loosely speaking, 
Euler inte\emph{grands}.
By defining this space we will discover concrete formulas for representing the Euler integral 
\eqref{Euler_integral} as a differential operator inside a $\mD$-module.

Twisted de Rham cohomology has recently garnered a lot of attention in the study of FIs
because it offers a novel and potentially more powerful framework for IBP reduction.
See e.g.~%
\cite{
        Mizera:2017rqa,
        Mastrolia:2018uzb,
        Frellesvig:2019kgj,
        Frellesvig:2020qot,
        Frellesvig:2021vem,
        Mizera:2020wdt,
        Mizera:2019vvs,
        Caron-Huot:2021xqj,
        Caron-Huot:2021iev,
        Chestnov:2022xsy,
        Gasparotto:2022mmp,
        Gasparotto:2023roh,
        De:2023xue,
        Gasparotto:2023cdl}
and references therein.

\subsection{"Traditional" de Rham cohomology}

Before discussing the "twisted" de Rham cohomology group,
it is instructive to first summarize the "traditional" theory.
The story begins with the celebrated theorem by Stokes:
\eq{
        \int_\G \dd \o =
        \int_{\p \G} \o
        \, .
        \label{Stokes_theorem}
}
$\o = h(x) \dd^k x$ is a differential $k$-form,
where $h(x)$ is some rational function,
and $\dd$ is the total derivative on integration variables acting as
\eq{
        \dd \o = 
        \frac{\p h(x)}{\p x_1} \dd x_1 \wedge \dd^k x + 
        \cdots + 
        \frac{\p h(x)}{\p x_n} \dd x_n \wedge \dd^k x
        \, .
} 
Note that $\dd$ sends a $k$-form to a $(k+1)$-form.

The notation $\p \G$ denotes the boundary of the $k$-dimensional integration contour $\G$
(see the classical textbook \cite{lee2003introduction} for a rigorous definition).
Note that the boundary is $(k-1)$-dimensional.

We focus on a particular class of integrals $\int_\G \o$ for which
\begin{align*}
        \begin{array}{cccccl}
                & \G \text{ is a \emph{cycle} }    & \iff & \hspace{0.5cm} \p \G  & = & 0
                \\
                & \o \text{ is a \emph{co-cycle} } & \iff & \hspace{0.5cm} \dd \o & = &  0 \, .
        \end{array}
\end{align*}
Consider now the following two scenarios:
\begin{enumerate}
        \item 
                Fix $\o$ and shift $\G$ by the boundary of another contour,
                $\G \to \G + \p \g$.
                It follows from Stokes' theorem that
                \eq{
                        \int_{\G + \p \g} \o =
                        \int_\G \o + \int_{\p \g} \o =
                        \int_\G \o + \cancelto{0}{\int_{\g} \dd \o} =
                        \int_\G \o
                        \, .
                }
                So,
                the value of the integral does not depend on any particular contour,
                but rather an \emph{equivalence class} $[\G]$ of contours,
                where $\G \sim \G'$ precisely when $\G = \G' + \p \g$.
        \item 
                Fix $\G$ and shift $\o$ by the total derivative of another differential form,
                $\o \to \o + \dd \psi$.
                Stokes' theorem implies that
                \eq{
                        \int_\G (\o + \dd \psi) =
                        \int_\G \o + \int_\G \dd \psi =
                        \int_\G \o + \cancelto{0}{\int_{\p \G} \psi} =
                        \int_\G \o
                        \, .
                }
                The numerical value of the integral is thus determined
                by an \emph{equivalence class} $[\o]$ of differential forms,
                where $\o \sim \o'$ iff $\o = \o' + \dd \psi$.
\end{enumerate}
It is therefore natural to study contours up to boundary terms,
and differential forms up to total derivatives.
Let's formalize these notions.

Define 
\eq{
        C_k = 
        \big \{ 
                \text{formal sums of $k$-dimensional cycles} \ \G
        \big\}
        \, .
}
The boundary map 
$
        \p_k : C_k \to C_{k-1} 
$
sends $k$-cycles to $(k-1)$-cycles.
An important result is that "the boundary of a boundary is zero":
\eq{
        \p_k \circ \p_{k+1} = 0 
        \quadit{\implies}
        \mathrm{Image}(\p_{k+1}) 
        \ \subset \
        \mathrm{Kernel}(\p_k) 
        \, .
}
It thus makes sense to define the $k$th \emph{homology group} as the quotient space
\eq{
        H_k = 
        \frac{\mathrm{Kernel}(\p_k)}{\mathrm{Image}(\p_{k+1})}
        \, .
}
The numerator indicates that we focus on $k$-cycles $\G$ for which $\p_k \G = 0$,
and the denominator means that we mod out by boundary terms of the form $\p_{k+1} \g$.
So this space indeed consists of the equivalence classes $[\G]$ discussed above.

Furthermore,
let
\eq{
        \O^k = 
        \big\{
                \text{formal sums of co-cycle $k$-forms} \ \o
        \big\}
        \, .
}
The total derivative
$
        \dd_k: \O^k \to \O^{k+1}
$
sends $k$-forms to $(k+1)$-forms.
It holds that
\eq{
        \dd_{k} \circ \dd_{k-1} = 0 
        \quadit{\implies}
        \mathrm{Image}(\dd_{k-1}) 
        \ \subset \
        \mathrm{Kernel}(\dd_k) 
        \, .
}
The $k$th \emph{de Rham cohomology group} is then defined by the quotient
\eq{
        H^k = 
        \frac{\mathrm{Kernel}(\dd_k)}{\mathrm{Image}(\dd_{k-1})}
        \, .
}
The numerator consists of co-cycles satisfying $\dd_k \o = 0$,
and the denominator mods out by total derivatives $\dd_{k-1} \psi$.
This space therefore consists of the equivalence classes $[\o]$ mentioned earlier.


\subsection{Twisted de Rham cohomology}
\label{sec:twisted_de_rham}

Now consider integrals of the form
\eq{
        \int_\G U \, \o_q
        \quad \text{where} \quad
        U = g(z|x)^{\b_0} \, x^{-\b'}
        \quad \text{and} \quad
        \o_q = g(z|x)^{-q_0} x^{q'} \frac{\dd x}{x}
        \, ,
        \label{U_omega_integral}
}
with 
$
        x^{\b'} = x_1^{\b_1} \cdots x_n^{\b_n}.
$
We collect the integer exponents of $\o_q$ into a vector
\eq{
        q = (q_0, \, q') \in \ZZ \times \ZZ^n
        \, .
}
In "traditional" de Rham cohomology,
we restricted ourselves to integrals of differential forms with rational coefficient functions.
Here there is a new ingredient:
the \emph{twist} $U$.
This function is multivalued because the exponents $\b = (\b_0, \b')$ are non-integer by stipulation.

In this setting,
Stokes' theorem \eqref{Stokes_theorem} says
\eq{
        \int_\G
        \dd \big[ U \, \o_q \big] =
        \int_{\p_U \G}
        U \, \o_q 
        \, .
        \label{U_omega_Stokes}
}
We have upgraded the usual boundary operator to $\p_U$ due to the multivaluedness of $U$.
$\p_U$ keeps track of the branch of $U$ because $\G$ might cross some branch cuts in its path;
we refer to \cite[Chapter 3.1]{Matsubara-Heo:2023ylc} for a proper definition.

On the LHS,
we can expand the integrand to get
\eq{
        \dd \big[ U \, \o_q \big] 
        &=
        U \dd \o_q + \dd U \wedge \o_q
        \\[3pt]&=
        U \underbrace{\left[ \dd  + \frac{\dd U}{U} \wedge \right]}_{\nabla} \o_q
        \, .
}
In the last line we defined the \emph{covariant derivative} $\nabla$.
Inserting the twist $U$ from \eqref{U_omega_integral},
it can be represented as
\eq{
        \nabla =
        \dd + 
        \b_0 \frac{\dd g(z|x)}{g(z|x)} \wedge -
        \sum_{i=1}^n \b_i \frac{\dd x_i}{x_i} \wedge
        \, .
}

We choose to focus on integrals $\int_\G U \o_q$ such that
\begin{align*}
        \begin{array}{cccccl}
                & \G \text{ is a \emph{twisted cycle}} & \iff & \hspace{0.5cm} \p_U \G & = & 0
                \\
                & \o_q \text{ is a \emph{twisted co-cycle}} & \iff & \hspace{0.5cm} \nabla \o_q & = &  0 \, .
        \end{array}
\end{align*}
Following the same logic as before,
we infer that the integral in question only depends on the equivalence classes $[\G]$ and $[\o_q]$,
where 
$\G \sim \G' \iff \G = \G' + \p_U \g$ 
and
$\o_q \sim \o_q' \iff \o_q = \o_q' + \nabla \psi$.

We won't need the twisted homology group in what follows,
so we skip that definition.
To define the \emph{twisted de Rham cohomology group},
we need to be careful in distinguishing between the $x$- and $z$-variables appearing in the integrand.
Let
\eq{
        X = 
        \left\{
                (z,x) \, \in \, \mathbb{A}^N \times (\CC^*)^n \ \big | \ g(z|x) \neq 0
        \right\}
        \quad \text{and} \quad
        Y = \mathbb{A}^N
        \, ,
}
where $\AA$ is the complex affine line 
(essentially $\CC$ without a distinguished choice of origin),
and $\CC^*$ is the complex torus
(which equals $\CC \setminus \{0\}$ as a set).
The space of holomorphic functions on $X$
(technically a \emph{sheaf})
is written as
\eq{
        \mO(X) = 
        \CC
        \Big[z_1, \ldots, z_N, x_1^{\pm 1}, \ldots, x_n^{\pm 1}, \frac{1}{g}\Big]
        \, .
        \label{holomorphic_functions_X}
}
For $1 \leq k \leq n$,
$\o_q$ from \eqref{U_omega_integral} should be thought of as a differential $k$-form in the $x$-variables
(i.e.~we allow for $\dd^k x$ but not $\dd^k z$),
with a coefficient function that lives in $\mO(X)$.
These objects are called \emph{relative $k$-forms} in the literature 
\cite[Section 1.5]{Hotta-Tanisaki-Takeuchi-2008} \cite[Section 4.1]{andre2020rham},
and they live in the space%
\footnote{
        The subscript $X/Y$ does not denote a quotient. 
        This is merely the standard notation for the space of relative $k$-forms.
}
\eq{
        \O^k_{X/Y} 
        \quad =
        \bigoplus_{\substack{K \, \subset \, \{1, \ldots,n\} \\ |K| \, = \, k}}
        \mO(X) \, \dd x^K
        \, .
}
The covariant derivative
$
        \nabla_k: \O^k_{X/Y} \to \O^{k+1}_{X/Y}
$
thus sends a relative $k$-form to a relative $(k+1)$-form.
A calculation shows that 
\eq{
        \nabla_{k} \circ \nabla_{k-1} = 0 
        \quadit{\implies}
        \mathrm{Image}(\nabla_{k-1}) 
        \ \subset \
        \mathrm{Kernel}(\nabla_k) 
        \, .
}
The $k$th \emph{twisted de Rham cohomology group} can at last be defined as
\eq{
        \deRham = 
        \frac{\mathrm{Kernel}(\nabla_k)}{\mathrm{Image}(\nabla_{k-1})}
        \, .
        \label{twisted_cohomology_group}
}
We have emphasized the dependence on the $A$-matrix and the $\b$-parameter vector,
as they are used to build the integrand \eqref{U_omega_integral}.

$\deRham$ is endowed with the structure of a finite-dimensional vector space \cite{AK}.
When $\b$ is generic enough 
(loosely speaking, it should not be integral),
then the dimension of $\deRham$ equals number of roots of the \emph{likelihood equations}
\cite{huh2013maximum, Matsubara-Heo:2023hmf}
\eq{
        \frac{\b_0}{g(z|x)} \frac{\p g(z|x)}{\p x_i} - \frac{\b_i}{x_i} = 0
        \quad \text{for} \quad
        i = 1, \ldots, n
        \, .
        \label{likelihood_equation}
}
The integer obtained in this fashion in fact equals the \emph{Euler characteristic} $\chi$ of the space
$
        (\CC^*)^n \setminus \{x_1 \cdots x_n \cdot g(z|x) = 0\}.
$
For FIs,
this integer even corresponds to the number of MIs \cite{Bitoun:2018afx}%
\footnote{
        Albeit without taking symmetry relations into account,
        which arise when different momentum routings result in the same FI.
}!
What is more,
this counting works whether or not the $z$'s take on special values.
In the GKZ case,
these variables are required to be independent and indeterminate.
But one is often interested in situations where some of the $z$'s are equal to each other,
and others equal unity.
The dimension of $\deRham$ will generally change when the $z$'s are fixed to special values,
but the counting via \eqref{likelihood_equation} will still be valid.

$\deRham$ can also be endowed with the structure of a $\mD$-module.
For generic $z$ and $\b$,
it is in fact isomorphic to the GKZ system as a $\mD$-module!
That is,
\eq{
        \mD_N / H_A(\b) \quad \cong \quad \deRham
        \, .
        \label{GKZ_de_Rham_isomorphism}
}
\appref{sec:GKZ_cohomology} presents a rough sketch for the proof of this claim. 
Here we simply take this fact as given,
and in the next subsection we leverage it to map elements from 
the RHS of the isomorphism to elements on the LHS.

\subsection{Representing Euler integrals inside \texorpdfstring{$\mD$-modules}{}}
\label{sec:Euler_integrals_as_operators}

The isomorphism \eqref{GKZ_de_Rham_isomorphism} sends the equivalence class of operators
$
        [1] \in \mD / H_A(\b) 
$
to the equivalence class of differential forms
$
        \left[ \frac{\dd x}{x} \right] \in \deRham.
$
Suppose we are given some other equivalence class $[\o(z)]$.
In this section we shall construct an operator $d$ with the property
\eq{
        d \bullet 
        \left[ \frac{\dd x}{x} \right] =
        [\o(z)]
        \, .
}

Given a vector
$
        q = (q_0, q') \in \ZZ \times \ZZ^n,
$
recall the expression for $\o_q$ from \eqref{U_omega_integral}.
For some fixed $\G$,
the Euler integral associated to this differential form is denoted by
\eq{
        \oq = \int_\G U \o_q
        \, .
}
By differentiating w.r.t.~$z_i$,
then $q$ gets shifted by the $i$th column vector of the $A$-matrix:
\eq{
        \p_i \bullet \oq =
        (\b_0 - q_0) \langle \o_{q+a_i} \rangle 
        \, .
        \label{shift_plus_a}
}
There exists an operator which shifts $q$ in the opposite direction as well,
i.e.~by $-a_i$.
To set this up,
the authors of \cite{saito_sturmfels_takayama_1999} construct a so-called \emph{creation operator} $C_i$
and a \emph{b-function} $b_i(\b)$ such that
\eq{
        C_i \, \p_i - b_i(\b) = 0
        \quad \text{mod} \quad
        H_A(\b)
        \, .
}
Using that $\oq$ is a solution to $H_A(\b-q)$,
one can then derive that
\eq{
        C_i \bullet \oq =
        b_i(\b-a_i) \langle \o_{q-a_i} \rangle 
        \, .
        \label{shift_minus_a}
}

Suppose that an integral $\oq$ is given.
Since the vectors $a_i$ span $\ZZ^{n+1}$ by stipulation,
it is possible to write the $q$-vector for $\oq$ as
\eq{
        q = \sum_{i=1}^N r_i \, a_i
        \, , \quad
        r_i \in \ZZ
        \, .
        \label{q_r_vector}
}
The integers $r_i$ allow us to balance shifts by $\pm a_i$ in the $q$-vector of $\oq$ via
\eqref{shift_plus_a} and \eqref{shift_minus_a}.
It thereby becomes possible to erect the operator
\eq{
        d_q = 
        \prod_{r_i < 0} C_i^{-r_i}
        \prod_{r_i > 0} \frac{1}{B(\b) B'(\b)} \p^{r_i}
        \label{shift_q_operator}
}
satisfying%
\footnote{
        Contrary to what was promised in the early remarks of this section,
        this equation is formulated as an action on a integral $\langle \o_q \rangle$
        rather than on a differential form $[\o(z)]$.
        The action of a differential operator on $[\o(z)]$ follows from \eqref{D_action_on_form},
        but the formula for $d_q$ remains the same in that case.
}
\eq{
        d_q \bullet
        \langle \o_0 \rangle =
        \oq
        \, , \quad
        \o_0  = \frac{\dd x}{x}
        \, .
        \label{shift_q_operator_action}
}
The two functions $B(\b)$ and $B'(\b)$ are built from the prefactors of
\eqref{shift_plus_a} and \eqref{shift_minus_a}.
Their explicit formulas are somewhat lengthy,
so we refer to \cite{Matsubara-Heo-Takayama-2020b} for details.
The formula \eqref{shift_q_operator} has been implemented in the \package{asir} package \package{mt\_gkz};
see \exref{ex:r_vector} below.

Note that the vector
$
        r = [r_1, \ldots, r_N]^T \in \ZZ^N
$
from \eqref{q_r_vector} is not unique for fixed $q$.
Given a collection of vectors $u_i \in \ZZ^N$ which lie in $\ker{A}$,
then it is the case that
$
        q = A \cdot \big(r + \sum_i c_i \, u_i\big)
$
for any choice of $c_i \in \ZZ$.
This freedom can be exploited to simplify the operator $d_q$ from \eqref{shift_q_operator}.
In particular,
the creation operators $C_i$ tend to have a large degrees in $\p_i$,
so for practical purposes it is useful to adjust the coefficients $c_i$ such that $r_i > 0$,
in which case the first product of \eqref{shift_q_operator} drops out.

Later on when we discuss applications to FIs,
the $q$-vector will be used to specify a particular integral within a family.
In the Lee-Pomeransky representation of a FI
(shown in the future equation \eqref{lee_pomeransky_representation}),
$q_0$ is related to the spacetime dimension,
and $q'$ contains the propagator powers.

\begin{ex}
\label{ex:r_vector}
Consider the GKZ system for the 4-fold Euler integral
\eq{
        f_\b(z) =
        \int_\G
        \big(
                z_1 x_1 + z_2 x_2 + z_3 x_3 + z_4 x_4 +
                z_5 x_1 x_3 + z_6 x_2 x_4
        \big)^{\b_0}
        x^{-\b'}
        \frac{\dd x}{x}
        \, .
}
The goal of this example is to build the operator $d_q$ corresponding to the differential form
\eq{
        \o_q = \left[ \frac{x_1 x_2}{g^2} \frac{\dd x}{x} \right]
        \quad \text{with} \quad
        q = (2,1,1,0,0)
        \, .
}
We shall gain more insight from deriving $d_q$ by hand,
rather than using the closed formula \eqref{shift_q_operator}.
The idea is to take suitable derivatives of $f_\b(z)$ w.r.t.~$z_i$.
Since $\o_q$ has a $g^2$ in the denominator,
we ought to take two derivatives to decrease $\b_0 \to \b_0 - 2$.
The powers of $x_1$ and $x_2$ in $\o_q$ are obtained by specifically choosing 
derivatives w.r.t.~$z_1$ and $z_2$.
So we have
\eq{
        \p_1 \p_2 \bullet f_\b(z) =
        \b_0(\b_0-1)
        \int_\G
        g(z|x)^{\b_0-2} x^{-\b'} x_1 x_2 \frac{\dd x}{x}
        \, .
}
Rearranging factors,
\eq{
        \frac{\p_1 \p_2}{\b_0(\b_0-1)} \bullet f_\b(z) =
        \int_\G 
        g(z|x)^{\b_0} x^{-\b'}
        \left[ \frac{x_1 x_2}{g^2} \frac{\dd x}{x} \right]
        \, .
}
The $\mD$-module representative for the differential form $\o_q$ is therefore
$
        d_q = 
        \frac{\p_1 \p_2}{\b_0(\b_0-1)}
        \, .
$

The same computation can be performed in \package{asir} via the following script:
\begin{lstlisting}[style=mystyle]
import("mt_gkz.rr");

A = [
        [1,1,1,1,1,1]
        [1,1,0,0,0,0]
        [0,0,1,1,0,0]
        [1,0,0,0,1,0]
        [0,0,1,0,0,1]
];
Ap = mt_gkz.make_Ap(A, 1); 
Ap = vtol(Ap);
Beta = [b0,b1,b2,b3,b4];

Rvec = [1,1,0,0,0,0];
mt_gkz.ff(Rvec, A, Ap, Beta); 
\end{lstlisting}
The first row of \soft{Ap} consists of $1$'s, 
and the remaining rows contain only $0$'s%
\footnote{
        \soft{Ap} is  built from the vectors called $\mathbf{a}_i'$ in \cite{Matsubara-Heo-Takayama-2020b}.
        See the documentation of \package{mt\_gkz} for additional details.
}.
The $r$-vector \soft{Rvec} satisfies $q = A \cdot r$.
The final command \soft{mt\_gkz.ff(...)} outputs the result \verb|(dx1*dx2)/(b0^2-b0)|,
as expected.
\end{ex}

To summarize,
we now have three ways of representing \emph{the same} object:
\begin{enumerate}
        \item
                The Euler integral 
                $
                        \langle \o_q \rangle = \int_\G U \o_q,
                $
                written in the notation of \eqref{U_omega_integral}.
                This lives in some suitable space of functions.
        \item 
                The equivalence class of differential forms $[\o_q]$.
                This lives in the twisted cohomology group \eqref{twisted_cohomology_group}.
        \item 
                The equivalence class of differential operators $[d_q]$
                defined by formula \eqref{shift_q_operator}.
                It lives inside the $\mD$-module $\mD_N / H_A(\b)$.
\end{enumerate}

        \section{Basis of standard monomials}
\label{sec:basis_of_stds}

As the GKZ $\mD$-module is holonomic,
it must have a finite collection of standard monomials serving as a basis.
There is a simple formula for the size of this basis
given a genericity condition on the parameter vector $\b = (\b_0, \ldots, \b_n)$.
We require some definitions from polyhedral geometry to state the formula.
\begin{itemize}
        \item 
                Interpret the column vectors $a_i$ of the $A$-matrix as points in $\ZZ^{n+1}$.
                The positive real \emph{cone} generated by these points is defined as
                \eq{
                        \mathbf{C}[A] =
                        \left\{
                                \sum_{i=1}^N \l_i \,  a_i
                                \ \Big | \
                                \l_i \geq 0
                        \right\}
                        \, .
                }
        \item 
                We say that $\b$ is \emph{non-resonant} if any integer shift 
                $
                        \b + \ZZ^{n+1}
                $
                contains no points on the \emph{facets} of $\mathbf{C}[A]$.
                Facets are defined as codimension-1 faces of $\mathbf{C}[A]$.

                Alternatively,
                we can say that $\b$ is resonant when some integer translate of the facet hyperplanes
                for $\mathbf{C}[A]$ can be made to "hit" $\b$.

                While this definition is somewhat technical,
                in practice it just means that $\b$ should be "generic enough",
                in the sense of not containing integer values.
        \item
                The \emph{Newton polytope} associated to $A$ is
                \eq{
                        \mathbf{N}[A] =
                        \left\{
                                \sum_{i=1}^N \l_i \,  a_i
                                \ \Big | \
                                \l_i \geq 0
                                \quad \text{and} \quad
                                \sum_{i=1}^N \l_i = 1
                        \right\}
                        \, .
                        \label{Newton_polytope}
                }
                In other words,
                $\mathbf{N}[A]$ is the convex hull of $A$'s column vectors $a_i \in \ZZ^{n+1}$.
\end{itemize}
When $\b$ is non-resonant,
it was proved in \cite{Adolphson} that the holonomic rank of the GKZ system,
and hence also the number of standard monomials,
is given by
\eq{
        R = n! \, \times \, \mathrm{vol}(\mathbf{N}[A])
        \, ,
        \label{GKZ_holonomic_rank}
}
where 
$
        \mathrm{vol}(\mathbf{N}[A])
$
is the volume of the Newton polytope
(this can be efficiently computed with software such as \package{Polymake} \cite{polymake}).
This formula is quite remarkable.
Recall that $R$ counts the number of independent solutions to the PDEs 
\eqref{E_i_annihilator}-\eqref{Box_u_annihilator}.
The latter is statement of complex analysis,
while the formula \eqref{GKZ_holonomic_rank} is purely combinatorial.

For GKZ systems,
not only is the number of basis elements known a priori,
there is even a fast algorithm to determine \emph{explicit} standard monomials due to 
Hibi, Nishiyama and Takayama \cite{Hibi_Nishi_Taka}.
The idea of the algorithm is to compute standard monomials for a simpler,
auxiliary ideal.
The construction can be summarized as follows.
Given the toric ideal of a GKZ system
\eq{
        \mI_\Box =
        \big\langle
                \Box_u
        \big\rangle_{\forall u \, \in \, \ker{A}}
        \, ,
}
one must first build its ideal of initial terms 
$
        \mathrm{in}_\prec(\mI_\Box)
$
(recall \eqref{initial_ideal}).
Any generator $d$ of
$
        \mathrm{in}_\prec(\mI_\Box)
$
can be written in the form
$
        d =
        \prod_{i=1}^N \p^{q_i}
        \text{ for }
        q_i \in \ZZ_{\geq 0}
$
The \emph{distraction} of $d$ is defined by
\eq{
        \mathrm{dist}(d) =
        \prod_{i=1}^N
        \p_i (\p_i -1 ) \cdots (\p_i - q_i - 1)
        \, .
}
The authors of \cite{Hibi_Nishi_Taka} then consider the ideal $\mI'$ generated as
\eq{
        \mI' \quad = \quad
        \sum_{i=0}^{n}
        \mD \, E_i \big |_{z_1 = \cdots = z_N = 1}
        \plus
        \sum_{\forall d \, \in \, \mathrm{in}_\prec(\mI_\Box)} 
        \mD \, \mathrm{dist}(d)
        \, ,
}
where $E_i$ are the homogeneity operators from \eqref{GKZ_operator_Euler}.
Surprisingly,
the set of standard monomials for $\mD / \mI'$ also constitutes a basis for the GKZ system $\mD / H_A(\b)$.
Hibi et al.~thus obtain standard monomials for the latter by a Gr\"obner basis computation in the former.
Crucially,
$\mI'$ is an ideal in the \emph{commutative} subring
$
        \CC[\p_1, \ldots, \p_N],
$
which makes the computation much more efficient than the 
execution of Buchberger's algorithm in a non-commutative Weyl algebra.

\begin{ex}
To illustrate the efficiency of this algorithm, 
consider the following $A$-matrix formed from a polynomial with $N=26$ terms:

{\scriptsize
\begin{align*}
A =
\arr{cccccccccccccccccccccccccc}{
1&  1&  1&  1&  1&  1&  1&  1&  1&  1&  1&  1&  1&  1&  1&  1&  1&  1&  1&  1&  1&  1&  1&  1&  1&  1 \\
1&  1&  1&  1& \mzero& \mzero& \mzero& \mzero& \mzero& \mzero& \mzero& \mzero& \mzero& \mzero& \mzero&  1&  1&  1&  1&  1&  1& \mzero& \mzero& \mzero& \mzero& \mzero \\
1& \mzero& \mzero& \mzero&  1&  1&  1& \mzero& \mzero& \mzero& \mzero& \mzero& \mzero& \mzero& \mzero&  1&  1& \mzero& \mzero& \mzero& \mzero&  1&  1&  1&  1& \mzero \\
\mzero&  1& \mzero& \mzero&  1& \mzero& \mzero&  1&  1&  1&  1& \mzero& \mzero& \mzero& \mzero& \mzero& \mzero&  1&  1& \mzero& \mzero&  1&  1& \mzero& \mzero&  1 \\
\mzero& \mzero& \mzero& \mzero& \mzero&  1& \mzero&  1& \mzero& \mzero& \mzero&  1&  1& \mzero& \mzero& \mzero& \mzero& \mzero& \mzero& \mzero& \mzero& \mzero& \mzero&  1& \mzero&  1 \\
\mzero& \mzero& \mzero& \mzero& \mzero& \mzero&  1& \mzero&  1& \mzero& \mzero& \mzero& \mzero&  1&  1&  1& \mzero&  1& \mzero&  1&  1&  1& \mzero& \mzero&  1& \mzero \\
\mzero& \mzero&  1& \mzero& \mzero& \mzero& \mzero& \mzero& \mzero&  1& \mzero&  1& \mzero&  1& \mzero& \mzero&  1& \mzero&  1&  1& \mzero& \mzero&  1&  1&  1& \mzero \\
\mzero& \mzero& \mzero&  1& \mzero& \mzero& \mzero& \mzero& \mzero& \mzero&  1& \mzero&  1& \mzero&  1& \mzero& \mzero& \mzero& \mzero& \mzero&  1& \mzero& \mzero& \mzero& \mzero&  1 \\
}
\, .
\end{align*}}

\noindent
The algorithm of \cite{Hibi_Nishi_Taka} has been implemented by Nobuki Takayama in \package{asir} 
via the command

\begin{lstlisting}[style=mystyle]
mt_gkz.cbase_by_euler(A);
\end{lstlisting}
In less than a minute on a laptop,
this command gives a list of $R=238$ standard monomials for the GKZ system.
The first few elements are
\eq{
        \Std =
        \big\{
                \p_{26} \p_{25}^4, \,
                \p_{26}^4 \p_{14}, \,
                \p_{25}^2 \p_{24}^2, \,
                \p_{25}^3 \p_8, \,
                \p_{25}^3 \p_{12}, \,
                \ldots
        \big\}
        \, .
}
\end{ex}

        \section{Integrand rescaling}
\label{sec:integrand_rescaling}

By cleverly rescaling the $z$-variables of the Euler integral \eqref{Euler_integral},
we show here how to reduce the number of variables from $N$ to $N-(n+1)$.
This reduction is useful for practical computations.

The construction follows from a homogeneity property satisfied by Euler integrals:
\eq{
        \label{Euler_integral_homogeneity}
        f_\b(t^{a_1} z_1, \, \ldots, \, t^{a_N} z_N) =
        t_0^{\b_0} \, t_1^{\b_1} \cdots t_n^{\b_n}
        f_\b(z)
        \, ,
}
where the vectors $a_i \in \ZZ^{n+1}$ come from the columns of the $A$-matrix.
By differentiating w.r.t.~$t$ and using the chain rule,
it can be shown that this property is equivalent to the collection of PDEs
$
        E_i \bullet f_\b(z) = 0
$
from \eqref{GKZ_operator_Euler}.
This is the reason for calling $E_i$ the "homogeneity operators".

We can freeze $(n+1)$ of the $z$-variables to unity
by judiciously choosing the rescaling parameters $t \in \CC^{n+1}$.
This is best illustrated by an example.

\begin{ex}
\label{ex:2_F_1_gkz}
This is a continuation of \exref{ex:2_F_1_macaulay}.
After relabeling $\b_i \to \b_{i-1}$ in the generators called $d_i$ in \eqref{2_F_1_generators},
we now see that this is the GKZ system built from $A$-matrix
\eq{
        A =
        \arr{cccc}{
                1 & 1 & 1 & 1 \\
                0 & 1 & 0 & 1 \\
                0 & 0 & 1 & 1 \\
        }
        \, .
}
The associated Euler integral is
\eq{
        f_\b(z) =
        \int_\Gamma
        \big(
                z_1 + z_2 x_1 + z_3 x_2 + z_4 x_1 x_2
        \big)^{\b_0}
        x_1^{-\b_1} x_2^{-\b_2}
        \frac{\dd x}{x}
        \, .
}
In this case,
the homogeneity property \eqref{Euler_integral_homogeneity} reads
\eq{
        \label{2_F_1_homogeneity}
        f_\b(t_0 z_1, \, t_0 t_1 z_2, \, t_0 t_2 z_3, \, t_0 t_1 t_2 z_4) =
        t_0^{\b_0} t_1^{\b_1} t_2^{\b_2} 
        f_\b(z_1, z_2, z_3, z_4)
        \, .
}
Choosing
\eq{
        t_0 = \frac{1}{z_1}
        \, , \quad
        t_1 = \frac{z_1}{z_2}
        \, , \quad
        t_2 = \frac{z_1}{z_3}
        \, ,
        \label{2_F_1_t_variables}
}
then \eqref{2_F_1_homogeneity} becomes
\eq{
        f_\b(1, \, 1, \, 1, w) =
        z_1^{-\b_0+\b_1+\b_2}
        z_2^{-\b_1}
        z_3^{-\b_2}
        f_\b(z_1,z_2,z_3,z_4)
        \quad \text{with} \quad
        w = \frac{z_1 \, z_4}{z_2 \, z_3}
        \, .
        \label{2_F_1_w_variable}
}
A $4$-variable problem has thereby been reduced to one that depends on a single cross-ratio variable $w$.

The three homogeneity operators $\{E_0, E_1, E_2\}$ have now been "gauge fixed" away by this rescaling.
The GKZ system then only depends on a single toric operator:
\eq{
        \Box_u \bullet f_\b(z) = 
        (\p_1 \p_4 - \p_2 \p_3) \bullet f_\b(z) = 
        0
        \, .
        \label{2_F_1_toric}
}
Inserting the representation of $f_\b$ from the RHS of \eqref{2_F_1_w_variable},
a short calculation shows that the two terms in $\Box_u$ get mapped to
\eq{
        \begin{array}{cccl}
                & \p_1 \p_4 
                \quad &\to \quad
                & -w^{-1} (\b_0-\b_1-\b_2-w\p_w)(w \p_w)
                \\
                & \p_2 \p_3 
                \quad &\to \quad
                & (\b_1 + w \p_w)(\b_2 + w \p_w)
                \, .
        \end{array}
}
The toric PDE \eqref{2_F_1_toric} thus turns into the second-order ODE
\eq{
        \label{2_F_1_w_operator}
        & L \bullet f_\b(1,1,1,w) = 0
        \quad \text{with} \\
        & L =
        w(1-w) \p_w^2 +
        \big[
                1 - \b_0 + \b_1 + \b_2 -
                (1 + \b_1 + \b_2)w
        \big]
        \p_w -
        \b_1 \b_2 
        \, ,
        \nonumber
}
which is recognized to be the DEQ for Gauss' hypergeometric ${}_2F_1$ function.
\end{ex}

The rescaling parameters \eqref{2_F_1_t_variables} were here pulled out of a hat.
In \appref{sec:general_rescaling_formula},
we present a general formula for how to choose these $t$'s.

        \section{Recurrence relations}
\label{sec:recurrence}

This section describes a method,
relying solely on products of Pfaffian matrices,
to determine recurrence relations among Euler integrals with different values of $\b$.
(We remark that these matrix products can be swiftly calculated using rational reconstruction.)
One can think of these recurrence relations as being a counterpart to IBP for FIs,
though in this case the IBPs come first and the Pfaffian systems last.
That logic is reversed by the method presented here,
as we obtain Pfaffian matrices first 
(via the Macaulay matrix method),
and then use those Pfaffians to derive recurrence relations/IBPs.

\subsection{Shifting \texorpdfstring{$\b$}{} by matrix factorials}

For convenience,
we shall work with a rescaled version of the Euler integral:
\eq{
        f_\b(z) \to
        f(\b) =
        \frac{1}{\Gamma(\b_0+1)} \int_\Gamma
        g(z|x)^{\b_0} x^{-\b'} \frac{\dd x}{x}
        \, ,
}
where we chose to only emphasize the $\b$-dependence of $f(\b)$ since $z$ will stay fixed throughout
the discussion.
The $\Gamma$-function prefactor provides a clean shift relation when a derivative is taken w.r.t.~$z_i$:
\eq{
        \p_i \bullet f(\b) = f(\b-a_i)
        \, .
        \label{beta_shift}
}

Let $\Std$ be a basis of standard monomials for a GKZ system $\mD / H_A(\b)$.
We convert this $\mD$-module basis into a basis of functions by writing
\eq{
        F(\b) =
        \arr{c}{
                \Std_1 \bullet f(\b) \\
                \vdots \\
                \Std_R \bullet f(\b)
        }
        \, .
}
Suppose we have computed the Pfaffian system for $F(\b)$ via the Macaulay matrix method.
Combining the relation
$
        \p_i \Std_j =
        \Std_j \p_i 
$
with \eqref{beta_shift},
we deduce that
\eq{
        \p_i \bullet F(\b) =
        P_i(\b) \cdot F(\b) =
        F(\b-a_i)
        \, .
}
In other words,
matrix multiplication by $P_i(\b)$ induces a difference equation for $F(\b)$.

It is possible to shift $\b$ in the opposite direction as well.
Assuming $\b$ is non-resonant,
then $P_i(\b+a_i)$ is invertible.
Setting
\eq{
        Q_i(\b) = P_i(\b+a_i)^{-1}
        \, ,
}
we thereby obtain an "inverse derivative operator" $\p_i^{-1}$ acting as
\eq{
        \p_i^{-1} \bullet F(\b) =
        Q_i(\b) \cdot F(\b) =
        F(\b+a_i)
        \, .
}

Let $b \in \ZZ_{\geq 0}$.
If $b > 0$,
we write the \emph{falling matrix factorial} as
\eq{
        \big[ P_i(\b) \big]_b =
        P_i\big( \b - (b-1)a_i \big)
        \cdot
        P_i\big( \b - (b-2)a_i \big)
        \cdots
        P_i(\b - a_i)
        \cdot 
        P_i(\b)
        \, ,
}
and the \emph{rising matrix factorial} as
\eq{
        \big( Q_i(\b) \big)_b =
        Q_i\big( \b + (b-1)a_i \big)
        \cdot
        Q_i\big( \b + (b-2)a_i \big)
        \cdots
        Q_i(\b + a_i)
        \cdot 
        Q_i(\b)
        \, .
}
If $b=0$,
then 
$
        \big[ P_i(\b) \big]_0 = 
        \big( Q_i(\b) \big)_0 = \mathbf{1}.
$

When the subscript $b$ is promoted to an integer vector $v \in \ZZ_{\geq 0}^N$,
these definitions are extended to
\eq{
        \big[ P(\b) \big]_v =&
        \prod_{i=1}^N
        \left[
                P_i \left( \b - {\textstyle\sum_{j=i+1}^N} \, v_j \, a_j \right)
        \right]_{v_i}
        \\
        \big( Q(\b) \big)_v =&
        \prod_{i=1}^N
        \left[
                Q_i \left( \b - {\textstyle\sum_{j=i+1}^N} \, v_j \, a_j \right)
        \right]_{v_i}
        \, ,
}
with the convention that 
$
        \sum_{j=i+1}^N v_j a_j = 0
$
when $i=N$.

Now, 
by induction over $b>0$,
it is short proof 
(see \cite[Lemma 6.3]{Chestnov:2022alh})
to show that
\eq{
        \begin{array}{ccccl}
                \p^b_i \bullet F(\b) &=&
                \big[ P_i(\b) \big]_b \cdot F(\b) &=&
                F(\b - b a_i)
                \\
                \p^{-b}_i \bullet F(\b) &=&
                \big( Q_i(\b) \big)_b \cdot F(\b) &=&
                F(\b + b a_i)
                \, .
        \end{array}
}
Repeatedly applying these relations for each $v_i$ in a given vector $v \in \ZZ_{\geq 0}^N$,
we obtain the general shift relations
\eq{
        \begin{array}{ccccl}
                \p^v_i \bullet F(\b) &=&
                \big[ P_i(\b) \big]_v \cdot F(\b) &=&
                F
                \left(\b - \sum_{i=1}^N v_i \, a_i \right)
                \\[5pt]
                \p^{-v}_i \bullet F(\b) &=&
                \big( Q_i(\b) \big)_v \cdot F(\b) &=&
                F
                \left(\b + \sum_{i=1}^N v_i \, a_i \right)
                \, .
        \end{array}
        \label{general_beta_shift}
}
Note that these matrix factorial relations are not uniquely represented because
\eq{
        \begin{array}{cccc}
                F\big( (\b-a_i) - a_j \big) &=&
                P_i(\b-a_j) \cdot P_j(\b) \cdot F(\b)
                \\ & \text{and} & \\
                F\big( (\b-a_j) - a_i \big) &=&
                P_j(\b-a_i) \cdot P_i(\b) \cdot F(\b)
        \end{array}
}
imply the commutation relation
\eq{
        P_i(\b-a_i) \cdot P_j(\b) = P_j(\b-a_i) \cdot P_i(\b)
        \, ,
}
and likewise for $Q_i$.

\subsection{General \texorpdfstring{$\b$}{} shift algorithm}

Here we employ the formulas derived in the previous subsection to state a general algorithm
for expressing a given Euler integral as a sum of basis integrals.
To begin,
write the standard monomials in the form
$
        \Std_i = \p^{q_i}
$
for some 
$
        q_i \in \ZZ_{\geq 0}^N.
$
Recalling that the $i$th component of $F(\b)$ is of the form
$
        \Std_i \bullet f(\b),
$
then \eqref{beta_shift} implies that
\eq{
        F(\b) =
        \arr{c}{
                f(\b - A \cdot q_1)
                \\ \vdots \\
                f(\b - A \cdot q_R)
        }
        \, .
}
Given an integer vector $q_0 \in \ZZ^N$,
we seek a recurrence relation for
$
        f(\b - A \cdot q_0)
$
of the form
\eq{
        f(\b - A \cdot q_0) = 
        \sum_{i=1}^R
        c_i(\b,z) \,
        f(\b - A \cdot q_i) 
        \quadit{\text{with}}
        c_i \in \QQ(\b,z)
        \, .
        \label{Euler_integral_recurrence_relation}
}

Observe that
$
        f(\b - A \cdot q_0)
$
is the first element of the vector
$
        F\big(\b - A \cdot (q_0-q_1) \big).
$
Combining this fact with the shift relations \eqref{general_beta_shift},
we are lead to the following algorithm.

\newpage

\begin{algorithm}
        \underline{Input}:
        \vspace{-0.25cm}
        \begin{itemize}
                \item
                        Vector $q_0 \in \ZZ^N$.
                \item 
                        Indeterminate vector $\beta$.
                \item 

                        Monomial basis 
                        $
                                \Std =
                                \big\{
                                        \p^{q_1},
                                        \ldots,
                                        \p^{q_R}
                                \big\}
                        $
                        for
                        $
                                q_i \in \ZZ_{\geq 0}^N.
                        $
        \end{itemize}
        \underline{Output}:
        The recurrence relation \eqref{Euler_integral_recurrence_relation}.
        \vspace{0.2cm}
    \begin{algorithmic}[1]
        \State Calculate $P_i(\beta)$, $i=1,\ldots,N,$ w.r.t.~the basis $\Std$ 
        by calling \algref{alg:Pfaffian_by_MM}.
        \vspace{0.2cm}
        \State Decompose 
        \eq{ 
                \nonumber
                q_0 - q_1 = q^+ - q^- = q
        }
        for $q^\pm \in \ZZ_{\geq 0}^N$.
        \vspace{0.2cm}
        \State Compute the matrix factorial
        \eq{
                \nonumber
                \Big[
                        P\Big(\b - {\textstyle \sum_{q_j < 0}} \, q_j \, a_j \Big)
                \Big]_{q^+}
                \ \cdot \
                \big( Q(\b) \big)_{q^-}
                \ \cdot \
                F(\b)
                \, .
        } 
        \\
        \Return the first element of step 3.
    \end{algorithmic}
    \caption{: Recurrence relations for Euler integrals}
    \label{alg:recurrence_by_MM}
\end{algorithm}

\noindent
Since this algorithm avoids the use of derivatives,
it is possible to fix the $z$-variables to special values in the Pfaffian matrices.
We give a simple example of how to use this algorithm in \secref{sec:recurrence_relation_example}.

        \chapter{Generalized Feynman Integrals}
\label{ch:gfi}

\vspace{-0.3cm}

In the physics literature,
Tullio Regge observed already in 1968 that FIs often evaluate to 
\emph{hypergeometric functions} \cite{Regge:1968rhi} 
(for instance the ${}_2F_1$ function that we saw in the previous chapter).
In the mathematics literature,
although a large zoo of hypergeometric%
\footnote{
        Let
        $F(z) = \sum_{n=0}^\infty c_n z^n$.
        We say that $F$ is \emph{hypergeometric} if $c_{n+1}/c_n$ is a rational function of $n$.
        The generalization to multivariate and logarithmic series is immediate.
}
functions 
(carrying famous names such as 
Gauss, 
Kummer, 
Mellin,
Barnes, 
Appell, 
Kampé de Fériet, 
Lauricella, 
Horn and Meijer)
were known in the last half of the previous century,
there was arguably no overarching framework that tied them all together.
The work of Gel'fand, Kapranov and Zelevinsky in the 1990s did exactly that:
it gave a general construction which subsumed the whole zoo of hypergeometric functions as special cases.

Following in the footsteps of Regge,
it is then natural to wonder whether GKZ systems can say something about FIs.
The first proper study in this direction appears to have been by Nasrollahpoursamami in 2016
\cite{Nasrollahpoursamami:2016,Nasrollahpoursamami:2017shc}.
There has since then been a flurry of works studying this topic
\cite{Pal:2021llg,
      Pal:2023kgu,
      Ananthanarayan:2022ntm,
      Feng:2022kgh,
      Feng:2022ude,
      Zhang:2023fil,
      walther2022feynman,
      Dlapa:2023cvx,
      Matsubara-Heo:2023hmf,
      Chestnov:2022alh,
      Chestnov:2023kww,
      Munch:2022ouq,
      Klemm:2019dbm,
      Bonisch:2020qmm,
      Bonisch:2021yfw,
      Vanhove:2018mto,
      vanhove-2021,
      Tellander:2021xdz,
      Klausen:2019hrg,
      Klausen:2021yrt,
      Klausen:2023gui,
      Henn:2023tbo,
      Feng:2019bdx,
      delaCruz:2019skx,
      Agostini:2022cgv,
      Srednyak:2023oyq}.
There is now even a practical program, 
named \package{FeynGKZ} \cite{Ananthanarayan:2022ntm},
that computes series expansions for FIs based on the GKZ framework.

The connection between FIs and GKZ systems is not so immediate in the momentum space representation.
The connection becomes more clear in parametric representations such as the 
\emph{Lee-Pomeransky representation} (LPr) \cite{Lee:2013hzt}.
The FI then looks more like the Euler integral defined in \eqref{Euler_integral},
which is a solution to the GKZ system.
The difference is that the $z$-variables take on special values in the LPr,
so the class of Euler integrals is too general.
Though the proper function space for FIs is unknown 
(cf.~\secref{sec:general_features}),
the Euler integral can thus be thought of as an "upper bound" on the complexity of this space.

In what follows,
we use the technology developed in the preceding chapters to study some simple FIs in the GKZ framework.
We especially focus on Pfaffian systems for MIs.

        \section{Lee-Pomeransky representation as a GKZ system}

Fix a Feynman graph $G$.
There are two special \emph{Symanzik polynomials} associated to $G$,
given by
\eq{
        \label{U_poly}
        \mU(x) &= 
        \sum_T \prod_{e \, \notin \, T} x_e
        \\
        \mF(x) &= 
        - \sum_F p(F)^2 \prod_{e \, \notin \, F}
        \plus
        \mU(x) \sum_{i=1}^n m_i^2 x_i
        \, .
        \label{F_poly}
}
$T$ and $F$ are sets of so-called \emph{spanning trees} and \emph{spanning forests} respectively,
and $p(F)^2$ is the Minkowski square of the momentum flowing into certain forest components.
We shall not dwell on these definitions here,
as there are many exhaustive expositions already,
see e.g.~\cite[Chapter 3]{Weinzierl:2022eaz}.
The parameters $m_i$ denote the internal masses associated to the $n$ internal edges of $G$.

Starting from the momentum space representation of a FI,
it is possible to derive an equivalent parametric representation called the 
\emph{Lee-Pomeransky representation} (LPr) \cite{Lee:2013hzt}
(see also \cite[Section 2.5]{Weinzierl:2022eaz}).
We denote it by
\eq{
       \label{lee_pomeransky_representation}
       I(\DD_0|\nu) =
       \frac
       { \G\lrbrk{ \frac{\DD}{2} } }
       { \G\lrbrk{ \frac{\DD}{2} - \o } \G\brk{\nu} }
       \int_0^\infty
       \mG\brk{x}^{-\frac{\DD}{2}} \,
       x^{\nu} \,
       \frac{\dd x}{x}
       \, ,
}
where 
\begin{itemize}
        \item
                $\mG$ is the \emph{Lee-Pomeransky} (LP) \emph{polynomial} 
                \eq{
                        \mG\brk{x} = \mU\brk{x} + \mF\brk{x}
                        \, .
                }
        \item 
                $\o$ is the \emph{superficial degree of divergence}
                \eq{
                        \o = \sum_{i=1}^n \nu_e - \DD L / 2
                }
                written in terms of edge weights $\nu = (\nu_1, \ldots, \nu_n) \in \ZZ^n$,
                the spacetime dimension $\DD = \DD_0 - 2\e$,
                and the number of loops $L$.
                We assume $\DD_0 \in 2 \cdot \ZZ_{>0}$ is even for simplicity.
        \item 
                $\G(\nu)$ is shorthand for $\prod_{i=1}^n \G(\nu_i)$.
\end{itemize}

\subsection{Generalized Feynman integral}

Let $0 < \d \ll 1$.
The two small parameters $(\e,\d)$ correspond to analytic regulators in the works of Speer 
\cite{speer1969theory,Speer1971}.
Fix the exponents $\b$ of the Euler integral $f_\b(z)$ from \eqref{Euler_integral} to%
\footnote{
        Writing $- \e \d$ instead of, say, $\d$ is merely a matter of convention.
        This convention was chosen so as to more easily compare results 
        from \secref{sec:MM_box} with \cite{Mizera:2019vvs}.
}
\eq{
        \b =
        (\e, \, -\e \d, \, \ldots, - \e \d)
        \minus
        (\DD_0/2, \, \nu_1, \ldots, \, \nu_n)
        \, .
        \label{gfi_beta}
}
The \emph{generalized FI} (GFI) is now defined to be
\eq{
        \gfi = \gfic \times f_\b(z)
        \label{generalized_feynman_integral}
        \, .
}
\begin{itemize}
        \item 
                Inserting the special choice of $\b$ from \eqref{gfi_beta},
                the Euler integral reads
                \eq{
                        f_\b(z) =
                        \int_{(0,\infty)^n}
                        \mG(z|x)^{\e - \DD_0/2} \,
                        x_1^{\nu_1 + \e \d} \cdots x_n^{\nu_n + \e \d} \,
                        \frac{\dd x}{x}
                        \, ,
                }
                where the integration contour has been fixed to
                $
                        \G = (0,\infty)^n.
                $
        \item 
                Rewriting the prefactor in \eqref{lee_pomeransky_representation} with this $\b$,
                we have
                \eq{
                        \gfic =
                        \frac
                        {\G \big( \DD_0/2 - \e \big)}
                        {
                                \G
                                \Big(
                                        (L+1)(\DD_0/2 - \e) -
                                        |\nu| -
                                        n \e \d
                                \Big)
                                \prod_{i=1}^n
                                \G\big(\nu_i + \e \d\big)
                        }
                        \, ,
                }
                with $|\nu| = \nu_1 + \ldots + \nu_n$.
        \item 
                The monomials $x^{\a_i}$ in $\mG$ come from the 
                LP polynomial \eqref{lee_pomeransky_representation}:
                \eq{
                        \mG(z|x) = \mU(z|x) + \mF(z|x) = \sum_{i=1}^N z_i \, x^{\a_i}
                        \, .
                        \label{LP_poly_GFI}
                }
                But, 
                in contrast to the original LP polynomial,
                each monomial coefficient $z_i$ is here taken to be indeterminate.
\end{itemize}

\noindent
There is a GKZ system $\mD/H_A(\b)$ associated to such a GFI.
According to the isomorphism \eqref{GKZ_de_Rham_isomorphism},
$\gfi$ is also represented by a twisted cohomology group $\deRham$.
In the notation of \secref{sec:Euler_integrals_as_operators},
given a twist
$
        U = \mG(z|x)^\e \, x^{\e \d}
$
and a differential form
\eq{
        [\o_{\DD_0/2, \nu}] =
        \gfic \times \mG^{-\DD_0/2} \, x^\nu \, \frac{\dd x}{x}
        \, ,
        \label{LPr_form}
}
we then have that
$
        \gfi = \langle \o_{\DD_0/2, \nu} \rangle.
$
The $r$-vector from \eqref{q_r_vector} therefore satisfies 
$
        A \cdot r = \big(\DD_0/2, \nu\big).
$
The columns $a_i = [1, \a_i]^T$ of this $A$-matrix come from the multi-indices $\a_i$ in \eqref{LP_poly_GFI}.

Now,
making the identifications
\begin{itemize}
        \item
                $\d \to 0$
        \item 
                $
                        z_i \to 
                        \ZZ_{>0}
                        \ \text{or} \
                        \big\{ \text{kinematic variables} \ (m_i^2, p_i^2, p_i \cdot p_j) \big\}
                $
\end{itemize}
then a GFI agrees exactly with the LPr.
The $\d \to 0$ limit is not so troublesome 
(this parameter is inserted to ensure non-resonance of the GKZ system).
The more pressing issue is how to deal with each $z_i$ being indeterminate in the GFI.
By the homogeneity rescaling described in \secref{sec:integrand_rescaling},
we can at least rescale $(n+1)$ of the $z$-variables to unity
in order to match with some or all of the monomial coefficients in the $\mU$-polynomial.
In doing so,
it is possible to exactly match the GFI with a large class of 1-loop graphs 
\cite[Section 5.1]{Chestnov:2022alh} -
in particular those which are fully massless,
or contain off-shell legs whose squared momenta are independent, 
$p_i^2 \neq p_j^2 \neq 0$.

In general,
however,
there are \emph{more} GKZ variables than there are independent monomial coefficients
in the proper LP polynomial.
This is especially true at $L=2$ loops and beyond.
In those cases one must take a kind of "limit",
called a \emph{restriction},
of the GKZ system.
We study such restrictions in detail in Chapters 
\eqref{ch:restrictions} and \eqref{ch:restrictions_examples}.
For the remaining part of this chapter though,
we only study examples where the number of GKZ variables can be made to match with the proper FI.

        \section{Macaulay matrix example: 1-loop massless box}
\label{sec:MM_box}

Here we compute the Pfaffian system for the 1-loop massless box diagram using the Macaulay matrix method
from \chapref{sec:macaulay_matrices}.
This example was studied in the LPr using twisted cohomology in \cite{Mizera:2019vvs}.

The kinematic configuration is
\vspace{-0.5cm}

\noindent
\begin{minipage}{0.78\textwidth}
\begin{align*}
        \sum_{i=1}^4 p_i = 0
        \, , \quad
        p_1^2 = \cdots = p_4^2 = 0
        \, , \quad
        s = 2 p_1 \cdot p_2
        \, , \quad
        t = 2 p_2 \cdot p_3 
        \, .
\end{align*}
\end{minipage}
\begin{minipage}{0.2\textwidth}
        \vspace{0.5cm}
        \includegraphics[scale=0.15]{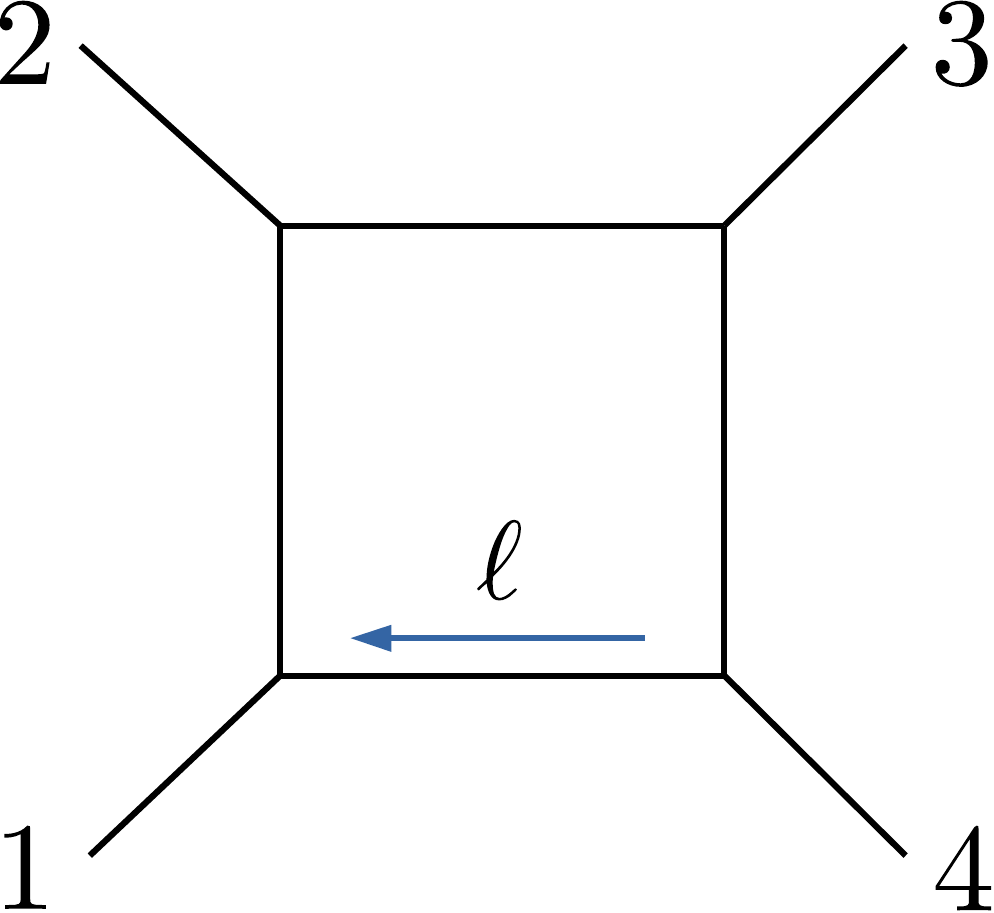}
\end{minipage}

\noindent
In momentum space,
one has the following four inverse propagators:
\eq{
        \begin{array}{lllllll}
                & D_1 &=& -\ell^2
                \, ,
                & D_2 &=& -(\ell-p_1)^2
                \, ,
                \\
                & D_3 &=& -(\ell-p_1-p_2)^2
                \, , 
                & D_4 &=& -(\ell-p_1-p_2-p_3)^2
                \, .
        \end{array}
}
The corresponding GFI reads 
\eq{
        \gfi =
        \gfic
        \int_{(0,\infty)^4}
        \mG(z|x)^{\e-\DD_0/2}
        \,
        x_1^{\nu_1+\e\d}
        \cdots
        x_4^{\nu_4+\e\d} 
        \,
        \frac{\dd x}{x}
        \, ,
}
where
\eq{
        \mG(z|x) =
        \sum_{i=1}^4 z_i x_i + z_5 x_1 x_3 + z_6 x_2 x_4
        \, .
        \label{1L_box_LP}
}

Upon rescaling the integration variables in the proper LPr as 
$
        x \to x/(-s),
$
then its monomial coefficients become
\eq{
        z_1 = \ldots = z_5 = 1
        \, , \quad
        z_6 = \frac{t}{s} = z
        \, .
        \label{1L_box_z1_z6}
}
So the proper LP has one independent coefficient $z$.
On the other hand, 
the GKZ system has $N=6$ variables.
Using homogeneity,
we can rescale $n+1 = 4+1 = 5$ of those variables to unity.
To match with \eqref{1L_box_z1_z6},
we choose to rescale $\{z_1, \ldots, z_5\}$.
Including an overall $s$-dependent prefactor because of the $x$-rescaling above,
\eq{
        \gfi \to
        (-s)^{\DD_0/2-\e-4\e\d-|\nu|} \times \gfi
        \, ,
}
then our GFI matches exactly with the LPr in the limit $\d \to 0$.

Having "gauged away" the Euler homogeneity operators,
the GKZ system now only contains the toric part.
Using \eqref{homogeneity_partial_derivative} or the \package{asir} command \soft{mt\_gkz.gkz\_b},
the single toric operator works out to
\eq{
        \label{1L_box_toric}
        \mI_\Box =& \
        z^3 [1 + z] \p^3  
        \plus
        z^2 \big[3 (1 + z) + \e (2 + z + 6 \d (1 + z))\big] \p^2 
        \\ \nonumber & \plus
        z \big[(1 + \e + 3 \d \e)^2 + z (1 + \e + 6 \d \e +  \e^2 \d (2 + 9 \d)) \big] \p 
        \plus
        z \e^3  \d^2 [1 + 4 \d] 
        \, ,
}
where $\p = \p_z$.

\subsection{Basis}

Since the number of GKZ variables matches with the proper LPr,
the formula $R = n! \times \mathrm{vol}(\mathbf{N}[A])$ ought to correctly count the number of MIs.
In \package{polymake},
this volume can be computed as
\begin{lstlisting}[style=mystyle]
$P = new Polytope(POINTS =>
        [[1,1,0,0,0],
         [1,0,1,0,0],
         [1,0,0,1,0],
         [1,0,0,0,1],
         [1,1,0,1,0],
         [1,0,1,0,1]]
);
print $P -> VOLUME;
\end{lstlisting}
The matrix above is the transpose of the $A$-matrix associated to the LP polynomial \eqref{1L_box_LP},
due to the conventions of the software.
The script outputs the volume $1/8$,
so we expect $R = 4! \times 1/8 = 3$ MIs.
This indeed holds true,
as can be verified by an IBP calculation.

Taking inspiration from \cite{Henn:2014qga},
we choose the following basis of canonical integrals near $4$ dimensions:
\eq{
        e\supbrk{I} =
        (-s)^{\e+1}
        \arr{c}{
                z \, I(4 | 0 1 0 2) \\
                I(4 | 1 0 2 0) \\
                \e \, z \, (-s) \, I(4 | 1 1 1 1)
        }
        \, .
        \label{box_integral_basis}
}
Using \eqref{LPr_form},
this corresponds to the basis of differential forms 
\eq{
        e\supbrk{\text{dR}} = 
        \mathrm{c} \cdot
        \arr{c}{
                \frac{x_4}{x_1 \, x_2 \, \mG^2} \dd x \\[4pt]
                \frac{x_3}{x_2 \, x_4 \, \mG^2} \dd x \\[4pt]
                \frac{1}{\mG^2} \dd x 
        }
        \, ,
}
where $\mathrm{c}$ is a $3 \times 3$ diagonal matrix containing the $\Gamma$-function constants
$
        \gfic
$
multiplied by the prefactors from \eqref{box_integral_basis}.
The formulas 
\eqref{shift_q_operator} and \eqref{homogeneity_partial_derivative}
turn these differential forms into a $\mD$-module basis
\eq{
        e\supbrk{\mD} =
        \mathrm{c} \cdot
        \arr{c}{
                \frac{\e \d - 1}{\e (\e - 1)} \p -
                \frac{z}{\e (\e - 1)} \p^2 \\[4pt]
                \frac{\e (1 - 3 \d) (4 \d - 1)}{\e - 1} +
                \frac{z ( 7 \e \d - 2 \e - 1)}{\e (\e - 1)} \p -
                \frac{z^2}{\e (\e - 1)} \p^2 \\[4pt]
                \frac{4 \e \d - \e - 1}{\e (\e - 1)} \p  -
                \frac{z}{\e (\e - 1)} \p^3
        }
        \, .
}

\subsection{Macaulay matrix}

Our goal is to obtain the Pfaffian system
$
        \p e\supbrk{\mD} = P \cdot e\supbrk{\mD}.
$
Using the Macaulay matrix method,
we first construct the Pfaffian system in the basis of standard monomials,
\eq{
        \p \Std =
        \p \,
        \arr{c}{
                \p^2 \\
                \p \\
                1
        }
        =
        \arr{c}{
                \p^3 \\
                \p^2 \\
                \p
        }
        =
        P\supbrk{\Std} 
        \cdot 
        \arr{c}{
                \p^2 \\
                \p \\
                1
        }
        \, ,
        \label{box_Std_system}
}
whereafter we gauge transform to the $e\supbrk{\mD}$ system.
With the toric operator \eqref{1L_box_toric} and the basis $\Std$ as input,
the Macaulay data works out to
\eq{
        \nonumber
        M_\Ext &=
        \arr{c}{ z^2(z+1) }
        \\[5pt] 
        M_\Std &=
        \arr{c}{
                z \big[\e(6\d+2) + z(6\e\d+\e+3) + 3\big] \\[3pt]
                \big(3\e\d+\e+1\big)^2 + z\big[ \e^2\d(9\d+2) +6\e\d + \e + 1\big] \\[3pt]
                \e^3\d^2(4\d+1)
        }^T
        \\[5pt] \nonumber
        C_\Ext &=
        \arr{ccc}{1 & 0 & 0}
        \, , \quad
        C_\Std =
        \arr{ccc}{0 & 0 & 0 \\ 1 & 0 & 0 \\ 0 & 1 & 0}
        \\[5pt] \nonumber
        \Mons &= \Ext \ \sqcup \ \Std = \{\p^3\} \ \sqcup \ \{\p^2, \, \p, \, 1\}
        \, .
}
Letting
$
        C = \big[ c_{11} \ c_{12} \ c_{13} \big]
$
be an unknown matrix,
we then solve 
$
        C_\Ext - C \cdot M_\Ext = 0
$
to obtain
\eq{
        C = \arr{ccc}{\frac{1}{z^2(z+1)} & 0 & 0}
        \, .
}
The Pfaffian matrix in the basis of standard monomials is then
\eq{
        P\supbrk{\Std} &= 
        C_\Std - C \cdot M_\Std \\&=
        \arr{ccc}{
                P\supbrk{\Std}_{11} & P\supbrk{\Std}_{12} & P\supbrk{\Std}_{13} \\
                1 & 0 & 0 \\
                0 & 1 & 0
        }
        \, ,
}
with
\eq{
        \nonumber
        P\supbrk{\Std}_{11} &=
        - \frac{\e\big[ 6\d(z+1) + z + 2 \big] + 3(z+1)}{z(z+1)}
        \\[2pt]
        P\supbrk{\Std}_{12} &=
        - \frac{\big( 3\e\d+\e+1 \big)^2 + z \big[ \e^2\d(9\d+2) + 6\e\d + \e + 1 \big]}{z^2(z+1)}
        \\[2pt]
        P\supbrk{\Std}_{13} &=
        - \frac{\e^3\d^2(4\d+1)}{z^2(z+1)}
        \, .
        \nonumber
}
Finally, 
using the basis change method described in \appref{sec:basis_change} we build a matrix $G$ such that 
$
        e\supbrk{\mD} = G \cdot e\supbrk{\Std},
$
leading to the gauge transformation
\eq{
        P &=
        \left(\p \bullet G + G \cdot P\supbrk{\Std}\right) \cdot G^{-1} \\ &=
        \e
        \arr{ccc}{
                - \frac{1}{z} & 0 & 0 \\
                0 & 0 & 0 \\
                - \frac{2}{z(z+1)} & \frac{2}{z+1} & - \frac{1}{z(z+1)}
        }
        \, .
}
The limit $\d \to 0$ has already been taken here.
This matrix is canonical as expected,
and is in agreement with an independent \package{LiteRed} computation.

        \section{Macaulay matrix example: 1-loop pentagon with off-shell leg}

Here we compute a Pfaffian system for the 1-loop massless pentagon integral with one massive leg.
The kinematics are
\vspace{-0.4cm}

\noindent
\begin{minipage}{0.65\textwidth}
\begin{align*}
        \nonumber
        \sum_{i=1}^5 p_i = 0
        \, , \quad
        p_1^2 = \cdots &= p_4^2 = 0
        \, , \quad
        p_5^2 = p^2
        \\
        \text{and} \quad
        s_{ij} &= 2 p_i \cdot p_j
        \, .
\end{align*}
\end{minipage}
\begin{minipage}{0.25\textwidth}
        \vspace{0.5cm}
        \includegraphics[scale=0.15]{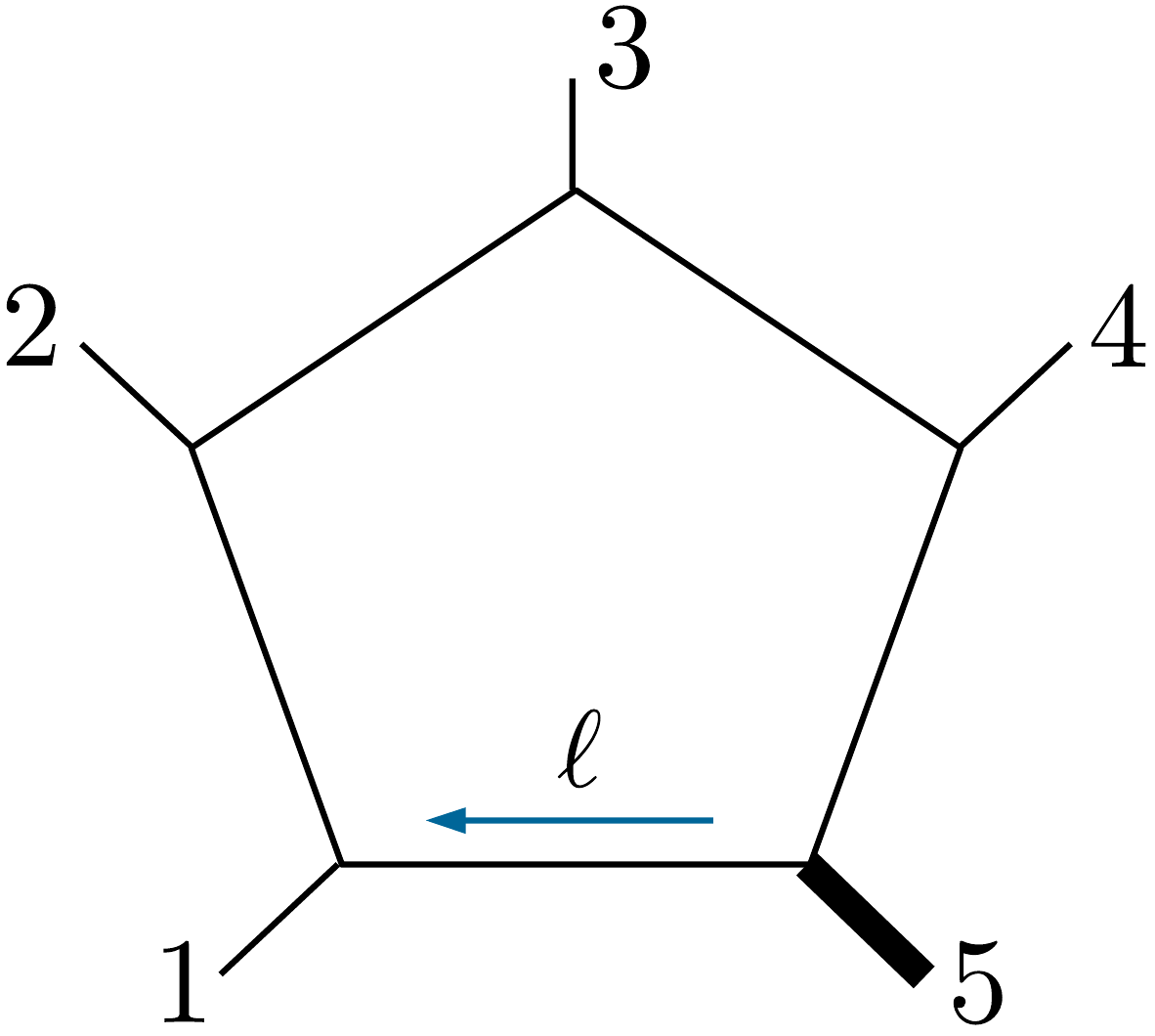}
\end{minipage}

\vspace{0.5cm}
\noindent
Note that the identity
$
        (p_1 + p_2 + p_3 + p_4)^2 = (-p_5)^2 = p^2
$
can be used to eliminate one of the Mandelstam variables $s_{ij}$.

The inverse propagators are given by
\eq{
        \begin{array}{llllll}
                D_1 &=& - \ell^2
                \, , \quad
                D_2 = -(\ell-p_1)^2
                \, , \quad
                & D_3 &=& -(\ell-p_1-p_2)^2 
                \\
                D_4 &=& -(\ell-p_1-p_2-p_3)^2 
                \, , \quad
                & D_5 &=& -(\ell-p_1-p_2-p_3-p_4)^2
                \, .
        \end{array}
}
The associated GFI is
\eq{
        \gfi = 
        \gfic (-s_{12})^{\DD_0/2 - \e - |\nu| - 5 \e \d}
        \int_{(0,\infty)^5}
        \mG(z|x)^{\e-\DD_0/2} \, 
        x_1^{\nu_1+\e\d} \cdots x_5^{\nu_5+\e\d} \,
        \frac{\dd x}{x}
        \, ,
        \label{GFI_pentagon}
}
with LP monomials given by
\eq{
        & \mG(z|x) = 
        \sum_{i=1}^5 z_i \, x_i +
        z_7    \, x_1 \, x_4 +
        z_8    \, x_1 \, x_5 +
        z_9    \, x_2 \, x_4 +
        z_{10} \, x_1 \, x_5 +
        z_{11} \, x_3 \, x_5 
        \, ,
        \\
        & \text{corresponding to} \quad 
        A = 
        \arr{ccccccccccc}{
                1 & 1 & 1 & 1 & 1 & 1 & 1 & 1 & 1 & 1 & 1 \\
                1 & 0 & 0 & 0 & 0 & 1 & 1 & 1 & 0 & 0 & 0 \\
                0 & 1 & 0 & 0 & 0 & 0 & 0 & 0 & 1 & 1 & 0 \\
                0 & 0 & 1 & 0 & 0 & 1 & 0 & 0 & 0 & 0 & 1 \\
                0 & 0 & 0 & 1 & 0 & 0 & 1 & 0 & 1 & 0 & 0 \\
                0 & 0 & 0 & 0 & 1 & 0 & 0 & 1 & 0 & 1 & 1 
        }
        \, .
        \nonumber
}
The $s_{12}$-dependent prefactor in \eqref{GFI_pentagon} results from rescaling integration variables by
$
        x_i \to x_i/(-s_{12}).
$

The proper LP polynomial has the monomial coefficients
\eq{
        \begin{array}{lllll}
        &z_1 &=& \ldots \ \ = \ \ z_6 \ \ = \ \ 1
        \\
        &z_7 &=& 1+y_2+y_4
        \\\
        &z_8 &=& y_1
        \\
        &z_9 &=& y_4 
        \\
        &z_{10} &=& -1+y_1-y_2-y_3
        \\
        &z_{11} &=& -1+y_1-y_2-y_3-y_4-y_5
        \end{array}
}
in terms of ratios
\eq{
        y_1 = \frac{p^2}{s_{12}}
        \ , \
        y_2 = \frac{s_{13}}{s_{12}}
        \ , \
        y_3 = \frac{s_{14}}{s_{12}}
        \ , \
        y_4 = \frac{s_{23}}{s_{12}}
        \ , \
        y_5 = \frac{s_{24}}{s_{12}}
        \ .
}
There are 5 monomial coefficients that differ from unity.
We can exactly match them with GKZ variables using homogeneity.
More precisely,
the GKZ system originally has $N=11$ variables,
but $n+1=5+1=6$ variables can be rescaled as 
$
        z_1 = \ldots = z_6 = 1,
$
leaving us with 5 variables $\{z_7, \ldots, z_{11}\}$.

\subsection{Basis}

Computing the volume of the Newton polytope with \package{polymake},
we expect
$
        R = 5! \times 13/120 = 13
$
MIs.
With the \package{asir} command \soft{mt\_gkz.cbase\_by\_euler},
we instantly identify a basis of $13$ standard monomials
\eq{
        e\supbrk{\Std} = 
        \arr{c}{
                \p_9 \p_{11}^2 \\
                \p_9^2 \\
                \p_{10}^2 \\
                \p_8 \p_{11} \\
                \p_9 \p_{11} \\
                \p_{10} \p_{11} \\
                \p_{11}^2 \\
                \p_7 \\
                \p_8 \\
                \p_9 \\
                \p_{10} \\
                \p_{11} \\
                1
        }
        \, .
        \label{pentagon_Std}
}
For the sake of verifying our results against an independent IBP calculation,
we also determine an integral basis using \package{LiteRed}:
\eq{
        e\supbrk{I} =
        (-s_{12})^\e
        \arr{c}{
                I(2|0,0,1,0,1) \\ 
                I(2|0,1,0,0,1) \\ 
                I(2|0,1,0,1,0) \\ 
                I(2|1,0,0,0,1) \\ 
                I(2|1,0,0,1,0) \\ 
                I(2|1,0,1,0,0) \\ 
                \e I(4|1,0,1,0,1) \\ 
                \e (-s_{12}) I(4|0,1,1,1,1) \\ 
                \e (2\e-1) I(6|1,0,1,1,1) \\ 
                \e (-s_{12}) I(4|1,1,0,1,1) \\ 
                \e (-s_{12}) I(4|1,1,1,0,1) \\
                \e (-s_{12}) I(4|1,1,1,1,0) \\ 
                \e^2 (-s_{12}) I(6|1,1,1,1,1) 
        }        
        \, .
        \label{pentagon_integral_basis}
}
This basis has been slightly dressed up with prefactors to simplify its Pfaffian system.
Moreover,
the dimension $\DD_0$ of each $I(\DD_0|\nu)$ is chosen to ensure that its associated $r$-vector 
only contains non-negative integers,
as per the remark above \exref{ex:r_vector}.
The $\mD$-module basis corresponding to \eqref{pentagon_integral_basis} is computed by inserting
these $r$-vectors into the \package{asir} command \soft{mt\_gkz.rvec\_red2}, 
with the result
\eq{
        e\supbrk{\mD} =
        \mathrm{c} \cdot
        \arr{c}{
                \p_{11} 
                \\[3pt]
                \p_{10} 
                \\[3pt]
                \p_9 
                \\[3pt]
                \p_8 
                \\[3pt]
                \p_7 
                \\[3pt]
                \e (5 \d +1)+
                z_7 \p_7+
                z_8 \p_8+
                z_9 \p_9+
                z_{10} \p_{10}+
                z_{11} \p_{11} 
                \\[3pt]
                (4 \e \d +\e +1) \p_{11} +
                z_{11} \p_{11}^2+
                z_9 \p_9 \p_{11}+
                z_{10} \p_{10} \p_{11} 
                \\[3pt]
                \p_9 \p_{11} 
                \\[3pt]
                e\supbrk{\mD}_9
                \\[3pt]
                \p_7 \p_{10} 
                \\[3pt]
                (5 \e  \d +\e +1) \p_{10} + 
                z_{10} \p_{10}^2+
                z_7 \p_7 \p_{10}+
                z_8 \p_8 \p_{10}+
                z_9 \p_9 \p_{10}+
                z_{11} \p_{11} \p_{10} 
                \\[3pt]
                (5 \e  \d +\e +1) \p_9 + 
                z_9 \p_9^2+
                z_7 \p_7 \p_9+
                z_8 \p_8 \p_9+
                z_{10} \p_{10} \p_9+
                z_{11} \p_{11} \p_9 
                \\[3pt]
                (4 \e  \d +\e +2) \p_{11} \p_9 + 
                z_9 \p_{11} \p_9^2+
                z_{10} \p_{10} \p_{11} \p_9+
                z_{11} \p_{11}^2 \p_9 
        }
        \, ,
        \label{pentagon_D_basis}
}
where
\eq{
        \nonumber
        e\supbrk{\mD}_9 =& \
        \e \d \left(4 \e \d +1\right)\p_{11} +
        z_{11} \e \d \p_{11}^2 +
        z_7 \left(4 \e \d + \e +1\right) \p_7 \p_{11} +
        z_9 \left(5 \e \d + \e +2\right) \p_9 \p_{11} 
        \\ & +
        z_{10} \e  \d  \p_{10} \p_{11} +
        z_7 z_{11} \p_7 \p_{11}^2 +
        z_9 z_{11} \p_9 \p_{11}^2 +
        z_9^2 \p_9^2 \p_{11}
        \\ & +
        z_7 z_9 \p_7 \p_9 \p_{11} +
        z_7 z_{10} \p_7 \p_{10} \p_{11} +
        z_9 z_{10} \p_9 \p_{10} \p_{11} 
        \nonumber
        \, ,
}
and $\mathrm{c}$ is a $13 \times 13$ diagonal matrix containing prefactors
(it can be downloaded from a \package{mathematica} notebook in \cite{url-mm-data}).

\subsection{Macaulay matrix}

The Macaulay matrix data is obtained in less than a second on a laptop via the \package{asir} command
\soft{mt\_mm.find\_macaulay}.
The matrix $M_\Ext$ has dimensions $189 \times 113$,
meaning that there are $113$ exterior monomials.
After fixing every parameter in $M_\Ext$ to a number,
row reduction reveals that only $133$ out of the $189$ rows are independent.
The matrix containing these independent rows is denoted by $M'_\Ext$.
Starting from the $189 \times 13$ matrix $M_\Std$,
we similarly define the $133 \times 13$ matrix $M'_\Std$.

For each 
$
        i \in \{7, \ldots, 11\}
$
we proceed to solve the equation
\eq{
        C\supbrk{i}_\Ext - C\supbrk{i} \cdot M'_\Ext = 0
}
for the unknown matrices $C\supbrk{i}$.
This only takes a few minutes on a laptop with \package{FiniteFlow}.
The Pfaffian matrix in direction $i$ is finally obtained from
\eq{
        C\supbrk{i}_\Std - C\supbrk{i} \cdot M'_\Std = P\supbrk{\Std}_i 
        \, .
}
To compare it with \package{LiteRed},
we perform a gauge transformation from the basis 
$e\supbrk{\Std}$ in \eqref{pentagon_Std} to $e\supbrk{\mD}$ in \eqref{pentagon_D_basis}.
After sending $\d \to 0$, 
we find perfect agreement.

        \section{Recurrence relation example: 1-loop bubble}
\label{sec:recurrence_relation_example}

Here we showcase \algref{alg:recurrence_by_MM} by computing an IBP relation for a simple bubble diagram.
The inverse propagators are

\noindent
\begin{minipage}{0.60\textwidth}
\begin{align*}
        D_1 = -\ell^2 + m^2
        \, , \quad
        D_2 = -(\ell+p)^2 
        \, ,
\end{align*}
\end{minipage}
\begin{minipage}{0.25\textwidth}
        \includegraphics[scale=0.15]{Figs/MM_bubble.pdf}
\end{minipage}

\vspace{0.5cm}
\noindent
where $p$ is an incoming external momentum.
The GFI takes the form
\eq{
        I(\DD_0|\nu_1,\nu_2) =
        c(\DD_0|\nu_1,\nu_2)
        \int_{(0,\infty)^2}
        \mG(z|x)^{\e-\DD_0/2} \,
        x_1^{\nu_1 + \e\d} \, x_2^{\nu_2 + \e\d} \,
        \frac{\dd x}{x}
        \, ,
}
with the LP monomials
\eq{
        \mG(z|x) = 
        z_1 \, x_1 + z_2 \, x_2 + z_3 \, x_1 \, x_2 + z_4 \, x_1^2
        \, , \quad
        A = 
        \arr{cccc}{
                1 & 1 & 1 & 1 \\
                1 & 0 & 1 & 2 \\
                0 & 1 & 1 & 0
        }
        \, .
}
We avoid the use of homogeneity in order to follow the steps of 
\algref{alg:recurrence_by_MM} as closely as possible%
\footnote{
        \algref{alg:recurrence_by_MM} assumes that the input $\Std$ consists of single monomials.
        Rewriting the basis in rescaled variables using
        \eqref{homogeneity_partial_derivative} would generally lead to operators with several terms,
        slightly modifying the algorithm.
}.
Instead, 
we simply specify
\eq{
        \label{z_i_bubble}
        z_1 = z_2 = 1
        \, , \quad
        z_3 = m^2 - p^2
        , \quad
        z_4 = m^2 
}
at the end of the computation to match with the physical LPr.

Fixing a basis of MIs
\eq{
        e =
        \arr{c}{
                I(4|1,1) \\
                I(4|2,0)
        }
        \, ,
}
we now proceed to find $c_{1,2}$ in the decomposition%
\footnote{
        Here we pick an example where all the $\nu_i$ are non-negative.
        For negative $\nu_i$,
        one has to slightly tweak the algorithm with the shift vector $\a$ from
        \cite[Equation 6.28]{Chestnov:2022alh}.
}
\eq{
        e_0 = 
        I(4|1,2) = 
        c_1 e_1 + c_2 e_2
}
using \algref{alg:recurrence_by_MM}.

\begin{paragraph}{Step 0: Input.}

Set
$
        \b = \{\e, -\e\d, -\e\d\} \in \CC^{3} \setminus \ZZ^{3}.
$
According to the formula \eqref{Euler_integral_recurrence_relation},
\algref{alg:recurrence_by_MM} pertains to Euler integrals of the form
\eq{
        f(\b - A \cdot q) = \gfi / \gfic
        \, .
}
By the definition of $\gfi$ from \eqref{generalized_feynman_integral},
we thus require that the argument of $f$ be
\eq{
        \b - A \cdot q = \b - \big(\DD_0/2,\nu)
        \, .
}
(Such a $q$-vector plays a similar role as the $r$-vector from \eqref{q_r_vector}.)
For the integrals in the example at hand,
we have
\eq{
        \begin{array}{c||c|c|c}
                         & q & A \cdot q & \Std_q = \p^q \\ \hline \hline
                I(4|1,2) & q_0 = (0,1,1,0) & (2,1,2) & \p_2 \p_3 \\ \hline
                I(4|1,1) & q_1 = (1,1,0,0) & (2,1,1) & \p_1 \p_2 \\ \hline
                I(4|2,0) & q_2 = (2,0,0,0) & (2,2,0) & \p_1^2
        \end{array}
}
In the final column,
we provided the $\mD$-module representatives of the integrals 
(up to prefactors).

\end{paragraph}

\begin{paragraph}{Step 1.}

Calling  \algref{alg:Pfaffian_by_MM},
we obtain the $2 \times 2$ Pfaffian matrices
$
        \{P_1, P_2, P_3, P_4\}
$
in the basis
$
        \Std = \{\p^{q_1}, \, \p^{q_2}\}.
$

\end{paragraph}

\begin{paragraph}{Step 2.}

We decompose
\eq{
        \nonumber
        q_0 - q_1 &= 
        (-1,0,1,0) \\&=
        (0,0,1,0) - (1,0,0,0) \\&=
        \nonumber
        q^+ - q^- \\&= q
        \, .
        \nonumber
}

\end{paragraph}

\begin{paragraph}{Step 3.}

This step instructs us to compute the matrix factorial corresponding to the operator
$
        \p_3^{+1} \, \p_1^{-1},
$
where $\p_3$ stems from $q^+$ and $\p_1$ stems from $q^-$.
Explicitly,
we compute
\eq{
        \label{matrix_factorial_bubble}
        P_3(\b+a_1) \cdot Q_1(\b) \cdot e
        \, .
}

\end{paragraph}

\begin{paragraph}{Step 4: Output.}

The sought after recurrence relation is encoded in the first element of \eqref{matrix_factorial_bubble}.
After sending $\d \to 0$ and replacing the $z_i$ variables as in \eqref{z_i_bubble},
we obtain the result
\eq{
        I(4|1,2) =
        \frac{(1-2\e)(p^2+m^2)}{(p^2-m^2)^2} I(4|1,1) -
        \frac{2m^2}{(p^2-m^2)^2} I(4|2,0)
        \, ,
}
which is in agreement with \package{LiteRed}.

\end{paragraph}

        \chapter{Restriction of \texorpdfstring{$\mD$-modules}{}}
\label{ch:restrictions}

As emphasized in the previous chapter,
the GKZ system does not perfectly capture the nature of FIs in the LPr.
The former requires all monomial coefficients $z_i$ of the integrand polynomial 
$\mG(z|x)$ to be indeterminate,
but the latter has some of the $z_i$ fixed to special values.
The question is then whether there exists a kind of \emph{limit} of the GKZ system,
at the level of $\mD$-modules,
which will exactly identify it with the annihilating ideal of a proper FI.

To motivate the need for a restriction procedure,
let us consider an example.
\begin{ex}
\label{ex:2L_nbox}
Let

\noindent
\begin{minipage}{0.75\textwidth}
\begin{align*}
        \gfi \propto
        \int_{(0,\infty)^5}
        \mG(z|x)^{\e-\DD_0/2} \,
        x_1^{\nu_1 + \e\d} \cdots x_5^{\nu_5 + \e\d} \,
        \frac{\dd x}{x}
\end{align*}
\end{minipage}
\begin{minipage}{0.24\textwidth}
        \vspace{0.2cm}
        \includegraphics[scale=1]{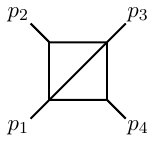}
\end{minipage}

\vspace{0.5cm}

\noindent
denote the GFI associated to the 2-loop massless N-box diagram.
The kinematics are
\eq{
        p_1^2 = \cdots = p_4^2 = 0
        \, \quad
        s = 2 p_1 \cdot p_2
        \, \quad
        t = 2 p_2 \cdot p_3
        \, ,
}
and the LP monomials read
\eq{
        \mG(z|x) &= 
        z_1 x_1 x_2 +
        z_2 x_1 x_4 +
        z_3 x_1 x_5 +
        z_4 x_2 x_3 +
        z_5 x_2 x_4 +
        z_6 x_2 x_5 
        \\ \nonumber & +
        z_7 x_3 x_4 +
        z_8 x_3 x_5 +
        z_9 x_1 x_2  x_4 +
        z_{10} x_2 x_3 x_5
        \, .
        \label{n_box_LP}
}
The holonomic rank of the GKZ system is $R = 9$,
as dictated by the formula \eqref{GKZ_holonomic_rank},
and it stays fixed when six of the $z_i$ variables are rescaled to unity via homogeneity.

After rescaling integration variables by $x \to x/s$,
the proper LPr has monomial coefficients confined to the space
\eq{
        Y' = 
        \big\{
                (z_1, \ldots, z_{10}) \in \AA^{10}
                \, \big | \,
                z_1 = \ldots = z_9 = 1
        \big\}
        \, ,
}
with $z_{10} = t/s$ being generic.
Counting the dimension of the twisted cohomology group $\deRham$ for $z \in Y'$
via the likelihood equation \eqref{likelihood_equation},
one finds that
$
        \dim(\deRham) = 3.
$
This is the correct number of MIs,
as as verified by an IBP calculation.

As $R = 9 > 3$,
we infer that the GKZ system is "too big".
We therefore seek some way to recover the space $\deRham$,
having special values for the $z$'s,
via some limiting procedure of the $\mD/H_A(\b)$.
\end{ex}

"Limits" of $\mD$-module are known under the name of \emph{restrictions} \cite[Chapter 5]{SST}.
Standard techniques for computing restrictions require huge computational resources,
and can generally not be applied to FIs of interest.
In this chapter,
we shall develop two new methods for computing restrictions that are more efficient.
The first one takes the Pfaffian system associated to a basis of Euler integrals as input.
The second one works at the level of the Macaulay matrix.
These two methods have different strengths and weaknesses,
so it depends on the context whether to apply one or the other.

        \section{Preliminaries}

We begin with some extra $\mD$-module formalism.
More details and examples besides what is presented here can be found in \cite[Appendix A]{Chestnov:2023kww}.

\subsection{Tensor products}

Tensor products of vector spaces are familiar to physicists from the study of quantum mechanics.
It is possible to define tensor products of $\mD$-modules as well.
Let $\mM$ and $\mN$ be two $\mD$-modules.
We proceed now to define the tensor product
$
        \mN \otimes_{\mD} \mM.
$

Let $n \otimes m$ denote the tuple $(n,m)$ where $n \in \mN$ and $m \in \mM$,
and define the space
$
        S(\mN \otimes \mM) = 
        \big\{
                \text{finite, formal sums} \
                \sum_i n_i \otimes m_i
        \big\}
        \, .
$
We specify an equivalence relation on $S$:
\begin{enumerate}
        \item
                Linearity in the left slot:
                \eq{
                        n_1 \otimes m + n_2 \otimes m 
                        \quad \sim \quad
                        (n_1 + n_2) \otimes m
                        \, .
                        \label{tensor_product_1}
                }
        \item 
                Linearity in the right slot:
                \eq{
                        n \otimes m_1 + n \otimes m_2
                        \quad \sim \quad
                        n \otimes (m_1 + m_2)
                        \, .
                        \label{tensor_product_2}
                }
        \item 
                Moving factors "across" the $\otimes$ sign:
                \eq{
                        (n \, d) \otimes m
                        \quad \sim \quad
                        n \otimes (d \, m)
                        \, ,
                        \label{tensor_product_3}
                }
                for $d \in \mD$.
                This is the reason for the subscript in $\otimes_{\mD}$.
\end{enumerate}
The $\mD$-module tensor product is then defined by the 
space $S(\mN \otimes \mM)$ modulo this equivalence relation,
that is
$
        \mN \otimes_{\mD} \mM := S(\mN \otimes \mM) / \sim.
$


\subsection{Definition of restriction}

The variables $z = (z_1, \ldots, z_N)$ parametrize an affine space $Y = \AA^N$.
We write $\mD_Y, \mR_Y$ etc.~for spaces of operators defined on $Y$.
Consider an affine hyperplane $Y ' \subset Y$ of lower dimension -
this will, later on, represent the proper space of kinematic variables.
By an affine linear change of coordinates,
we can define $Y'$ as the vanishing locus
\eq{
        Y' = \{z \in Y \, | \, z_1 = \cdots = z_{N'} = 0\} 
        \, .
}
\begin{definition}
Denote by $\iota: Y' \hookrightarrow Y$ the natural inclusion map.
Let $\mM$ be a $\mD_Y$-module.
The \emph{restriction module} of $\mM$ is defined as the $\mD_{Y'}$-module
\eq{
        \iota^* \mM := \frac{\mM}{z_1 \mM + \ldots + z_{N'} \mM}
        \, .
        \label{iota*_2}
}
The operation $\iota^*$ can be thought of as a pullback from $Y$ to $Y'$.
See \cite[Chapter 5]{SST} \cite[Chapter 6]{dojo} for details.
\end{definition}

A typical use case for the restriction module is the following.
Suppose the $\mD_Y$-module $\mM = \mD_Y / \mI$ is constructed from the annihilating ideal $\mI$
of some function $f(z)$ for $z \in Y$.
By computing $\iota^* \mM = \mD_{Y'} / \mI'$ for some $\mI'$,
then $\mI'$ is the annihilating ideal for $f(z)$ where now $z \in Y'$.
Naturally,
it is not necessary that $\mM$ be given by a quotient of operator spaces.
As in \exref{ex:function_space_as_D_module},
the $\mD_Y$-module $\mM$ can also be equal, as a set, to a suitable space of functions.

\begin{ex}
Consider the GKZ system associated to the Euler integral
\eq{
        f_\b(z_1,z_2,z_3) = 
        \int_{(0,\infty)^2} 
        (z_1 x_1 + z_2 x_2 + z_3 x_1 x_2)^{\b_0}
        x_1^{-\b_1} x_2^{-\b_2} 
        \frac{\dd x}{x}
        \, .
}
The annihilating ideal $\mI = H_A(\b)$ is solely generated by the homogeneity operators
\eq{
        \nonumber
        E_0 &= z_1 \p_1 + z_2 \p_2 + z_3 \p_3 - \b_0 \\
        E_1 &= z_1 \p_1 + z_3 \p_3 - \b_1 \\
        E_2 &= z_2 \p_2 + z_3 \p_3 - \b_2 
        \nonumber
        \, .
}
We are interested in the annihilating ideal for the function $f_\b(1,1,z_3)$.
By an affine linear coordinate change
$
        \{
        z_1 \to z_1 + 1
        \, ,
        z_2 \to z_2 + 1
        \, ,
        z_3 \to z_3
        \},
$
we hence have the affine subspace given by
$
        Y' = \{z_1 = z_2 = 0 \}.
$

A variant of the restriction algorithm presented in Oaku's seminal work \cite{OAKU1997495} 
has been implemented in the \package{asir} package \package{nk\_restriction} \cite{nakayama2010algorithm}.
Inputting the data above,
one finds the annihilating ideal
\eq{
       \mI' = \langle -z_3 \p_3 - \b_0 + \b_1 + \b_3 \rangle
       \, .
}
One can check that this indeed annihilates the function
\eq{
        f_\b(1,1,z_3) = 
        \frac{\Gamma(-\b_0+\b_1)\Gamma(-\b_0+\b_2)\Gamma(\b_0-\b_1-\b_2)}{\Gamma(-\b_0)}
        z_3^{-\b_0+\b_1+\b_2}
        \, .
}
\end{ex}

The discussion above was directed at $\mD$-modules,
i.e.~when the action is given by operators with polynomial coefficient functions.
We are,
more broadly,
interested in $\mR$-modules having rational function coefficients,
because this is what appears in Pfaffian systems.
An $\mR$-module naturally arises from a given $\mD$-module $\mM$ by taking a tensor product:
$
        \mR \otimes_{\mD_Y} \mM.
$
\begin{definition}
The \emph{rational restriction} of a $\mD_Y$-module $\mM$ onto $Y'$ is defined by
\eq{
        \mR_{Y'} \otimes_{\mD_{Y'}} \iota^* \mM
        \, ,
        \label{rational_restriction}
}
where $\iota^* \mM$ is the restriction module \eqref{iota*_2}.
\end{definition}
The restriction algorithms presented in this chapter will output Pfaffian systems
that live in the space \eqref{rational_restriction}.

\subsection{Normal form}
\label{sec:normal_form}

Let $\p_i \bullet e(z) = P_i \cdot e(z)$ be a Pfaffian system that is \emph{regular singular}
(cf.~\secref{sec:regular_holonomic}).
Loosely speaking,
this means that the solutions vector $e(z)$ is free of essential singularities such
$
        \exp(1/z)
$
at $z=0$.
It is expected that FIs fall into this category,
as the poles in $\e$-factorized Pfaffian systems always appear to be simple.

We focus on a single pole of the system,
which can be written as $Y' = \{z_1 = 0\}$ by a change of coordinates.
Theorems by Deligne \cite{deligne1970equations,deligneurl} ensure 
that the Pfaffian system can be brought into \emph{normal form} w.r.t.~$Y'$.
Namely,
there exists another set of Pfaffian matrices $\widetilde{P}_i(z)$ such that
\begin{enumerate}
        \item 
                $\widetilde{P}_1$ has a simple pole on $Y'$
                and
                $\{\widetilde{P}_2, \ldots, \widetilde{P}_N\}$ are finite on $Y'$: 
                \eq{
                        \label{normal_form_1}
                        \widetilde{P}_1(z) &= 
                        \sum_{n=-1}^\infty \widetilde{P}_{1,n}(z') z_1^n
                        \, , \quad \text{and}
                        \\[5pt]
                        \widetilde{P}_i(z) &= 
                        \sum_{n=0}^\infty \widetilde{P}_{i,n}(z') z_1^n
                        \, , \hspace{0.3cm} \quad \text{for} \quad
                        i = 2, \ldots, N \, ,
                        \label{normal_form_2}
                }
                where the truncated list of variables is
                \eq{
                        z' = (z_2, \ldots, z_N)
                        \, .
                }
                Note that the expansion of $\widetilde{P}_1$ starts at $n=-1$,
                and the expansions for $\widetilde{P}_{i \, = \, 2,\ldots,N}$ start at $n=0$.
        \item 
                The spectrum of the \emph{residue matrix} $\widetilde{P}_{1,-1}(z')$ is \emph{non-resonant},
                meaning that 
                1) no two distinct eigenvalues of $\widetilde{P}_{1,-1}(z')$ have an integral difference, 
                and
                2) the only integral eigenvalue, if it exists, is $0$.
\end{enumerate}
The matrices $\widetilde{P}_i(z)$ are related to the original ones $P_i(z)$ via a gauge transformation.
By the work of Moser \cite{Moser},
and later Barkatou et al.~\cite{barkatou,BARKATOU201741},
there exist explicit \emph{Moser reduction} algorithms for constructing such a gauge transformation
(see \cite[Appendix B]{Chestnov:2023kww} for a review).
Moser reduction has been implemented in software such as
\package{Fuchsia} \cite{Gituliar:2017vzm} and \package{Isolde} \cite{barkatou2013isolde}.

It can be computationally expensive to enforce a non-resonant spectrum on a Pfaffian system,
as this step of the algorithm requires the calculation of Jordan decompositions%
\footnote{
        In particular,
        one is required to know the sizes of the Jordan cells \cite[Appendix B.2]{Chestnov:2023kww}.
}.
However,
if we only seek to perform Moser reduction w.r.t.~a single pole $z_1 = 0$ rather than many,
then the algorithm can readily be applied to rather large systems.
Moreover,
since the Jordan decomposed matrix does not depend on the $z$-variables
\cite[Theorem 12.1]{haraoka2020linear},
it is possible speed up the computation by fixing the variables to generic numbers.

        \section{Restriction at the level of Pfaffian systems}
\label{sec:restriction_pfaffian_level}

In this section,
we present a restriction algorithm which takes a Pfaffian system on $Y = \AA^N$ as input
and outputs a new Pfaffian system that holds true on an affine subspace $Y' \subset Y$.
As has been mentioned,
the original motivation was to calculate restrictions of GKZ systems.
However,
the algorithm turns out to work for any regular holonomic $\mD$-module.
So it can,
rather serendipitously,
also be applied to Pfaffian systems for FIs obtained by conventional means,
such as IBPs in momentum space.

As it will turn out,
the algorithm can also be used to compute logarithmic series solutions to Pfaffian systems.
This is helpful for calculating asymptotic expansions for FIs,
such as threshold and small-mass expansions.

\subsection{Restriction of a logarithmic connection}
\label{sec:restriction_connection}

Suppose we are given a Pfaffian system on the "big" space $Y$.
We want to restrict it onto a "smaller" space $Y' = \{z_1 = 0\}$.
Further,
$z_1=0$ should constitute a pole in the Pfaffian system.
There might poles elsewhere on $Y$,
but we imagine working "locally" around this single pole%
\footnote{
        The notion of "local" is here meant w.r.t.~the Zariski topology.
        So to be rigorous,
        we ought to formulate everything in terms of 
        Zariski open subsets $Y_0$ of $Y$ as in \cite[Section 3]{Chestnov:2023kww}.
        Many such mathematical details are neglected here in order to take the shortest 
        path towards defining the practical algorithms.
}.
Assume that the Pfaffian system is in normal form w.r.t.~$Y'$
(cf. \secref{sec:normal_form}).
Then it is possible to write it in a peculiar way:
\eq{
        \label{peculiar_pfaffian_1}
        z_1 \p_1 \bullet e(z) &= P_1 \cdot e(z)
        \\
        \p_i \bullet e(z) &= P_i \cdot e(z)
        \, , \quad
        i = 2, \ldots, N 
        \, ,
        \label{peculiar_pfaffian_2}
}
where $e(z)$ is an $R$-dimensional vector of functions.
We would like to use this Pfaffian system to construct a certain
$\mD$-module $\mM$ consisting of \emph{regular} functions.
The action of the derivative $\p_i$ should be well-defined on such a space,
in the sense of not creating unwanted singularities.
We have therefore guarded ourselves in two ways to ensure well-definedness:
1) by working locally around $Y'$ we can thereby ignore singularities elsewhere, 
and 2) by moving $1/z_1$ from the RHS of \eqref{peculiar_pfaffian_1} to get $z_1 \p_1$ on the LHS.
We are thence motivated to define a subring of $\mD_Y$ by
\eq{
        \mD_Y(\log Y') = 
        \CC \langle z_1, \ldots, z_N, {\color{red} z_1 \p_1}, \p_2, \ldots, \p_N \rangle
        \, ,
}
and the abelian group
\eq{
        \mM = 
        \big\{
                R\text{-dimensional vectors of functions on} 
                \ Y \
                \text{that are regular near}
                \ Y'
        \big\}
        \, .
}
The action of the ring $\mD_Y(\log Y')$ onto $\mM$ is well-defined.
The generators $z_i$ act by scalar multiplication,
and the generators $\{z_1 \p_1, \p_2, \ldots, \p_N\}$ act according to the RHSs of the 
Pfaffian system \eqref{peculiar_pfaffian_1}-\eqref{peculiar_pfaffian_1}.
This turns $\mM$ into a $\mD_Y(\log Y')$-module.
In technical jargon, 
such an $\mM$ is called a \emph{logarithmic connection} 
\cite[Section 11.4]{andre2020rham} \cite[Section 5.2.2]{Hotta-Tanisaki-Takeuchi-2008}.
We now proceed to study the rational restriction
$
        \mR_{Y'} \otimes_{\mD_{Y'}} \iota^* \mM
$
of the logarithmic connection $\mM$.

To start,
let $P_{1,-1}(z')$ be the residue matrix 
(cf.~\eqref{normal_form_1}) 
associated to $P_1(z)$ defined in \eqref{peculiar_pfaffian_1},
where $z' = (z_2, \ldots, z_N)$.
The entries of $P_{1,-1}(z')$ live in the field $\mK_{Y'}$ consisting of rational functions on $Y'$.
The $R$ rows of $P_{1,-1}(z')$ generate a subspace of $\mK_{Y'}^R$ which we call $\mN'$.
The following quotient space is important for our construction of the rational restriction:
\eq{
        \mM' = \mK_{Y'}^R \, / \, \mN'
        \, .
        \label{mod_out_by_residue_matrix}
}
As a set,
this consists of $R$-dimensional vectors of rational functions on $Y'$,
and we identify two vectors if they differ by another vector lying in the row space of $P_{1,-1}(z')$.
$\mM'$ can actually be turned into an $\mR_{Y'}$-module.
The action of the derivatives $\{\p_2, \ldots, \p_N\}$ on an 
$R$-dimensional equivalence class $[m] \in \mM'$ is
\eq{
        \p_i \bullet [m] = P_{i,0}(z') \cdot [m]
        \quadit{\text{for}}
        i = 2, \ldots, N
        \, ,
}
where the matrices $P_{i,0}(z')$ come from the leading terms in the $z_1$-expansion \eqref{normal_form_2}.

With all of this build-up,
we can finally state the following crucial isomorphism of $\mR_{Y'}$-modules%
\footnote{
        The formula \eqref{restriction_module_isomorphism}
        shows the rational restriction of $\mM$ when it equals a space of functions as a set.
        Recalling the discussion from \secref{sec:Pfaffian_systems},
        this isomorphism can also be employed when $\mM$ is of the form
        $
                \mD_{Y} / \mI
        $
        for some annihilating ideal of differential operators $\mI$.
}:
\eq{
        \mR_{Y'} \otimes_{\mD_{Y'}} \iota^* \mM
        \ \simeq \
        \mM'
        \, .
        \label{restriction_module_isomorphism}
}
We refer to \cite[Proposition 3.4]{Chestnov:2023kww} for many more mathematical details as well as a proof.

The takeaway message up until now is this.
We are given some Pfaffian matrices $\{P_1(z), \ldots, P_N(z)\}$ on $Y$ which define $\mM$.
We would like to know the Pfaffian matrices $\{Q_2(z'), \ldots, Q_N(z')\}$ on the smaller space $Y'$.
The matrices $Q_i(z')$ are associated to the rational restriction module of $\mM$.
Because of the isomorphism \eqref{restriction_module_isomorphism},
the matrices $Q_i(z')$ are equally well associated to the quotient space $\mM'$.
Fortunately,
we have a lot of information about $\mM'$:
it is completely specified by the residue matrix $P_{1,-1}(z')$!
In the following subsection,
we use this information to provide an explicit formula for the Pfaffian matrices $Q_i(z')$.

\subsection{Constructing the restricted Pfaffian system}

Let $R'$ denote the holonomic rank of $\mM'$.
In general $R' < R$,
i.e.~the holonomic rank drops when we fix $z$-variables to special values.
Given an element $m \in \mK^R_{Y'}$,
we write $[m]$ for its equivalence class in $\mM'$.

We define three sets of Pfaffian systems:
\begin{enumerate}
        \item 
                We suppose the existence of a \emph{special} basis
                $
                        f = f(z') = \{f_1, \ldots, f_R\} 
                $
                obeying the Pfaffian system
                \eq{
                        \p_i \bullet f = \widetilde{Q}_i(z') \cdot f
                        \quadit{\text{for}}
                        i = 2, \ldots, N
                        \, .
                        \label{1st_pfaffian_system}
                }
                We do not yet know $f$ nor $\widetilde{Q}_i(z')$.
        \item 
                The basis $f$ is special for the following reason:
                we assume that
                $
                        \{f_{R'+1}, \ldots, f_R\}
                $
                is a basis of $\mN'$.
                The set 
                $
                        [f] = \{[f_1], \ldots, [f_{R'}]\}
                $
                is then a basis for $\mM'$,
                with an associated Pfaffian system
                \eq{
                        \p_i \bullet [f] = Q_i(z') \cdot [f]
                        \quadit{\text{for}}
                        i = 2, \ldots, N
                        \, .
                        \label{2nd_pfaffian_system}
                }
                By stipulation,
                the first $R' \times R'$ block of $\widetilde{Q}_i(z')$ equals $Q_i(z')$.
                We do not yet know $[f]$ nor $Q_i(z')$.
        \item 
                Let
                $
                        e = e(z')= \{e_1, \ldots, e_R\} 
                $
                be the basis associated to the Pfaffian system
                \eq{
                        \p_i \bullet e = P_{i,0}(z') \cdot e
                        \quadit{\text{for}}
                        i = 2, \ldots, N
                        \, .
                        \label{3rd_pfaffian_system}
                }
                The matrices $P_{i,0}(z')$ \emph{are} assumed to be 
                known via the formula \eqref{normal_form_2}.
                In contrast to $\widetilde{Q}_i(z')$,
                the matrices $P_{i,0}(z')$ have no nice block structure.
\end{enumerate}
We are hence in the following situation:
we know $P_{i,0}(z')$ (for the basis $e$),
but we do not know $\widetilde{Q}_i(z')$ (for the special basis $f$).
We would like to know $\widetilde{Q}_i(z')$,
because then we would automatically get the Pfaffian matrices $Q_i(z')$ 
for the rational restriction module $\mM'$.
We shall achieve this goal by building a gauge transformation mapping $e$ to the special basis $f$.
This is done by through three new matrices $\{\Bmat, \Rmat, \Mmat\}$,
which we define now.
\begin{itemize}
        \item 
                Recall the basis 
                $
                        [f] = \{[f_1], \ldots, [f_{R'}]\}
                $
                of $\mM'$.
                Each representative of these equivalence classes can be expanded in the basis $e$ via some
                $
                        R' \times R
                $ 
                matrix $\Bmat$ having entries in $\mK_{Y'}$:
                \eq{
                        f_i = \sum_{j=1}^R \Bmat_{ij} \, e_j
                        \quadit{\text{for}}
                        i = 1, \ldots, R'
                        \, .
                        \label{B_matrix}
                }
                The entries of $\Bmat$ are not known at this point.
        \item 
                By the definition of $\mM'$,
                the residue matrix $P_{1,-1}(z')$ has exactly $R-R'$ independent rows.
                Define an $(R-R') \times R$ matrix by
                \eq{
                        \Rmat = \rowred{ P_{1,-1}(z') }
                        \, ,
                        \label{R_matrix}
                }
                where \soft{RowReduce} includes the deletion of zero-rows.
                Then
                \eq{
                        f_i = \sum_{j=1}^R \Rmat_{ij} \, e_j
                        \quadit{\text{for}}
                        i = R'+1, \ldots, R
                }
                forms a basis of $\mN'$.
        \item 
                Aligning the row of $\Bmat$ and $\Rmat$ produces an invertible $R \times R$ matrix $\Mmat$:
                \eq{
                    \Mmat =
                    \arr{c}{
                            {\color{red1} \Bmat} \\ 
                            \hline 
                            {\color{purple1} \Rmat}
                    }
                    =
                    \vcenter{\hbox{ \includegraphics[scale=1]{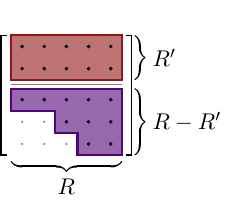} }}
                    \ .
                    \label{M_matrix}
                }
                Note that $\Rmat$ is known by stipulation,
                and the matrix $\Bmat$ can be anything as long as $\Mmat$ becomes invertible.
                We shall give a simple recipe for generating $\Bmat$ later on.
\end{itemize}
In vector notation,
we now have that $f = \Mmat \cdot e$.
This means that the unknown matrices $\widetilde{Q}_i(z')$ 
(associated to $f$) 
are obtained from the known matrices $P_{i,0}(z')$ 
(associated to $e$)
by gauge transforming with $\Mmat^{-1}$.
Because the sought after matrix $Q_i(z')$ is given by the first $R' \times R'$ block of $\widetilde{Q}_i(z')$,
we can finally write the key formula
\begin{empheq}[box=\fbox]{align}
        \widetilde{Q}_i(z')
        \quadit{=}
        \big[
                \p_i \bullet \Mmat
                \plus
                \Mmat \cdot P_{i,0}(z')
        \big]
        \cdot
        \Mmat^{-1}
        \quadit{=}
        \arr{c|c}{
                Q_i(z') & \star \\
                \hline
                \mathbf{0} & \star
        }
        \, .
        \label{M_gauge_transformation}
\end{empheq}
See 
\cite[Appendix C.2.1]{Chestnov:2023kww} 
for an explanation of the zero block in the lower left corner.
Obtaining this block is a useful check for practical computations.

Before closing this section,
let us observe that the integer $N$ does not have to be the "$N$" of the GKZ system.
In fact,
we did not use any notions related to the GKZ system to derive \eqref{M_gauge_transformation},
so the preceding discussion holds true for any (regular singular) Pfaffian system!

        \subsection{Holomorphic restriction of solutions}
\label{sec:holomorphic_restriction}

The previous section focused on the matrices associated to Pfaffian systems.
In the following two subsections,
we draw our attention to the \emph{solution vectors} of Pfaffian systems,
in particular how to obtain actual series representations using restriction.
This topic can be viewed as singular boundary value problem 
(see \cite{haraoka2020linear} \cite{takayama1992propagation} and references therein).

To emphasize the connection to MIs,
let us use the notation $\vec{I}(z)$ for the solution vector of a Pfaffian system
\eq{
        \p_i \bullet \vec{I}(z) = P_i(z) \cdot \vec{I}(z)
        \, , \quad
        i = 1, \ldots, N
        \, .
        \label{I_Pfaffian_system}
}
Suppose that we are interested in solutions that are \emph{holomorphic} on $Y' = \{z_1 = 0 \}$.
This means that $\vec{I}(z)$ enjoys a Taylor expansion
\eq{
        \vec{I}(z) =
        \sum_{n=0}^\infty \In (z') \, z_1^n
        \, ,
        \label{I_holomorphic_solution}
}
where as usual $z' = (z_2, \ldots, z_N)$.
The issue at hand is how compute the expansion coefficient functions $\In$ up to some appropriate order 
$
        \mO(z_1^k).
$

Assume that the matrices $P_i(z)$ are in normal form.
Insert the $z_1$-expansions of $P_i(z)$ and $\vec{I}(z)$,
given by  
\eqref{normal_form_1}-\eqref{normal_form_2} and \eqref{I_holomorphic_solution},
into the Pfaffian system \eqref{I_Pfaffian_system}.
Relations among the $\In$ are then obtained by comparing powers of $z_1^n$ on both sides of the equality
\cite[Chapter 4]{haraoka2020linear} \cite[Appendix D.6]{Chestnov:2023kww}.
Comparing the leading-order terms in $z_1$,
the following relations are found:
\eq{
        \begin{array}{rl}
                & \p_i \bullet \Iz(z') \ = \ P_{i,0}(z') \cdot \Iz(z')
                \, , \quad
                i = 2, \ldots, N
                \\[10pt]
                & \text{Rank jump constraint:} \quad P_{1,-1}(z') \cdot \Iz (z') \ = \ 0
                \, .
        \end{array}
        \label{holomorphic_restriction}
}
The higher-order terms in $z_1$ yield the recursion relation%
\footnote{
        The matrix in front of the sum is indeed invertible because the spectrum of 
        $P_{1,-1}(z')$ is non-resonant.
        More precisely,
        the only potential integral eigenvalue of $P_{1,-1}(z')$ is $0$,
        and the diagonal is shifted by some $n>0$,
        wherefore the determinant of the full expression is non-zero.
}
\eq{
        \In(z') =
        \big[ n\mathbf{1} - P_{1,-1}(z') \big]^{-1}
        \ \cdot \
        \sum_{i=0}^{n-1} P_{1,n-i-1} \cdot \vec{I} \, \supbrk{i} (z')
        \, , \quad
        n > 0
        \, .
        \label{holomorphic_recursion}
}

The set of PDEs \eqref{holomorphic_restriction} is dubbed the 
\emph{holomorphic restriction} 
of the Pfaffian system \eqref{I_Pfaffian_system}.
The first equation of \eqref{holomorphic_restriction} shows that $\Iz$ plays 
the role of the basis $e$ in \eqref{3rd_pfaffian_system}.
The second equation of \eqref{holomorphic_restriction} shows why the residue matrix $P_{1,-1}(z')$ 
played such a central role in the construction of \secref{sec:restriction_connection},
wherein we modded out by the rows of $P_{1,-1}(z')$ to obtain the rational restriction module.

At the level of solutions for Pfaffian systems,
we now have a clear interpretation for the \emph{rank jump} from $R$ to $R'$:
the equation
$
        P_{1,-1}(z') \cdot \Iz (z') = 0
$
imposes constraints among the entries of $\Iz(z')$ in the limit $z_1 \to 0$,
meaning that some of them become linearly dependent on each other.
For FIs,
these relations turn out to be the new IBP relations among MIs which arise when the limit 
$
        z_1 \to 0
$
is taken at the \emph{integrand} level.
An example of this phenomenon is shown in \secref{sec:1L_bhabha_scattering}.

Observe that once $\Iz$ is determined from the PDEs \eqref{holomorphic_restriction},
then \eqref{holomorphic_recursion} fixes all higher-order terms $\In$ via simple matrix multiplication
(one can even use finite field reconstruction in this step for efficiency).
The PDEs \eqref{holomorphic_restriction} are,
however,
written in a redundant way due to the residue matrix constraint.
We thus seek to "mod out" by this constraint,
in order to write the system in a minimal basis.
To set this up,
begin by writing $\Iz$ in terms of some dummy symbols, 
say
\eq{
        \Iz = \arr{c}{I_1 \\ \vdots \\ I_R}
        \, .
}
Recall the definition of the $\Rmat$-matrix from \eqref{R_matrix}.
Because of the second equation in \eqref{holomorphic_restriction},
we have that
\eq{
        \Rmat \cdot \Iz = 0 .
}
This equation can be solved to determine which of the $I_i$ symbols are linearly dependent on each other.
Removing those linearly dependent entries from $\Iz$,
the remaining ones are collected into a new $R'$-dimensional basis
\eq{
        \vec{J}
        \quadit{=}
        \arr{c}{ I_{i_1} \\ \vdots \\ I_{i_{R'}} }
        \quadit{=}
        \Bmat \cdot \Iz
        \, .
}
The newly introduced $\Bmat$-matrix has dimension $R' \times R$ and entries $\Bmat_{ij} \in \{0,1\}$.
In fact,
this is precisely the $\Bmat$-matrix from \eqref{B_matrix}.
Therefore,
both blocks of the $\Mmat$-matrix \eqref{M_matrix} are now known.
By construction,
it holds that
\eq{
        \Mmat \cdot \Iz =
        \arr{c}{\Bmat \\ \hline \Rmat} \cdot \Iz =
        \arr{c}{\vec{J} \\ \hline \mathbf{0}}
        \, .
        \label{M_matrix_multiplication}
}
The Pfaffian matrices $Q_i(z')$ for the basis $\vec{J}$ 
now immediately follow from \eqref{M_gauge_transformation},
giving the desired, restricted Pfaffian system
\eq{
        \p_i \bullet \vec{J}(z') = Q_i(z') \cdot \vec{J}(z')
        \quadit{\text{for}}
        i = 2, \ldots, N
        \, .
}
This system is \emph{simpler} to solve than the original Pfaffian system 
\eqref{I_Pfaffian_system} for $\vec{I}(z)$ because
1) it has a smaller rank $R' < R$,
and 2) it contains one less variable.
Once $\vec{J}$ is solved for,
then $\Iz$ is determined by multiplying equation \eqref{M_matrix_multiplication} with $\Mmat^{-1}$.

\algref{alg:holomorphic_restriction} summarizes the preceding discussion.
Note that an initial Moser reduction step is required if the input is not in normal form.

\begin{algorithm}[H]
        \underline{Input}: 
        \begin{itemize}
                \item 
                Pfaffian system in normal form of rank $R$:
                $
                        \p_i \bullet \vec{I}(z) =
                        P_i(z) \cdot \vec{I}(z)
                $
                for 
                $
                        i = 1, \ldots, N.
                $
                \item 
                Integer $k>0$.
        \end{itemize}
        \underline{Output}: 
        The first $k$ coefficient functions of
        $
                \vec{I}(z) = 
                \sum_{n=0}^\infty 
                \In(z') \, z_1^n .
        $
        \vspace{0.2cm}
    \begin{algorithmic}[1]
            \State 
            Set 
            $\Rmat = \soft{RowReduce}[P_{1,-1}]$
            and
            $R' = R - \rank{\Rmat}$.
            \vspace{0.4cm}
            \State 
            Write $\Iz = \big[I_1, \ldots, I_R \big]^T$.
            Eliminate linearly dependent dummy symbols $I_i$ by solving $\Rmat \cdot \Iz = 0$.    
            Collect linearly independent ones into $\vec{J}$.
            Fix $\Bmat$ via
            $
                \vec{J} = \Bmat \cdot \Iz.
            $
            \vspace{0.4cm}
            \State 
            Set
            $
                \Mmat = \arr{c}{\Bmat \\ \hline \Rmat}.
            $
            \vspace{0.4cm}
            \State
            For each $i = 2, \ldots, N$,
            compute 
            $
                \big( \p_i \bullet \Mmat + \Mmat \cdot P_{i,0} \big) \cdot \Mmat^{-1}
            $
            and save the upper left $R' \times R'$ block matrices dubbed $Q_i$.
            \vspace{0.4cm}
            \State
            Solve 
            $
                \p_i \bullet \vec{J} = Q_i \cdot \vec{J}
            $
            for $\vec{J}$ and set
            $
                \Iz = \Mmat^{-1} \cdot \arr{c}{\vec{J} \\ \hline \mathbf{0}}.
            $
            \vspace{0.4cm}
            \State
            For each $n = 1,\ldots,k$, calculate $\In$ using the recursion \eqref{holomorphic_recursion}.
            \\
            \vspace{0.4cm}
            \Return $\{\vec{I} \, \supbrk{0}, \ldots, \vec{I} \, \supbrk{k}\}$.
    \end{algorithmic}
    \caption{: Holomorphic series expansion}
    \label{alg:holomorphic_restriction}
\end{algorithm}

\noindent

\begin{figure}[H]
        \centering
        \includegraphics[scale=0.8]{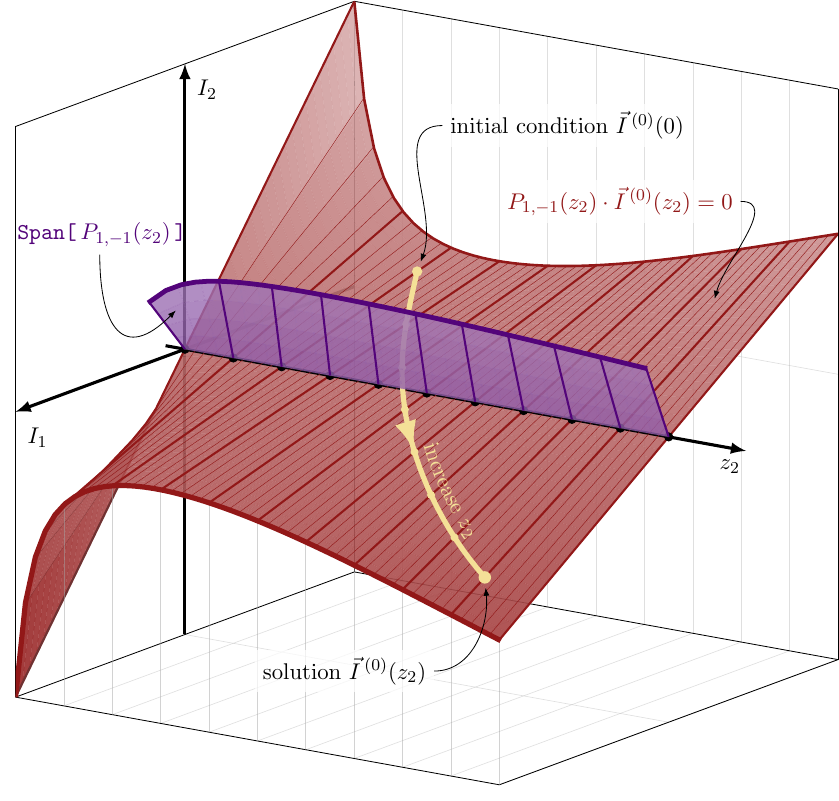}
        \caption{
                The holomorphic restriction of a rank $R=2$ solution vector 
                $
                \vec{I}(z_1,z_2) = \arr{c}{I_1(z_1,z_2) \\ I_2(z_1,z_2)}
                $
                as $z_1 \to 0$.
                The {\color{purple1} purple} and {\color{red1} red} 
                surfaces depict $\Rmat$ and $\Bmat$ respectively.
        }
        \label{fig:holomorphic_restriction}
\end{figure}

A geometric interpretation of the holomorphic restriction is illustrated in 
\figref{fig:holomorphic_restriction}.
There are three axes,
one for the variable $z_2$ and two for the vector components of
$
        \Iz(z') = [I_1(0,z_2), \, I_2(0,z_2)]^T,
$
where $I_1(z_1,z_2)$ and $I_2(z_1,z_2)$ represent a basis before restriction.

The {\color{red1} red} surface depicts all possible solutions to the rank jump constraint,
i.e.~the nullspace of the residue matrix $P_{1,-1}(z_2)$.
For fixed $z_2$,
this nullspace spans a 1-dimensional vector space shown as the {\color{red1} red} lines.
As $z_2$ varies,
the lines trace out the whole {\color{red1} red} surface%
\footnote{Such a family of vector spaces is naturally formalized in terms of vector bundles.}.
The solutions living on the surface constitute a basis,
determined by the row space of the $\Bmat$-matrix \eqref{B_matrix},
for the restricted Pfaffian system.
Since the nullspace is 1-dimensional,
the size of this basis is $R' = R - 1 = 1$.

For fixed $z_2$,
each {\color{purple1} purple} line represents the 1-dimensional span of the row vectors in $P_{1,-1}(z_2)$.
Varying $z_2$,
then these lines trace out the {\color{purple1} purple} surface.
This surface is thus a depiction of the $\Rmat$-matrix \eqref{R_matrix}.

Since the {\color{red1} red} and {\color{purple1} purple} surfaces are orthogonal to each other,
together they span a 2-dimensional vector space.
This implies that the $\Mmat$-matrix \eqref{M_matrix} is full rank and therefore invertible.

The {\color{yellow1} yellow} curve illustrates a \emph{particular} solution vector $\Iz(z_2)$,
given some initial condition $\Iz(0)$.
Observe that this solution is constrained to lie on the {\color{red1} red} surface -
this is analogous to the situation in classical mechanics where one finds the solution to
some equations of motion given a holonomic constraint.
According to \eqref{holomorphic_restriction},
the "flow" of the solution vector is governed by the Pfaffian matrix $P_{2,0}(z_2)$:
given some point $\Iz(\zcirc{2})$ on the curve,
then the solution is evolved a small step in $z_2$ by
\eq{
        \nonumber
        \Iz(\zcirc{2} + \Delta z_2) 
        & \ \simeq \
        \Iz(\zcirc{2}) \ + \ \Delta z_2 \Big[ \p_2 \bullet \Iz(z_2) \Big] \Big|_{z_2 = \zcirc{2}}
        \\& \ = \
        \Iz(\zcirc{2}) \ + \ \Delta z_2 \Big[ P_{2,0}(\zcirc{2}) \cdot \Iz(\zcirc{2}) \Big] 
        \, .
}

        \subsection{Logarithmic restriction of solutions}
\label{sec:logarithmic_restriction}

Now suppose that the solution vector $\vec{I}(z)$ to the Pfaffian system \eqref{I_Pfaffian_system} is
logarithmically singular rather than holomorphic on $Y' = \{z_1 = 0\}$.
Our goal is to state an algorithm for calculating the logarithmic series expansion of $\vec{I}(z)$.

The first step is to write down the most general ansatz to a Pfaffian system with regular singularities.
When the Pfaffian system is in normal form,
it is known that the ansatz decomposes into a sum over unique eigenvalues of the residue matrix
\cite[Chapter 4]{haraoka2020linear}.
Write the spectrum of the residue matrix as
\eq{
        \spec{P_{1,-1}(z')} =
        \Big\{
                \underbrace{\l_1, \ldots, \l_1}_{\L_1}
                \ , \
                \underbrace{\l_2, \ldots, \l_2}_{\L_2}
                \ , \ldots
        \Big\}
        \, ,
}
where $\L_i$ is the multiplicity of eigenvalue $\l_i$.
Even though $P_{1,-1}(z')$ depends on $z'$,
the integrability relations \eqref{integrability} actually dictate that the 
eigenvalues $\l_i$ are independent of $z'$ \cite[Theorem 12.1]{haraoka2020linear}.
For FIs in DR,
it means that $\l_i$ can only depend on $\e$.
In all examples known to us,
the dependence is even linear in $\e$:
\eq{
        \l_i = a_i + b_i \e
        \quadit{\text{for}}
        a_i, b_i \in \QQ
        \, .
}
We are not aware of a proof of the above statement.

The ansatz for a logarithmically divergent solution vector can now be represented as
\eq{
        \vec{I}(z) 
        \quadit{=}
        \sum_{ \l \ \in ! \ \spec{P_{1,-1}} } 
        z_1^\l
        \ \times \
        \sum_{n=0}^\infty 
        \ z_1^n
        \ \times \
        \sum_{m=0}^{M_\l}
        \ \Ilnm(z') 
        \ \times \
        \log^m(z_1)
        \label{I_asymptotic_series}
        \, ,
}
where "$\in !$" means "unique element of".
In DR,
$\Ilnm$ can be further expanded in $\e$.
Note that there are two sources of logarithms in that case:
from the $\e$-expansion of
$
        z_1^{\l_i} = z_1^{a_i+b_i\e}
$
and from the finite sum over $m$.

The integer $M_\l \in \ZZ_{\geq 0}$ denotes the maximal power of $\log^m(z_1)$.
Its value depends on the Jordan form of the residue matrix.
Decomposing this matrix into blocks labeled by unique eigenvalues
\eq{
        \jord{P_{1,-1}} = 
        \bigoplus_{ \l \ \in ! \ \spec{P_{1,-1}} }
        \, 
        \mathsf{J} \supl
        \, ,
        \label{jordan_decomposition_residue_matrix}
}
then%
\footnote{
        Recall that a Jordan block has the eigenvalue $\l$ on the diagonal,
        and $1$'s on the superdiagonal.
        $0$'s on the superdiagonal correspond to Jordan blocks of size $1$.
}
\eq{
        M_\l =
        \big[
                \text{size of the largest Jordan block of} \
                \mathsf{J}\supl
        \big]
        - 1
        \, .
        \label{max_log_power}
}

The ansatz \eqref{I_asymptotic_series} can now be inserted into the Pfaffian system 
\eqref{I_Pfaffian_system}
together with the $z_1$-expansions of $P_i(z)$ from 
\eqref{normal_form_1}-\eqref{normal_form_2}.
One thereby obtains relations among the coefficient functions $\Ilnm$ by comparing the coefficients of 
$
        z_1^n \, \log^m(z_1)
$ 
on both sides of the equality \cite[Chapter 4]{haraoka2020linear}.
For fixed $\l$,
among these relations one finds 
\eq{
        \begin{array}{ll}
                & \p_i \bullet \Ilzz (z') \ = \
                P_{i,0}(z') \ \cdot \ \Ilzz (z')
                \qquad \text{for} \qquad
                i = 2, \ldots, N
                \\[5pt]
                & \text{Rank jump constraint}: \quad
                \big[ P_{1,-1}(z') - \l \mathbf{1} \big]^{M_\l+1} \cdot \ \Ilzz (z') \ = \ 0
                \, . 
        \end{array}
        \label{logarithmic_restriction}
}
We call this the \emph{logarithmic restriction} of the solution vector.
If $\l = 0$,
which is the unique integral eigenvalue for a system in normal form,
then \eqref{logarithmic_restriction} reduces to the holomorphic restriction \eqref{holomorphic_restriction}.

The second equation of \eqref{logarithmic_restriction} represents a constraint on the vector $\Ilzz$;
it must be a generalized eigenvector for $P_{1,-1}(z')$.
Recall that we modded out by the rows of $P_{1,-1}(z')$ in the holomorphic case because of the constraint
$
        P_{1,-1}(z') \cdot \Iz = 0
$
(cf.~\eqref{mod_out_by_residue_matrix}).
For fixed $\l$,
in the logarithmic case we instead mod out the rows of
$
        \big[ P_{1,-1}(z') - \l \mathbf{1} \big]^{M_\l+1}
$
(this would therefore supplant the $\mN'$ in \eqref{mod_out_by_residue_matrix}).

The vector $\Ilzz$ hence plays the role of the basis $e$ in \eqref{3rd_pfaffian_system},
and we can repeat the construction of \secref{sec:restriction_connection} by replacing $\Rmat$ with
\eq{
        \Rmat\supl =
        \rowred{
                \big( P_{1,-1}(z') - \l \mathbf{1} \big)^{M_\l+1}
        }
        \, .
        \label{R_matrix_l}
}
This matrix can be inserted into the formula \eqref{M_gauge_transformation} 
together with some basis matrix $\Bmat \supl$.
The result is a restricted Pfaffian system of size
$
        {R\,'}\supl 
$
for a basis $\Jl(z')$:
\eq{
        \p_i \bullet \Jl(z') \ = \ Q_i\supl(z') \ \cdot \ \Jl(z')
        \quadit{\text{for}}
        i = 2, \ldots, N
        \, .
}
The rank of this system is precisely the eigenvalue multiplicity of $\l$,
which also equals the size of the block matrix $\mathsf{J}\supl$ in 
\eqref{jordan_decomposition_residue_matrix}.
That is,
\eq{
        {R\,'}\supl = \L 
        \, .
}

The higher-order terms $\Ilnm$ follow from recursion relations given $\Ilzz$ as input.
\begin{itemize}
        \item 
                \underline{Shifting powers of $\log^m(z_1)$}.
                Set
                \eq{
                        \Phi\supbrk{\l,m} = 
                        \frac{P_{1,-1} - \l \mathbf{1}}{m}
                        \, .
                }
                Then
                \eq{
                        \vec{I} \, \supbrk{\l,0,m} 
                        \quadit{=}
                        \Phi\supbrk{\l,m}
                        \ \cdot \
                        \vec{I} \, \supbrk{\l,0,m-1} 
                        \quadit{\text{for}}
                        0 < m \leq M_\l
                        \, .
                        \label{m_shift}
                        \tag{A}
                }
                The terminating condition is
                $
                        \vec{I} \, \supbrk{\l,0,M_\l+1} = 0.
                $
        \item 
                \underline{Shifting powers of $z_1^n$}.
                Define
                \eq{
                        \Psi\supbrk{\l,n} =
                        \big[
                                (\l+n)\mathbf{1} - P_{1,-1}
                        \big]^{-1}
                        \, .
                }
                It holds that
                \eq{
                        \Ilnm
                        \quadit{=}
                        \Psi\supbrk{\l,n}
                        \ \cdot \
                        \left(
                                \sum_{\nu=0}^{n-1}
                                P_{1,n-\nu-1} 
                                \ \cdot \
                                \vec{I} \, \supbrk{\l,\nu,m}
                                \minus
                                (m+1) \vec{I} \, \supbrk{\l,n,m+1}
                        \right)
                        \, .
                        \label{n_shift}
                        \tag{B}
                }
                The last term drops out when $m = M_\l$,
                leading to%
                \footnote{
                        \eqref{n_shift} in fact also reduces to \eqref{m_shift} when $n=0$.
                }
                \eq{
                        \vec{I} \, \supbrk{\l,n,M_\l}
                        \quadit{=}
                        \Psi\supbrk{\l,n}
                        \ \cdot \
                        \sum_{\nu=0}^{n-1}
                        P_{1,n-\nu-1} 
                        \ \cdot \
                        \vec{I} \, \supbrk{\l,\nu,M_\l}
                        \, .
                        \label{n_shift_simple}
                        \tag{C}
                }
                The index $n$ runs up to $\infty$,
                so this recursion ought to be truncated based on a
                numerical accuracy condition for the full solution vector $\vec{I}(z)$.
\end{itemize}

\noindent
Starting from $\Ilzz$,
we present a flowchart showing how to use the recursion relations 
\eqref{m_shift}, \eqref{n_shift} and \eqref{n_shift_simple} 
to compute $\Ilnm$ for all
$
        0 \leq n \leq N_\l
$
and 
$
        0 \leq m \leq M_\l.
$
The flowchart is adapted from \cite[Page 42]{haraoka2020linear},
and the tuple $(n,m)$ represents $\Ilnm$.
\eq{
        \label{recursion_flowchart}
        \begin{array}{ccccccccc}
          (0,0)
        & \overset{\eqref{m_shift}}{\longrightarrow}
        & (0,1)
        & \overset{\eqref{m_shift}}{\longrightarrow}
        & \cdots
        & \overset{\eqref{m_shift}}{\longrightarrow}
        & (0, M_\lambda-1)
        & \overset{\eqref{m_shift}}{\longrightarrow}
        & (0, M_\lambda)
        \\[10pt]
        &&&&&&&
        & \hspace{0.3cm} \rotatebox[origin=c]{270}{$ \overset{\eqref{n_shift_simple}}{\longrightarrow} $}
        \\[10pt]
          (1,0)
        & \overset{\eqref{n_shift}}{\longleftarrow}
        & (1,1)
        & \overset{\eqref{n_shift}}{\longleftarrow}
        & \cdots
        & \overset{\eqref{n_shift}}{\longleftarrow}
        & (1, M_\lambda-1)
        & \overset{\eqref{n_shift}}{\longleftarrow}
        & (1, M_\lambda)
        \\[10pt]
        &&&&&&&
        & \hspace{0.3cm} \rotatebox[origin=c]{270}{$ \overset{\eqref{n_shift_simple}}{\longrightarrow} $}
        \\
        &&&&&&&
        & \vdots
        \\[10pt]
        &&&&&&&
        & \hspace{0.3cm} \rotatebox[origin=c]{270}{$ \overset{\eqref{n_shift_simple}}{\longrightarrow} $}
        \\[10pt]
          (N_\lambda ,0)
        & \overset{\eqref{n_shift}}{\longleftarrow}
        & (N_\lambda ,1)
        & \overset{\eqref{n_shift}}{\longleftarrow}
        & \cdots
        & \overset{\eqref{n_shift}}{\longleftarrow}
        & (N_\lambda,M_\lambda-1)
        & \overset{\eqref{n_shift}}{\longleftarrow}
        & (N_\lambda,M_\lambda)
        \end{array}
}

\begin{ex}
Fix $\l$ and drop the $\l$-index from $\Ilnm$, $\Phi \supbrk{\l,m}$ and $\Psi \supbrk{\l,n}$.
We showcase the recursion for $0 \leq n \leq 2$ and $0 \leq m \leq 2$.
\begin{enumerate}
        \item 
                $n=0$: Using \eqref{m_shift} twice,
                \eq{
                        \Iab{0}{1} &= \Phia{1} \cdot \Iab{0}{0}
                        \\
                        \Iab{0}{2} &= \Phia{2} \cdot \Iab{0}{1}
                        \, .
                }
        \item 
                $n=1$: Using \eqref{n_shift_simple} once,
                \eq{
                        \Iab{1}{2} = \Psia{1} \cdot P_{1,0} \cdot \Iab{0}{2}
                        \, .
                }
                Using \eqref{n_shift} twice,
                \eq{
                        \Iab{1}{1} &= \Psia{1} \cdot \Big( P_{1,0} \cdot \Iab{0}{1} - 2 \Iab{1}{2} \Big)
                        \\
                        \Iab{1}{0} &= \Psia{1} \cdot \Big( P_{1,0} \cdot \Iab{0}{0} - 1 \Iab{1}{1} \Big)
                        \, .
                }
        \item 
                $n=2$: Using \eqref{n_shift_simple} once,
                \eq{
                        \Iab{2}{2} = 
                        \Psia{2} \cdot 
                        \Big( P_{1,1} \cdot \Iab{0}{2} + P_{1,0} \cdot \Iab{1}{2} \Big)
                        \, .
                }
                Using \eqref{n_shift} twice,
                \eq{
                        \Iab{2}{1} &= 
                        \Psia{2} \cdot 
                        \Big( P_{1,1} \cdot \Iab{0}{1} + P_{1,0} \cdot \Iab{1}{1} - 2 \Iab{2}{2} \Big)
                        \\
                        \Iab{2}{0} &= 
                        \Psia{2} \cdot 
                        \Big( P_{1,1} \cdot \Iab{0}{0} + P_{1,0} \cdot \Iab{1}{0} - 1 \Iab{2}{1} \Big)
                        \, .
                }
\end{enumerate}
Note that this recursion can be performed over a finite field for the sake of efficiency.
\end{ex}

The discussion above is summarized in the following algorithm.

\begin{algorithm}[H]
        \underline{Input}: 
        \begin{itemize}
                \item 
                Pfaffian system in normal form:
                $
                        \p_i \bullet \vec{I}(z) =
                        P_i(z) \cdot \vec{I}(z)
                $
                for 
                $
                        i = 1, \ldots, N.
                $
                \item 
                Integer $k>0$.
        \end{itemize}
        \underline{Output}: 
        The series expansion coefficients $\Ilnm$ from \eqref{I_asymptotic_series} up to order $\mO(z_1^k)$.
        \vspace{0.2cm}
    \begin{algorithmic}[1]
            \State
            Compute
            $
                \jord{P_{1,-1}(z')}
            $
            with $z'$ fixed to generic numbers.
            Save 
            \begin{itemize}
                \item
                        unique eigenvalues 
                        $
                            \{\l_1, \l_2, \ldots\}
                        $
                \item 
                        eigenvalue multiplicities
                        $
                            \{\L_1, \L_2, \ldots\}
                        $
                \item 
                        integers 
                        $
                            \{M_{\l_1}, M_{\l_2}, \ldots\}
                        $
                        from \eqref{max_log_power}.
            \end{itemize}
            \vspace{0.2cm}
            \For{each unique $\l$}
            \vspace{0.2cm}
            \State 
            Repeat steps 1.~to 5.~in \algref{alg:holomorphic_restriction} with the replacements
            \begin{itemize}
            \addtolength{\itemindent}{0.6cm}
                \item
                        $
                            \Rmat \leftarrow 
                            \Rmat\supl =
                            \rowred{
                                    \big( P_{1,-1}(z') - \l \mathbf{1} \big)^{M_\l+1}
                            }
                        $
                \item 
                        $R' \leftarrow \L$
                \item 
                        $\Iz \leftarrow \Ilzz$
            \end{itemize}
            \vspace{0.2cm}
            \hspace{0.5cm} and save the result for $\Ilzz$.
            \vspace{0.2cm}
            \For{$0 \leq n \leq k$ and $0 \leq m \leq M_\l$}
                \vspace{0.2cm}
                \State
                Input $\Ilzz$ into the recursion relations
                \eqref{m_shift}, \eqref{n_shift}, \eqref{n_shift_simple} 
                to compute $\Ilnm$.
                \vspace{0.2cm}
            \EndFor
            \vspace{0.2cm}
            \EndFor 
            \\
            \vspace{0.4cm}
            \Return The coefficient functions
            $
                \bigcup_{\l} \ 
                \bigcup_{0 \, \leq \, n \, \leq k} \ 
                \bigcup_{0 \, \leq \, m \, \leq M_\l} \
                \Ilnm
            $
    \end{algorithmic}
    \caption{: Logarithmic series expansion}
    \label{alg:logarithmic_restriction}
\end{algorithm}

We expect this to be an efficient method for computing series expansions of MIs.
Its efficiency stems from trading a "big" Pfaffian system
(satisfied by the full solution vector $\vec{I}(z)$)
for several "small" Pfaffian systems
(satisfied by the restricted bases $\Jl(z')$),
each one containing one less variable.
For instance,
if the "big" system includes unrationalizable square roots and elliptic functions,
then it is likely that the "small" systems only contain simple GPLs.

The catch is that the algorithm does not produce results valid in the entire phase space;
there must exist a variable $z_1 \ll 1$ that can be used for the expansion.
Luckily,
there are many such situations in high-energy physics:
$z_1$ could be a small mass, 
a threshold variable (e.g.~$s-4m^2$),
an angle between two collinear momenta (e.g.~in Regge theory),
a small or large transverse momentum ($z_1=p_T$ or $z_1=1/p_T$),
a soft momentum variable in Post-Newtonian/Minkowskian gravity,
and so forth.

The \emph{method of regions} is the classic approach 
to asymptotic expansions \cite{Beneke:1997zp,Smirnov:2002pj,Jantzen:2011nz}.
More recent approaches include 
\cite{Moriello:2019yhu,Egner:2023kxw,Lee:2017qql,Lee:2018ojn,Kudashkin:2017skd}.
It would be interesting to import some of the wisdom from these studies 
in order to enhance the logarithmic restriction approach presented here.

        \section{Restriction at the level of Macaulay matrices}

While the protocol of the previous section is applicable to any regularly singular Pfaffian system,
it does require an unrestricted system as input,
which might be hard to obtain.
It should also be mentioned that this method is only conjectured to yield the rational restriction
of the twisted cohomology group \eqref{twisted_cohomology_group} when it is viewed as a $\mD$-module
\cite[Remark 3.5]{Chestnov:2023kww}.

Building on the Macaulay matrix method introduced in \chapref{sec:macaulay_matrices},
this section presents a holomorphic restriction protocol which works entirely at the level of
$\mD$-modules.
Theorems from \cite[Section 4]{Chestnov:2023kww} ensure that the method yields correct results.
Some steps of the construction are technical,
so we refer to the previous reference for details.

\subsection{Restriction to a hyperplane}

Let $\mI$ be a holonomic ideal in $\mD_Y$.
This could e.g.~be the annihilating ideal for the GKZ system.
The goal is to compute a Pfaffian system for the rational restriction module onto the hyperplane given by
$
        Y' = \{z_1 = \cdots = z_{N'} = 0\}.
$

The first question to answer is that of a basis for the restriction module.
In \cite[Algorithm 4.7]{Chestnov:2023kww},
a systematic search algorithm is formulated which succeeds when the dimension 
of the basis matches the expected holonomic rank $R'$ of the restriction module.
The algorithm requires $R'$ as input,
but this is easily obtained for the case of Euler integrals by computing the 
Euler characteristic $\chi$ via the likelihood equations \eqref{likelihood_equation}.
Moreover,
this algorithm relies entirely on numerical linear algebra,
with all parameters fixed to generic numbers,
meaning that it will terminate with modest computing power even for relatively large Pfaffian systems.

The resulting standard monomials for the restriction module,
call them $\RStd$,
now serve as input for the Macaulay matrix $M$ in \algref{alg:Pfaffian_by_MM}.
Recall that $M$ is built by acting with "seed" monomials on the generators of $\mI$.
Since $\mI$ contains unrestricted variables defined on $Y$,
so will $M$.
The key observation is this:
all the entries of $M$ are polynomial in $z \in Y$,
so we can simply fix $z_1 = \ldots = z_{N'} = 0$ in $M$,
and then proceed with the usual Macaulay matrix algorithm.

\begin{algorithm}[H]
        \underline{Input}: 
        \begin{itemize}
                \item 
                        The restricted basis $\RStd$.
                \item 
                        Generators $\{d_1, \ldots, d_D\}$ of a holonomic ideal $\mI$ in $\mD_Y$.
                \item 
                        Direction $i$.
        \end{itemize}
        \underline{Output}: 
        Pfaffian matrix $Q_i(z')$ for the restriction
        \eq{
                \mR_{Y'} 
                \otimes_{\mD_{Y'}}
                \mD_Y / \big( \mI + z_1 \mD_Y + \ldots + z_{N'} \mD_Y \big)
                \, .
        }
    \begin{algorithmic}[1]
            \vspace{-0.2cm}
            \State Call \algref{alg:Pfaffian_by_MM} with the replacements
            \begin{itemize}
                \item 
                        $\Std \leftarrow \RStd$
                \item
                        $
                                M \leftarrow 
                                \arr{c|c}{M_\Ext & M_\RStd} \Big|_{z_1 = \ldots = z_{N'} = 0}
                        $
            \end{itemize}
            \vspace{0.1cm}
            \vspace{-0.2cm}
            \State \Return $Q_i(z') = C_\RStd \supbrk{i} - C \supbrk{i} \cdot M_\RStd$
    \end{algorithmic}
    \caption{: Rationally restricted Pfaffian system by the Macaulay matrix}
    \label{alg:macaulay_restriction}
\end{algorithm}

\noindent
Let us remark that the basis $\RStd$ can also be guessed,
assuming $R'$ is known,
by executing \algref{alg:macaulay_restriction} several times with all parameters set to numbers
until it succeeds.

\algref{alg:macaulay_restriction} has been implemented in the \package{asir} package
\package{mt\_mm} by Nobuki Takayama,
as showcased in the following example.

\begin{ex}
\label{ex:mm_restriction}
Let 
$
        \mI = \langle d_1, d_2 \rangle
$ 
be the ideal generated by
\eq{
        d_1 &= 
        z_1 (1-z_1) \p_1^2 - z_1 z_2 \p_1 \p_2 + (c_1 - z_1 (a+b_1+1)) \p_1 - b_1 z_2 \p_2 - a b_1
        \\[2pt]
        d_2 &= 
        z_2 (1-z_2) \p_2^2 - z_1 z_2 \p_1 \p_2 + (c_2 - z_2 (a+b_2+1)) \p_2 - b_2 z_1 \p_1 - a b_2
        \, ,
}
which annihilates Appell's $F_2$ function 
\eq{
        \nonumber
        F_2(a,b_1,&b_2,c_1,c_2 | z_1, z_2) 
        =
        \frac{\Gamma(c_1)\Gamma(c_2)}{\Gamma(b_1)\Gamma(b_2)\Gamma(c_1-b_1)\Gamma(c_2-b_2)}
        \ \times
        \\&
        \int_{(0,1)^2}
        (1-x_1)^{c_1-b_1-1} 
        (1-x_2)^{c_2-b_2-1}
        (1-z_1x_1-z_2x_2)^{-a}
        x_1^{b_1} 
        x_2^{b_2}
        \frac{\dd x}{x}
        \label{Appell_F2}
        \, .
}
This function appears often in the FI calculus \cite{Ananthanarayan:2021bqz,Ananthanarayan:2021yar}.

The following \package{asir} script computes the holomorphic restriction to $Y'=\{z_1 = 0\}$.

\begin{lstlisting}[style=mystyle]
import("mt_mm.rr")$

Vars = [z1, z2]$
DVars = [dz1, dz2]$

D1 = z1*(1-z1)*dz1^2 - z1*z2*dz1*dz2 + (c1-z1*(a+b1+1))*dz1 - b1*z2*dz2 - a*b1;
D2 = z2*(1-z2)*dz2^2 - z1*z2*dz1*dz2 + (c2-z2*(a+b2+1))*dz2 - b2*z1*dz1 - a*b2;
Ideal = [D1, D2];

NumericRule = [[z2,z2+1/3], [a,1/2], [b1,1/3], [b2,1/5], [c1,1/7], [c2,1/11]];
IdealNumeric = base_replace(Ideal, NumericRule);

RStd = mt_mm.restriction_to_pt_(IdealNumeric, Gamma = 2, KK = 4, Vars | p = 10^8);
RStd = reverse(map(dp_ptod, RStd[0], DVars));

Ideal = map(dp_ptod, Ideal, DVars);
MMData = mt_mm.find_macaulay(Ideal, RStd, Vars | restriction_var = [z1]);

Q2 = mt_mm.find_pfaffian(MMData, Vars, 2 | use_orig = 1);
\end{lstlisting}

\noindent
The inputs of \soft{mt\_mm.restriction\_to\_pt\_()} merit explanation.
The list \soft{IdealNumeric} contains generators with $z_2$ and exponent parameters fixed to generic numbers.
The integers \soft{Gamma} and \soft{KK} refer to the $\g$ and $k$ of 
the basis search algorithm \cite[Algorithm 4.7]{Chestnov:2023kww}.
The value \soft{p = 10\string^8} means that the function is executed over a finite field of order 
$p = 100000007$,
which is more efficient than a symbolic calculation.

The result of this script is the Pfaffian system
\eq{
        \p_2 \, \RStd = Q_2 \cdot \RStd
}
given by
\eq{
        \p_2 \arr{c}{1 \\ \p_2} =
        \arr{cc}{
                0 & 1 \\ 
                \frac{-a b_2}{z_2 (z_2-1)} & \frac{c_2-(a+b_2+1)}{z_2(z_2-1)}
        }
        \cdot
        \arr{c}{1 \\ \p_2} 
        \, .
}
\end{ex}

\subsection{Restriction to a hypersurface}

This chapter has,
up until now,
only dealt with restrictions onto hyperplanes.
Namely,
we have assumed that 
$
        Y' = \{z_1 = \ldots = z_{N'} = 0\}
$
is defined by an affine \emph{linear} coordinate change.
To date,
there is no known algorithm for restrictions of $\mD$-modules onto general hypersurfaces.
Surprisingly,
the Macaulay matrix restriction can be generalized to this scenario.

Suppose that 
$
        Y' = \{ f = 0 \}
$
is the vanishing locus of an irreducible polynomial $f = f(z) \in \CC[z_1, \ldots, z_N]$.
The idea is,
essentially,
just to solve the Macaulay matrix equation \eqref{MM_Ext_eq} for $C \supbrk{i}$ modulo $f$:
\eq{
        C_\Ext \supbrk{i} \ - \ C\supbrk{i} \cdot M_\Ext \equiv 0
        \quad \text{mod} \ f
        \, .
        \label{MM_Ext_eq_mod_f}
}
We give one example below to illustrate this procedure.
See \cite{url-restriction-data} for further results on restrictions onto hypersurface singularities of
Horn's hypergeometric functions $\{H_1, \ldots, H_7\}$.

\begin{ex}
We compute the restriction of Appell's $F_4$ function
\eq{
        \nonumber
        & F_4(a,b,c_1,c_2 | z_1, z_2) =
        \frac{\Gamma(c_1) \Gamma(c_2)}{\Gamma(a) \Gamma(b) \Gamma(c_1-a) \Gamma(c_2-b)}
        \ \times
        \\ &
        \int_{(0,1)^2}
        (1-x_1)^{c_1-a-1}
        (1-x_2)^{c_2-b-1}
        (1-z_1 x_1)^{a-c_1-c_2+1}
        \ \times
        \\ & \nonumber \hspace{1.1cm}
        (1-z_2 x_2)^{b-c_1-c_2+1}
        (1-z_1 x_1 - z_2 x_1)^{c_1+c_2-a-b-1}
        x_1^a
        x_2^b
        \frac{\dd x}{x}
}
onto its singular locus defined by
\eq{
        L = z_1 z_2 f = 0
        \quadit{\text{with}}
        f = (z_1-z_2)^2 - 2(z_1+z_2) + 1 
        \, .
}
In terms of Euler operators $\t_i = z_i \, \p_i$,
the annihilating ideal 
$
        \mI = \langle d_1, d_2 \rangle
$
is generated by
\eq{
        d_1 &=
        \t_1 (\t_1 + c_1 - 1) - z_1(\t_1 + \t_2 + a)(\t_1 + \t_2 + b)
        \\
        d_2 &=
        \t_2 (\t_2 + c_2 - 1) - z_2(\t_1 + \t_2 + a)(\t_1 + \t_2 + b)
        \, .
}
The holonomic rank outside the singular locus is $R = 4$.
On $Y' = \{ f = 0 \}$, 
the rank drops to $R' = 3$.
Both of these ranks assume that the exponent parameters $(a,b,c_1,c_2)$ are generic.

We search for a restricted basis by executing \algref{alg:macaulay_restriction}
with different guesses for $\RStd$ until it succeeds.
The parameters are set to generic numbers in this search.
The result is 
\eq{
        \RStd = \big[1, \p_1, \p_2\big]^T
        \, .
}
Next we build a $6 \times 10$ Macaulay matrix 
$
        M_1 = \big[ M_\Ext \big|  M_\RStd \big]
$
of degree $1$.
The left block contains $7$ columns labeled by
$
        \Ext = 
        \big[ 
                \p_1^3, \, \p_1^2 \p_2, \, \p_1 \p_2^2, \, \p_2^3, \, \p_1^2, \, \p_1 \, \p_2, \, \p_2^2 
        \big]^T
        \, .
$

The goal is now to solve \eqref{MM_Ext_eq_mod_f} modulo $f$.
Since $f$ is quadratic in $z_1$,
we can eliminate all instances of $z_1^p$ for $p \geq 2$ via polynomial division by $f$.
It is therefore enough to consider the ansatz
\eq{
        C\supbrk{i} = C_0\supbrk{i} + C_1\supbrk{i} z_1
        \, ,
}
where $C_{1,2}\supbrk{i}$ are independent of $z_1$.
The reduction modulo $f$ can be performed via a sequence of replacements
$
        z_1^p \to g_p(z_2) + h_p(z_2) z_1.
$
For instance,
$f=0$ immediately gives
\eq{
        z_1^2 \to (-1+2z_2-z_2^2) + 2(1+z_2)z_1
        \, .
}
Inserting this into $z_1 f = 0$,
we further have
\eq{
        z_1^3 \to 2(-1+z_2+z_2^2-z_2^3) + (3+10z_2+3z_2^2) z_1
        \, ,
}
and so forth.

After solving for $C_{1,2}\supbrk{i}$,
the Pfaffian matrices are computed as usual by
\eq{
        Q_i = C_\RStd \supbrk{i} - C \supbrk{i} \cdot M_\RStd
        \, .
}
Setting
$
        (a , b , c_1 , c_2) = (-2/3 , \, 1/3 , \, 1/3 , \, 1/3)
$
to simplify the output,
we finally get
\eq{
        \nonumber
        Q_1 &=
        \arr{ccc}{
                 0 & 1 & 0 \\
                 -\frac{5 (3 z_1 z_2+z_1-5 z_2 (z_2+2)-1)}{162 (z_2-1)^3 z_2} & 
                 \frac{z_1 (41 z_2-5)-z_2 (87 z_2+62)+5}{108 (z_2-1)^2 z_2} & 
                 \frac{5 (3 z_1 z_2+z_1-5 z_2 (z_2+2)-1)}{108 (z_2-1)^3} \\
                 \frac{4 (-z_1+2 z_2+2)}{81 (z_2-1)^2 z_2} & -\frac{2}{27 z_2} & 
                 \frac{2 (z_1-2 (z_2+1))}{27 (z_2-1)^2} 
        }
        \\
        Q_2 &=
        \arr{ccc}{
                 0 & 0 & 1 \\
                 \frac{4 (-z_1+2 z_2+2)}{81 (z_2-1)^2 z_2} & 
                 -\frac{2}{27 z_2} & \frac{2 (z_1-2 (z_2+1))}{27 (z_2-1)^2} \\
                 \frac{5 (-z_1+3 z_2+1)}{162 (z_2-1) z_2^2} & 
                 -\frac{5 (z_1+z_2-1)}{108 z_2^2} & \frac{5 z_1-51 z_2+31}{108 (z_2-1) z_2} 
        }
}
See the script 
\cite[\soft{2023-01-26-F4-on-sing.py}]{url-restriction-data} 
for comparisons of known analytic results with numerical solutions to this Pfaffian system.
\end{ex}

        \chapter{Restriction Examples}
\label{ch:restrictions_examples}

This chapter gives examples of how to compute Pfaffian systems for restriction modules related to FIs.
Sections \eqref{sec:1L_bhabha_scattering} and \eqref{sec:bhabha_2L_scattering}
also include partial results for logarithmic series expansions related to 1- and 2-loop Bhabha scattering.

        \section{Holomorphic restriction: 2-loop N-box}

This is a continuation of \exref{ex:2L_nbox}.
The goal is to apply the holomorphic restriction protocol of 
\secref{sec:holomorphic_restriction} to that GKZ system.
Schematically,
we restrict 
\eq{
    A = 
    \arr{ccccc ccccc}{
        1 & 1 & 1 & 1 & 1 & 1 & 1 & 1 & 1 & 1\\
        1 & 1 & 1 & \mzero & \mzero & \mzero & \mzero & \mzero & 1 & \mzero\\
        1 & \mzero & \mzero & 1 & 1 & 1 & \mzero & \mzero & 1 & 1\\
        \mzero & \mzero & \mzero & 1 & \mzero & \mzero & 1 & 1 & \mzero & 1\\
        \mzero & 1 & \mzero & \mzero & 1 & \mzero & 1 & \mzero & 1 & \mzero\\
        \mzero & \mzero & 1 & \mzero & \mzero & 1 & \mzero & 1 & \mzero & 1
    }
    \quad
    \xrightarrow[i \, \neq \, 10]{z_i = 1}
    \quad
    \includegraphicsbox{Figs/2L_nbox.pdf}
    \, .
}
We begin with a rank $R=9$ Pfaffian system,
obtained via the Macaulay matrix method,
in terms of variables 
$
        z \in Y = \AA^{10}.
$
After a coordinate shift
$
        (z_1,\ldots,z_9,z_{10}) \to (z_1,\ldots,z_9,z_{10}) + (1,\ldots,1,0),
$
the restriction surface is
$
        Y' = \{z_1 = \ldots = z_9 = 0\}.
$
Six variables can immediately be restricted using homogeneity
(cf.~\secref{sec:integrand_rescaling}),
say
$
        \{z_1,z_2,z_3,z_4,z_5,z_9\}.
$
So the restriction protocol need only be applied to $\{z_6, z_7, z_8\}$.
By calculating the Euler characteristic on $Y'$,
we expect a rank $R' = 3$ Pfaffian system in the end.

\subsection{Basis}

The method of \cite{Hibi_Nishi_Taka} 
(implemented in the \package{asir} command \soft{mt\_gkz.cbase\_by\_euler(A)})
gives a basis of 9 standard monomials
\eq{
        \Std = 
        \arr{c}{
                \p_8^2 \\
                \p_6 \p_{10} \\
                \p_7 \p_{10} \\
                \p_{10}^2 \\
                \p_6 \\
                \p_7 \\
                \p_8 \\
                \p_{10} \\
                1
        }
        \, .
}
The Macaulay matrix method then outputs a Pfaffian system
\eq{
        \p_i \, \Std = P_i \cdot \Std
        \quadit{\text{for}}
        i = 6,7,8,10
        \, .
        \label{n_box_pfaffian_system}
}
For the sake of verifying our results,
an independent IBP calculation is performed for the following MIs in $\DD = 4 - 2\e$ dimensions:
\eq{
        \arr{c}{
                I_{02202} \\
                I_{22020} \\
                I_{11111}
        }
        =
        G_0 \cdot e\supbrk{I}
        \, .
        \label{n_box_integral_basis}
}
The subscripts on the integrals label powers of the inverse propagators
\eq{
        D_1 = \ell_1^2
        \, , \
        D_2 = \ell_2^2
        \, , \
        D_3 = (p_1-\ell_1)^2
        \, , \
        D_4 = (p_{12}-\ell_{12})^2
        \, , \
        D_5 = (-p_{123}+\ell_{12})^2
        \, ,
}
and $G_0$ is a matrix containing powers of $s = 2 p_1 \cdot p_2$ that render $e\supbrk{I}$ dimensionless.

We would like to land on the basis \eqref{n_box_integral_basis}
after computing the restriction of \eqref{n_box_pfaffian_system},
in order to compare our Pfaffian matrices with those coming from IBPs.
By the formulas of \secref{sec:Euler_integrals_as_operators},
we therefore translate the integrals \eqref{n_box_pfaffian_system} into $\mD$-module elements:
\eq{
        e\supbrk{I} 
        \to
        e\supbrk{\mD} 
        =
        \mathrm{c} \cdot
        \arr{c}{
                \p_{10}^2
                \\
                \e(2+5\d)(1+2\e+5\e\d) +
                2z_{10}(1+2\e+5\e\d)\p_{10} +
                z_{10}^2\p_{10}
                \\
                (1+\e+2\e\d)\p_{10} +
                z_{10}\p_{10}^2 -
                z_7\p_7\p_{10} +
                z_6\p_6\p_{10}
        }
        \, .
}
The $3 \times 3$ diagonal matrix $\mathrm{c}$ contains prefactors depending on $\{\e,\d,s\}$.
Thus, there is a $3 \times 9$ $\Bmat$-matrix \eqref{B_matrix}
with rows and columns labeled by $e\supbrk{\mD}$ and $\Std$ respectively:
\eq{
        \Bmat =
        \arr{ccccc ccccc}{
                \mzero & \mzero & \mzero & 1 & \mzero & \mzero & \mzero & \mzero & \mzero
                \\[4pt]
                \mzero & \mzero & \mzero & z_{10}^2 & \mzero & \mzero & \mzero & 
                2z_{10} (1 + 2 \e + 5 \e \d) & \e(2 + 5 \d) (1 + 2 \e + 5 \e \d)
                \\[4pt]
                \mzero & z_6 & -z_7 & z_{10} & \mzero & \mzero & \mzero & 1 + \e + 2 \e \d & \mzero
        }
        \, .
        \label{nbox_B}
}
This matrix satisfies%
\footnote{
        One may employ \eqref{G_G1_G2} in cases where $e\supbrk{\mD}$ contains 
        some monomials that are not contained in $\Std$.
}
$
        e\supbrk{\mD} = \mathrm{c} \cdot \Bmat \cdot \Std.
$

\subsection{Normal form}

The Pfaffian system \eqref{n_box_pfaffian_system} must be brought to normal form 
before the restriction procedure can be applied
(cf.~\secref{sec:normal_form}).
The system turns out to be in normal form w.r.t.~the variables $z_6$ and $z_7$,
but $P_8$ has a second-order pole in $z_8$:
\eq{
        P_8 \big |_{z_6, \, z_7 \to 1} =
        (z_8-1)^2 \, P_{8,-2} \big |_{z_6, \, z_7 \to 1} + \ldots
        \, .
}
The second-order pole can be cured by Moser reduction,
namely a gauge transformation by
\eq{
        G_1 = \diag{(z_8-1)^{-1}, \, 1, \ldots, 1}
        \, .
}
The notation
\eq{
        G[P_i] = G^{-1} \cdot ( P_i \cdot G - \p_i \bullet G)
}
is used throughout this chapter to denote gauge transformations.
In addition to curing the second-order pole,
the new residue matrices $G[P_i]_{-1}, i \in \{6,7,8\},$ also obtain non-resonant spectra.
Consequently,
the system is in normal form.
To simplify notation, let us
\eq{
        \text{replace} \ G_1[P_i] \ \to \ P_i \ \text{in the rest of this example}.
}

\subsection{Restriction}

The Pfaffian-level restriction protocol implores us to build the matrix
\eq{
        \Mmat =
        \arr{c}{\Bmat \\ \hline \Rmat}
}
coming from \eqref{M_matrix}.
The $\Bmat$-matrix was already found in \eqref{nbox_B}.
The $\Rmat$-matrix is constructed from the independent rows of all the (gauged transformed) residue matrices:
\eq{
        \Rmat =
        \rowred{
                \begin{array}{c}
                        P_{6,-1} \\ \hline P_{7,-1} \\ \hline P_{8,-1}
                \end{array}
        }
        \, .
}
We remind that the $\rowred{...}$ operation includes the deletion of zero-rows.

The $\Mmat$-matrix can now be inserted into \eqref{M_gauge_transformation}, 
giving
\eq{
        \big(
                \p_{10} \bullet \Mmat + \Mmat \cdot P_{10,0}
        \big)
        \cdot
        \Mmat^{-1}
        =
        \arr{c|c}{
                Q_{10} & \star \\ \hline
                \mathbf{0}_{6 \times 3} & \star
        }
        \, ,
}
where 
\eq{
        P_{10,0} = P_{10} \big |_{z_6, \,z_7, \, z_8 \to 0}
        \, ,
}
which is well-defined because the system is in normal form.
Sending the regulator $\d \to 0$ in $Q_{10}$,
the final result is
\eq{
        Q_{10} =
        \arr{ccc}{
                -\frac{2 \brk{1 + \e}}{z_{10}} & \mzero & \mzero \\
                \mzero & \mzero & \mzero \\
                -\frac{z_{10} \e}{6 \brk{1 + z_{10}}} &
                \frac{\e}{6 z_{10} \brk{1 + z_{10}}} &
                -\frac{z_{10} + 2 \e}{z_{10} \brk{1 + z_{10}}}
        }
        \, .
}
This Pfaffian matrix agrees with an independent IBP calculation for the basis 
$e \supbrk{I}$ in \eqref{n_box_integral_basis},
where $z_{10} = t/s$ is a ratio of Mandelstam variables.

In conclusion,
the $\mD$-module associated to the massless 2-loop N-box diagram 
is isomorphic to the rational restriction of a GKZ system.

        \section{Logarithmic restriction: 1-loop Bhabha scattering}
\label{sec:1L_bhabha_scattering}

This example uses \algref{alg:logarithmic_restriction} to compute a logarithmic restriction.
Our goal is the small-mass expansion of the 1-loop box topology for Bhabha scattering:
\eq{
    \nonumber
    \includegraphicsbox{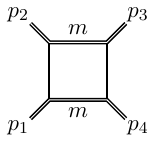}
    \quad
    \xrightarrow{m \to 0}
    \quad
    \includegraphicsbox{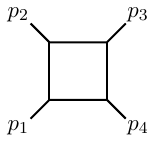}
    \, .
}
The momentum space integral family in $\DD = 4 - 2\e$ dimensions is
\eq{
        I_{\nu_1 \ldots \nu_5}(m) =
        \int
        \frac
        {\dd^\DD \ell}
        {
                \big[\ell^2-m^2]^{\nu_1}
                \big[(\ell+p_{1})^2]^{\nu_2}
                \big[(\ell+p_{12})^2-m^2]^{\nu_3}
                \big[(\ell+p_{123})^2]^{\nu_4}
        }
        \, ,
        \label{1L_bhabha_family}
}
and the kinematic variables are
\eq{
        p_1^2 = p_2^2 = p_3^2 = p_4^2 = m^2
           , \quad
        s = p_{12}^2
        \, , \quad
        t = p_{23}^2
        \, .
}
IBP reduction reveals a basis of 5 MIs:
\eq{
        \arr{c}{
                I_{0010}(m) \\
                I_{0101}(m) \\
                I_{1010}(m) \\
                I_{0111}(m) \\
                I_{1111}(m) 
        }
        =
        G_0 \cdot \vec{I}
        \, , \quad
        G_0 =
        (-s)^{-\e} \,
        \diag{
                (-s), \, 1, \, 1, \, (-s)^{-1}, \, (-s)^{-2}
        }
        \, .
        \label{1L_bhabha_masters}
}
The purpose of $G_0$ is to factor out the mass dimension of the integrals.
Using IBPs,
the MIs $\vec{I}$ are found to obey a Pfaffian system
\eq{
        \p_i \bullet \vec{I} =
        P_i \cdot \vec{I}
        \quadit{\text{with}}
        z_1 = \frac{m^2}{-s}
        \quadit{\text{and}}
        z_2 = \frac{t}{s}
        \, .
        \label{1L_bhabha_system_1}
}
Using the method of \secref{sec:logarithmic_restriction},
we now compute the logarithmic restriction of solutions onto
$
        Y' = \{z_1 = 0\},
$
i.e.~for $m \sim 0$.

\subsection{Normal form}

The matrices $P_{1,2}$ in \eqref{1L_bhabha_system_1} are not in normal form w.r.t.~$Y'$.
In particular,
$P_1$ has a second-order pole there,
and $P_2$ a first-order pole.
This is cured via Moser reduction:
\eq{
        \vec{I} \to G_1 \cdot \vec{I}
        \quadit{\text{with}}
        G_1 = \diag{z_1, \, 1, \, 1, \, 1, \, 1}
        \, .
}
To simplify notation,
we
\eq{
        \text{replace} \ G_1[P_i] \ \to \ P_i \ \text{in this example}.
}
The gauge transformed Pfaffian matrices now have the desired pole structure:
\eq{
        \begin{array}{lllllll}
                P_1 & = & z_1^{-1} & \times & P_{1,0}(z_2)  & + & \mO(z_1^0)
                \\[2pt]
                P_2 & = & z_1^0    & \times & P_{2,0}(z_2)  & + & \mO(z_1)
                \, .
        \end{array}
        \label{1L_bhabha_normal}
}
In addition,
the spectrum of the residue matrix $P_{1,-1}$ is non-resonant,
meaning that the system is now in normal form.

From 
\eq{
        \jord{P_{1,-1}} =
        \arr{ccccc}{
                0      & \mzero & \mzero & \mzero & \mzero \\
                \mzero & 0      & \mzero & \mzero & \mzero \\
                \mzero & \mzero & 0      & \mzero & \mzero \\
                \mzero & \mzero & \mzero & -\e    & 1      \\
                \mzero & \mzero & \mzero & \mzero & -\e 
        }
        \label{1L_bhabha_jordan}
}
we read off the data
\eq{
        \begin{array}{ccl}
                \{ \l_1, \, \l_2 \} &=& \{0, \, -\e\} \\
                \{ \L_1, \, \L_2 \} &=& \{3, \,   2\} \\
                \{ M_{\l_1}, \, M_{\l_2} \} &=& \{0, \, 1\}
                \, ,
        \end{array}
}
in the notation of step 1.~of \algref{alg:logarithmic_restriction}.
The ansatz for the logarithmic expansion \eqref{I_asymptotic_series} thus takes the form
\eq{
        \label{1L_bhabha_ansatz}
        \vec{I} 
        \quad &= \quad
        z_1^{\l_1}
        \sum_{n=0}^\infty \Iab{\l_1}{n}(z_2) z_1^n
        \\ & + \quad
        z_1^{\l_2}
        \sum_{n=0}^\infty
        \big[
                \Iabc{\l_2}{n}{0}(z_2)
                \plus
                \Iabc{\l_2}{n}{1}(z_2) \times \log(z_1)
        \big]
        z_1^n
        \, .
        \nonumber
}
The logarithm appears due to the "$1$" in the $\l_2$-eigenvalue block of \eqref{1L_bhabha_jordan}.
Recall that the vector $\Iabc{\l_2}{0}{1}$ is immediately obtained from $\Iabc{\l_2}{0}{0}$ 
via the recursion relation \eqref{m_shift}.
Higher-order terms in $z_1^n$ are obtained from the recursions \eqref{n_shift} and \eqref{n_shift_simple}.
If what follows,
we determine $\Iab{\l_1}{0}$ from the solution to a Pfaffian system of rank $\L_1 = 3$,
and $\Iabc{\l_2}{0}{0}$ from the solution to a system of rank $\L_2 = 2$.
Both systems are univariate.

\subsection{Eigenvalue \texorpdfstring{$\l_1 = 0$}{}}

There is no logarithm for the eigenvalue $\l_1$,
so this case can in fact be treated with the holomorphic restriction of \secref{sec:holomorphic_restriction}.

\subsubsection{Rank jump}
Recall the constraint from the second equation of \eqref{holomorphic_restriction}:
\eq{
        \Rmat\supla{1} \cdot \Iab{\l_1}{0}
        &=
        \rowred{P_{1,-1}} \cdot \Iab{\l_1}{0}
        \\&=
        \arr{ccccc}{
                1 & \mzero & \mzero & \mzero & \mzero \\
                \mzero & 1 & \mzero & \frac{z_2 \e}{1-2\e} & \mzero
        }
        \cdot
        \Iab{\l_1}{0}
        \label{1L_bhabha_l1_constraint}
        \\&=
        0
        \, .
}
Two constraints are here imposed on the entries of the $5$-dimensional vector $\Iab{\l_1}{0}$.
This is the reason why the rank drops to $5-2=3$.

The entries of $\Iab{\l_1}{0}$ apparently correspond to the MIs \eqref{1L_bhabha_masters} with the limit
$z_1 \to 0$ taken at the \emph{integrand} level%
\footnote{
        This is called the \emph{hard region} in the method of regions \cite{Beneke:1997zp,Smirnov:2002pj}.
},
i.e.~setting $m^2=0$ inside the propagators of \eqref{1L_bhabha_family}.
The interpretation of the two constraints from \eqref{1L_bhabha_l1_constraint} is now clear:
\begin{enumerate}
        \item 
                The first constraint
                $
                        I\supbrk{\l_1,0}_1 = 0
                $ 
                means that the massless tadpole $I_{0010}(m=0)$ should vanish,
                which is indeed true in DR.
        \item 
                The second constraint
                $
                        I\supbrk{\l_1,0}_2 
                        +
                        \frac{z_2 \e}{1-2\e} I\supbrk{\l_1,0}_4
                        = 0
                $
                is an IBP relation between the massless $t$-channel bubble 
                $I_{0101}(m=0)$ and the massless triangle $I_{0111}(m=0)$.
                We have verified this relation using \package{LiteRed}.
\end{enumerate}

\subsubsection{Basis}

Following step 2. of \algref{alg:holomorphic_restriction},
a basis for the restricted Pfaffian system can be found by simply eliminating linearly dependent entries of
$\Iab{\l_1}{0}$.
We should eliminate two integrals according to the above discussion,
so we have%
\footnote{
        We choose the bubble instead of the triangle for the first basis element
        since it is preferred to have basis elements with the fewest number of distinct propagators.
}
\eq{
        \Ja{\l_1} 
        =
        \arr{c}{
                I\supbrk{\l_1,0}_2 \\
                I\supbrk{\l_1,0}_3 \\
                I\supbrk{\l_1,0}_5 
        }
        =
        (-s)^\e
        \arr{c}{
                I_{0101}(0) \\
                I_{1010}(0) \\
                (-s)^2 I_{1111}(0) 
        }
        \, .
        \label{1L_bhabha_J}
}
The factors of $(-s)$ stem from the matrix $G_0$ in \eqref{1L_bhabha_masters}.
The $\Bmat$-matrix \eqref{B_matrix} is therefore
\eq{
        \Bmat \supla{1}
        =
        \arr{ccccc}{
                \mzero & 1      & \mzero & \mzero & \mzero \\
                \mzero & \mzero & 1      & \mzero & \mzero \\
                \mzero & \mzero & \mzero & \mzero & 1
        }
        \quadit{\text{such that}}
        \Ja{\l_1} = \Bmat \supla{1} \cdot \Iab{\l_1}{0}
        \, .
        \label{1L_bhabha_B_l1}
}

\subsubsection{Restriction}

The $\Rmat$- and $\Bmat$-matrices from \eqref{1L_bhabha_l1_constraint} and \eqref{1L_bhabha_B_l1} combine to
\eq{
        \Mmat \supla{1} =
        \arr{c}{\Bmat \supla{1} \\ \hline \Rmat \supla{1}}
        \quadit{\text{such that}}
        \Mmat \supla{1} \cdot \Iab{\l_1}{0} =
        \arr{c}{\Ja{\l_1} \\ \hline 0 \\ 0}
        \, .
}
The $3 \times 3$ Pfaffian matrix $Q_2\supla{1}$ associated to $\Ja{\l_1}$ is now computed with 
\eqref{M_gauge_transformation}.
Namely,
by gauge transforming $P_{2,0}$ from \eqref{1L_bhabha_normal} with
$
        \left(\Mmat \supla{1} \right)^{-1},
$
we get
\eq{
        \left(
                \p_2 \bullet \Mmat \supla{1}
                +
                \Mmat \supla{1} \cdot P_{2,0}
        \right)
        \cdot
        \left(\Mmat \supla{1} \right)^{-1}
        =
        \arr{c|c}{
                Q_2 \supla{1} & \star \\ \hline
                \mathbf{0}_{2 \times 3} & \star
        }
        \, .
}
Explicitly,
we have derived the Pfaffian system
\eq{
        \p_2 \bullet \Ja{\l_1} = Q_2 \supla{1} \cdot \Ja{\l_1}
        \quadit{\text{with}}
        Q_2 \supla{1} =
        \arr{ccc}{
                -\frac{\e}{z_2}  & \mzero & \mzero \\
                \mzero           & \mzero & \mzero \\
                \frac{2(1-2\e)}{z_2^2(z_2+1)} &
                \frac{2(2\e-1)}{z_2  (z_2+1)} &
                \frac{-z_2-\e-1}{z_2 (z_2+1)}
        }
        \, .
        \label{1L_bhabha_Q2_l1}
}
This system agrees with an independent IBP calculation performed directly on the basis
of massless integrals \eqref{1L_bhabha_J}.
See \cite{Henn:2014qga} for a pedagogical exposition on how to solve 
these equations by passing to a canonical basis.

\subsection{Eigenvalue \texorpdfstring{$\l_2 = -\e$}{}}

The Pfaffian system associated to the eigenvalue $\l_2 = -\e$ turns out to be even simpler
than the one for $\l_1 = 0$.
Let's build it.

\subsubsection{Rank jump}

The rank jump constraint from \eqref{logarithmic_restriction} given $M_{\l_2} = 1$ reads
\eq{
        \Rmat\supla{2} \cdot \Iabc{\l_2}{0}{0}
        &=
        \rowred{(P_{1,-1} - \l_2 \mathbf{1} )^2} \cdot \Iabc{\l_2}{0}{0}
        \\&=
        \arr{ccccc}{
                1      & \mzero & \mzero & \frac{z_2 \e}{1-\e}   & \frac{z_2 \e}{2(1-\e)} \\
                \mzero & 1      & \mzero & \mzero                & \mzero                   \\
                \mzero & \mzero & 1      & \mzero                & \mzero
        }
        \cdot
        \Iabc{\l_2}{0}{0}
        \label{1L_bhabha_l2_constraint}
        \\&=
        0
        \, .
}
There are 3 equations imposed on the entries of the $5$-dimensional vector $\Iabc{\l_2}{0}{0}$,
wherefore the rank drops to $2$.

In this case,
it is less clear what the relations \eqref{1L_bhabha_l2_constraint} mean!
The present author does not know what the momentum space representations looks like for other
eigenvalues besides $\l = 0$.
This is worth clarifying, 
since it would help with the determination of boundary constants via direct integration.

\subsubsection{Basis}

The constraints \eqref{1L_bhabha_l2_constraint} fix
$
        I_2\supbrk{\l_2,0,0} = I_3\supbrk{\l_2,0,0} = 0,
$
and 
$I_5\supbrk{\l_2,0,0}$ 
becomes a linear combination of 
$I_1\supbrk{\l_2,0,0}$ and $I_4\supbrk{\l_2,0,0}$.
So 
\eq{
        \Ja{\l_2} =
        \arr{c}{
                I\supbrk{\l_2,0,0}_1 \\
                I\supbrk{\l_2,0,0}_4 
        }
}
is a linearly independent basis.
The $\Bmat$-matrix is therefore given by
\eq{
        \Bmat \supla{2} =
        \arr{ccccc}{
                1 & \mzero & \mzero & \mzero & \mzero \\
                \mzero & \mzero & \mzero & 1 & \mzero
        }
        \quadit{\text{such that}}
        \Ja{\l_2} = \Bmat \supla{2} \cdot \Iabc{\l_2}{0}{0}
        \, .
        \label{1L_bhabha_B_l2}
}

\subsubsection{Restriction}

Stacking the $\Rmat$- and $\Bmat$-matrices from
\eqref{1L_bhabha_l2_constraint} and \eqref{1L_bhabha_B_l2}
on top of each other,
we get
\eq{
        \Mmat \supla{2} =
        \arr{c}{\Bmat \supla{2} \\ \hline \Rmat \supla{2}}
        \quadit{\text{such that}}
        \Mmat \supla{2} \cdot \Iabc{\l_2}{0}{0} =
        \arr{c}{\Ja{\l_2} \\ \hline 0 \\ 0 \\ 0}
        \, .
}
The Pfaffian system for $\Ja{\l_2}$ now follows from \eqref{M_gauge_transformation}:
\eq{
        \left(
                \p_2 \bullet \Mmat \supla{2}
                +
                \Mmat \supla{2} \cdot P_{2,0}
        \right)
        \cdot
        \left(\Mmat \supla{2} \right)^{-1}
        =
        \arr{c|c}{
                Q_2 \supla{2} & \star \\ \hline
                \mathbf{0}_{3 \times 2} & \star
        }
        \, .
}
Explicitly,
we have that
\eq{
        \p_2 \bullet \Ja{\l_2} = Q_2 \supla{2} \cdot \Ja{\l_2}
        \quadit{\text{with}}
        Q_2 \supla{2} =
        \arr{cc}{0 & 0 \\ \frac{\e-1}{z_2^2} & \frac{-1}{z_2}}
        \, .
        \label{1L_bhabha_Q2_l2}
}
The general solution is swiftly found to be
\eq{
        \Ja{\l_2}
        =
        \arr{c}{
                c_1(\e) \\
                \frac{c_1(\e)(1-\e)\log(z_2)}{z_2} + \frac{c_2(\e)}{z_2}
        }
        \, ,
        \label{1L_bhabha_J_l2_sol}
}
where $c_{1,2}(\e)$ are two boundary constants.

\subsection{Result for the logarithmic series}

The solution vectors in \eqref{1L_bhabha_ansatz} at $0$th order in $z_1$ take the form
\eq{
        \Iab{\l_1}{0} &=
        \left( \Mmat \supla{1} \right)^{-1}
        \cdot
        \arr{c}{\Ja{\l_1} \\ \hline 0 \\ 0}
        \\[3pt]
        \Iabc{\l_2}{0}{0} &=
        \left( \Mmat \supla{2} \right)^{-1}
        \cdot
        \arr{c}{\Ja{\l_2} \\ \hline 0 \\ 0 \\ 0}
        \\[3pt]
        \Iabc{\l_2}{0}{1} &= (P_{1,-1} - \l_2 \mathbf{1}) \cdot \Iabc{\l_2}{0}{0}
        \, ,
}
where the last equation comes from the recursion \eqref{m_shift}.
The vector $\Ja{\l_1}$ comes from solving the massless 1-loop box system \eqref{1L_bhabha_Q2_l1},
and $\Ja{\l_2}$ was given in \eqref{1L_bhabha_J_l2_sol}.
We do not include higher-order terms in $z_1^n$ 
(coming from the recursions \eqref{n_shift} and \eqref{n_shift_simple})
in order to show a compact,
final result.

Inserting the solution vectors above into \eqref{1L_bhabha_ansatz} we get%
\footnote{We are neglecting the DR scale $\mu$ in factors such as $(-s/\mu^2)^{-\e}$.}
\eq{{\scriptstyle
        \arr{c}{
                I_{0010}(m) \\[3pt] 
                I_{0101}(m) \\[3pt] 
                I_{1010}(m) \\[3pt] 
                I_{0111}(m) \\[3pt] 
                I_{1111}(m) 
        }
        \overset{m \to 0}{=}
        \arr{c}{
                0 \\[3pt]
                (-s)^{-\e}I_{0101}(0) \\[3pt]
                (-s)^{-\e}I_{1010}(0) \\[3pt]
                \frac{(-s)^{1-\e}(1-2\e)}{t \e} I_{0101}(0) \\[3pt]
                (-s)^{-2-\e} I_{1111}(0) 
        }
        +
        \arr{c}{
                \frac{c_1}{\e-1} (m^2)^{1-\e} \\[3pt]
                0 \\[3pt]
                0 \\[3pt]
                \frac{-(m^2)^{-\e}}{t}
                \big[ c_1 \log(-t/m^2) + c_2 \big]
                \\[3pt]
                \frac{-2(m^2)^{-\e}}{s t \e}
                \big[ c_1(\e \log(-t/m^2) - 1) + \e c_2 \big]
        }
        \label{1L_bhabha_result}
}}
The first and second terms on the LHS come from $\l_1$ and $\l_2$ respectively.
Here we reinstated all prefactors,
wrote $z_1 = m^2/(-s)$ and $z_2 = t/s$,
and lastly rescaled $c_1 \to \frac{c_1}{\e-1}$ to simplify the expression.
Note that only the triangle and the box integrals receive logarithmic corrections.
The bubbles would naturally receive power corrections in $z_1$,
which we are neglecting. 

The two boundary constants are determined as follows.

\begin{itemize}
        \item
                $c_1$ is fixed by comparing with the exact expression
                $
                        I_{0010}(m) = - \Gamma(1-\e)(m^2)^{1-\e}
                $
                for the tadpole on the LHS.
                This gives
                $
                        c_1 = (1-\e) \Gamma(\e-1).
                $
        \item 
                $c_2=c_2(\e)$ can be fixed order-by-order in $\e$ by numerically matching with
                an integral on the LHS,
                say the 5th element $I_{1111}(m)$.
                On the LHS we insert the ansatz
                $
                        c_2 = \sum_{i=-2}^2 c\supbrk{i}_2 \e^i,
                $
                which starts at $i = -2$ 
                because the massless box $I_{1111}(0)$ has a double pole in $\e$.
                On the RHS we numerically evaluate $I_{1111}(m)$ at the point
                $
                        (s,t,m^2) = (-1, -9, 10^{-3})
                $
                with \package{AMFlow} \cite{Liu:2022chg}.
                By comparing the LHS and RHS we get
                \eq{
                        \Big[
                                c\supbrk{-2}_2, c\supbrk{-1}_2, c\supbrk{0}_2, c\supbrk{1}_2, c\supbrk{2}_2
                        \Big] 
                         =
                        \big[ 1, −0.588997, −7.12155, −4.69031, −20.4068 \big]
                        \, .
                        \nonumber
                }
                \package{AMFlow} is easily able to compute $\mO(10^2)$ digits for this topology.
                Analytical expressions for these constants could therefore,
                if needed,
                be found with \soft{PSLQ} \cite{PSLQ}.
\end{itemize}
Evaluating at other phase space points with small $m^2$,
we find that the LHS of \eqref{1L_bhabha_result} is a fair approximation to the
$\e$-expansions of $I_{0111}(m)$ and $I_{1111}(m)$ up to $\mO(\e^3)$.
For instance,
at $(s,t,m^2) = (-1,-8,10^{-3})$ we find agreement within $0.01\%$ accuracy for the triangle,
and within $0.001\%$ accuracy for the box.
The non-leading terms in $\e$ for the bubbles $I_{0101}(m)$ and $I_{1010}(m)$ 
give errors between $5-20\%$ because we have neglected power corrections.

        \section{Logarithmic restriction: 2-loop Bhabha scattering}
\label{sec:bhabha_2L_scattering}

This is another example of logarithmic restriction using \algref{alg:logarithmic_restriction}.
We provide partial results on the small-mass
logarithmic expansion of the 2-loop Bhabha double-box topology:
\eq{
        \includegraphicsbox{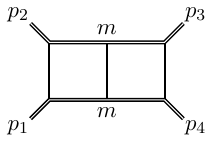}
        \quad
        \xrightarrow{m \to 0}
        \quad
        \includegraphicsbox{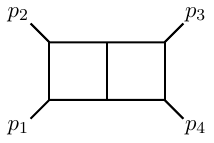}
        \, .
}
The momentum space integral family in $\DD = 4-2\e$ dimensions reads
\eq{
        I_{\nu_1 \ldots \nu_9}(m) =
        \int
        \frac{\dd^\DD \ell_1 \dd^\DD \ell_2}
        {
            D_1^{\nu_1}
            \cdots
            D_9^{\nu_9}
        }
        \, ,
        \label{2L_bhabha_family}
}
with
\eq{
        \begin{array}{llllllllll}
                &D_1 &=& \ell_1^2  - m^2
                &D_2 &=& (\ell_1 + p_1 + p_2)^2 - m^2
                &D_3 &=& \ell_2^2 - m^2 
                \\
                &D_4 &=& (\ell_2 + p_1 + p_2)^2 - m^2
                &D_5 &=& (\ell_1+p_1)^2
                &D_6 &=& (\ell_1-\ell_2)^2  
                \\
                &D_7 &=& (\ell_2-p_3)^2
                &D_8 &=& (\ell_2+p_1)^2
                &D_9 &=& (\ell_1-p_3)^2 \, ,
        \end{array}
        \nonumber
}
where $D_8$ and $D_9$ denote ISPs.
Solutions for this integral family were found in \cite{Henn:2013woa}
(see also \cite{Duhr:2021fhk}).

Inspired by \cite{Henn:2013woa},
we choose the following basis for the unrestricted Pfaffian system:
\eq{
        \hspace{-0.5cm}
        \begin{array}{lllllll}
        G_0 \cdot \vec{I} = 
        \big[
        & I_{2 0 2 0 0 0 0 0 0} ,
        & I_{0 0 0 0 2 2 1 0 0} ,
        & I_{0 0 1 0 2 2 0 0 0} ,
        & I_{1 2 2 0 0 0 0 0 0} ,
        & I_{0 1 2 0 0 2 0 0 0} ,
        & I_{0 2 2 0 0 1 0 0 0} ,
        \\
        & I_{0 0 1 1 1 2 0 0 0} ,
        & I_{0 1 2 0 1 1 0 0 0} ,
        & I_{1 0 0 0 1 1 2 0 0} ,
        & I_{1 2 1 2 0 0 0 0 0} ,
        & I_{0 1 1 0 1 1 1 0 0} ,
        & I_{0 1 1 0 1 2 1 0 0} ,
        \\ 
        & I_{0 2 1 0 1 1 1 0 0} ,
        & I_{0 2 2 0 1 1 1 0 0} ,
        & I_{1 0 1 1 1 1 0 0 0} ,
        & I_{1 0 1 2 1 1 0 0 0} ,
        & I_{1 0 2 0 1 1 1 0 0} ,
        & I_{1 0 1 0 1 1 2 0 0} ,
        \\ 
        & I_{1 1 0 0 1 1 2 0 0} ,
        & I_{1 1 1 0 1 1 1 0 0} ,
        & I_{1 1 1 0 1 1 1 -1 0},
        & I_{1 1 1 1 1 1 1 0 0} ,
        & I_{1 1 1 1 1 1 1 -1 0} & \big]^T \, .
        \end{array}
        \label{2L_bhabha_basis}
}
The $23 \times 23$ diagonal matrix $G_0$ factors out the mass dimension of each integral via
powers of $(-s)$.
IBP reduction with \package{kira} \cite{Klappert:2020nbg} then provides a Pfaffian system
$
        \p_i \bullet \vec{I} = P_i \cdot \vec{I}
$
in terms of the kinematic variables 
\eq{
        z_1 = \frac{m^2}{-s}
        \quadit{\text{and}}
        z_2 = \frac{t}{s}
        \, .
}
Our goal is to restrict onto the surface $Y' = \{z_1=0\}$.

\subsection{Normal form}

The Pfaffian system associated to the basis \eqref{2L_bhabha_basis} is not in normal form.
The first issue to resolve is the higher-order singularities on $Y'$:
\eq{
        \begin{array}{lllllll}
                P_1 & = & z_1^{-3} & \times & P_{1,-3}(z_2)  & + & \mO(z_1^{-2})
                \\[2pt]
                P_2 & = & z_1^{-2} & \times & P_{2,-2}(z_2)  & + & \mO(z_1^{-1})
                \, .
        \end{array}
        \label{2L_bhabha_normal}
}
Moser reduction suggests to gauge transform with
\eq{
        G_1 =
        \soft{Diagonal}\big[\underbrace{ z_1, \ldots, z_1}_{8 \ \text{times}}, \, 1, \ldots, \, 1\big]
        \quadit{\text{and}}
        G_2 =
        \soft{Diagonal}\big[\underbrace{ z_1, \ldots, z_1}_{12 \ \text{times}}, \, 1, \ldots, \, 1\big]
        \, .
        \nonumber
}
Introducing the shorthand notation
$
G_{i \ldots j} = G_i \cdot \ldots \cdot G_j,
$
then the new pole structure becomes
\eq{
        \begin{array}{lllllll}
                G_{21}[P_1] & = & z_1^{-1} & \times & G_{21}[P_1]_{-1}  & + & \mO(z_1^0)
                \\[2pt]
                G_{21}[P_2] & = & z_1^{-1} & \times & G_{21}[P_2]_{-1}  & + & \mO(z_1^0)
                \, .
        \end{array}
}
So we have cured the higher-order poles in $P_1$.
But $P_2$ is still not finite on $Y'$ because the residue matrix $G_{21}[P_1]_{-1}$ has a resonant spectrum%
\footnote{
        See \cite[Appendix B]{Chestnov:2023kww} for a proof of why non-resonance of the 
        residue matrix $P_{1,-1}$ implies finiteness of the remaining Pfaffian matrices
        $\{P_2, \ldots P_N\}$ on $\{z_1=0\}$.
}.
In \cite[Appendix B]{Chestnov:2023kww},
it is shown how to construct a sequence of gauge transformations
(involving the Jordan decomposition of the residue matrix) 
to cure the spectrum%
\footnote{
        The \package{Mathematica} scripts in \cite{url-restriction-data} contain
        the gauge transformations for this particular example.
}.
It takes a matter of seconds on a laptop to compute these gauge transformations,
even without fixing $z$-variables to generic numbers
(recall that the Jordan decomposition is independent of $z$).
Writing $G_3$ for the product of said gauge matrices,
we get
\eq{
        G_{321}[P_2] = G_{321}[P_2]_0 + \mO(z_1)
        \, ,
        \label{2L_bhabha_G3}
}
wherefore the system is now in normal form.
For brevity,
we
\eq{
        \text{replace} \ G_{321}[P_i] \ \to \ P_i \ \text{in the rest of this example}.
}

$P_{1,-1}$ denotes the residue matrix in this newfound basis.
Its Jordan form enjoys an eigenvalue decomposition
\eq{
        \jord{P_{1,-1}} =
        \mathsf{J} \supla{1} 
        \oplus
        \mathsf{J} \supla{2} 
        \oplus
        \mathsf{J} \supla{3} 
        \oplus
        \mathsf{J} \supla{4} 
}
in terms of block matrices
\eq{
        \nonumber
        \mathsf{J} \supla{1} &= 
        \mathbf{0}_{8 \times 8}
        \, , \quad
        \mathsf{J} \supla{2} = 
        \arr{c}{\l_2}
        \\
        \mathsf{J} \supla{3} &= 
        \arr{ccccccc}{
                \l_3   & \mzero & \mzero & \mzero & \mzero & \mzero & \mzero \\
                \mzero & \l_3   & \mzero & \mzero & \mzero & \mzero & \mzero \\
                \mzero & \mzero & \l_3   & 1      & \mzero & \mzero & \mzero \\
                \mzero & \mzero & \mzero & \l_3   & \mzero & \mzero & \mzero \\
                \mzero & \mzero & \mzero & \mzero & \l_3   & 1      & \mzero \\
                \mzero & \mzero & \mzero & \mzero & \mzero & \l_3   & \mzero \\
                \mzero & \mzero & \mzero & \mzero & \mzero & \mzero & \l_3 
        }
        \\ \nonumber
        \mathsf{J} \supla{4} &= 
        \arr{ccccccc}{
                \l_4   & \mzero & \mzero & \mzero & \mzero & \mzero & \mzero \\
                \mzero & \l_4   & \mzero & \mzero & \mzero & \mzero & \mzero \\
                \mzero & \mzero & \l_4   & \mzero & \mzero & \mzero & \mzero \\
                \mzero & \mzero & \mzero & \l_4   & \mzero & \mzero & \mzero \\
                \mzero & \mzero & \mzero & \mzero & \l_4   & 1      & \mzero \\
                \mzero & \mzero & \mzero & \mzero & \mzero & \l_4   & 1 \\
                \mzero & \mzero & \mzero & \mzero & \mzero & \mzero & \l_4 
        }
        \, ,
}
where the eigenvalues are
\eq{
        \{ \l_1, \, \l_2, \, \l_3, \, \l_4 \}
        =
        \left\{0, \, 2(\DD-5)\, , \frac{\DD-4}{2}, \, \DD-4 \right\}
        \, .
}
Their associated multiplicities
\eq{
        \{ \L_1, \, \L_2, \, \L_3, \, \L_4 \}
        =
        \{8, \, 1, \, 7, \, 7\}
}
count the sizes of the four different Pfaffian systems that will be obtained via logarithmic restriction.
The Jordan blocks above specify the maximum logarithm powers
\eq{
        \{ M_{\l_1}, \, M_{\l_2}, \, M_{\l_3}, \, M_{\l_3} \}
        =
        \{0, \, 0, \, 1, \, 2\}
}
for the series ansatz \eqref{I_asymptotic_series}.
The vector of MIs $\vec{I}$ can therefore be represented as
\eq{
        \begin{array}{llllllll}
                G_{321} \cdot \vec{I}(z) &=&
                & z_1^{\l_1} \Iab{\l_1}{0} (z) 
                & + &
                z_1^{\l_2} \Iab{\l_2}{0} (z) & &
                \\[6pt] 
                & + &
                & z_1^{\l_3} 
                \big[ 
                        \Iab{\l_3}{0} (z) 
                        & + &
                        \Iab{\l_3}{1} (z) \log(z_1)
                \big] 
                & &
                \\[6pt]
                & + &
                & z_1^{\l_4} 
                \big[ 
                        \Iab{\l_4}{0} (z) 
                        & + &
                        \Iab{\l_4}{1} (z) \log(z_1)
                        & + &
                        \Iab{\l_4}{2} (z) \log^2(z_1)
                \big] 
                \, ,
        \end{array}
        \label{2L_bhabha_ansatz}
        \raisetag{2.5\baselineskip}
}
where $z = (z_1,z_2)$.
The logarithm-free terms above all have holomorphic expansions in $z_1$ of the form
\eq{
        \Iab{\l}{0}(z) =
        \sum_{n=0}^\infty \Iabc{\l}{n}{0}(z_2) \, z_1^n
        \, .
}
Recall that all higher-order terms $\Ilnm$ in 
$
        z_1^n \, \log^m(z_1)
$
are uniquely specified given $\Ilzz$ as input via the recursion relations 
\eqref{m_shift}, \eqref{n_shift} and \eqref{n_shift_simple}.

\subsection{Restriction}

We proceed to compute the $\L \times \L$ Pfaffian matrices $Q_2 \supl(z_2)$ associated to each $\Ilzz$.
The full expression \eqref{2L_bhabha_ansatz} has not yet been mounted
because the boundary constants are missing,
so this step is left for future study.

\subsubsection*{Eigenvalue \texorpdfstring{$\l_1$}{}}

The $\Rmat$-matrix for $\l_1 = 0$ splits into a $0$-block and an identity matrix block:
\eq{
        \Rmat \supla{1} =
        \rowred{P_{1,-1}} =
        \arr{c|c}{ \mathbf{0}_{15 \times 8} & \mathbf{1}_{15} }
        \, .
}
Since $\Rmat \supla{1} \cdot \Iabc{\l_1}{0}{0} = 0$,
this means that 
$
        I \supbrk{\l_1,0,0}_9 = \ldots = I \supbrk{\l_1,0,0}_{23} = 0.
$
The $8$ remaining entries of $\Iabc{\l_1}{0}{0}$ thus make up a basis for the restricted Pfaffian system:
\eq{
        \Ja{\l_1} =
        \arr{c}{I \supbrk{\l_1,0,0}_1 \\ \vdots \\ I \supbrk{\l_1,0,0}_8}
        \, .
}
This corresponds to the $\Bmat$-matrix
\eq{
        \Bmat \supla{1} = 
        \arr{c|c}{ \reflectbox{$\mathbf{1}$}_8 & \mathbf{0}_{8 \times 15} }
        \quadit{\text{such that}}
        \Ja{\l_1} = \Bmat \supla{1} \cdot \Iabc{\l_1}{0}{0}
        \, ,
}
where $\reflectbox{$\mathbf{1}$}_8$ represents the "reversed identity matrix",
i.e.~the $8 \times 8$ matrix with $1$'s on the anti-diagonal.
Setting
\eq{
        \Mmat \supla{1} = 
        \arr{c}{\Bmat \supla{1} \\ \hline \Rmat \supla{1}}
        \, ,
}
then the Pfaffian matrix associated to $\Ja{\l_1}$ is
\eq{
        \left(
                \p_2 \bullet \Mmat \supla{1}
                +
                \Mmat \supla{1} \cdot P_{2,0}
        \right)
        \cdot
        \left(\Mmat \supla{1} \right)^{-1}
        =
        \arr{c|c}{
                Q_2 \supla{1} & \star \\ \hline
                \mathbf{0}_{15 \times 8} & \star
        }
        \, .
}
We find that
\eq{
        \p_2 \bullet \Ja{\l_1} = 
        Q_2 \supla{1} \cdot \Ja{\l_1}
}
is indeed a Pfaffian system for the massless double-box topology,
as verified by an independent IBP computation.
Each entry of $\Ja{\l_1}$ is a linear combination%
\footnote{
        It is a linear combination because the gauge transformation matrix 
        $G_3$ from \eqref{2L_bhabha_G3} is non-diagonal.
}
of integrals in the family \eqref{2L_bhabha_family} with $m^2=0$ at the integrand level.
The matrix $Q_2 \supla{1}$ has the form
$
        \frac{Q_a(\e)}{z_1^2} + \frac{Q_b(\e)}{z_1} + \frac{Q_c(\e)}{z_1+1} 
$
where $Q_{a,b,c}(\e)$ contain rational functions in $\e$ with maximal degree $\e^2$.
Using the software \package{canonica} \cite{Meyer:2017joq},
a canonical form for $Q_2 \supla{1}$ is found within seconds.

\subsubsection*{Eigenvalue \texorpdfstring{$\l_2$}{}}

The system for the eigenvalue
$
        \l_2 = 2(\DD-5) = -2 - 4\e
$
is peculiar.
The rank jump constraint
\eq{
        \Rmat \supla{2} \cdot \Iabc{\l_2}{0}{0} = 
        \rowred{P_{1,-1} - \l_2 \mathbf{1}} \cdot \Iabc{\l_2}{0}{0} =
        0
}
sets every entry of $\Iabc{\l_2}{0}{0}$ to zero except for the $15$th entry
$
        J\supla{2} = I \, \supbrk{\l_2,0,0}_{15}
$.
The Pfaffian system for this single, 
non-trivial function
works out to
\eq{
        \p_2 \bullet J\supla{2} = - \frac{J\supla{2}}{z_2} 
        \quadit{\implies}
        J\supla{2} = \frac{c\supla{2}(\e)}{z_2}
        \, .
}

\subsubsection*{Eigenvalue \texorpdfstring{$\l_3$}{}}

The system associated to eigenvalue 
$
        \l_3 = \frac{\DD-4}{2} = -\e
$
has rank $\L_3 = 7$.
Working out the $\Rmat$- and $\Bmat$-matrices as usual,
one finds a basis
\eq{
        \Ja{l_3} =
        \arr{c}{
                I\supbrk{\l_3,0,0}_{16} \\[2pt]
                \vdots \\[2pt]
                I\supbrk{\l_3,0,0}_{22} 
        }
        \, .
}
The associated Pfaffian matrix $Q_2\supla{3}$ has the form
$
        \frac{Q_a(\e)}{z_2^2} +
        \frac{Q_b(\e)}{z_2} + 
        \frac{Q_c(\e)}{(z_2+1)^2} + 
        \frac{Q_d(\e)}{z_2-1},
$
where $Q_{a,b,c,d}(\e)$ contain rational functions in $\e$ of maximal degree $\e^3$.
This system is swiftly $\e$-factorized with \package{canonica}.

\subsubsection*{Eigenvalue \texorpdfstring{$\l_4$}{}}

The last eigenvalue
$
        \l_4 = \DD-4 = -2\e
$
also has rank $\L_4 = 7$.
Its associated rank jump constraint 
$
        \Rmat \supla{4} \cdot \Iabc{\l_4}{0}{0} = 0
$
contains linear relations with rather complicated coefficients.
This is in contrast to the rank jump constraints for the other eigenvalues,
which were all of the form 
$
        I\supbrk{\l,0,0}_i = 0.
$
We do not know why this is the case.

Regardless,
we are easily able to calculate the Pfaffian system associated to the basis
\eq{
        \Ja{\l_4} = 
        \arr{c}{
                I\supbrk{\l_4,0,0}_{5}  \\[5pt]               
                I\supbrk{\l_4,0,0}_{17} \\[5pt]               
                I\supbrk{\l_4,0,0}_{19} \\[5pt]               
                I\supbrk{\l_4,0,0}_{20} \\[5pt]               
                I\supbrk{\l_4,0,0}_{21} \\[5pt]               
                I\supbrk{\l_4,0,0}_{22} \\[5pt]               
                I\supbrk{\l_4,0,0}_{23} 
        }
        \, .
}
The Pfaffian matrix has the singularity structure
$
        Q_2\supla{3} =
        \frac{Q_a(\e)}{z_2^2} +
        \frac{Q_b(\e)}{z_2} + 
        \frac{Q_c(\e)}{(z_2+1)^2} + 
        \frac{Q_d(\e)}{z_2-1},
$
where $Q_{a,b,c,d}(\e)$ have maximal degree $\e^3$.
\package{Canonica} brings this system into canonical form with ease.

\subsection{Discussion}

General solutions to the Pfaffian systems above are easily obtained 
in terms of HPLs in the variable $z_2 = t/s$.
What is more,
every step in the construction of those systems can be automatized.

It is worth comparing the present method with previous studies.
The seminal calculation carried out in \cite{Henn:2013woa} 
was complicated by an unrationalizable square root.
For another 2-loop Bhabha scattering topology \cite{Duhr:2021fhk},
the authors had to deal with elliptic functions.
Because our restriction protocol involves smaller, univariate Pfaffian systems,
such complications are more easily avoided.
The cost is that a smaller part of phase space is covered,
since $m^2$ is assumed to be small
(it could e.g.~be the electron mass $m_e \sim 0.511 \ \text{MeV}$).

As mentioned,
the final expression for the logarithmic series \eqref{2L_bhabha_ansatz} still lacks the determination
of boundary constants.
These could be fixed by matching with the full set of two-variable 
MIs $\vec{I}$ from \eqref{2L_bhabha_basis} after imposing physically motivated constraints.
For instance,
one could impose real-valuedness in the Euclidean region,
or regularity at certain points in phase space.
Another way is just to numerically evaluate the two-variable integrals $\vec{I}$ and match
this with the logarithmic series.
Both of these approaches are,
however,
unsatisfactory from the point of view of the restriction protocol.
It would be preferable to somehow fix the boundary constants solely within each univariate system
$
        \p_2 \bullet \Ja{\l} = Q_2\supl \cdot \Ja{\l},
$
rather than comparing with the two-variate integrals.
This is indeed possible for the eigenvalue $\l = 0$,
since in this case we know the momentum space representation of $\Ja{\l}$.
A speculative suggestion is to perform an \emph{additional} restriction on these univariate systems,
in which case the rank jump constraints might provide relations among boundary constants.

        \section{Macaulay-level restriction: 2-loop diagonal box}

This example concerns the rational restriction of the GKZ system 
associated to a 2-loop massless diagonal-box diagram.
We employ \algref{alg:macaulay_restriction},
which is based on the Macaulay matrix.

\subsection{Setup}

The LP monomials are given by
\eq{
    \nonumber
    \mG(z|x) = \ &
    z_{1} x_1 x_4 +
    z_{2} x_1 x_5 +
    z_{3} x_1 x_6 +
    z_{4} x_2 x_4 +
    z_{5} x_2 x_5 +
    z_{6} x_2 x_6 + \\ \nonumber &
    z_{7} x_3 x_4 +
    z_{8} x_3 x_5 +
    z_{9} x_3 x_6 +
    z_{10} x_4 x_5 +
    z_{11} x_4 x_6 +
    z_{12} x_4 x_7 + \\ &
    z_{13} x_5 x_7 +
    z_{14} x_6 x_7 +
    z_{15} x_1 x_3 x_4 +
    z_{16} x_1 x_3 x_5 +
    z_{17} x_1 x_3 x_6 + \\ \nonumber &
    z_{18} x_1 x_4 x_5 +
    z_{19} x_2 x_4 x_6 +
    z_{20} x_2 x_4 x_7 +
    z_{21} x_2 x_5 x_7 +
    z_{22} x_2 x_6 x_7 \, .
}
Recall that this enters the integrand of the GFI \eqref{generalized_feynman_integral}.
$\mG(z|x)$ gives rise to to a GKZ system with $A$-matrix
\eq{A = \arr{cccccccccccccccccccccc}{
1 & 1 & 1 & 1 & 1 & 1 & 1 & 1 & 1 & 1 & 1 & 1 & 1 & 1 & 1 & 1 & 1 & 1 & 1 & 1 & 1 & 1 \\
1 & 1 & 1 & \mzero & \mzero & \mzero & \mzero & \mzero & \mzero & \mzero & \mzero & \mzero & \mzero & \mzero & 1 & 1 & 1 & 1 & \mzero & \mzero & \mzero & \mzero \\
\mzero & \mzero & \mzero & 1 & 1 & 1 & \mzero & \mzero & \mzero & \mzero & \mzero & \mzero & \mzero & \mzero & \mzero & \mzero & \mzero & \mzero & 1 & 1 & 1 & 1 \\
\mzero & \mzero & \mzero & \mzero & \mzero & \mzero & 1 & 1 & 1 & \mzero & \mzero & \mzero & \mzero & \mzero & 1 & 1 & 1 & \mzero & \mzero & \mzero & \mzero & \mzero \\
1 & \mzero & \mzero & 1 & \mzero & \mzero & 1 & \mzero & \mzero & 1 & 1 & 1 & \mzero & \mzero & 1 & \mzero & \mzero & 1 & 1 & 1 & \mzero & \mzero \\
\mzero & 1 & \mzero & \mzero & 1 & \mzero & \mzero & 1 & \mzero & 1 & \mzero & \mzero & 1 & \mzero & \mzero & 1 & \mzero & 1 & \mzero & \mzero & 1 & \mzero \\
\mzero & \mzero & 1 & \mzero & \mzero & 1 & \mzero & \mzero & 1 & \mzero & 1 & \mzero & \mzero & 1 & \mzero & \mzero & 1 & \mzero & 1 & \mzero & \mzero & 1 \\
\mzero & \mzero & \mzero & \mzero & \mzero & \mzero & \mzero & \mzero & \mzero & \mzero & \mzero & 1 & 1 & 1 & \mzero & \mzero & \mzero & \mzero & \mzero & 1 & 1 & 1 \\
} \, . \nonumber}
The restriction we compute is
\eq{
        A
        \quad
        \xrightarrow
        [1 \leq i \, \leq \, 18 \ \text{and} \ 19 \, \leq \, j \, \leq \, 22]
        {z_i = 1 \ \text{and} \ z_j = t / s}
        \quad
        \includegraphicsbox{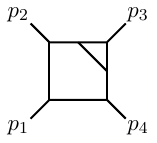}
        \, ,
}
where $s$ and $t$ stand for Mandelstam variables.
To be clear,
after a linear shift of coordinates,
this is a restriction onto the affine hyperplane
\eq{
        \nonumber
        Y' = 
        \big\{ 
                (z_1, \ldots, z_{22}) \in \AA^{22} 
                \, \big | \, 
                z_1 = \ldots = z_{18} = 0 
                \ \text{and} \
                z_{19} = z_{20} = z_{21} = z_{22} = z 
                \ \text{is generic} 
        \big\}
        \, .
}
Using homogeneity
(cf.~\secref{sec:integrand_rescaling})
we can immediately restrict $z_i = 0$ for $i \in \{1,2,5,7,9,11,13,18\}$.
This means that \algref{alg:macaulay_restriction} should be applied to restrict
$z_i = 0$ for $i \in \{3,4,6,8,10,12,14,15,16,17\}$ and $z_i = z = t/s$ for $i \in \{19,20,21,22\}$.

\subsection{Restriction}

For generic values of the analytic regulators $\e$ and $\d$,
the GKZ system is of rank $R = 115$.
Computing the Euler characteristic by counting the number of 
critical points of \eqref{likelihood_equation} on $Y'$,
we find a rank drop to $R' = 7$.
This number is verified by counting the number of MIs using IBPs.

We must find a basis of the restriction module before calling \algref{alg:macaulay_restriction}.
Executing \cite[Algorithm 4.7]{Chestnov:2023kww} as in \exref{ex:mm_restriction} with parameters 
$\gamma =3, \, k = 3, \, p = 100000007$ and $z = 1/7$,
we obtain the $7$-dimensional basis%
\footnote{
        This took 40.56s on a machine with the following specs:
        \soft{Intel(R) Core(TM) i7-10700K CPU @ 3.80GHz, 8G memory}.
}
\eq{
        \RStd = 
        \arr{c}{
                \p_{21} \p_{22} \\
                \p_{22}^2 \\
                \p_6 \\
                \p_{17} \\
                \p_{21} \\
                \p_{22} \\
                1
        }
        \, .
}
A valid Macaulay matrix $M_1 = \arr{c|c}{M_\Ext & M_\RStd}$
is then found%
\footnote{
        It took 65.76s on the same machine to compute this Macaulay matrix.
}
for degree $Q=1$,
meaning that the equation \eqref{MM_Ext_eq} given by
$
        C_\Ext - C \cdot M_\Ext = 0
$
has a solution for the matrix $C$.
The matrix $M_\Ext$ has dimension $1305 \times 2763$,
so there are $2763$ exterior monomials $\Ext$.
The $1305$ rows are not all independent;
by row reducing $M_\Ext$ with $(\e,\d,z)$ fixed to generic numbers,
we identify $912$ linearly independent equations.
It then takes less than $3$ minutes on a laptop to determine $C$ via \package{FiniteFlow}.
The $7 \times 7$ Pfaffian matrix $P_z$ finally results from
$
        P_z = C_\RStd - C \cdot M_\RStd.
$
This becomes a Pfaffian matrix for the proper 2-loop diagonal-box topology after multiplying by
the LP prefactors from \eqref{generalized_feynman_integral} and sending $\d \to 0$.

As an aside,
note how large the discrepancy is between the ranks 
$R = 115$ and $R' = 7$ for the GKZ system and the proper FI.
So even though we benefit from being able to immediately write down an 
annihilating ideal and a basis in the GKZ setting,
the additional complexity brought about by unspecified $z$-variables may quickly become too much to handle.

        \part{Numerical Integration}

        \chapter{Tropical Integration}
\label{ch:tropical}

This thesis has,
so far,
predominantly been an \emph{analytic} study of Euler and Feynman integrals.
In particular,
we have focused on the DEQs obeyed by said integrals.
The ideal situation is that 
1) these DEQs can be built with reasonable computer resources, 
and
2) they can be solved symbolically in terms of special functions.
For sufficiently difficult problems -
e.g.~the evaluation of 2- or 3-loop diagrams with several kinematic scales -
either or both of these steps become insurmountable with the presently available analytical methods.
These difficulties are in many cases unavoidable if the accuracy of theoretical predictions 
is made to compete with the experimental precision at the LHC.
For instance,
the slow convergence rate of the QCD perturbative series
(stemming from the coupling constant being $\a_s(M_Z) \sim 0.1$ near the weak scale)
often requires 3-loop results to get cross sections at $\%$-level accuracy.
Furthermore,
the study of Higgs phenomenology typically requires the calculation 
of diagrams with many internal masses that do not evaluate to GPLs,
which complicates the determination of canonical bases in the method of DEQs for FIs.

In cases where analytical methods fall short,
\emph{numerical} methods may instead offer a way forward.
The past few years have indeed seen a surge in the development of 
public codes tailored to the numerical evaluation of FIs.
Using series of expansions of Pfaffian systems,
the codes
\package{AMFlow} \cite{Liu:2022chg},
\package{DiffExp} \cite{Hidding:2020ytt,Dubovyk:2022frj}
and
\package{SeaSyde} \cite{Armadillo:2022ugh}
are able to to achieve phenomenal accuracy.
They derive Pfaffian systems by fixing parameters to numbers \emph{before} solving the IBPs,
which yields a significant speed boost.
The \package{pySecDec} codebase \cite{Borowka:2017idc},
employing \emph{Monte Carlo} integration in combination with \emph{sector decomposition},
has been a staple of high-energy physics calculations for many years,
and was recently upgraded in \cite{Heinrich:2023til}.
Sophisticated manipulations of momentum space integrals via the \emph{Loop-tree duality} 
led to efficient and stable numerical integration schemes in \cite{Capatti:2019ypt,Capatti:2020ytd}.
The authors of \cite{Lairez:2023nih} presented a high-precision 
numerical algorithm by interpreting FIs as \emph{period integrals}.
Further promising studies were carried out in \cite{Pittau:2021jbs,Zeng:2023jek,Jinno:2022sbr},
though so far without publicly available codes.

\textbf{Part III} of this thesis is a contribution to this flourishing research program.
The present chapter describes a Monte Carlo integration scheme called \emph{tropical integration}.
It was established by Borinsky in \cite{Borinsky:2020rqs} 
(see also \cite{Borinsky:2022qcf}), 
and subsequently generalized by Borinsky, Tellander and the present author in \cite{Borinsky:2023jdv}.
A public code named \ft was developed alongside the aforementioned article.
\ft is able to swiftly numerically evaluate FIs at high loop-order with many mass scales
at \textperthousand-level accuracy given modest computer resources.
The next chapter gives a tutorial on how to use the code,
and showcases several state-of-the-art examples.
The present chapter explicates the theoretical background.

        \section{Tropical Monte Carlo sampling, in brief}
\label{sec:tropica_MC_brief}

Tropical integration is a special instance of Monte Carlo (MC) integration.
MC methods deal with integrals of the form
\eq{
        I_f = \int_\Gamma f(x) \mu
        \, ,
}
where $f(x)$ is a square-integrable function integrated over an $n$-dimensional domain $\Gamma$,
and $\mu$ is a probability measure that is positive on $\Gamma$ and normalized to unity,
$
        \int_\Gamma \mu = 1.
$
By sampling many independent random points
$
        \{x\supbrk{1}, \ldots, x\supbrk{N}\}
$
from $\mu$,
it follows from the central limit theorem that
\eq{
        I_f \simeq I_f\supbrk{N}
        \quadit{\text{where}}
        I_f\supbrk{N} = \frac{1}{N} \sum_{i=1}^N f\big( x\supbrk{i} \big)
        \, .
}
Here $I_f\supbrk{N}$ is simply the average of the integrand $f$ evaluated at the random points $x\supbrk{i}$.
For sufficiently large $N$,
the expected error of this approximation is
\eq{
        \s_f = \sqrt{ \frac{I_{f^2} - I_f^2 }{N} }
        \quadit{\text{where}}
        I_{f^2} = \int_\Gamma f(x)^2 \mu
        \, .
}
This error can itself be estimated by
\eq{
        \s_f \simeq \s_f\supbrk{N}
        \quadit{\text{where}}
        \s_f\supbrk{N} =
        \sqrt{ 
                \frac
                {I_{f^2}\supbrk{N} - \left[ I_f\supbrk{N} \right]^2}
                {N-1} 
        }
        \quadit{\text{and}}
        I_{f^2}\supbrk{N} = 
        \frac{1}{N} \sum_{i=1}^N f\big( x\supbrk{i} \big)^2
        \, .
        \label{sampling_variance}
}

A common choice for $\mu$ is just the uniform measure 
$
        \mu = \dd^n x
$
on the $n$-dimensional unit cube $\Gamma = [0,1]^n$.
Tropical integration,
instead,
makes use of polytopal and tropical geometry 
(see \cite{MaclaganSturmfels,ziegler2012lectures} for seminal textbooks on these topics)
to build a more sophisticated \emph{tropical measure} $\mu\trop$.
This measure is tailored to estimating integrals of functions 
$
        f(x) = \frac{g(x)^a}{h(x)^b}
$
given by homogeneous polynomials $g,h \in \CC[x]$ and rational numbers $a,b \in \QQ$.
A good example hereof is the evaluation of the integral
$\int_0^1 \int_0^1 \frac{\dd x \wedge \dd y}{x+y} = 2 \log 2$.
"Ordinary" MC fails because the variance of $\frac{1}{x+y}$ is infinite w.r.t.~the uniform measure.
It is, however, easily evaluated with the tropical measure.
See \cite[Section 3.1.1]{favorito} for a detailed discussion.

\subsection{Tropical approximation}

As we are ultimately interested in integrating homogeneous polynomials,
let us now define the notion of \emph{real projective space} $\RR\PP^{n-1}$;
this which will serve as an integration domain%
\footnote{In practice, this domain can be parametrized as a simplex polytope.}
later on.
$\RR\PP^{n-1}$ is defined by the set of points in 
$
        \RR^n \setminus \brc{0}
$
modulo the equivalence relation
\eq{
        \hspace{-2cm}
        \begin{array}{rcl}
        (x_0, \ldots, x_{n-1})
        & \sim &
        (y_0, \ldots, y_{n-1})
        \\
        & \Updownarrow &
        \\
        \exists \l \in \RR \setminus \{0\} \, : \,
        \brk{x_0, \ldots, x_{n-1}}
        & = &
        \brk{\l y_0, \ldots, \l y_{n-1}} \, .
        \end{array}
}
We denote the coordinates on $\RR\PP^{n-1}$ by
$
        \sbrk{x_0 : \ldots : x_{n-1}}.
$
Two points such as
$
        \sbrk{1 : 0 : 2}
$
and
$
        \sbrk{\tfrac{1}{2} : 0 : 1}
$
would be equivalent in $\RR\PP^2$ because they are related by
$
        \brk{1 \pcomma 0 \pcomma 2}
        = 
        \brk{2 \cdot \tfrac{1}{2} \pcomma 2 \cdot 0 \pcomma 2 \cdot 1} .
$
\emph{Positive} projective space is further given by
\eq{
        \PP^{n-1}_+ =
        \{x = [x_0: \cdots : x_{n-1}] \in \RR \PP^{n-1} \, | \, x_i > 0\}
        \, .
        \label{positive_projective_space}
}

Above we made the unusual choice of starting the index at $0$ rather than $1$.
This convention allows for seamless interoperability with the \package{python} programming language,
to be used in the next chapter.

Let $p$ be a homogeneous polynomial.
In multi-index notation,
\eq{
        p(x) = 
        \sum_{k \in \supp{p}}
        c_k \, x^k
        \ ,
}
where
$
        x^k = x_0^{k_0} \cdots x_{n-1}^{k_{n-1}}
$
and $|k| = k_0 + \ldots + k_{n-1}$ is equal for all multi-indices $k$.
The \emph{tropical approximation} \cite{Panzer:2019yxl} to $p$ is 
the central concept in tropical integration:
\eq{
        p \trop (x) = \max_{k \in \supp{p}} x^k
        \, .
}
If,
for instance,
$
        p(x) = 
        x_0 x_1 x_2^2 - 2 x_1^4 + 42 i x_0^2 x_1^2,
$
then
$
        p \trop (x) = 
        \max \big\{ x_0 x_1 x_2^2, \, x_1^4, \, x_0^2 x_1^2 \big\}.
$
In other words,
the tropical approximation neglects the monomial coefficients,
and only cares about the largest monomial.
The set of monomial exponents $\supp{p}$ is hence the essential data for $p \trop$.
Interpreting each multi-index in $\supp{p}$ as a vector in $\ZZ^n_{\geq 0}$ and taking the convex hull,
we see that $p \trop$ is simply a function avatar of the \emph{Newton polytope} $\mathbf{N}[p]$ from
\eqref{Newton_polytope}.

Suppose that $\mathbf{N}[p]$ lives in $\RR^n$ as parametrized by $x$.
By virtue of being a polytope,
$\mathbf{N}[p]$ is given by a finite intersection of \emph{half-spaces},
each of the form
$
        \{v \in \RR^n \, | \, v \cdot x \leq c\}
$
for some scalar $c \in \RR$.
A subset $F \subset \mathbf{N}[p]$ is called a \emph{face} of $\mathbf{N}[p]$ iff
\begin{enumerate}
        \item 
                There exist $v \in \RR^n$ and $c \in \RR$ such that $\mathbf{N}[p]$ is contained in
                $
                        \{v \in \RR^n \, | \, v \cdot x \leq c\}.
                $
        \item 
                $F$ is the intersection of $\mathbf{N}[p]$ with
                $
                        \{v \in \RR^n \, | \, v \cdot x = c\}.
                $
\end{enumerate}
The polynomial $p$ truncated to a face $F$ is given by
\eq{
        p_F(x) = 
        \sum_{k \, \in \, F \, \cap \, \supp{p}}
        c_k \, x^k
        \, .
}
$p$ is called \emph{completely non-vanishing} on a domain $X$ 
if for each face $F$ of the Newton polytope we have
$
        p_F (x) \neq 0
        , \,
        x \in X.
$

\begin{figure}[t!]
        \centering
        \includegraphics[scale=0.8]{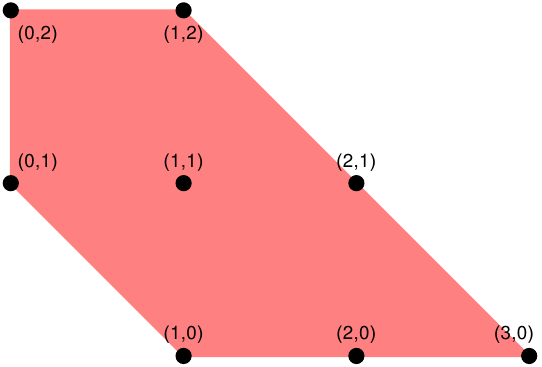}
        \caption{
                Newton polytope for the polynomial $p(x)$ in \eqref{polynomial_for_newton_polyope}.
                The black dots represent the monomial exponents of $p(x)$,
                and the pink area is their convex hull in $\RR^2$ as parametrized by $(x_0,x_1)$.
        }
        \label{fig:newton_polytope_example}
\end{figure}

\begin{ex}
Consider the polynomial
\eq{
        p(x) = 
        x_0 + x_1 + 2 x_0^2 + x_1^2 + 3 x_0 x_1 + 2 x_0^2 x_1 + x_0 x_1^2 + x_0^3 
        \label{polynomial_for_newton_polyope}
        \, .
}
It is non-homogeneous,
but that is inconsequential for the purpose of this example.
Its associated Newton polytope $\mathbf{N}[p]$ is shown in \figref{fig:newton_polytope_example}.
As an intersection of half-spaces,
the Newton polytope is represented by
\eq{
        \mathbf{N}[p] = 
        \big\{ x_0 \geq 0\big\}
        \, \cap \, 
        \big\{ x_1 \geq 0 \big\}
        \, \cap \, 
        \big\{ x_1 \leq 2 \big\}
        \, \cap \, 
        \big\{ x_0 + x_1 \geq 1 \big\}
        \, \cap \, 
        \big\{ x_0 + x_1 \leq 3 \big\}
        \, .
}

Consider a face $F$ of $\mathbf{N}[p]$,
for example the bottom line segment touching the $x_0$-axis given by
$
        F = \{(x_0,x_1) \, | \, 1 \leq x_0 \leq 3 \ \text{and} \ x_1 =0 \}.
$
$F$ is indeed a face since
1) $\mathbf{N}[p]$ is contained in the half-space
$
        \{x_1 \geq 0\},
$
and
2)
$F$ is given by an intersection
$
        F = 
        \mathbf{N}[p] 
        \, \cap \, 
        \{x_1 = 0 \}.
$
The polynomial truncated to this face is given by
\eq{
        p_F(x) = x_0 + 2 x_0^2 + x_0^3 
        \, .
}

Because each monomial coefficient of $p$ is positive,
we have that $p$ is completely non-vanishing on, 
say,
$X = \RR^2_{>0}$.
\end{ex}

Borinsky proved that there is really a sense in which $p\trop(x)$ \emph{approximates} the values of $p(x)$;
this is the backbone behind tropical integration.
Take a homogeneous polynomial $p \in \CC[x]$ that is completely non-vanishing on 
positive projective space $\PP^{n-1}_+$.
Then there exist two constants $C_1, C_2 > 0$ such that \cite[Theorem 8]{Borinsky:2020rqs}
\eq{
        C_1 
        \ \leq \
        \frac{|p(x)|}{p \trop (x)}
        \ \leq \
        C_2
        \quadit{\text{for all}}
        x \in \PP^{n-1}_+
        \, .
        \label{tropical_approximation_property}
}
In other words,
the tropical approximation of $p$ bounds the original polynomial from above and from below.

We can now describe the intuition behind tropical integration 
(a precise algorithm is given in \secref{sec:tropical_sampling_algorithm}).
Consider a convergent integral
\eq{
        I 
        = 
        \int_{\PP^{n-1}_+} \frac{g(x)^a}{h(x)^b} \, \O
}
given by two homogeneous polynomials $g$ and $h$ that are completely non-vanishing on the integration domain,
and $a,b \in \QQ$.
We insert a clever way of writing the number "$1$" into the integrand:
\eq{
        \begin{array}{lllll}
        I 
        &=&
        \displaystyle
        \int_{\PP^{n-1}_+} \frac{g(x)^a}{h(x)^b} \, \O
        &\times&
        \displaystyle
        \left( \frac{g \trop(x)}{g \trop (x)} \right)^a
        \displaystyle
        \left( \frac{h \trop(x)}{h \trop (x)} \right)^b
        \vspace{0.3cm}
        \\
        &=&
        \displaystyle
        \underbrace{
                \int_{\PP^{n-1}_+} \frac{g\trop(x)^a}{h\trop(x)^b} \, \O
        }_{\propto \ \text{tropical measure} \ \mu\trop}
        &\times&
        \underbrace{
                \displaystyle
                \left( \frac{g(x)}{g \trop (x)} \right)^a
                \displaystyle
                \left( \frac{h\trop(x)}{h(x)} \right)^b
        }_{\text{bounded due to } \eqref{tropical_approximation_property} }
        \, .
        \end{array}
        \label{tropical_factorization}
}
The integrand has been factored into something resembling a probability measure
(after being suitably normalized)
multiplied by a function that is square-integrable as a consequence of being bounded.
This is now a suitable setup for MC integration:
we sample a large number of random points $x\supbrk{i}$ from $\mu \trop$,
and then take the average of the function $(g/g\trop)^a \, (h\trop/h)^b$ evaluated at these points.

        \section{Projective Feynman integrals}

As we will present a numerical code later on,
let us be explicit about our conventions for FIs.
Take $G$ to be a one-particle irreducible Feynman diagram with edge set $E$ and vertex set $V$.
With our index convention,
we have
$
        E = \{0,1,\ldots,|E|-1\}
$
and
$
        V = \{0,1,\ldots,|V|-1\}.
$
Every edge $e \in E$ is equipped with a mass $m_e$ and an edge weight $\nu_e$.
Every vertex $v \in V$ comes with an incoming momentum vector $p_v$.
Internal vertices are distinguished by having vanishing incoming momenta;
i.e.,
writing
$
        V = V_\ext \, \sqcup \, V_\Int,
$
then $p_v = 0$ if $v \in V_\Int$.

The \emph{incidence matrix} $\mE$ of $G$ is defined by choosing an arbitrary orientation for the edges,
and setting
$
        \mE_{v,e} = \pm 1
$
if edge $e$ points to/from vertex $v$ and
$
        \mE_{v,e} = 0
$
if $e$ is not incident to $v$.

In terms of this data,
we write the FI associated to $G$ as
\eq{
        I_G =
        \int
        \prod_{e \in E}
        \frac{\dd^\DD \ell_e}{i \pi^{\DD/2}}
        \left( \frac{-1}{\ell_e^2 - m_e^2 + i\vare} \right)^{\nu_e}
        \prod_{v \in V \setminus \{v_0\}} 
        i \pi^{\DD/2}
        \d \supbrk{\DD} \Big( p_v + \sum_{e \in E} \mE_{v,e} \, \ell_e \Big)
        \, .
        \label{momentum_space_integral_conventions}
}
The loop momenta $\ell_e$ are integrated over $\DD$-dimensional Minkowski space with mostly-minus signature,
$
        \ell_e^2 = 
        \left( \ell^0_e \right)^2 - \left( \ell^1_e \right)^2 - \left( \ell^2_e \right)^2 - \ldots,
$
and an overall momentum-conserving $\d$-function 
$
        \d(p_{v_0} + \ldots + p_{v_{|V|-1}})
$
has been extracted by removing a vertex $v_0 \in V$.

To evaluate $I_G$ numerically,
we use an equivalent parametric representation
(see \cite[Section 2.5]{Weinzierl:2022eaz} for a derivation starting from momentum space):
\eq{
        \label{projective_feynman_integral}
        I_G &= 
        \Gamma(\o) \int_{\PPS} \phi
        \qquad \text{with}
        \\[3pt]
        \phi &=
        \left( \prod_{e \in E} \frac{x_e^{\nu_e}}{\Gamma(\nu_e)} \right)
        \frac{1}{\mU(x)^{\DD/2}}
        \left( \frac{1}{\mV(x) - i\vare \sum_{e \in E} x_e} \right)^\o \O
        \, .
        \nonumber
}
This is an integral over positive projective space $\PPS$,
defined in \eqref{positive_projective_space},
w.r.t.~the volume form
\eq{
        \O =
        \sum_{e=0}^{|E|-1}
        \brk{-1}^{|E|-e-1} \,
        \frac
        { \dd x_0 \wedge \cdots \wedge \widehat{\dd x_e} \wedge \cdots \wedge \dd x_{|E|-1} }
        { x_1 \cdots \widehat{x}_e \cdots x_{|E|-1} } 
        \, ,
}
where hats denote omission.
The superficial degree of divergence of the graph $G$ is given by
$
        \o = \sum_{e \in E} \nu_e - \DD \cdot L / 2
$
where $L = |E|-|V|+1$ is the number of loops.

$\mU(x)$ is the first Symanzik polynomial,
previously defined in \eqref{U_poly}.
Recalling also the second Symanzik polynomial $\mF(x)$ from \eqref{F_poly},
we set 
\eq{
        \mV(x) = \frac{\mF(x)}{\mU(x)}
        \, .
}
Since
$
        \mU(\l x) = \l^L \mU(x)
$
and
$
        \mF(\l x) = \l^{L+1} \mF(x),
$
one can quickly verify that the integrand $\phi$ is invariant under the coordinate rescaling
$
        x \to \l x,
$
wherefore it is appropriate to call \eqref{projective_feynman_integral} a 
\emph{projective integral representation}.

The graph-theoretic formulas 
\eqref{U_poly} and \eqref{F_poly} for $\mU$ and $\mF$ 
are in fact not so efficient for numerically evaluating millions of points during MC integration.
The reason is that the number of monomials grows quickly as a function of $|E|$ and $L$,
making it slow to add up all terms during each numerical evaluation.
Let us therefore present more efficient formulas in terms of matrices 
that scale linearly with the complexity of $G$.
To start,
define the $(|V|-1) \times (|V|-1)$ \emph{reduced Laplacian matrix} component-wise by 
\eq{
        \mL_{uv} = \sum_{e \in E} \frac{\mE_{u,e} \, \mE_{v,e}}{x_e}
        \quadit{\text{for}}
        u,v \, \in \, V \setminus \{v_0\}
        \, .
        \label{laplacian_matrix}
}
As this matrix is symmetric,
it is only necessary to explicitly calculate the diagonal and the upper triangular block.
The lower triangular block can then be copied from the upper one.
It turns out that $\mL$ is positive definite.

Let 
$
        \mP_{uv} = p_u \cdot p_v
$
be the $(|V|-1) \times (|V|-1)$ matrix containing Minkowski scalar products between all independent momenta.

We then have the identities
\eq{
        \mU(x) &= 
        \det \big[ \mL(x) \big]
        \prod_{e \in E} x_e
        \\[2pt]
        \mF(x) &=
        \mU(x)
        \left[
                - \sum_{u,v \, \in \, V \setminus \{v_0\}} 
                \big( \mL^{-1}(x) \big)_{uv} \, \mP_{uv} 
                + \sum_{e \in E} m_e^2 \, x_e
        \right]
        \, .
        \label{U_F_formulas}
}
Note that the first sum in $\mF$ can be expressed as 
$
        \tr\left[ \mL^{-1} \cdot \mP \right].
$
Determinants and traces are computationally cheap,
but the matrix inverse $\mL^{-1}$ appears to be expensive at first glance.
However,
notice that we are only interested in the combination
$
        \mathsf{X} = \mL^{-1} \cdot \mP,
$
not $\mL^{-1}$ itself.
Owing to the fact that $\mL$ is positive definite,
then $\mathsf{X}$ can be swiftly evaluated using \emph{Cholesky decomposition}.
More precisely,
in $O(|V|^3)$ time,
it is possible to decompose
$
        \mL = 
        \mathsf{L} \cdot \mathsf{D} \cdot \mathsf{L}^T
$
in terms of a lower triangular matrix $\mathsf{L}$ and a diagonal matrix $\mathsf{D}$.
Afterwards,
the solution $\mathsf{X}$ to the linear system
$
        \mL \cdot \mathsf{X} = \mP
$
is immediately obtained by solving
\eq{
        \mathsf{L} \cdot \mathsf{Y}_1 = \mP 
        \, , \quad
        \mathsf{D} \cdot \mathsf{Y}_2 = \mathsf{Y}_1
        \, , \quad
        \mathsf{L}^T \cdot \mathsf{X} = \mathsf{Y}_2
        \label{cholesky_solve}
}
in cascade.

Later on,
we will actually require $\mL$ to take on complex values rather than real ones.
Then $\mL$ is no longer Hermitian,
wherefore Cholesky decomposition fails.
In this case,
one must use the less efficient LU (lower-upper) decomposition.

\subsection{\texorpdfstring{$i\vare$-prescription}{} via contour deformation}
\label{sec:ieps_deformation}

For the purposes of this thesis,
say that a FI computation is in the \emph{Euclidean regime} if
$
        \mF(x) \geq 0
$
for all
$
        x \in \PPS
$
(a more refined notion of the Euclidean regime is given in \cite[Section 2.2]{Borinsky:2023jdv}).
Then it is only the case that $\mF(x) = 0$ on the boundary $x \to 0$ of the integration domain,
which is assumed to be an integrable singularity.
In our convention,
being in the Euclidean regime is analogous to having a FI with scalar products computed in the
Euclidean all-minus metric.
Life is simple in the Euclidean regime because one avoids the singularities in 
\eqref{momentum_space_integral_conventions} coming from setting the time-components equal to
$
        \ell^0 = \pm \sqrt{\vec{\ell} \cdot \vec{\ell} + m^2}
$
in the propagators.
Consequently, 
Feynman's $i\vare$-prescription can be dropped,
and the resulting FI will be purely real.

Alas,
life is not so simple in Minkowski space.
Let us say that we are in the \emph{Minkowski regime} when we are not in the Euclidean regime.
In this case,
the monomial coefficients of $\mF(x)$ can be both positive and negative.
The $i\vare$ regulates singularities coming from having $\mF(x) = 0$ inside the integration domain.
In momentum space,
the $i\vare$ regulates the $\ell^0$-singularities by pushing them slightly away from
the real axis.
An alternative description is to deform the integration contour given by
$
        \ell^0 \in (-\infty,\infty)
$
to one that goes around the poles%
\footnote{
        With two poles there are four possible choices for how to go around them,
        but only one is consistent with the notion of causality.
        This contour corresponds to a propagator
        $\ell^2 - m^2 + i\vare$.
        Two non-casual examples are
        $(\ell^0+i\vare)^2 - \vec{\ell} \cdot \vec{\ell} - m^2$
        and
        $\ell^2 - m^2 - i\vare$.
        The correct contour goes
        below the singularity at
        $
                -\sqrt{\vec{\ell} \cdot \vec{\ell} + m^2}
        $
        and above the one at
        $
                +\sqrt{\vec{\ell} \cdot \vec{\ell} + m^2}.
        $
        I.e.
        \includegraphics[scale=0.05]{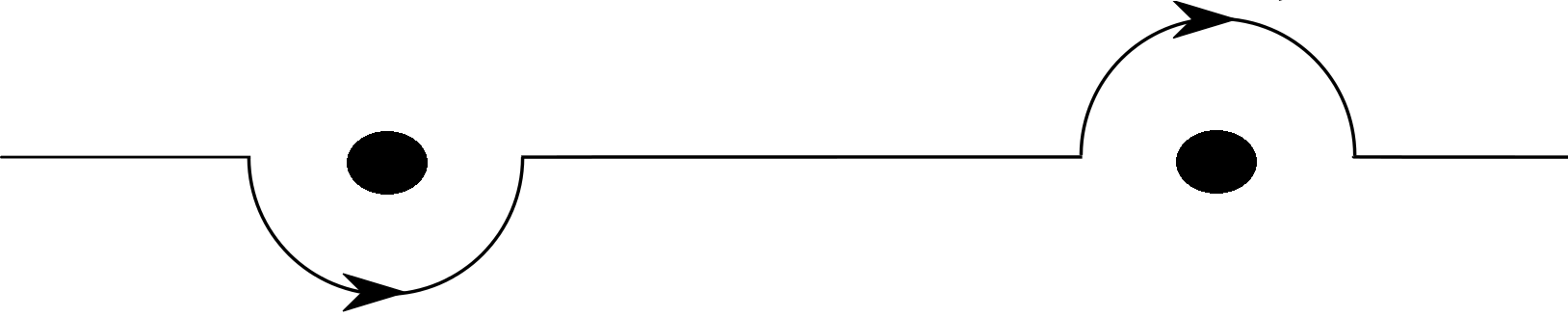}
        in the complex $\ell^0$-plane.
}.

The $i\vare$-prescription only dictates on which side of the poles we integrate.
But for the purpose of numerical integration we must make an \emph{explicit} choice for the contour,
preferably with decent numerical stability.
Moreover,
the deformation ought to preserve the projective symmetry of the integral
in order to play well with tropical MC sampling.
We find that the following deformation into the complex plane proposed in 
\cite{Mizera:2021icv,Hannesdottir:2022bmo} 
fulfils both of these criteria:
\begin{empheq}[box=\fbox]{align}
        X_e(x) = x_e \exp \left[ -i \l \frac{\p \mV(x)}{\p x_e} \right]
        \quadit{\text{for}}
        \l \in \RR_{>0}
        \, .
        \label{contour_deformation}
\end{empheq}
Because $\mU$ and $\mF$ are homogeneous polynomials of degrees $L$ and $L+1$ respectively,
then $\mV = \mF / \mU$ is a rational function of degree $1$,
wherefore $\p_e \mV = \p \mV(x) / \p x_e$ has homogeneous degree $0$.
Thus,
\eqref{contour_deformation} preserves projective symmetry.

The value of parameter $\l$ is finite but small.
This is in contrast to the infinitesimal $\vare$ of the $i\vare$-prescription.
As long as $\l$ is is small enough that no poles of the integrand $\phi$ are hit,
then Cauchy's theorem guarantees that the deformation is valid.
Note that $\l$ must have a mass dimension in order to cancel that of $\p_e \mV$ in the exponential.
In practice,
we shall choose $\l$ to be of order $\mO(1 / \L^2)$,
where $\L^2$ is the largest kinematic scale in $\mV$.

To see why \eqref{contour_deformation} is consistent with the $i\vare$-prescription,
let us consider the \emph{Schwinger representation} of a FI \cite[Section 2.5.2]{Weinzierl:2022eaz}:
\eq{
        I = 
        \int_0^\infty 
        \ldots \exp \left[ i \left( - \mV(x) + i\vare \sum_{e \in E} x_e \right) \right]
        \, .
}
Here we ignored factors in the integrand which are irrelevant for the present argument.
Observe that the $i\vare$ term cures the divergence at $x_e \to \infty$ by gifting an exponential damping
$
        \sim 
        \exp\left[- \vare x_e \right].
$
Notice that a similar damping is achieved by adding a 
\emph{positive} imaginary part to the first term $-\mV(x)$.
In that case the $i\vare$ could be dropped.
Indeed,
changing coordinates according to \eqref{contour_deformation},
then a Taylor expansion in $\l$ shows that
\eq{
        \label{taylor_in_lambda}
        - \mV(X) = 
        - \mV(x) + i \l \sum_{e \in E} x_e \left[ \frac{\p \mV(x)}{\p x_e} \right]^2 
        + \mO(\l^2)
        \, .
}
Assuming that 1) there are no solutions to the \emph{Landau equations}
\eq{
       x_e \frac{\p \mV(x)}{\p x_e} = 0
       \quadit{\text{for each}}
       e \in E
       \quadit{\text{and for any}}
       x \in \PPS
       \, ,
}
and 2) $\l$ is small enough such that the $\mO(\l^2)$ terms to not change the sign
of the $\mO(\l)$ term,
then \eqref{taylor_in_lambda} has a positive imaginary part,
as desired.
For the remaining part of this thesis,
it is assumed that the Landau equations have no solutions.

A short calculation shows that the $|E| \times |E|$
Jacobian matrix associated to \eqref{contour_deformation} is
\eq{
        (\mJ_\l(x))_{eh} =
        \d_{eh} - i \l x_e \frac{\p^2 \mV(x)}{\p x_e \p x_h}
        \quadit{\text{for}}
        e,h \in E
        \, .
        \label{tropical_jacobian}
}
Coming back to the projective FI \eqref{projective_feynman_integral},
its deformed version now reads
\eq{
        I_G =
        \Gamma(\o) \int_{\PPS}
        \left( \prod_{e \in E} \frac{X_e^{\nu_e}}{\Gamma(\nu_e)} \right)
        \frac
        {\det \mJ_\l(x)}
        {\mU(X)^{\DD/2} \ \times \ \mV(X)^\o} \,
        \O
        \, ,
        \label{deformed_feynman_integral}
}
where
$
        X = (X_0, \ldots, X_{|E|-1})
$
with $X_e$ given by \eqref{contour_deformation}.

\subsection{\texorpdfstring{$\e$-expansion}{} for quasi-finite integrals}
\label{sec:quasi_finite}

We have so far not made any assumptions regarding the finiteness of the FI
\eqref{deformed_feynman_integral}
as the DR parameter $\e$ tends to zero.
Recall that $\e$ appears inside
$
        \DD = \DD_0 - 2\e
$
and 
$
        \o = \sum_{e \in E} \nu_e - \DD L /2
$ 
in this representation.

We shall restrict our attention to the category of integrals dubbed as 
\emph{quasi-finite} in \cite{vonManteuffel:2014qoa}.
This means that $I_G$ in \eqref{deformed_feynman_integral} can have overall $\e$-poles coming
from the prefactor $\Gamma(\o)$,
but the integral $\int_{\PPS} (\ldots)$ is itself finite as $\e \to 0$.
Note that this is less restrictive than requiring the whole FI to be finite.

Given this,
the $\e$-expansion of $I_G$ is derived by merely Taylor expanding the integrand.
Writing
$
        \o_0 = \sum_{e \in E} \nu_e - \DD_0 L /2,
$
the expansion becomes
\eq{
        \label{deformed_eps_expansion}
        I_G &= 
        \Gamma(\o_0 + \e L)
        \sum_{k=0}^\infty
        \frac{\e^k}{k!} \ \times
        \\ &
        \int_{\PPS}
        \left( \prod_{e \in E} \frac{X_e^{\nu_e}}{\Gamma(\nu_e)} \right)
        \frac
        {\det \mJ_\l(x)}
        {\mU(X)^{\DD_0/2} \ \times \ \mV(X)^{\o_0}}
        \log^k \left[ \frac{\mU(X)}{\mV(X)^L} \right]
        \O
        \, .
        \nonumber
}
If the integral is finite for $k=0$,
then the $\log^k$ factors cannot spoil the convergence for higher orders in $k$.

Our ultimate goal is to numerically evaluate the integrals appearing in the $\e$-expansion above.
To this end,
it is important that the tropical approximation property \eqref{tropical_approximation_property} holds true,
but this is now complicated by the fact that $X$ depends on $\l$.
So we make the following \emph{assumption}:
there exist $\l$-dependent constants $C_1(\l), C_2(\l) > 0$ such that for small enough $\l > 0$,
\eq{
        C_1(\l) 
        \ \leq \
        \left|
                \left( \frac{\mU \trop (x)}{\mU(X)} \right)^{\DD_0/2}
                \left( \frac{\mV \trop (x)}{\mV(X)} \right)^{\o_0}
        \right|
        \ \leq \
        C_2(\l) 
        \quadit{\text{for all}}
        x \in \PPS
        \, .
        \label{tropical_approximation_property_lambda}
}
We have checked this assumption through extensive numerical testing,
but lack a mathematical argument for its validity.

The statement \eqref{tropical_approximation_property_lambda} suggest an optimal value of $\l$:
choose it such that the bounds are as tight as possible.
This is an interesting research direction,
as it would constitute the first \emph{canonical} contour deformation without any free parameters.

\subsection{Convergence of Euler integrals}
\label{sec:convergence_of_euler_integrals}

It was proven by Berkesch, Forsg{\aa}rd and Passare 
that a quasi-finite representation can always be found \cite[Theorem 2.4]{berkesch2014euler}.
The proof of this theorem gives a concrete prescription for factoring out $\e$-poles in front of
finite integrals,
but we have not yet implemented this protocol%
\footnote{
        Their method requires expanding out derivatives of $\mU$ and $\mV$,
        which can lead to a huge proliferation of terms if done symbolically.
        This problem is naturally avoided in the setting of numerical MC integration.
        Indeed,
        we shall provide efficient formulas in \secref{sec:evaluating_deformed}
        for derivatives of $\mU$ and $\mV$ that could be used in a future implementation.
}.

Assuming a quasi-finite basis of MIs is known,
one can alternatively employ a combination of dimensional recurrence relations and IBPs 
as in \cite{vonManteuffel:2014qoa} to write a given FI as linear combination of said basis.

To find a quasi-finite basis,
one ought to have test for finiteness of Euler integrals
(in contrast to the GKZ setting studied earlier in this thesis,
we are now allowing the monomial coefficients of the Euler integrands
to be fixed to special values).
Convergence theorems for Euler integrals were first discovered in the mathematics literature by
Berkesch, Forsg{\aa}rd, Nilsson and Passare 
\cite[Theorem 1]{nilsson2013mellin} \cite[Theorem 2.2]{berkesch2014euler}.
In the physics literature,
convergence criteria were later established by Panzer, Schabinger and von Manteuffel in 
\cite{vonManteuffel:2014qoa},
followed by works of Arkani-Hamed et al.~%
\cite[Claim 1]{Arkani-Hamed:2019mrd}
and Borinsky 
\cite[Theorem 3]{Borinsky:2020rqs}.

The most general, proven statement is roughly as follows.
Suppose 
$
        \{f_1(x), \ldots, f_l(x)\}
$
are completely non-vanishing polynomials on $\RR^n_{>0}$
(an equivalent statement can be formulated in projective space).
Let 
$
        s = (s_1, \ldots, s_l) \in \CC^l
$
and
$
        \nu = (\nu_1, \ldots, \nu_n) \in \CC^n
$
be complex parameters.
Assume $\Re{s_i} >0 $.
The integral
\eq{
        I =
        \int_{\RR^n_{>0}}
        \frac{x_1^{\nu_1} \cdots x_n^{\nu_n}}{f_1^{s_1} \cdots f_l^{s_l}}
        \frac{\dd x}{x}
}
is convergent iff%
\footnote{
        An additional, weak assumption is that the Minkowski sum of Newton polytopes
        $
                \mathbf{N}[f_1] + \ldots + \mathbf{N}[f_l]
        $
        should have dimension $n$.
}
\eq{
        \label{convergence_criteria}
        &\arr{c}{
                \Re{\nu_1} \\
                \vdots \\
                \Re{\nu_n} 
        }
        \ \in \
        \mathrm{int}(\mathbf{N}(s))
        \quadit{\text{where}}
        \\[6pt]
        & \mathbf{N}(s) = \Re{s_1} \cdot \mathbf{N}[f_1] + \cdots + \Re{s_l} \cdot \mathbf{N}[f_l]
        \, .
        \nonumber
}

\begin{itemize}
        \item
                The notation 
                $
                        s \cdot \mathbf{N}[f]
                $ 
                means that the Newton polytope associated to $f$ is rescaled by a constant 
                $s \in \RR_{\geq 0}$.
                When $\mathbf{N}[f]$ is determined by a collection half-spaces
                $
                        \{v_i \cdot x \leq c_i\},
                $
                then $s \cdot \mathbf{N}[f]$ has the effect of rescaling the $c$'s,
                that is
                $
                        \{v_i \cdot x \leq s \, c_i\}.
                $
        \item 
                The \emph{Minkowski sum} of two polytopes $\mathbf{N}[f_A]$ and $\mathbf{N}[f_B]$ is given by
                $
                        \mathbf{N}[f_A] + \mathbf{N}[f_B] =
                        \big\{ 
                                a + b \, \big| \, a \in \mathbf{N}[f_A] \ \text{and} \ b \in \mathbf{N}[f_B] 
                        \big\}.
                $
        \item 
                A point $x$ is in the \emph{interior} of $\mathbf{N}(s)$, 
                $x \in \mathrm{int}(\mathbf{N}(s))$,
                if there exists a small ball 
                (w.r.t.~the Euclidean metric) 
                centered at $x$ that is completely contained in $\mathbf{N}(s)$.
\end{itemize}
The condition \eqref{convergence_criteria} can be checked in practice as follows.
First compute the half-space representation of $\mathbf{N}(s)$
(for example with \package{sage} or \package{polymake}).
This gives a set of inequalities of the form
$
        \{v \cdot x \leq c\}
$
for the hyperplanes that cut out $\mathbf{N}(s)$.
For FIs we have $s_1 = \DD/2$ and $s_2 = \o$,
so these inequalities would depend parametrically on the DR parameter $\e$.
A vector 
$
        v = \left[ \Re{\nu_1}, \ldots, \Re{\nu_n} \right]^T
$ 
is then in the interior of $\mathbf{N}(s)$ if the inequalities are strict,
namely
$
        \{v \cdot x < c\}.
$

This suggests the following heuristic algorithm for determining a quasi-finite basis:
check whether the criterion \eqref{convergence_criteria} is fulfilled during a 
bottom-up search in the space $(\nu,\DD)$ of propagator powers and spacetime dimensions%
\footnote{
        The search space for $\nu$ is cut down drastically in case \emph{any} other basis is known,
        so as to pinpoint a list of relevant integral sectors.
}.

We remark that the alternative quasi-finite basis search 
algorithm from \cite{vonManteuffel:2014qoa} has been 
automated in \package{Reduze2} \cite{vonManteuffel:2012np} via the command \soft{find\_finite\_masters}.

        \section{Tropical Monte Carlo sampling, in detail}

We have already briefly explained the idea behind tropical MC sampling in \secref{sec:tropica_MC_brief}.
This algorithm is here described in detail.
The goal is to tropically integrate the terms in the $\e$-expansion \eqref{deformed_eps_expansion},
so let us factorize the integrals in a way analogous to \eqref{tropical_factorization}:
\eq{
        I_G =
        I\trop \
        \frac{\Gamma(\o_0 + \e L)}{\prod_{e \in E} \Gamma(\nu_e)} \
        \sum_{k=0}^\infty \frac{\e^k}{k!} I\supbrk{k}
        \, ,
}
where the $k$th order integral reads as
\eq{
        I\supbrk{k} = 
        \int_{\PPS}
        \frac
        {
                \det \mJ_\l(x)
                \ \times \
                \prod_{e \in E} (X_e / x_e)^{\nu_e}
        }
        {
                \big( \mU(X) / \mU\trop(x) \big)^{\DD_0/2}
                \ \times \
                \big( \mV(X) / \mV\trop(x) \big)^{\o_0}
        }
        \log^k \left[ \frac{\mU(X)}{\mV(X)^L} \right]
        \ \times \
        \mu\trop
        \, ,
        \label{tropical_factorization_U_V}
}
and the \emph{tropical probability measure} is
\eq{
        \mu\trop =
        \frac{1}{I\trop} \
        \frac
        {\prod_{e \in E} x_e^{\nu_e}}
        {\mU\trop(x)^{\DD_0/2} \ \times \ \mV\trop(x)^{\o_0}} \
        \O
        \, .
        \label{tropical_measure}
}
The normalization factor $I\trop$ ensures that
$
        \int_{\PPS} \mu\trop = 1
$
(a formula for $I\trop$ is given further down in \eqref{Itrop}).

The idea of tropical MC is to sample $N$ points 
$
        \{x\supbrk{1}, \ldots, x\supbrk{N}\}
$
from the measure $\mu\trop$,
then evaluate the integrand of \eqref{tropical_factorization_U_V} at those points,
and finally take the average of these evaluations.
Borinsky introduced two different algorithms that sample from $\mu\trop$ in \cite{Borinsky:2020rqs}.
The first algorithm \cite[Section 5]{Borinsky:2020rqs} 
works for any projective integral over rational functions.
A key step in the algorithm requires the triangulation of a certain \emph{normal fans}%
\footnote{
        A normal fan is essentially a collection of cones that
        can be associated to the normal directions of each face of a given polytope 
        \cite{ziegler2012lectures}.
}
associated to the Newton polytopes of $\mU$ and $\mF$.
The computation of such triangulations unfortunately 
becomes a bottleneck when a Feynman diagram has many edges.

The second algorithm \cite[Section 6]{Borinsky:2020rqs}
takes special features of $\mU$ and $\mF$ into account,
thereby avoiding the triangulation step.
Namely,
the Newton polytope of $\mU$ is always a so-called \emph{generalized permutahedron},
and the same holds true for the Newton polytope of $\mF$ given sufficiently generic kinematics.
The triangulation step of the previous algorithm gets traded for a much simpler preprocessing step,
involving the calculation of a certain recursive function called $J(\g)$ below.

\subsection{Generalized permutahedra}
\label{sec:GP}

\begin{figure}[t]
        \centering
        \includegraphics[scale=0.15]{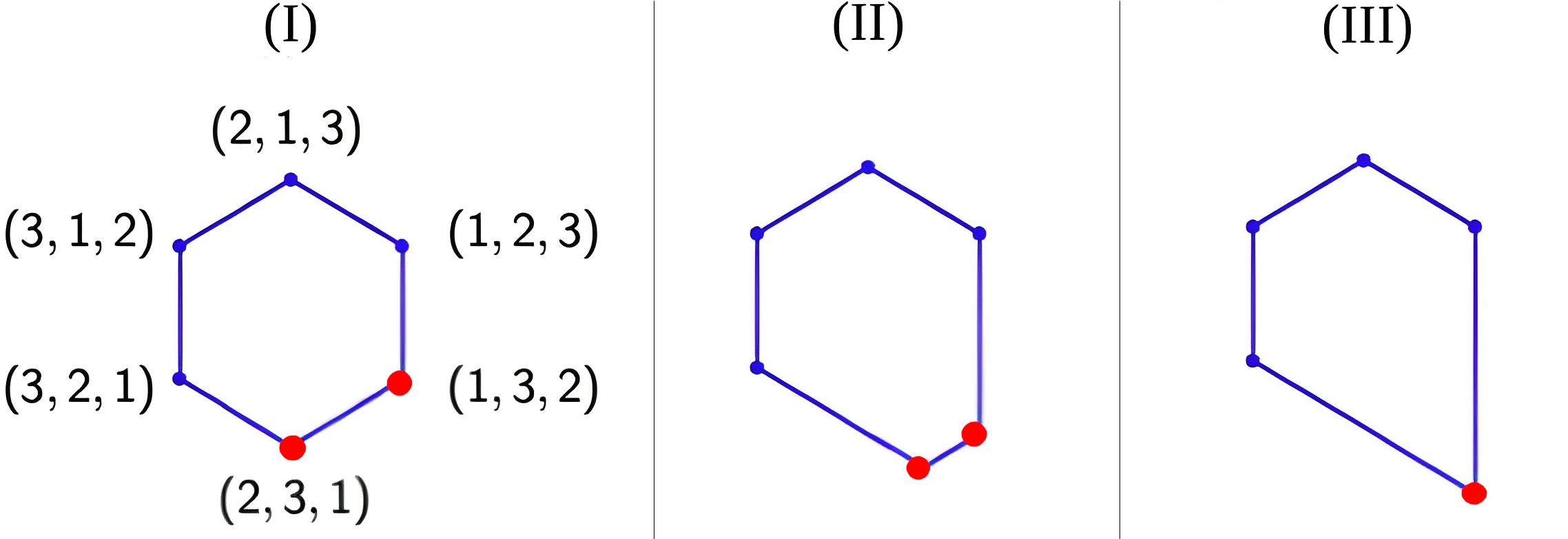}
        \caption{
                \textbf{(I)}
                The standard permutahedron $\Pi_3$. Each vertex is labeled by a permutation of $(1,2,3)$.
                Since the vector components $(x_1,x_2,x_3)$ of each vertex satisfies $x_1+x_2+x_3=6$,
                the dimension of the polytope drops from 3 to 2.
                \textbf{(II)}
                Deforming the two red vertices while preserving edge direction gives a GP.
                \textbf{(III)}
                Continuing to deform the vertices creates another GP.
                A facet has now degenerated to a point.
        }
        \label{fig:GP}
\end{figure}

Let $E$ be any finite set.
Consider a function%
\footnote{The notation $2^E$ stands for the power set of $E$.}
$
        z: 2^E \to \RR
$
that assigns a number to each subset $\g$ of $E$.
Given such a $z$,
the \emph{base polytope} $\mathbf{P}[z] \subset \RR^{|E|}$ is defined by the set of points
$
        (a_0, \ldots, a_{|E|-1}) \in \RR^{|E|}
$
satisfying
\eq{
        \sum_{e \in E} a_e &= z(E)
        \quadit{\text{and}}
        \\
        \sum_{e \in \g} a_e &\geq z(\g) 
        \quadit{\text{for all}}
        \g \subsetneq E
        \nonumber
        \, .
}
The function $z$ is called \emph{supermodular} if it satisfies the inequalities
\eq{
        z(\g) + z(\d)
        \ \leq \
        z(\g \, \cap \, \d) + z(\g \, \cup \, \d)
        \quadit{\text{for all pairs of subsets}}
        \g, \d \subset E
        \, .
        \label{supermodular}
}
Given a supermodular function $z$,
Aguiar and Ardila proved that the base polytope $\mathbf{P}[z]$ is always a 
\emph{generalized permutahedron} (GP) \cite[Theorem 12.3]{Aguiar2017HopfMA}
(as a shorthand, we shall often say that a polytope "has the GP property", or simply "is GP").
This is a wonderful class of polytopes defined by Postnikov in \cite{postnikov2009permutohedra}.
For the purposes of this discussion,
we shall simply \emph{define} a GP to a supermodular base polytope.
To build some intuition for these polytopes,
let us nevertheless give a geometric description of Postnikov's original construction.
To start,
define the \emph{standard permutahedron} by
\eq{
       \Pi_n =
       \soft{ConvexHull}\!
       \left[
                \text{permutations of} \ (1,\ldots,n)
       \right]
       \, \subset \, \RR^n
       \, .
}
In words,
we take the convex hull in $\RR^n$ of the vector
$
        \mathrm{v} = (1,\ldots,n)
$
and all its $n!$ permutations
$
        \mathrm{v}_\s = (\s(1), \ldots, \s(n)).
$
Equivalently,
each permutation of $\mathrm{v}$ corresponds to a vertex.
The dimension of $\Pi_n$ is $n-1$ rather than $n$
because its points are all constrained to lie on the hyperplane
$
        1 + 2 +\ldots + n = n(n+1)/2.
$
Postnikov then defines a GP as a special deformation of a standard permutahedron:
one is allowed to move any facet along its normal ray as long as it does not pass through any vertices.
Equivalently,
one can move any vertex as long as the directions are preserved.
Examples of generalized permutahedra are shown in \figref{fig:GP}.

\subsubsection{GP property of \texorpdfstring{$\mU$}{}}

Let us now see how this connects to FIs.
Take the set $E$ to be the edge set of a Feynman graph $G$.
A subset of $E$ corresponds to a subgraph $\g$ of $G$, 
so $z$ can be thought of as a collection of numbers for each subgraph of $G$.
The integer $L_\g$ denotes the number of loops of $\g$.
It was shown by Schultka \cite{Schultka:2018nrs} that the function
\eq{
        z_\mU(\g) = L_\g
        \label{z_U}
}
is supermodular,
and Newton polytope $\newtU$ of the $\mU$-polynomial equals the base polytope $\baseU$.
According to the theorem by Aguiar and Ardila,
this means that $\newtU$ is GP.

\subsubsection{GP property of \texorpdfstring{$\mF$}{}}

The situation is much more complicated for the $\mF$-polynomial.
Following Brown \cite{Brown:2015fyf},
call a subgraph $\g \subset E$ 
\emph{mass-momentum spanning} 
if the $\mF$-polynomial restricted to the cograph $G \setminus \g$ vanishes identically,
$\mF_{G \setminus \g} = 0$.
Since the $\mF$-polynomial contains all the kinematic scales ($s,t,m^2$ etc.) of a given FI,
an equivalent definition is to say that $\mF_{G \setminus \g}$ is scaleless.

We further define the notion of \emph{generic kinematics}.
Consider the $\mF$-polynomial of a 1-loop bubble graph with two different internal masses:
$
        \mF_\text{bubble}(x) = (p^2-m_0^2-m_1^2) x_0 x_1 + m_0^2 x_0^2 + m_1^2 x_1^2.
$
Notice that if the kinematics are tuned to enforce $p^2 - m_0^2 - m_1^2 = 0$,
then the first monomial drops out.
Special choices of kinematics can therefore change the Newton polytope $\newtF$,
as its vertices stem from the monomial exponents of $\mF$.
The question is now how to avoid this degeneration of $\newtF$ by
setting a criterion which prevents cancellation among momentum and mass terms in $\mF$.
To this end,
begin by recalling the scalar product matrix $\mP_{uv} = p_u \cdot p_v$,
where $u,v \in V$ are vertices of $G$.
Call a vertex $v$ \emph{internal} if $\mP_{uv} = 0$ for all $u \in V$ and \emph{external} otherwise.
We then say that a kinematic configuration is \emph{generic} if
\eq{
        \sum_{u,v \, \in \, V'} \mP_{uv}
        \ \neq \
        \sum_{e \, \in \, E'} m_e^2
}
for each proper subset of external vertices $V' \subsetneq V_\ext$ 
and each non-empty subset of edges $E' \subset E$.
When the kinematics are not generic,
they are called \emph{exceptional}.
For example,
kinematics are always generic in the Euclidean regime if 
1) $m_e > 0$ for all $e \in E$,
or 2) $\sum_{u,v \, \in \, V'} \mP_{uv} < 0$ for all $V' \subsetneq V_\ext$.
See \cite[Section 2.2]{Borinsky:2023jdv} for additional details.

Now,
assuming generic kinematics,
it was shown by Schultka \cite{Schultka:2018nrs}%
\footnote{
        Schultka technically proved this theorem with kinematics in the Euclidean regime.
        The same result also holds in the Minkowski regime
        because the Newton polytopes coincide in both regimes given the genericity assumption.
}
that the function
\eq{
        z_\mF =
        \begin{cases}
                L_\g +1 & \text{if $\g$ is mass-momentum spanning} 
                \\
                L_\g    & \text{else} \, .
        \end{cases}
        \label{z_F}
}
is supermodular,
and the Newton polytope $\newtF$ equals the base polytope $\baseF$.
Therefore,
for generic kinematics,
the Newton polytope $\newtF$ is GP.
There are reasons to expect that $\newtF$ is also GP even for 
exceptional kinematics in the Euclidean regime \cite[Conjecture 3.7]{Borinsky:2023jdv}.

\begin{figure}
\captionsetup[subfigure]{justification=centering}
        \centering
        \begin{subfigure}[t]{0.22\textwidth}
                \centering
                \includegraphics[width=\textwidth]{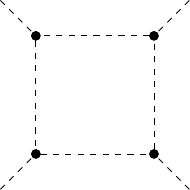}
                \caption{On-shell box. \\ Not GP.}
                \label{fig:1L_on_shell_box}
        \end{subfigure}
        \hspace{0.2cm}
        \begin{subfigure}[t]{0.22\textwidth}
                \centering
                \includegraphics[width=\textwidth]{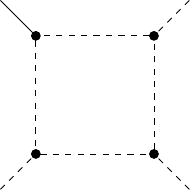}
                \caption{One off-shell leg. \\ Not GP.}
                \label{fig:1L_1leg_box}
        \end{subfigure}
        \hspace{0.2cm}
        \begin{subfigure}[t]{0.22\textwidth}
                \centering
                \includegraphics[width=\textwidth]{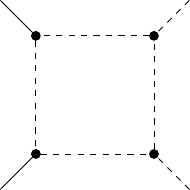}
                \caption{Two off-shell legs (adjacent). Not GP.}
                \label{fig:1L_2leg_adj_box}
        \end{subfigure}
        \hspace{0.2cm}
        \begin{subfigure}[t]{0.22\textwidth}
                \centering
                \includegraphics[width=\textwidth]{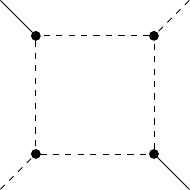}
                \caption{Two off-shell legs (crossed). Is GP.}
                \label{fig:1L_2leg_cross_box}
        \end{subfigure}
        \caption{
                1-loop box diagrams with massless internal propagators and different combinations 
                of on- or off-shell external legs.
                On-shell ($p^2 = 0$) legs are depicted by dashed lines,
                and off-shell ($p^2 \neq 0$) legs are drawn as solid lines.
                In all four cases we have $\mathbf{N}[\mF] = \mathbf{P}[z_\mF]$,
                but only \eqref{fig:1L_2leg_cross_box} is GP.
        }
        \label{fig:1L_boxes}
\end{figure}

It should be emphasized that $\newtF$ is generally \emph{not} GP outside of the Euclidean regime.
Indeed, 
the polytopes for three box diagrams in Figures
\eqref{fig:1L_on_shell_box}, \eqref{fig:1L_1leg_box} and \eqref{fig:1L_2leg_adj_box}
are \emph{not} GP.
The GP property \emph{is} fulfilled for the polytope associated to the diagram 
\eqref{fig:1L_2leg_cross_box} 
though.

The story is further complicated by an interesting observation of Arkani-Hamed, Hillman and Mizera 
\cite[Section IV]{Arkani-Hamed:2022cqe}:
even for exceptional kinematics,
the Newton polytope of $\mF$ is often equal to the base polytope $\baseF$
with $z_\mF$ given by \eqref{z_F}!
In fact,
for all four examples in \figref{fig:1L_boxes} we have that
$
        \newtF = \baseF.
$
This is expected for the polytope of the last graph \eqref{fig:1L_2leg_cross_box} since it is GP,
but how is this consistent with the fact that the first three cases
\eqref{fig:1L_on_shell_box}, \eqref{fig:1L_1leg_box} and \eqref{fig:1L_2leg_adj_box}
are not GP?
The answer is that $z_\mF$ is not supermodular for those graphs.
So the function $z_\mF$ does \emph{not} fulfil 
the inequalities \eqref{supermodular} for the first three boxes,
but it \emph{does} so for the last box.

While $\newtF = \baseF$ for generic kinematics,
we are not aware of a set of necessary and sufficient conditions that specify 
when $\newtF$ equals $\baseF$ for exceptional kinematics.
It is at least the case that $\newtF \subset \baseF$,
because we can only lose monomials in the exceptional case.
A counter example%
\footnote{We thank Erik Panzer for sharing this example with us.}
to equality is given by the following 1-loop triangle:
\eq{
       \mF_\text{triangle} =
       \mF \Big( \includegraphicsbox{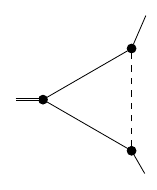} \Big) =
       m^2 (x_0^2 + x_1^2) + (2m^2 - p^2) x_0 x_1
       \, .
       \label{exceptional_triangle}
}
The dashed line is massless, 
the double line has off-shell momentum $p^2 \neq 0$,
and every single solid line has mass $m$.
This is an exceptional kinematic configuration.
The Newton polytope of $\mF_\text{triangle}$ is the convex hull of the points
$(2,0), (0,2)$ and $(1,1)$.
Since they all lie on a line,
the dimension of $\mathbf{N}[\mF_\text{triangle}]$ equals 1.
However,
the base polytope $\mathbf{P}[z_{\mF_\text{triangle}}]$ can be shown to have dimension 2.
Hence 
$
        \mathbf{P}[z_{\mF_\text{triangle}}] \neq \mathbf{N}[\mF_\text{triangle}].
$

Why is this discussion relevant for tropical MC sampling?
The answer is that the sampling algorithm of \ft assumes that $\newtF$ is GP.
Curiously,
although we have highlighted the complications with this assumption above,
it appears not to matter that much in practice.
Indeed, 
by evaluating the examples from this section with \ft
(i.e.~quasi-finite versions of the four boxes in \figref{fig:1L_boxes} and the triangle from 
\eqref{exceptional_triangle}), 
we find agreement with other numerical codes in all cases.
The catch is that numerical convergence for \ft becomes slower in the non-GP cases.
We suspect that \ft will fail for more complicated diagrams with exceptional kinematics,
but we did not yet find such an example.

\subsection{Tropical sampling algorithm}
\label{sec:tropical_sampling_algorithm}

\begin{table}[t]
        \centering
        \begin{tabular}{l|l|l} 
                    & Euclidean                  & Minkowski  \\ [1ex] 
        \hline
        Generic     & no deformation / always GP & deformation / always GP  \\ \hline
        Exceptional & no deformation / always GP & deformation / not always GP  \\ [1ex] 
        \end{tabular}
        \caption{
                Summary of whether contour deformation is required
                and whether the GP property of $\newtF$ is fulfilled for each kinematic regime.
        }
        \label{tab:regime}
\end{table}

\ft employs a modification of \cite[Algorithm 4]{Borinsky:2020rqs} to sample from the tropical measure
$\mu \trop$ given in \eqref{tropical_measure},
under the assumption that $\newtF$ is GP.
The algorithm includes a preprocessing step,
to be performed once,
followed by a sampling step.
The preprocessing step scales as $\mO(n 2^n)$,
where $n = |E|$ is the number of edges in a Feynman diagram.
This scaling is determined from the preparation of a certain 
table of rational numbers $J(\g)$ for each subgraph $\g \subset E$.
The sampling step scales as $\mO\left(n^2 \d^{-2}\right)$,
where $\d$ is the demanded relative accuracy of the result.
This is just ordinary MC scaling.

Because of the preprocessing step,
the time complexity of this algorithm is not fully polynomial.
It is a very interesting question whether a polynomial 
time algorithm exists for the numerical evaluation of FIs.

\subsubsection{Preprocessing}

The first task is to determine the kinematic regime,
namely Euclidean vs.~Minkowski and generic vs.~exceptional.

The regime is Euclidean if the scalar product matrix $P_{uv}$ is negative semi-definite,
and Minkowski otherwise.
If the kinematics are in the Minkowski regime,
then contour deformation needs to be switched on.

If the kinematic point is exceptional and Minkowski,
then $\newtF$ might not be GP,
and it might not even be equal to $\baseF$.
In this case,
\ft prints a warning saying that the integration might give the wrong result
(though, as mentioned, we have not yet witnessed erroneous numerics).
The program then bravely continues under the assumption that $\newtF = \baseF$.
For generic kinematics,
regardless of whether we are in the Euclidean or Minkowski regimes,
the sampling algorithm is guaranteed to converge to the correct result 
\cite[Proposition 31]{Borinsky:2020rqs}.

A summary of the preceding paragraphs is given in \tabref{tab:regime}.
For brevity,
we have left out a technicality in that table regarding the Euclidean regime:
there is in fact a larger regime, 
containing the Euclidean one, 
called the \emph{pseudo-Euclidean regime};
it does not require contour deformation,
but might witness a failure of the GP property for exceptional kinematics.
See \cite[Section 2.2]{Borinsky:2023jdv} for details.

The next task is to prepare data associated to each of the $2^{|E|}$ subgraphs $\g \subset E$:
\begin{enumerate}
        \item 
                We calculate the loop number 
                $
                        L_\g = |\g| - |V_\g| + 1,
                $
                where $V_\g$ is the vertex set associated to $\g$.
        \item 
                We check whether or not $\g$ is mass-momentum spanning,
                i.e.~whether $\mF_{G \setminus \g} = 0$.
                By definition,
                $\d_\g^\mathrm{m.m} = 1$ if $\g$ is mass-momentum spanning and 
                $\d_\g^\mathrm{m.m} = 0$ otherwise.
        \item 
                We calculate the \emph{generalized degree of divergence} \cite[Section 7.2]{Borinsky:2020rqs}
                \eq{
                        \o(\g) = 
                        \sum_{e \in \g} \nu_e - \DD_0 L_\g/2 - \o_0 \, \d_\g^\mathrm{m.m}
                        \, ,
                }
                where $\DD_0$ is the integer part of the spacetime dimension and
                $
                        \o_0 = \sum_{e \in E} \nu_e - \DD_0 L / 2.
                $
                If $\o(\g) \leq 0$ for any proper subgraph (i.e.~not the whole graph $\g = E$),
                then $G$ has a subdivergence.
                It implies that every term of the $\e$-expansion \eqref{deformed_eps_expansion} is divergent,
                wherefore \ft prints an error and terminates.
                In this case,
                one has to shift the integrand exponents $(\nu,\DD)$ to obtain a quasi-finite integral
                (cf.~\secref{sec:convergence_of_euler_integrals}).
        \item 
                If $\o(\g) > 0$ for all $\g \subset E$,
                then we \emph{recursively} compute a table of numbers $J(\g)$ via the formulas
                \eq{
                        \nonumber
                        J(\emptyset) &= 
                        \o(\emptyset) = 1 
                        \\[5pt]
                        J(\g) &= 
                        \sum_{e \in \g} \frac{J(\g \setminus e)}{\o(\g \setminus e)}
                        \quadit{\text{for all}}
                        \g \subset E
                        \, ,
                        \label{J_recursion}
                }
                where $\g \setminus e$ means that we delete the edge $e$ from $\g$.
                The last step of the recursion is for $\g = E$.
                In \cite[Proposition 29]{Borinsky:2020rqs}, 
                Borinsky proved the surprising identity
                \eq{
                        J(E) = I\trop
                        \label{Itrop}
                        \, ,
                }
                where $I\trop$ is the tropical normalization factor from \eqref{tropical_measure}.
                The rational numbers $J(\g)$ are intimately related to the values $z_\mF(\g)$
                \cite[Section 6.1]{Borinsky:2020rqs}.
                Indeed,
                this is the step during preprocessing wherein the GP assumption is employed.
\end{enumerate}
The data $(L_\g, \, \d_\g^\mathrm{m.m})$ feeds into the computation of $z_\mF(\g)$ via \eqref{z_F}.
For an exceptional kinematic point in the Minkowski regime,
\ft checks whether $z_\mF$ is supermodular by testing a 
(more efficient)
variant of the inequalities \eqref{supermodular},
see \cite[Equation (26)]{Borinsky:2023jdv}.
\ft prints a message if the inequalities are satisfied to notify that $\baseF$ is GP,
and a warning otherwise.

The output of this preprocessing is the data
$
        \bigcup_{\g \subset E} 
        \big\{ 
                L_\g , \, \d_\g^\mathrm{m.m} , \, \o(\g) , \, J(\g) \big\} \bigcup \big\{ I\trop 
        \big\},
$
which is stored in computer memory.

\subsubsection{MC sampling}

The validity of this tropical sampling algorithm was proven in 
\cite[Proposition 31]{Borinsky:2020rqs}.

\begin{algorithm}[H]
\vspace{0.2cm}
Initialize the variables $\g = E$ and $\kappa = U = 1$.
\vspace{0.2cm}
\begin{algorithmic}[1]
\While{$\g \neq \emptyset$}
\vspace{0.1cm}
\State{Pick a random edge $e \in \g$ with probability $$p_e^\g = \frac{1}{J(\g)} \frac{J(\g\setminus e)}{\o(\g\setminus e)}.$$}
\vspace{0.1cm}
\State{Set $x_e = \kappa$.}
\vspace{0.1cm}
\State{If $\g$ is mass-momentum spanning but $\g\setminus e$ is not, set $V = x_e$.}
\vspace{0.1cm}
\State{If $L_{\g\setminus e} < L_\g$, multiply $U$ with $x_e$ and store the result in $U$, i.e.~set $U \gets x_e \cdot U $.}
\vspace{0.1cm}
\State{Remove the edge $e$ from $\g$, i.e.~set $\g \gets \g \setminus e$.}
\vspace{0.1cm}
\State{Pick a uniformly distributed random number $\xi \in [0,1]$.}
\vspace{0.1cm}
\State{Multiply $\kappa$ with $\xi^{1/\o(\g)}$ and store the result in $\kappa$, i.e.~set $\kappa \gets \kappa \, \xi^{1/\o(\g)}$.}
\vspace{0.1cm}
\EndWhile
\vspace{0.2cm}
\end{algorithmic}
\textbf{return} $x = [x_0,\ldots,x_{|E|-1}] \in \PPS$, $\mU\trop(x) = U$ and $\mV\trop(x) = V$.
\caption{Generating a sample $x$ distributed as $\mu\trop$ from \eqref{tropical_measure}}
\label{alg:tropical_sampling}
\end{algorithm}

\noindent
The main idea of this algorithm is to interpret $p_e^\g$ as a probability distribution 
over the edges $e$ of a given subgraph $\g$.
Indeed,
it follows from the definition of $J(\g)$ in \eqref{J_recursion} that
$
        p_e^\g \geq 0
$
and
$
        \sum_{e \in \g} p_e^\g = 1.
$
Starting from $\g = E$,
the algorithm proceeds to \emph{cut} an edge $e$ of the graph $G$ with probability $p_e^\g$.
This yields a new graph $\g \setminus e$ with the edge removed.
The cutting process repeats until all edges are gone.

We refer to \cite[Section 6.1]{Borinsky:2020rqs} for explanations on why \algref{alg:tropical_sampling}
yields correct values for the triple $\big\{x, \, \mU\trop(x), \, \mV\trop(x)\big\}$.

\subsection{Evaluating deformed Feynman integrands}
\label{sec:evaluating_deformed}

\algref{alg:tropical_sampling} explains how to efficiently sample points $x$ from $\mu\trop$,
but it does not take care of all the extra factors in the integrand of \eqref{deformed_eps_expansion}
that arose due to contour deformation.
In particular,
we would like to efficiently evaluate the factors
$
        X_e = x_e \exp \left[ -i \l \p_e \mV(x) \right]
        , \,
        \mU(X)
        , \,
        \mV(X)
$
and $\mJ_\l(x)$ for any $ x \in \PPS$.

The non-trivial step in evaluating $X_e$ is the derivative $\p_e \mV(x)$.
Moreover,
the Jacobian $\mJ_\l(x)$ includes the second-order derivative 
$
        \p_e \p_h \mV(x)
$
according to \eqref{tropical_jacobian}.
For the sake of efficiency,
it would be nice to have formulas for these derivatives based only on linear algebra.
To this end,
begin by defining the $(|V|-1) \times (|V|-1)$ matrix
\eq{
        \mM(x) = \mL^{-1}(x) \cdot \mP \cdot \mL^{-1}(x)
        \, ,
}
where the Laplacian matrix $\mL(x)$ and the scalar product matrix $\mP$ come from \eqref{U_F_formulas}.
To avoid confusion,
we clarify that $\mL^{-1}_{uv}$ means the $(u,v)$th component of the inverse matrix $\mL^{-1}$.
Let $e$ and $h$ be two edges that respectively connect the vertices
$
        (u_e,v_e) \text{ and } (u_h,v_h).
$
Furthermore,
define two matrices component-wise by
\eq{
        \begin{array}{llllllllll}
                \mA_{eh}(x) &=&
                \dfrac{1}{x_e x_h}
                &\Big[
                        \mM_{u_e \, u_h}(x) &+&
                        \mM_{v_e \, v_h}(x) &-&
                        \mM_{u_e \, v_h}(x) &-&
                        \mM_{v_e \, u_h}(x) 
                \Big]
        \\[9pt]
                \mB_{eh}(x) &=&
                \dfrac{1}{x_e x_h}
                &\Big[
                        \mL^{-1}_{u_e \, u_h}(x) &+&
                        \mL^{-1}_{v_e \, v_h}(x) &-&
                        \mL^{-1}_{u_e \, v_h}(x) &-&
                        \mL^{-1}_{v_e \, u_h}(x) 
                \hspace{0.2cm} \Big]
                \, ,
        \end{array}
        \nonumber
}
with the convention that 
$
        \mL^{-1}_{uv}(x) = \mM_{uv}(x) = 0
$
if either $u$ or $v$ equal the vertex $v_0$ that was removed in the FI
\eqref{momentum_space_integral_conventions}
by momentum conservation.
Using the identity
\eq{
        \frac{\p}{\p x_e} \mL^{-1}_{uv}(x) =
        \left(
                \mL^{-1}(x) \cdot \frac{\p \mL(x)}{\p x_e} \cdot \mL^{-1}(x)
        \right)_{uv}
        \, ,
}
a tedious but straightforward calculation gives the sought after derivative formulas
\eq{
        \label{1st_order_V_derivative}
        \frac{\p \mV(x)}{\p x_e} &= 
        - \mA_{ee}(x) + m_e^2
        \\[9pt]
        \frac{\p^2 \mV(x)}{\p x_e \p x_h} &=
        \frac{2 \d_{eh} \mA_{ee}(x)}{x_e} -
        2 \big( \mA(x) \circ \mB(x) \big)_{eh}
        \, ,
        \label{2nd_order_V_derivative}
}
where the last term denotes the element-wise matrix product (also called the Hadamard product):
$
        \big( \mA(x) \circ \mB(x) \big)_{eh} = \mA(x)_{eh} \times \mB(x)_{eh}.
$

\subsubsection*{Summary}

Let us summarize all the non-trivial steps that must be carried out to evaluate
the integrand of \eqref{deformed_eps_expansion} at a single point.
Cholesky decomposition is used to evaluate determinants and inverses whenever possible,
that is when the matrix in question is Hermitian and positive semi-definite
(see the discussion around equation \eqref{cholesky_solve}).
When those two conditions are invalid,
the less efficient LU decomposition is opted for instead.
\begin{enumerate}
        \item
                Use \algref{alg:tropical_sampling} to output the triple
                $
                        \big\{x, \, \mU\trop(x), \, \mV\trop(x) \big\}.
                $
        \item 
                Compute the Laplacian matrix $\mL(x)$ via \eqref{laplacian_matrix}.
        \item 
                Compute the inverse $\mL^{-1}(x)$ via Cholesky decomposition.
        \item 
                Use $\mL(x)$ and $\mL^{-1}(x)$ to evaluate derivatives of $\mV(x)$ using
                \eqref{1st_order_V_derivative} and \eqref{2nd_order_V_derivative}.
        \item 
                Compute the deformed $X$ parameters by
                $
                        X_e = x_e \exp \left[ -i \l \p_e \mV(x) \right].
                $
        \item 
                Compute the Jacobian $\mJ_\l(x)$ using \eqref{tropical_jacobian}.
        \item 
                Evaluate $\det[\mJ_\l(x)]$ using LU decomposition.
        \item 
                Compute the deformed Laplacian $\mL(X)$.
        \item 
                Compute $\mL^{-1}(X)$ and $\det[\mL(X)]$ using LU decomposition.
        \item 
                Employ the formulas \eqref{U_F_formulas} to evaluate
                $
                        \mU(X), \, \mF(X) \text{ and } \mV(X) = \mF(X) / \mU(X).
                $
\end{enumerate}

\noindent
For a kinematic point in the Euclidean regime we are allowed to set $\l = 0$ implying $X = x$,
in which case \ft only executes step 1.

        \chapter{The Program \package{feyntrop}}
\label{ch:feyntrop}

We have implemented the contour-deformed tropical integration algorithm described in the previous chapter
in a \package{C++} module named \ft\!\!.
The module is equipped with a \package{python} interface.
The \package{C++} codebase is an upgrade to the one developed by Borinsky in \cite{Borinsky:2020rqs}.
\ft relies on \package{Eigen3} \cite{eigenweb} for optimized linear algebra routines,
on \package{OpenMP} \cite{chandra2001parallel} for parallelization of MC sampling,
and on \package{xoshiro256+} \cite{blackman2021scrambled} for pseudorandom number generation.

\ft has been verified against 
\package{AMFlow} \cite{Liu:2022chg} and \package{pySecDec} \cite{Borowka:2017idc}
for roughly 15 different 2- to 5-point Feynman diagrams between 1 and 3 loops at varying kinematic points,
both in the Euclidean and Minkowski regimes.
Numerical agreement was found in all cases within the given uncertainty bounds.
In the Euclidean regime,
the code was checked up to $\mO(10)$ loop-orders in \cite{Borinsky:2020rqs} against analytic results
for conformal 4-point functions from $\varphi^4$-theory.

In this chapter we explain how to use the code,
and thereafter present state-of-the-art examples of numerical evaluations of FIs.

        \section{Basic usage of \ft}

Source code for \ft is available in the \package{github} repository
$$
        \href{https://github.com/michibo/feyntrop}{\soft{https://github.com/michibo/feyntrop}}
        \, .
$$
To download and install it,
the following commands can be run in a Linux terminal:
\begin{lstlisting}[style=mystyle]
git clone https://github.com/michibo/feyntrop.git
cd feyntrop
make clean && make
\end{lstlisting}
(It should also work in a \package{macOS} environment,
although \package{C++} compiler issues may ensue.
See the \soft{README.txt} document on \package{github}).

\ft is loaded in a \package{python} environment by importing the file \soft{py\_feyntrop.py}
located in the top directory.
To test whether the installation was successful,
the commands
\begin{lstlisting}[style=mystyle]
cd tests
python test_suite.py
\end{lstlisting}
will compute six examples with 1-2 loops and 2-5 legs and compare the results with precomputed values.

The \package{python} interface serves three purposes:
\begin{enumerate}
        \item 
                It simplifies the specification of vertices $V$ and edges $E$ of a Feynman diagram,
                in comparison with the more verbose data structures of \package{C++}.
        \item 
                It allows for self-chosen momentum variables given by a list of replacement rules,
                rather than having to manually input the full matrix of scalar products 
                $
                        \mP_{uv}
                $
                from \eqref{U_F_formulas}.
        \item 
                The numerical results for the $\e$-expansion are printed in an easily readable format.
\end{enumerate}

\noindent
It is also possible to directly interface with the \package{C++} module via \soft{.json} files.
See the \soft{README.txt} document in the \package{github} repository for more details on this.

\subsection{Tutorial}

\begin{figure}[!ht]
        \centering
        \includegraphicsbox{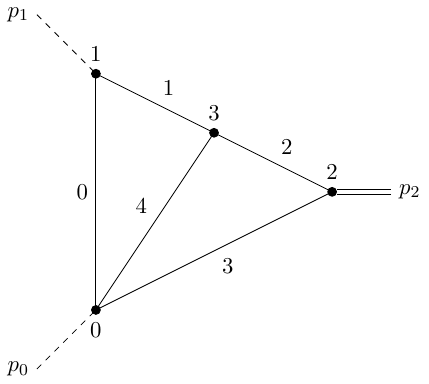}
        \caption{2-loop 3-point function in the \ft tutorial.}
        \label{fig:feyntrop_tutorial}
\end{figure}

The basic workflow of \ft is illustrated here by means of an example.
The code can be inspected with \package{jupyter} \cite{Kluyver2016jupyter} by calling
\begin{lstlisting}[style=mystyle]
jupyter notebook tutorial_2L_3pt.ipynb
\end{lstlisting}
in the top directory of the \ft package.
We will evaluate the 2-loop 3-point function shown in 
\figref{fig:feyntrop_tutorial}
in $\DD = 2 - 2\e$ dimensions.
The dashed lines represent massless on-shell external momenta:
$p_0^2 = p_1^2 = 0$.
The solid,
internal lines each have mass $m$.
The double line denotes an off-shell external momentum $p_2^2 \neq 0$.
The four vertices 
$
        V = V_\ext \sqcup V_\Int = \{0,1,2\} \sqcup \{3\}
$
and four edges
$
        E = \{0,1,2,3\}
$
have been explicitly labeled.
The convention of \ft requires one to label external vertices \emph{before} internal ones.

By the conventions of \eqref{momentum_space_integral_conventions},
we have the following momentum space representation with edge weights $\nu_0 = \ldots = \nu_4 = 1$:
\eq{
       \nonumber
       \pi^{-2+2\e} \int
       \frac
       {\dd^{2-2\e} \ell_0 \, \dd^{2-2\e} \ell_1 }
       { 
                (q_0^2 - m^2 + i\vare) 
                (q_1^2 - m^2 + i\vare) 
                (q_2^2 - m^2 + i\vare) 
                (q_3^2 - m^2 + i\vare) 
                (q_4^2 - m^2 + i\vare) 
       }
       \, ,
}
where the $\d\supbrk{D}$-functions of 
\eqref{momentum_space_integral_conventions} 
have been integrated out by setting
\eq{
       q_0 = \ell_0
       \, , \quad
       q_1 = \ell_0 + p_1
       \, , \quad
       q_2 = \ell_0 + \ell_1 + p_1
       \, , \quad
       q_3 = p_0 - \ell_0 - \ell_1
       \, , \quad
       q_4 = \ell_1
       \, . 
}
We evaluate the integral at the phase space point
\eq{
        m^2 = 0.2
       \, , \quad
       p_0^2 = p_1^2 = 0
       \, , \quad
       p_2^2 = 1
       \, ,
}
which is in the Minkowski regime given that $p_2^2 > 0$.

The \ft computation begins by importing \soft{py\_feyntrop.py} in a \package{python} environment:
\begin{lstlisting}[style=mystyle]
from py_feyntrop import *
\end{lstlisting}
Here it is assumed that the files 
\soft{feyntrop.so} and \soft{py\_feyntrop.py} 
are both in the working directory.
The graph in \figref{fig:feyntrop_tutorial} is defined by providing a list of edges,
edge weights $\nu_e$ and squared masses $m_e^2$ in the format
\eq{
        \Big[
        \big
                (\big( u_0, v_0 \big)
                \, , \, 
                \nu_0
                \, , \, 
                m_0^2
        \big)
        \, , \, \ldots \, , \,
        \big(
                \big( u_{|E|-1}, v_{|E|-1} \big)
                \, , \, 
                \nu_{|E|-1}
                \, , \, 
                m_{|E|-1}^2
        \big)
        \Big]
        \, .
}
The notation $(u_e, \, v_e)$ means that there is an edge $e$ connecting the vertices $u_e$ and $v_e$.
The vertex ordering $(u_e, \, v_e)$ is insignificant.
For the example at hand,
we thus write
\begin{lstlisting}[style=mystyle]
edges =  [((0,1), 1, 'mm'), ((1,3), 1, 'mm'), ((2,3), 1, 'mm'), 
          ((2,0), 1, 'mm'), ((0,3), 1, 'mm')]
\end{lstlisting}
The edge weights were here set to $\nu_e = 1$ for all $e$.
The chosen symbol \soft{mm} stands for $m^2$,
and will be replaced by its value $0.2$ later on%
\footnote{
        It is also allowed to input numerical values for the masses already in the \soft{edges} list.
        The first entry,
        for instance,  
        would then be \soft{((0,1), 1, '0.2')}.
}.

The next step is to fix the momentum variables.
By the zero-indexing convention,
we have external momenta
$
        \big\{p_0, \ldots, p_{|V_\ext|-1} \big\}.
$
But the last momentum $p_{|V_\ext|-1}$ is automatically inferred by \ft using momentum conservation,
leaving 
$
        \big\{p_0, \ldots, p_{|V_\ext|-2} \big\}
$
to be fixed by the user.
A momentum configuration is then specified by providing a collection of scalar products
\eq{
        p_u \cdot p_v
        \quadit{\text{for all}}
        0 \leq u \leq v \leq |V_\ext| - 2
        \, .
}
In the code,
we must give a replacement rule for each scalar product in terms of some variables of choice.
In the present example there are $|V_\ext|=3$ external vertices,
so we ought to provide replacement rules for
$
        \{p_0^2, p_1^2, p_0 \cdot p_1\} .
$
We have $p_0^2 = p_1^2 = 0$,
and by momentum conservation,
\eq{
        p_2^2 = (-p_0-p_1)^2 = 2 p_0 \cdot p_1 \implies p_0 \cdot p_1 = p_2^2/2
        \, .
}
In the syntax of \ft we then write
\begin{lstlisting}[style=mystyle]
replacement_rules = [(sp[0,0], '0'), (sp[1,1], '0'), (sp[0,1], 'pp2/2')]
\end{lstlisting}
The \soft{s}calar \soft{p}roduct between $p_u$ and $p_v$ is represented by the symbol \soft{sp[u,v]},
and \soft{pp2} corresponds to $p_2^2$.
Since we will eventually substitute values for the two parameters \soft{pp2} and \soft{mm},
we further define
\begin{lstlisting}[style=mystyle]
phase_space_point = [('mm', 0.2), ('pp2', 1)]
\end{lstlisting}
which fixes $m^2 = 0.2$ and $p_2^2 = 1$.

The final data to be provided are
\begin{lstlisting}[style=mystyle]
D0 = 2
eps_order = 5
Lambda = 7.6
N = int(1e7)
\end{lstlisting}
\soft{D0} is the integer part of $\DD = \DD_0 - 2\e$.
We expand up to,
but not including,
\soft{eps\_order}.
\soft{Lambda} is the contour deformation parameter in \eqref{contour_deformation}.
\soft{N} denotes the number of MC sampling points,
so increasing this value will yield better numerical accuracy
(the error on MC sampling scales as $\mO(1/\sqrt{\soft{N}})$ according to \eqref{sampling_variance}).

Tropical Monte Carlo integration is finally performed by calling
\begin{lstlisting}[style=mystyle]
trop_res, Itr = tropical_integration( 
        N, 
        D0, 
        Lambda, 
        eps_order, 
        edges, 
        replacement_rules, 
        phase_space_point)
\end{lstlisting}
If no error is printed,
then \soft{trop\_res} will contain the $\e$-expansion \eqref{deformed_eps_expansion}
\emph{without} the prefactor
$
        \Gamma(\o) / \big( \Gamma(\nu_0) \cdots \Gamma(\nu_{|E|-1}) \big)
        =
        \Gamma(2\e + 3).
$
\soft{Itr} is the value of the tropical normalization factor from \eqref{Itrop}.
Running this code on a laptop gives the output
\begin{lstlisting}[style=mystyle]
Prefactor: gamma(2*eps + 3).
(Effective) kinematic regime: Minkowski (generic).
Generalized permutahedron property: fulfilled.
Analytic continuation: activated. Lambda = 7.6
Started integrating using 8 threads and N = 1e+07 points.
Finished in 6.00369 seconds = 0.00166769 hours.

-- eps^0: [-46.59  +/- 0.13] + i * [ 87.19  +/- 0.12]
-- eps^1: [-274.46 +/- 0.55] + i * [111.26  +/- 0.55]
-- eps^2: [-435.06 +/- 1.30] + i * [-174.47 +/- 1.33]
-- eps^3: [-191.72 +/- 2.15] + i * [-494.69 +/- 2.14]
-- eps^4: [219.15  +/- 2.68] + i * [-431.96 +/- 2.67]
\end{lstlisting}
\soft{trop\_res} contains the printed $\e$-expansion coefficients in the format
$$
        \big[
                \left( 
                        \left( \mathrm{re}_0, \, \sigma_0^\mathrm{re} \right)
                        \, , \,
                        \left( \mathrm{im}_0, \, \sigma_0^\mathrm{im} \right)
                \right)
                \, , \, \ldots \, , \,
                \left( 
                        \left( \mathrm{re}_4, \, \sigma_4^\mathrm{re} \right)
                        \, , \,
                        \left( \mathrm{im}_4, \, \sigma_4^\mathrm{im} \right)
                \right)
        \big]
        \, ,
$$
where 
$
        \mathrm{re}_0 \pm \sigma_0^\mathrm{re} 
$
is the real part of the $0$th order term,
and so forth.
To obtain the $\e$-expansion of the integral with the prefactor included,
one may call
\begin{lstlisting}[style=mystyle]
eps_expansion(trop_res, edges, D0)
\end{lstlisting}
which gives
\begin{lstlisting}[style=mystyle]
174.3842115*i - 93.17486662 + eps*(-720.8731714 + 544.3677186*i) +
eps**2*(-2115.45025 + 496.490128*i) + eps**3*(-3571.990969 - 677.5254794*i) +
eps**4*(-3872.475723 - 2726.965026*i) + O(eps**5)
\end{lstlisting}

\subsubsection*{On the choice of \texorpdfstring{$\l$}{}}

A few comments are in order regarding the choice of the deformation parameter $\l$
(what was called \soft{Lambda} in the code).
The analytic expression for the integral \eqref{deformed_feynman_integral} is independent of $\l$
(so long as the deformed contour doesn't hit any poles),
but the numerical error coming from MC sampling depends rather sensitively on its value.
Moreover,
the optimal value for $\l$ will generally vary depending on the phase space point.
It is currently up to the user to find a suitable value by trial and error,
e.g.~by performing several integrations with low sampling numbers $N$.
A heuristic starting value is $\mO(1/\L^2)$,
where $\L^2$ is the largest kinematic scale in the $\mF$-polynomial
(for instance some mass $m^2$).

In the future,
it would be beneficial to automate the choice of $\l$ either by
1) minimizing the sampling variance $\sigma_f$, 
given in \eqref{sampling_variance},
by solving for $\l$ in
\eq{
        \p_\l \sigma_f = 0
        \, ,
}
or 2) by tightening the tropical approximation bounds in \eqref{tropical_approximation_property_lambda}.

        \section{Examples of Feynman integral evaluations}

In this section,
we use \ft to numerically evaluate many FIs of interest.
Each script is available in the \soft{/examples} folder in the \package{gitub} repository.
When possible,
the results have been verified against \package{AMFlow} and/or \package{pySecDec}%
\footnote{
        We thank Vitaly Magerya for help with verifying results using \package{pySecDec}.
}.
The exception is the example from \secref{sec:5L_2pt},
as we were not been able to re-compute these values with any other software in reasonable time -
though it might be possible now with the new release of \package{pySecDec} \cite{Heinrich:2023til}.

Each example can be computed on a laptop within a few minutes.
Though to properly showcase the potential of \ft\!\!,
we shall only include results obtained from a high-performance machine:
a single \soft{AMD EPYC 7H12} 64-core processor using all cores.
With $N = 10^8$ MC sampling points each example only took a few seconds to finish.
The relative accuracies of the results fall between $10^{-2}$ to $10^{-4}$.

The phase space points are chosen to be away from any kinematic thresholds to ensure numerical convergence.
In all but the last example, the $\e$-expansion goes up to and including $\mO(\e^{2L})$.

\newpage

\subsection{5-loop 2-point zigzag diagram}
\label{sec:5L_2pt}

Here we evaluate a 5-loop 2-point function with all masses different in $\DD = 3-2\e$ dimensions:

\hspace{4.5cm} \includegraphicsbox{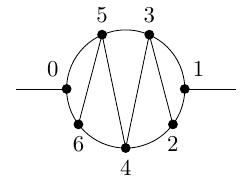}

\noindent
The edge set is
\begin{lstlisting}[style=mystyle]
edges = [((0,6), 1, '1') , ((0,5), 1, '2'), ((5,6), 1, '3'), 
         ((6,4), 1, '4') , ((5,3), 1, '5'), ((5,4), 1, '6'), 
         ((4,3), 1, '7') , ((4,2), 1, '8'), ((3,2), 1, '9'), 
         ((3,1), 1, '10'), ((2,1), 1, '11')]
\end{lstlisting}
Above we already input the chosen mass values
\eq{
        m_0^2 = 1
        \, , \quad
        m_1^2 = 2
        \, , \quadit{\ldots,}
        m_{10}^2 = 11
        \, .
}
The remaining kinematic variable to be specified is the incoming momentum.
We set $p_0^2 = 100$ in \ft by writing
\begin{lstlisting}[style=mystyle]
replacement_rules = [(sp[0,0], 'pp0')]
phase_space_point = [('pp0', 100)]
\end{lstlisting}

\noindent
The value $\l = 0.02$ turns out to yield small errors.
This is of order $\mO(1/p_0^2)$,
in accordance with the heuristic described near the end of the previous section.
With $N = 10^8$ MC sampling points,
the command \soft{tropical\_integration} in \ft gives

\begin{lstlisting}[style=mystyle]
Prefactor: gamma(5*eps + 7/2).
(Effective) kinematic regime: Minkowski (generic).
Finished in 9.62 seconds.
-- eps^0: [0.0001976 +/- 0.0000016]  +  i * [0.0001415 +/- 0.0000018]
-- eps^1: [-0.004961 +/- 0.000023 ]  +  i * [-0.000802 +/- 0.000024 ]
-- eps^2: [ 0.04943  +/-  0.00017 ]  +  i * [-0.01552  +/-  0.00017 ]
-- eps^3: [-0.25468  +/-  0.00083 ]  +  i * [ 0.24778  +/-  0.00093 ]
-- eps^4: [ 0.5909   +/-  0.0033  ]  +  i * [ -1.7261  +/-  0.0038  ]
-- eps^5: [  1.048   +/-   0.012  ]  +  i * [  7.410   +/-   0.013  ]
-- eps^6: [ -14.652  +/-   0.037  ]  +  i * [ -20.933  +/-   0.038  ]
-- eps^7: [  65.87   +/-   0.10   ]  +  i * [  35.25   +/-   0.11   ]
-- eps^8: [ -190.90  +/-   0.27   ]  +  i * [  -4.91   +/-   0.26   ]
-- eps^9: [ 393.08   +/-   0.70   ]  +  i * [ -182.56  +/-   0.59   ]
-- eps^10:[ -558.01  +/-   1.64   ]  +  i * [ 685.62   +/-   1.29   ]
\end{lstlisting}

\noindent
For fun,
the same computation with $N = 10^{12}$ samples results in

\begin{lstlisting}[style=mystyle]
Finished in 20 hours.
-- eps^0: [0.000196885 +/- 0.000000032]  +  i * [0.000140824 +/- 0.000000034]
-- eps^1: [-0.00493791 +/- 0.00000040 ]  +  i * [-0.00079691 +/- 0.00000038 ]
-- eps^2: [ 0.0491933  +/-  0.0000025 ]  +  i * [-0.0154647  +/-  0.0000025 ]
-- eps^3: [ -0.253458  +/-  0.000012  ]  +  i * [ 0.246827   +/-  0.000012  ]
-- eps^4: [ 0.587258   +/-  0.000046  ]  +  i * [ -1.720213  +/-  0.000046  ]
-- eps^5: [  1.05452   +/-   0.00015  ]  +  i * [  7.38725   +/-   0.00015  ]
-- eps^6: [ -14.66144  +/-   0.00047  ]  +  i * [ -20.86779  +/-   0.00046  ]
-- eps^7: [  65.8924   +/-   0.0013   ]  +  i * [  35.0793   +/-   0.0013   ]
-- eps^8: [ -190.9702  +/-   0.0036   ]  +  i * [  -4.4620   +/-   0.0034   ]
-- eps^9: [ 393.2522   +/-   0.0092   ]  +  i * [ -183.7431  +/-   0.0087   ]
-- eps^10:[ -558.202   +/-    0.023   ]  +  i * [  688.556   +/-    0.021   ]
\end{lstlisting}

\subsection{3-loop 4-point envelope diagram}

This example is a non-planar, $3$-loop $4$-point diagram in $\DD = 4 - 2 \e$ dimensions:

\vspace{0.5cm} \hspace{4.5cm} \includegraphicsbox{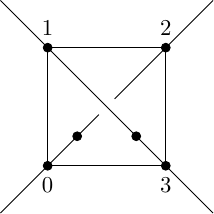} \vspace{0.5cm}

\noindent
The dots on the crossed lines represent squared propagators rather than vertices;
this choice of edge weights makes the diagram quasi-finite.
The edge data are
\begin{lstlisting}[style=mystyle]
edges = [((0,1), 1, 'mm0'), ((1,2), 1, 'mm1'), ((2,3), 1, 'mm2'), 
         ((3,0), 1, 'mm3'), ((0,2), 2, 'mm4'), ((1,3), 2, 'mm5')]
\end{lstlisting}
Note that the edge weights $\nu_4$ and $\nu_5$ equal 2.

We define two-index Mandelstam variables by
$
        s_{ij} = (p_i + p_j)^2.
$
They are input into replacement rules for \ft by writing
\soft{(sp[i,j], '(sij - ppi - ppj)/2')}
with
$
        0 \leq i \leq j \leq 2.
$
The chosen phase space point is 
\eq{
        \begin{array}{cccccccccc}
                & p_0^2 &=& 1.1
                \, , \quad
                & p_1^2 &=& 1.2
                \, , \quad
                & p_2^2 &=& 1.3
                \, , 
                \\
                & s_{01} &=& 2.1
                \, , \quad
                & s_{02} &=& 2.2
                \, , \quad
                & s_{12} &=& 2.3
                \, ,
                \\
                & m_0^2 &=& 0.05
                \, , \quad
                & m_1^2 &=& 0.06
                \, , \quad
                & m_2^2 &=& 0.07
                \, ,
                \\
                & m_3^2 &=& 0.08
                \, , \quad
                & m_4^2 &=& 0.09
                \, , \quad
                & m_5^2 &=& 0.1
                \, .
        \end{array}
}
Setting $\l = 1.24$ and $N = 10^8$,
we get the result

\begin{lstlisting}[style=mystyle]
Prefactor: gamma(3*eps + 2).
(Effective) kinematic regime: Minkowski (generic).
Finished in 5.12 seconds.
-- eps^0: [-10.8335 +/- 0.0084]  +  i * [-12.7145 +/- 0.0083]
-- eps^1: [ 47.971  +/- 0.059 ]  +  i * [-105.057 +/- 0.059 ]
-- eps^2: [ 413.05  +/-  0.23 ]  +  i * [  7.29   +/-  0.23 ]
-- eps^3: [ 372.07  +/-  0.65 ]  +  i * [ 947.82  +/-  0.65 ]
-- eps^4: [-1412.36 +/-  1.45 ]  +  i * [1325.74  +/-  1.45 ]
-- eps^5: [-2726.00 +/-  2.67 ]  +  i * [-1295.36 +/-  2.69 ]
-- eps^6: [ 287.25  +/-  4.28 ]  +  i * [-3982.04 +/-  4.30 ]
\end{lstlisting}

\subsection{2-loop 4-point \texorpdfstring{$\mu e$}{}-scattering diagram}

We evaluate a non-planar, $2$-loop $4$-point diagram that contributes to muon-electron scattering.
It is quasi-finite in $\DD = 6 - 2\e$ dimensions.
The diagram was evaluated for vanishing electron mass in \cite{DiVita:2018nnh},
but here we keep the electron massive.

\vspace{0.4cm} \hspace{4.2cm} \includegraphicsbox{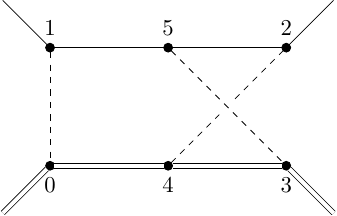} \vspace{0.4cm}

\noindent
The dashed lines represent photons.
The solid lines are electrons with mass $m$.
The double lines denote muons with mass $M \sim 200 m$.
The set of edges is
\begin{lstlisting}[style=mystyle]
edges = [((0,1), 1, '0'), ((0,4), 1, 'MM'), ((1,5), 1, 'mm'), ((5,2), 1, 'mm'),
         ((5,3), 1, '0'), ((4,3), 1, 'MM'), ((4,2), 1, '0')]
\end{lstlisting}
where \soft{mm} and \soft{MM} stand for $m^2$ and $M^2$ respectively.
We choose a phase space point similar to that of \cite[Section 4.1.2]{DiVita:2018nnh}:
\eq{
        p_0^2 = M^2 = 1
        \, , \quad
        p_1^2 = p_2^2 = m^2 = 1/200
        \, , \quad
        s_{01} = -1/7
        \, , \quad
        s_{12} = -1/3
        \, .
}
By momentum conservation,
this fixes
\eq{
        s_{02} = 2M^2 - 2m^2 - s_{01} - s_{12} = 2.49
        \, .
}
With additional settings $\l = 1.29$ and $N = 10^8$,
the result comes out to

\newpage

\begin{lstlisting}[style=mystyle]
Prefactor: gamma(2*eps + 1).
(Effective) kinematic regime: Minkowski (exceptional).
Finished in 6.53 seconds.
-- eps^0: [1.16483 +/- 0.00083]  +  i * [0.24155 +/- 0.00074]
-- eps^1: [5.5387  +/- 0.0086 ]  +  i * [2.2818  +/- 0.0093 ]
-- eps^2: [15.171  +/-  0.058 ]  +  i * [10.079  +/-  0.064 ]
-- eps^3: [ 28.02  +/-  0.32  ]  +  i * [ 28.17  +/-  0.28  ]
-- eps^4: [ 38.20  +/-  1.42  ]  +  i * [ 56.94  +/-  0.85  ]
\end{lstlisting}
Because the momentum configuration is exceptional,
we cannot be sure that the GP property holds 
(cf.~\secref{sec:GP}).
In spite of that,
we are able to confirm the values above with both \package{AMFlow} and \package{pySecDec}.

The leading-order term in $\e$ differs from \cite[Equation (4.20)]{DiVita:2018nnh} by roughly 10\% 
because we have included the electron mass.
If we set $m^2 = 0$ in \ft\!\!,
then we are able to reproduce the value from that reference
(even though this is an even more exceptional configuration!).

\subsection{2-loop 5-point QCD-like diagram}

Here we evaluate a QCD-like, $(6-2\e)$-dimensional graph with 2 loops and 5 legs:

\vspace{0.4cm} \hspace{3cm} \includegraphicsbox{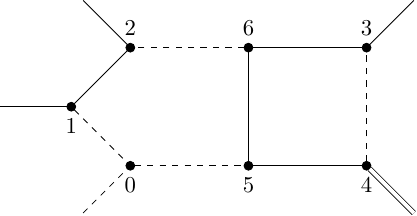} \vspace{0.4cm}

\noindent
The dashed lines represent on-shell gluons.
The solid lines are quarks each having mass $m$.
The double line denotes some off-shell leg $p_4^2 \neq 0$ fixed by momentum conservation.

The edge list is
\begin{lstlisting}[style=mystyle]
edges = [((0,1), 1, '0'), ((1,2), 1, 'mm'), ((2,6), 1, '0'), ((6,3), 1, 'mm'),
         ((3,4), 1, '0'), ((4,5), 1, 'mm'), ((5,0), 1, '0'), ((5,6), 1, 'mm')]
\end{lstlisting}
where \soft{mm} is the symbol standing for $m^2$.
Our chosen phase space point is
\eq{
        \begin{array}{lllll}
        & s_{01} = 2.2
        \, , 
        & s_{02} = 2.3 
        \, ,
        & s_{03} = 2.4
        \, , 
        & s_{12} = 2.5
        \, , 
        \\
        & s_{13} = 2.6
        \, , 
        & s_{23} = 2.7 
        \, ,
        & p_0^2 \hspace{0.15cm} = \hspace{0.15cm} 0
        \, , 
        & p_1^2 = p_2^2 = p_3^2 = m^2 = 1/2
        \, ,
        \end{array}
}
with $s_{ij} = (p_i + p_j)^2$ as before.
With settings $\l = 0.28$ and $N = 10^8$,
\ft gives the result
\begin{lstlisting}[style=mystyle]
Prefactor: gamma(2*eps + 2).
(Effective) kinematic regime: Minkowski (exceptional).
Finished in 8.20 seconds.
-- eps^0: [0.06480 +/- 0.00078]  +  i * [-0.08150 +/- 0.00098]
-- eps^1: [0.4036  +/- 0.0045 ]  +  i * [ 0.3257  +/- 0.0035 ]
-- eps^2: [-0.7889 +/- 0.0060 ]  +  i * [ 0.957   +/-  0.016 ]
-- eps^3: [-1.373  +/-  0.030 ]  +  i * [ -1.181  +/-  0.034 ]
-- eps^4: [ 1.258  +/-  0.088 ]  +  i * [ -1.205  +/-  0.036 ]
\end{lstlisting}
The kinematic configuration is once again exceptional.
The result is nevertheless verified with \package{pySecDec}.

\subsection{2-loop 5-point Higgs production diagram}

This example contributes to triple Higgs production via gluon fusion,
$gg \to HHH$,
in $\DD = 4 - 2 \e$ dimensions:

\hspace{4.0cm} \includegraphicsbox{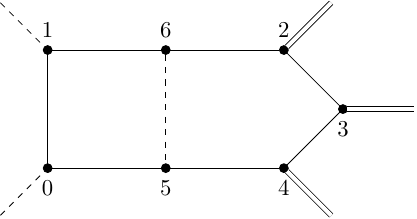} \vspace{0.4cm}

\noindent
The dashed lines represent massless, on-shell gluons.
The single solid lines are propagators containing the top quark mass.
The three external double lines are put on-shell to the Higgs mass.

The edge data are
\begin{lstlisting}[style=mystyle]
edges = [((0,1), 1, 'mm_top'), ((1,6), 1, 'mm_top'), ((5,6), 1, '0'),  
         ((6,2), 1, 'mm_top'), ((2,3), 1, 'mm_top'), ((3,4), 1, 'mm_top'),
         ((4,5), 1, 'mm_top'), ((5,0), 1, 'mm_top')]
\end{lstlisting}
with \soft{mm\_top} standing for the square of top quark mass, $m_t^2$.
The kinematic setup is given by
\eq{
        & p_0^2 = p_1^2 = 0
        \, , \quad
          p_2^2 = p_3^2 = p_4^2 = m_H^2
        \, , 
        \\
        & s_{01} = 5 m_H^2 - s_{02} - s_{03} - s_{12} - s_{13} - s_{23}
        \, ,
        \nonumber
}
with the last relation being a consequence of momentum conservation.
So the space of kinematic variables is parametrized by
$
        (s_{02}, s_{03}, s_{12}, s_{23}, m_t^2, m_H^2),
$
and we choose the point
\eq{
        \begin{array}{lllll}
        & m_t^2 = 1.8995
        \, , \quad
        & m_H^2 = 1
        \, , \quad
        & s_{02} = -4.4
        \, , \quad
        & s_{03} = -0.5
        \, , 
        \\
        & s_{12} = -0.6
        \, , \quad
        & s_{13} = -0.7
        \, , \quad
        & s_{23} = 1.8
        \, .
        & 
        \end{array}
}
This point lies in the physical region,
and it contains the physically relevant mass ratio $m_t^2 / m_H^2 = 1.8995$.
For $\l = 0.64$ and $N = 10^8$,
the FI then evaluates to
\begin{lstlisting}[style=mystyle]
Prefactor: gamma(2*eps + 4).
(Effective) kinematic regime: Minkowski (generic).
Finished in 8.12 seconds.
-- eps^0: [-0.0114757 +/- 0.0000082]  +  i * [0.0035991 +/- 0.0000068]
-- eps^1: [ 0.003250  +/- 0.000031 ]  +  i * [-0.035808 +/- 0.000041 ]
-- eps^2: [ 0.046575  +/- 0.000098 ]  +  i * [0.016143  +/- 0.000088 ]
-- eps^3: [ -0.01637  +/-  0.00017 ]  +  i * [ 0.03969  +/-  0.00016 ]
-- eps^4: [ -0.02831  +/-  0.00023 ]  +  i * [-0.00823  +/-  0.00024 ]
\end{lstlisting}

\subsection{4-loop 0-point QED-like diagram}

Next we evaluate a QED-like, $4$-loop vacuum diagram in $\DD = 4 - 2\e$ dimensions:

\vspace{0.4cm} \hspace{4.5cm} \includegraphicsbox{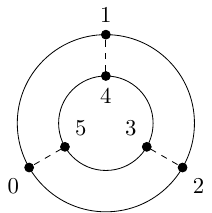} \vspace{0.4cm}

\noindent
The dashed lines are photons,
and the solid lines are fermions with mass $m$.
Since this diagram has no external momenta,
it is not necessary to perform any contour deformation.
This means that we can set $\l = 0$,
and result is expected to be purely real.
In the \ft script,
we write \soft{replacement\_rules = []} to specify that no external momenta are present.

The edge set is given by
\begin{lstlisting}[style=mystyle]
edges = [((0,1), 1, 'mm'), ((1,2), 1, 'mm'), ((2,0), 1, 'mm'),
         ((0,5), 1, '0' ), ((1,4), 1, '0' ), ((2,3), 1, '0' ),
         ((3,4), 1, 'mm'), ((4,5), 1, 'mm'), ((5,3), 1, 'mm')]
\end{lstlisting}
with \soft{mm} standing for $m^2$.
Setting this mass to unity by
\begin{lstlisting}[style=mystyle]
phase_space_point = [('mm', 1)]
\end{lstlisting}
the result for $N = 10^8$ sampling points becomes
\begin{lstlisting}[style=mystyle]
Prefactor: gamma(4*eps + 1).
(Effective) kinematic regime: Euclidean (generic).
Finished in 3.58 seconds.
-- eps^0: [3.01913 +/- 0.00047]  +  i * [0.0 +/- 0.0]
-- eps^1: [-7.0679 +/- 0.0021 ]  +  i * [0.0 +/- 0.0]
-- eps^2: [20.5399 +/- 0.0074 ]  +  i * [0.0 +/- 0.0]
-- eps^3: [-27.895 +/-  0.024 ]  +  i * [0.0 +/- 0.0]
-- eps^4: [62.043  +/-  0.074 ]  +  i * [0.0 +/- 0.0]
-- eps^5: [-59.46  +/-  0.23  ]  +  i * [0.0 +/- 0.0]
-- eps^6: [155.27  +/-  0.73  ]  +  i * [0.0 +/- 0.0]
-- eps^7: [-90.81  +/-  2.26  ]  +  i * [0.0 +/- 0.0]
-- eps^8: [403.78  +/-  6.71  ]  +  i * [0.0 +/- 0.0]
\end{lstlisting}

\subsection{An elliptic, conformal, 4-point integral}

The final example is a 1-loop 4-point conformally invariant integral \cite{Henn:2009bd}.
The analytic result for this diagram with edge weights
$
        \nu_0 = \nu_1 = \nu_2 = \nu_3 = 1/2
$
and dimension $\DD = 2$ was calculated in \cite[Section 7.2]{Corcoran:2021gda}:
\eq{
        \includegraphicsbox{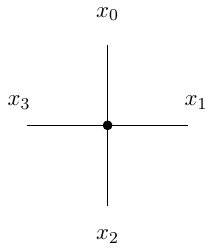}
        \ = \
        \frac{4}{\sqrt{-p_2^2}} 
        \big[
                K(z) K(1-\bar{z}) + K(\bar{z}) K(1-z)
        \big]
        \, .
        \label{conformal_integral}
}
This result features the elliptic integral $K$ given by
\eq{
        K(z) = \int_0^1 \frac{\dd x}{ \sqrt{(1-x^2) (1 - z^2 x^2)} }
}
and kinematic variables
\eq{
        z \bar{z} = \frac{p_0^2}{p_2^2}
        \, , \quad
        (1-z)(1-\bar{z}) = \frac{p_1^2}{p_2^2}
        \, .
}
The denominator of \eqref{conformal_integral} differs from \cite[Equation (7.6)]{Corcoran:2021gda}
because we have used conformal symmetry to send $x_3 \to \infty$,
where $x_3$ is a dual momentum variable
(defined by $p_i = x_i - x_{i+1}$).
This reduces the kinematic space to that of a 3-point function.

In \ft we thus specify an edge list for a 1-loop graph with 3 external legs:
\begin{lstlisting}[style=mystyle]
edges = [((0,1), 1/2, '0'), ((1,2), 1/2, '0'), ((2,0), 1/2, '0')]
\end{lstlisting}
All internal masses are set to zero,
otherwise conformal symmetry would be broken,
and every edge weight has been set to $1/2$.
It is most natural to study conformal integrals in the Euclidean regime,
so let us pick
\eq{
        p_0^2 = -2
        \, , \quad
        p_1^2 = -3
        \, , \quad
        p_2^2 = -5
        \, . 
}
With $N = 10^8$ sampling points,
we obtain
\begin{lstlisting}[style=mystyle]
(Effective) kinematic regime: Euclidean (generic).
Finished in 1.34 seconds.
-- eps^0: [9.97192 +/- 0.00027]  +  i * [0.0 +/- 0.0]
\end{lstlisting}
This result agrees with the analytic formula \eqref{conformal_integral}.
We hope that our code can furnish further study into conformal integrals,
as they are important for $\mN = 4$ SYM and the cosmological bootstrap.

        \chapter{Conclusion and Outlook}
\label{ch:conclusion}

\section*{Conclusion}

This thesis has been an investigation into the mathematical structure of Feynman integrals.
Its overarching goal has been to develop novel techniques for evaluating these integrals.

\vspace{0.5cm} 

In \textbf{Part I} of this thesis,
we studied integration-by-parts identities and the method of Pfaffian DEQs for evaluating Feynman integrals.
When the Feynman integrals in question admit a solution in terms of multiple polylogarithms,
we showed how the Magnus expansion can bring the Pfaffian system into canonical form,
whose solution is readily obtained via the path-ordered exponential.

These techniques were used to compute form factors for a leptophillic dark matter model.
The calculation involved a class of 2-loop 3-point Feynman integrals with massive internal propagators,
which were solved for in terms of multiple polylogarithms.

\vspace{0.5cm} 

In \textbf{Part II},
we delved deeper into the mathematics of DEQs by formalizing the notion of a $\mD$-module.
A $\mD$-module encapsulates the algebraic properties of differential operators 
that annihilate a given function such as a Feynman integral.

We studied a broad class of functions called Euler integrals containing Feynman integrals as a special case.
Euler integrals are annihilated by a known collection of 
higher-order differential operators called the GKZ system.
We showed how to connect the GKZ system to the more 
manageable first-order Pfaffian system through the Macaulay matrix.

In some cases,
the Euler integral matched exactly with a Feynman integral.
In other cases,
the Euler integral had too many unfixed parameters.
The latter case was treated by developing two new $\mD$-module restriction methods.
One of those methods drove us to establish an algorithm for 
obtaining logarithmic series solutions to Pfaffian systems.

\vspace{0.5cm} 

In \textbf{Part III},
we formulated a numerical integration scheme for the evaluation of Feynman integrals
which combined Monte Carlo integration with tropical geometry.
The algorithm relied on the assumption that the Newton polytope of the 
$\mF$ Symanzik polynomial is a generalized permutahedron,
and we gave a detailed discussion of when this can be expected to hold.
The $i\vare$-prescription was incorporated into the algorithm via contour deformation,
which allowed us to evaluate Feynman integrals in physical regions of kinematic space.
The algorithm was implemented in a publicly available program named \package{feyntrop},
whose validity was verified against existing numerical codes.
We used \ft to numerically evaluate Feynman integrals between 1-5 loops and 0-5 legs. 

\section*{Outlook}

There are several avenues of research to pursue based on the results of this thesis.

\begin{paragraph}{Applying the Macaulay matrix to other ideals.}

        \algref{alg:Pfaffian_by_MM} takes a holonomic ideal $\mI$
        together with a $\mD$-module basis as input,
        feeds it through the Macaulay matrix,
        and outputs a Pfaffian system.
        In all examples of this thesis,
        we worked with the GKZ system $\mI = H_A(\b)$ simply 
        because $H_A(\b)$ and a basis of standard monomials 
        are easily determined given an Euler integral.
        However,
        there might very well be other situations where an annihilating ideal is known.
        If a basis is not known,
        it could be guessed by executing \algref{alg:Pfaffian_by_MM} 
        (with parameters set to numbers)
        for different choices of standard monomials until it succeeds.

        A non-GKZ example is for instance the conformally invariant, 
        massless, 
        1-loop 3-point function studied in \cite{Henn:2023tbo}.
        An annihilating ideal was there determined from the generators of conformal symmetry.
        It would be interesting to study these operators with the Macaulay matrix,
        as well as those coming from the larger Yangian symmetry group \cite{Loebbert:2022nfu}.

        Another route would be to employ the algorithms from 
        \cite{Lairez:2022zkj,Lairez:2023nih,Doran:2023yzu} 
        to compute Picard-Fuchs ideals for Feynman integrals;
        these algorithms were effective for rather complicated diagrams,
        e.g.~at 2 loops and 7 points.
        As an aside,
        it would be interesting to see whether a canonical Feynman integral has any nice
        structure when it is represented as an operator inside the $\mD$-module $\mD/\mI$,
        where $\mI$ is the Picard-Fuchs ideal.
\end{paragraph}

\begin{paragraph}{More complicated examples of logarithmic restriction.}

        An algorithm for the calculation of logarithmic series solutions to
        Pfaffian systems was presented in \secref{sec:logarithmic_restriction}.
        We gave results for 1-loop Bhabha scattering in \secref{sec:1L_bhabha_scattering},
        and partial results for the 2-loop Bhabha double-box in \secref{sec:bhabha_2L_scattering}.
        It would be instructive to first complete the 2-loop calculation;
        this involves the determination of boundary constants and power corrections $(m^2/s)^n$.
        
        Afterwards, it would be interesting to see how far this approach can go.
        The complexity of the restricted Pfaffian systems for the 
        2-loop Bhabha double-box was miniscule compared to today's standards, 
        indicating that more complicated diagrams could potentially be evaluated with this method.

\end{paragraph}

\begin{paragraph}{Expansion around singular locus.}

        Recall the form of the logarithmic expansion ansatz \eqref{I_asymptotic_series}.
        There we chose a distinguished singularity $z_1 = 0$ to expand around.
        It would be interesting to reformulate the logarithmic restriction procedure
        to instead expand around the entire singular locus $L(z)$ of a Pfaffian system,
        where $L(z)$ is a product of irreducible hypersurfaces.
        For instance,
        the singular locus for Appell's $F_4$ function is
        $L(z) = z_1 z_2 [ (z_1-z_2)^2 - 2(z_1+z_2)+ 1 ]$.
        For Feynman integrals,
        $L(z)$ would correspond to the \emph{Landau variety} \cite{Mizera:2021icv}.

        The series ansatz is now roughly represented as 
        (see \cite[Equation (12.15)]{haraoka2020linear} for details)
        \eq{
                \vec{I}(z) 
                \quadit{\sim}
                L(z)^\l \times
                \sum_{n=0}^\infty \sum_{m=0}^M \Inm(z) \times L(z)^n \times \log[L(z)]^m
                \, ,
        }
        where $\l \in \RR$ and $M \in \ZZ_{\geq 0}$.
        Analytical solutions to Feynman integrals are often hard to numerically evaluate near Landau singularities,
        but a solution written in this fashion could potentially have better numerical stability.

\end{paragraph}

\begin{paragraph}{Applications in algebraic statistics.}

        Looking beyond physics,
        one finds that GKZ systems and Euler integrals also 
        frequently appear in the course of Bayesian inference 
        in algebraic statistics \cite[Section 3]{sattelberger2019d}.

        The article \cite{mano2021algebraic} gave an initial 
        application of the Macaulay matrix in this context;
        it is natural to expect that this research direction can be pushed even further 
        with the more sophisticated Macaulay matrix method developed here.
        Further, it appears that the Pfaffian-level restriction protocol 
        from \secref{sec:restriction_pfaffian_level} can be used to study
        contingency tables%
        \footnote{
                These tables are a standard tool in statistics
                to record interrelations between random variables.
        }
        in statistics \cite{saiei-talk}.

\end{paragraph}

\begin{paragraph}{Numerators and $\e$-divergent integrals in \ft\!\!.}

        The program \ft presented in \chapref{ch:feyntrop} has two main 
        shortcomings which would be beneficial to eliminate in future versions:
        \begin{enumerate}
                \item
                        It only works for scalar Feynman integrals.
                        But in gauge theories such as QED and QCD,
                        one encounters numerators in momentum space representation of the form
                        $(p \cdot \ell)^\nu$,
                        where $p$ and $\ell$ are external and loop momenta respectively.
                \item 
                        It assumes a quasi-finite integral representation (cf. \secref{sec:quasi_finite}).
        \end{enumerate}

        By the formulas of \cite[Appendix A]{Hannesdottir:2022bmo},
        numerators can be written in the Symanzik representation by including an extra
        (homogeneous) polynomial $\mN$ inside the integrand.
        There is no reason to expect that this polynomial would 
        enjoy the generalized permutahedron property though,
        so it might not be straightforward to include it in 
        the MC sampling of \algref{alg:tropical_sampling}.

        The quasi-finite condition can be lifted by following the protocol in 
        \cite[Page 5]{berkesch2014euler},
        where explicit formulas are given for factoring out $\e$-poles in front of finite integrals.
        These formulas would involve derivatives of the $\mU$ and $\mF$ Symanzik polynomials,
        and we fortunately have efficient expressions for those (cf. \secref{sec:evaluating_deformed}).

        With these two additions,
        we envision that \ft could be used in a wide range of phenomenological calculations.
        Especially pertaining to scattering amplitudes involving many mass scales,
        which are the hardest ones to treat analytically with DEQs due 
        to the proliferation of functions beyond the class of multiple polylogarithms%
        \footnote{
                \ft has already been used to check results for a non-planar 3-loop 
                calculation related to Higgs+jet production \cite{Henn:2023vbd}.
        }.
\end{paragraph}

\begin{paragraph}{A canonical contour deformation.}
        In \secref{sec:ieps_deformation},
        we implemented the $i\vare$-prescription via a contour deformation of the Schwinger parameters
        $
                x_e \to x_e \exp\left[-i\l \p_e \mV\right].
        $
        The small, but finite, number $\l \in \RR_{> 0}$ is a free parameter.
        An explicit value has to be specified by the user in \ft\!\!,
        and the accuracy of the result depends quite sensitively on this choice.

        It would be a significant development to determine a \emph{canonical} choice for $\l$,
        depending on the data of the Feynman integral in question as well as the given phase space point.
        This would not only improve the user experience in \ft\!\!,
        but also constitute the first example of a 
        contour deformation for Feynman integrals without any free parameters.

\end{paragraph}

	\appendix
	
	\chapter{Details on \texorpdfstring{$\mD$-modules}{}} 

Here we collect some additional details on $\mD$-modules that are referred to in the main text.

\section{Basis change in a holonomic \texorpdfstring{$\mD$-module}{}}
\label{sec:basis_change}

The goal in this section is to construct the Pfaffian system for the basis $e$
\eq{
        \p_i e = P_i\supbrk{e} \cdot e
}
given a known system in the standard monomial basis
\eq{
        \p_i \Std = P_i\supbrk{\Std} \cdot \Std
        \label{pfaffian_system_std}
        \, .
}
$e \in (\mR/\mI)^R$ is here an $R$-dimensional vector of 
(equivalence classes of) differential operators.

Observe that,
if we could find a matrix $G$ relating the two bases by
\eq{
        e = G \cdot \Std
        \, ,
        \label{e_G_Std}
}
then the problem would be solved by a gauge transformation%
\footnote{Note that $G$ is invertible under the assumption that $e$ and $\Std$ are both bases.}
\eq{
        P_i\supbrk{e} =
        \left[
                \p_i \bullet G \plus G \cdot P_i\supbrk{\Std}
        \right]
        \cdot
        G^{-1}
        \, .
}
To construct $G$,
we begin by writing the decomposition
\eq{
        e =
        \sum_q G\supbrk{1}_q \cdot \left( \p^q \, \Std \right)
        \, ,
}
where 
$
        G_q \supbrk{1} \in \CC(z)^{R \times R}
$.
In other words,
we write each element of $e$ as a sum of rational functions 
times derivatives $\p^q$ that act on $\Std$.
The idea is now to transform the expression $\p^q \, \Std$ into the form
$
        G_q\supbrk{2} \cdot \Std,
$
for some 
$
        G_q\supbrk{2} \in \CC(z)^{R \times R},
$
by repeated application of \eqref{pfaffian_system_std}.
This is best illustrated by means of an example.
\begin{ex}
Let
$
        q = \{1,1,0,0,\ldots,0\} .
$
Then $\p^q = \p_1 \p_2$.
Applying \eqref{pfaffian_system_std} twice,
the expression $\p^q \, \Std$ becomes
\eq{
        \p_1 \p_2 \, \Std &=
        \p_1 \left[ P_2\supbrk{\Std} \cdot \Std \right] \\&=
        \left[ \p_1 \bullet P_2\supbrk{\Std} \right] \cdot \Std
        \plus
        P_2\supbrk{\Std} \cdot \left[ \p_1 \, \Std \right] \\&=
        \left[ \p_1 \bullet P_2\supbrk{\Std} \right] \cdot \Std
        \plus
        P_2\supbrk{\Std} \cdot P_1\supbrk{\Std} \cdot \Std
        \, .
}
The matrix
\eq{
        G_q\supbrk{2} = 
        \p_1 \bullet P_2\supbrk{\Std} 
        \plus 
        P_2\supbrk{\Std} \cdot P_1\supbrk{\Std}
        \, ,
}
hence satisfies the sought after relation 
$
        \p^q \, \Std = G_q\supbrk{2} \cdot \Std
$
.
\end{ex}

We conclude that the gauge transformation matrix $G$ in \eqref{e_G_Std} takes the form
\eq{
        G = 
        \sum_q G_q\supbrk{1} \cdot G_q\supbrk{2}
        \label{G_G1_G2}
}
where $G_q\supbrk{2}$ is built from matrix products between Pfaffian matrices $P_i\supbrk{\Std}$
and derivatives thereof.

\section{Basis change without derivatives}

We concluded above that the gauge transformation matrix \eqref{G_G1_G2}
is constructable from derivatives of Pfaffian matrices $P_i(z)$.
In the GKZ setting we might,
however,
like to fix some of the $z$-variables to numbers.
Then we no longer have the luxury of taking derivatives.
As a remedy to this,
let us derive a formula for $\p_i \bullet P_j$ which is based solely on matrix multiplication.

Using the same setup as in \secref{sec:recurrence},
we set
\eq{
        f(\b) =
        \frac{1}{\Gamma(\b_0+1)} \int_\Gamma
        g(z|x)^{\b_0} x^{-\b'} \frac{\dd x}{x}
        \quadit{\text{and}}
        F(\b) =
        \arr{c}{
                \Std_1 \bullet f(\b) \\
                \vdots \\
                \Std_R \bullet f(\b)
        }
        \, .
}
As described in \secref{sec:recurrence},
there are two ways of expressing the derivative of $F(\b)$:
\eq{
        \label{pj_F(b)_1}
        \p_j \bullet F(\b) &= P_j(\b) \cdot F(\b) \\
        \p_j \bullet F(\b) &= F(\b-a_j)
        \, .
        \label{pj_F(b)_2}
}
Differentiating \eqref{pj_F(b)_1} w.r.t.~$z_i$,
we have
\eq{
        \p_i \, \p_j \bullet F(\b) &=
        (\p_j \bullet P_i) \cdot F +
        P_j \cdot (\p_i \bullet F) \\&=
        (\p_j \bullet P_i) \cdot F +
        P_j \cdot P_i \cdot F
        \, ,
        \label{pi_pj_F(b)_1}
}
where we omitted the $\b$-dependence for clarity.
On the other hand,
differentiating \eqref{pj_F(b)_2} yields
\eq{
        \p_i \, \p_j \bullet F(\b) &=
        \p_i \bullet F(\b-a_j) \\&=
        F(\b-a_j-a_i) \\&=
        P_j(\b-a_i) \cdot P_i(\b) \cdot F(\b)
        \, ,
        \label{pi_pj_F(b)_2}
}
where the identity
$
        P_k(\b) \cdot F(\b) = F(\b-a_k)
$
was applied twice in the final step.
Equating \eqref{pi_pj_F(b)_1} with \eqref{pi_pj_F(b)_2} and isolating $(\p_i \bullet P_j)$,
we conclude that
\eq{
        (\p_i \bullet P_j)(\b) = 
        \big[
                P_j(\b-a_i) - P_j(\b)
        \big]
        \cdot
        P_i(\b)
        \, .
}
The RHS of this equation can then be inserted into the $G$-matrix of \eqref{G_G1_G2}.

\section{General formula for homogeneity rescaling}
\label{sec:general_rescaling_formula}

We gave an example in \secref{sec:integrand_rescaling} of how to rescale
the GKZ variables $z = (z_1, \ldots, z_N)$ so as to fix $n+1$ of them to unity
(cf.~equation \eqref{2_F_1_t_variables}).
This section presents a general formula.

To start,
given an $(n+1) \times N$ matrix
$
        A = \big[a_1 \ | \ \ldots \ | \ a_N \big]
$
and a list of $n+1$ rescaling parameters
$
        t = \{t_1, \ldots, t_{n+1}\},
$
we construct a list of $N$ exponentials
\eq{
        t^A := \{t^{a_1}, \ldots, t^{a_N}\}
        \quadit{\text{where}}
        t^{a_i} := \prod_{j=1}^{n+1} t_j^{(a_i)_j}
        \, .
}
This matrix exponentiation notation obeys the rule
$
        (t^A)^B = t^{A \cdot B} .
$
For two sets $r$ and $s$,
both of length $N$,
the product $r \times s$ is defined to be
\eq{
        r \times s := \{r_1 s_1, \ldots, r_N s_N\}
        \, .
}
In this language,
the homogeneity property \eqref{Euler_integral_homogeneity} reads
\eq{
        f_\b(t^A \times z) = t^\b f_\b(z)
        \, .
        \label{Euler_integral_homogeneity_set_product}
}
Now fix two subsets
\eq{
        \s \subset \{1, \ldots, N\}
        \quadit{\text{and}}
        \eta = \{1, \ldots, N\} \setminus \s
        \, ,
}
where $\s$ has size $|\s| = n+1$,
and $\eta$ is its complement.
Let $A_\s$ denote the submatrix of $A$ whose columns are labeled by $\s$.
We assume that $\s$ is chosen such that $A_\s$ is invertible,
i.e.~$\det(A_\s) \neq 0$.

Returning to \eqref{Euler_integral_homogeneity_set_product},
a judicial choice for $t$ turns out to be
\eq{
        t = z_\s^{-A_\s^{-1}}
        \, ,
}
where $z_\s$ picks out the variables of $z = \{z_1, \ldots, z_N\}$ labeled by $\s$.
The homogeneity relation then becomes
\eq{
        f_\b
        \big(z_\s^{-A_\s^{-1} \cdot \, A} \times z\big) =
        z_\s^{-A_\s^{-1} \cdot \, \b}
        f_\b(z)
        \, .
        \label{2_F_1_homogeneity_z_sigma}
}
For this choice of $t$,
all the arguments on the LHS labeled by $\s$ are set to unity:
\eq{
        \Big\{
                z_\s^{-A_\s^{-1} \cdot \, A} \times z  
        \Big\}_i
        =
        \begin{cases}
                1 \, , & i \in \s \\
                z_\s^{-A_\s^{-1} \cdot \, a_i} \, z_i \, , & i \in \eta \, .
        \end{cases}
}
As desired,
the function on the RHS of \eqref{2_F_1_homogeneity_z_sigma}
\eq{
        g_\b(w) := z_\s^{-A_\s^{-1} \cdot \, \b}f_\b(z)
}
now effectively only depends on $N-(n+1)$ cross ratios
\eq{
        w := 
        \{w_1, \ldots, w_{N-(n+1)}\} :=
        z_\s^{-A_\s^{-1} \cdot \, A_\eta} \times z_\eta
        \, .
        \label{w_cross_ratios}
}

What's left is to rewrite the GKZ system in terms of the $w$-variables.
To this end,
we summarize a result which is proved in \cite[Appendix A]{Chestnov:2022alh}.
Given two positive integers $a,b \in \ZZ_{>0}$,
we first define the \emph{rising} and \emph{falling factorials} by
\eq{
        (a)_b &:= a(a+1)(a+2)\cdots(a+b+1)
        \\[3pt]
        [a]_b &:= a(a-1)(a-2)\cdots(a-b+1)
        \, .
        \label{scalar_rising_falling_factorials}
}
Given an $N$-dimensional vector $u$ and an integer vector $v \in \ZZ^N_{\geq 0}$,
we extend these definitions to
\eq{
        (u)_v &:= (u_1)_{v_1} \cdots (u_N)_{v_N}
        \\[3pt]
        [u]_v &:= [u_1]_{v_1} \cdots [u_N]_{v_N}
        \, .
        \label{vector_rising_falling_factorials}
}
In what follows,
we shall set 
\eq{
        u =
        \theta_w = 
        \big[\theta_{w_1}, \, \ldots, \, \theta_{ w_{N-(n+1)} }\big]^T
        \, ,
}
where $\theta_i = z_i \p_i$.

To rewrite the GKZ system in terms of $w$,
it suffices to find a relation between
$
        \p_z^v \bullet f_\b(z)
$
and $g_\b(w)$ for $v \in \ZZ^N_{\geq 0}$,
The master formula works out to
\eq{
        \label{homogeneity_partial_derivative}
        & \p_z^v \bullet f_\b(z) = \\[6pt]
        & (-1)^{|v_\s|}
        z_\s^{-A_\s^{-1} \cdot \big(\b + A \cdot v\big)}
        \left\{
                w^{-v_\eta} 
                \
                \big(
                        A_\s^{-1} \cdot \b +
                        A_\s^{-1} \cdot A_\eta \cdot \theta_w
                \big)_{v_\s}
                \
                [\theta_w]_{v_\eta}
                \ \bullet \
                g_\b(w)
        \right\}
        \
        \Big |_{w = \text{\eqref{w_cross_ratios}}}
        \nonumber
        \, ,
}
where $|v_\s| = \sum_{i \in \s} v_i$.
Although this expression looks somewhat daunting,
it has fortunately been implemented in the \package{asir} package \package{mt\_gkz}.

\begin{ex}
For the ${}_2F_1$ case of \exref{ex:2_F_1_gkz},
the following script will output a GKZ system in rescaled variables:
\begin{lstlisting}[style=mystyle]
A = [
        [1,1,1,1],
        [0,1,0,1],
        [0,0,1,1]
];
Beta = [b1,b2,b3];
Sigma = [1,2,3];
Ideal = mt_gkz.gkz_b(A, Beta, Sigma | partial = 1);
\end{lstlisting}
The choice of \soft{Sigma} means that we choose to rescale $z_1 = z_2 = z_3 = 1$.
The option \soft{partial = 1} indicates that the ideal is to be written in terms of 
$\p_i$ rather than $\theta_i$ symbols.
The list \soft{Ideal} will contain the single operator \eqref{2_F_1_w_operator}
(albeit in variables $x_i$ rather than $z_i$, as this is the convention for GKZ variables in \package{asir}).
\end{ex}

The formula \eqref{shift_q_operator} for the operator $d_q$,
which maps a differential form $[\o_q]$ to its corresponding $\mD$-module element,
has also been implemented with rescaled variables via the command \soft{mt\_gkz.rvec\_red2(...)}.

\section{GKZ systems are isomorphic to twisted cohomology groups}
\label{sec:GKZ_cohomology}

\secref{sec:twisted_de_rham} mentioned an isomorphism between the 
GKZ system $\mD_N / H_A(\b)$ and the twisted de Rham cohomology group $\deRham$.
This section provides a very rough sketch of the proof behind this theorem.
Heavy machinery from category theory is required to prove this in detail,
which is outside the scope of this thesis,
so we will content ourselves with summarizing the key steps.
The idea is to show that the GKZ system and the twisted cohomology group 
are both isomorphic to another $\mD$-module $\int_\pi \mL$,
to be described shortly,
wherefore $\mD/H_A(\b)$ and $\deRham$ must also be isomorphic.

Begin by recalling the space $\mO(X)$ from \eqref{holomorphic_functions_X}.
It is equipped with the structure of an abelian group under the addition of functions.
As mentioned in \secref{sec:what_is_a_D_module},
we can turn an abelian group into a $\mD$-module by specifying an action of the Weyl algebra $\mD$.
Given a function $h \in \mO(X)$,
the action is here
\eq{
        \p_{z_i} \bullet h &=
        \frac{\p h}{\p z_i} + 
        \frac{\b_0}{g(z|x)}
        \frac{\p g(z|x)}{\p z_i} h
        \\[3pt]
        \p_{x_i} \bullet h &=
        \frac{\p h}{\p x_i} + 
        \frac{\b_0}{g(z|x)}
        \frac{\p g(z|x)}{\p x_i} h -
        \frac{\b_i}{x_i} h
        \, .
        \label{O(X)_action}
}
The resulting $\mD$-module is denoted by the symbol $\mL$ instead of $\mO(X)$.

Let 
$
        \pi: (z,x) \to z
$
be a projection map.
The \emph{direct image} $\mD$-module
$
        \int_\pi \mL,
$
has an abstract categorical definition%
\footnote{
        In categorical terms,
        given a morphism
        $
                f: X \to Y
        $
        between two complex manifolds $X$ and $Y$,
        the direct image $\int_f$ is a \emph{functor} depending on $f$.
        Loosely speaking,
        it acts as a certain \emph{pushforward} from sheaves on $X$ to sheaves on $Y$.
        See \cite[Section 1.3]{Hotta-Tanisaki-Takeuchi-2008} for details.
},
but for the case of a projection it can be concretely realized as a quotient space that we proceed to build.
Write
\eq{
        \mD_{N+n} = \CC
        \langle 
                z_1, \ldots, z_N, x_1, \ldots, x_n, 
                \p_{z_1}, \ldots, \p_{z_N}, 
                \p_{x_1}, \ldots, \p_{x_n}
        \rangle
}
for the ring of differential operators in $N+n$ variables.
The subring
$
        \mD_N 
$
contains operators in the $z$-variables only.
It is possible to write 
$
        \mL =\mD_{N+n} / \mI,
$
where $\mI$ is the annihilating ideal of the Euler inte\emph{grand}
\eq{
        U = g(z|x)^{\b_0} x^{\b'}
        \, .
}
By differentiating $U$ w.r.t.~$(z,x)$ and recalling the expression for $g=g(z|x)$ from 
\eqref{g_Euler_integrand},
a short calculation shows that $\mI$ is generated by 
\eq{
        \mI =
        \sum_{i=1}^N \mD 
        \big[
                g \, \p_{z_i} - \b_0 \, x^{\a_i}
        \big]
        \plus
        \sum_{i=1}^n \mD 
        \Big[
                x_i \, g \, \p_{x_i} -
                \b_0 \sum_{j=1}^N (\a_j)_i \, z_j \, x^{\a_j} +
                \b_i \, g 
        \Big]
        \, ,
}
where $(\a_j)_i$ denotes the $i$th component of the multi-index $\a_j \in \ZZ^n$.
The direct image $\mD$-module then turns out to be represented by the quotient 
\eq{
        \int_\pi \mL =
        \frac{\mD_{N+n}}{\mI + \sum_{i=1}^n \p_{x_i} \mD_{N+n}}
        \, .
        \label{direct_image_functor}
}
This is now a $\mD_N$-module,
as the $x$-variables have been "projected out" by the map $\pi$.
The RHS of \eqref{direct_image_functor} is also called the \emph{integration module} \cite[Chapter 6]{dojo}%
\footnote{
        There are several \package{asir} packages for computing annihilation ideals of Euler integrands
        \cite{url-annihilation-ideal}.
        An algorithm for computing integration modules has been implemented in the package
        \package{nk\_restriction} \cite{nakayama2010algorithm}.
}.
It is an algebraic counterpart to the familiar integration operation in calculus for the following reason:
writing
\eq{
        \int_\pi \mL = \mD_N / \mJ
}
for some $\mJ$,
then the generators of $\mathcal{J}$ can be shown to annihilate the Euler inte\emph{gral} itself
\cite[Chapter 5]{SST}.
It is rather surprising that the integration of a function,
which is generally a highly transcendental operation,
can be captured in purely algebraic terms!

Now, 
it is a classical result that the twisted cohomology group
$
        \deRham
$
is isomorphic to the direct image $\int_\pi \mL$ as a $\mD$-module
\cite[Chapter 4]{Hotta-Tanisaki-Takeuchi-2008} \cite{adolphson2012A}.
The addition of differential forms forms an abelian group,
and $\deRham$ indeed becomes a $\mD$-module under the action
\eq{
        \p_{z_i} \bullet [\o(z)] =
        \left[
                \frac{\p \o(z)}{\p z_i} + \b_0 \frac{x^{\a_i}}{g(z|x)} \o(z)
        \right]
        \, ,
        \label{D_action_on_form}
}
where
$
        [\o(z)] \in \deRham
$
denotes an equivalence class of differential forms.
What is more,
Gel'fand, Kapranov and Zelevinsky 
proved that the annihilating ideal $\mJ$ defined above precisely equals $H_A(\b)$ \cite{GKZ_2}.
In other words,
$\int_\pi \mL$ is isomorphic to the GKZ system.

Since $\int_\pi \mL$ is isomorphic to both the GKZ system \emph{and} the twisted cohomology group,
we finally infer that
\eq{
        \mD_N / H_A(\b) \quad \cong \quad \deRham
        \, .
}
We emphasize that this isomorphism only holds true when the $z$-variables
are indeterminate and the $\b$-parameters are non-resonant.

        \bibliography{references}
        \bibliographystyle{JHEP}
\end{document}